\documentclass[useAMS]{mn2e}

\def\spose#1{\hbox to 0pt{#1\hss}}
\def\lta{\mathrel{\spose{\lower 3pt\hbox{$\mathchar"218$}}\raise 2.0pt\hbox{$\mathchar"13C$}}}
\def\gta{\mathrel{\spose{\lower 3pt\hbox{$\mathchar"218$}}\raise 2.0pt\hbox{$\mathchar"13E$}}}

\title[2Jy radio galaxies]{A near-IR study of the host galaxies of 2Jy radio sources at $0.03 \lta z \lta 0.5$: I -- the data\thanks{Based on observations collected at the European
    Southern Observatory, Chile  (programs 074.B-0296(A), 074.B-0296(B),
    075.B-0674(A) and 078.B-0500(A))}}
\author[K.\,J. Inskip et al ]{K. J. Inskip$^{1}$\thanks{E-mail:
inskip@mpia-hd.mpg.de}, C. N. Tadhunter$^{2}$,   R. Morganti$^{3,4}$,  
 J. Holt$^{5}$, C. Ramos Almeida$^{2,6}$ \&
\newauthor D. Dicken$^{7}$\\
$^{1}$ Max-Planck-Institut f\"{u}r Astronomie, K\"{o}nigstuhl 17,
    D-69117 Heidelberg, Germany\\
$^{2}$ Department of Physics \& Astronomy, University of Sheffield, Sheffield S3 7RH\\
$^{3}$ Netherlands Foundation for Research in Astronomy, Postbus 2,
  7990 AA Dwingeloo, The Netherlands\\
$^{4}$ Kapteyn Astronomical Institute, University of Groningen, P.O. Box 800,
9700 AV Groningen, The Netherlands\\
$^{5}$ Leiden Observatory, Leiden University, Niels Bohrweg 2, NL-2333 CA Leiden, The Netherlands\\
$^{6}$ Instituto de Astrof\' i sica de Canarias (IAC), C/V\' i a L\'actea, s/n,
E-38205, La Laguna, Tenerife, Spain\\
$^{7}$ Rochester Institute of Technology, 85 Lomb Memorial Drive, Rochester NY 14623, USA}
\begin{document}

\date{}

\pagerange{\pageref{firstpage}--\pageref{lastpage}} \pubyear{2010}

\maketitle

\label{firstpage}

\begin{abstract} 
We present the results of a program of $K-$ and $K_S-$band imaging of a sample of 2Jy radio galaxies with redshifts $0.03 \lta z \lta 0.5$, for which the host galaxy morphologies and structural parameters (effective radius, S\'ersic index and unresolved nuclear point source contribution) have been determined using \textsc{galfit}.  Two-thirds of our sample are best modelled as being hosted by massive elliptical galaxies with S\'ersic indices of $n=4-6$, with the remainder being better suited either by a mixture of morphological components (usually a bulge plus a small, less luminous, disk component) or by more disky galaxy models with $n=1-2$. Our measured galaxy sizes are generally in very good agreement with other imaging programs, both space- and ground-based.  We also determine a slightly higher average nuclear point source contribution than similar HST-based programs. This is due to our inability to separate the AGN emission from compact circum-nuclear stellar emission, but does not bias our modelling of the remainder of the host galaxies and our results remain robust.  We also observe that roughly half of the objects in our sample are either undergoing major or minor merger activity or are clearly morphologically disturbed.

\end{abstract}


\section{Introduction}
Over the last decade, it has become increasingly apparent that not only do supermassive black holes reside in the centres of all massive galactic bulges, but also that their ongoing evolution is strongly coupled, despite nine orders of magnitude difference between the relevant spatial scales of the black hole and its galactic host (Kormendy 1993; Magorrian et al 1998; Ferrarese \& Merritt 2000; Gebhardt et al 2000; Merritt \& Ferrarese 2001; Cattaneo \& Bernardi 2003; Marconi \& Hunt 2003; H\"aring \& Rix 2004). Regardless of the physical driving mechanisms for these correlations, it is certain that understanding the periods of AGN activity, when the central black hole is most able to grow via accretion, are of crucial importance for building up a cogent theoretical framework for galaxy evolution in general (e.g. Yu \& Tremaine 2002, Shankar et al 2004, Marconi et al 2004 and references therein). This problem poses two central, interlinked questions: what is the nature of an active galaxy, and how do they differ from the inactive population of galaxies from which they are (presumably) drawn?

Extragalactic radio sources play a particularly important role in such studies. Historically, they have been amongst the best studied distant AGN, due to the ease with which large samples of such objects could be assembled out to large redshifts. Under orientation-based unification schemes for radio galaxies and radio loud quasars (e.g. Barthel 1989; Urry \& Padovani 1995), they provided a perfect laboratory for studying the host galaxies of radio loud AGN without the complicating presence of an unobscured bright quasar nucleus.  Additionally, the presence of extensive powerful radio emission directly links the behaviour of these systems from the small scales of the central accreting black hole out to many hundreds or thousands of kiloparsecs into the surrounding intergalactic medium. Early studies of the proto-typical 3CR sources suggested that the host galaxies of these radio sources were members of a remarkably uniform population, and seemed to be ``standard candles'' to within $\sim 0.3$mag (Lilly \& Longair 1984), with characteristic effective radii of up to 15kpc and IR magnitudes dominated by emission from an old stellar population formed at $z > 5$ (Best et al 1998). However, subsequent revisions to the most widely accepted cosmological model (e.g. Hinshaw et al 2009) and an improved understanding of structure formation in the universe (e.g. Springel et al 2005) add considerable complexity to this picture.

While a radio source {\it may} undergo several active phases (e.g. Haehnelt \& Rees  1993; Saikia \& Jamrozy 2009 and references therein), this does not imply that any given radio source would necessarily have been active in the past, or will be so again in the future.  This opens up the scenario whereby radio galaxies at different epochs could be drawn from quite different galaxy populations: might this indeed be the case?  The tightness of the  $K-z$ relation (e.g. Inskip et al 2002; Willott et al 2003; Rocca-Volmerange et al 2004) and the total  near-IR luminosities of their host galaxies (Seymour et al 2007) suggest that radio sources are hosted by a relatively homogeneous, high mass  (up to $\sim 10^{12} \rm M_{\odot}$) population of galaxies over a wide redshift range. Even at higher redshifts, while morphological studies have shown that some sources are still undergoing active assembly, radio source host galaxies at redshifts of $z \sim 2-3$ are in general similar in size to those nearby (e.g. van Breugel et al 1998, Pentericci et al 2001). However, the increased luminosity of high-redshift radio galaxies relative to passive evolution models goes against the simplest picture of closed-box evolution of a single population of galaxies.

The high mass of radio source host galaxies at all epochs provides an additional constraint.  Since $z \sim 1$, the most massive members of the elliptical galaxy population have shown little evolution in mass (e.g. Heavens et al 2004; Thomas et al 2005). 
Although subsequent episodes of AGN activity in massive galaxies could be triggered by minor mergers or galaxy interactions which lead to little additional change in galaxy mass, the links between periods of black hole accretion and galaxy growth (as evidenced by the black hole mass vs. bulge mass relation) limits the ability of such objects to repeatedly host powerful, high accretion-rate AGN in more recent cosmic epochs. On top of this, it has been theorised that the very presence of a powerful radio source and/or AGN can limit the potential for further galaxy growth, via heating of the local environment (e.g. Fabian et al 2000; McNamara \& Nulsen 2007; Croton \& Farrar 2008; De Young 2010).  Equally, the actively accreting lower-redshift systems have likely been assembled from their less-massive constituent progenitor objects too recently to have hosted powerful AGN at high redshifts.

Similarly, at different redshifts the powerful radio source population samples different environments (e.g. Lilly \& Prestage 1987; Yates et al 1989; Hill \& Lilly 1991; Best 2000), with a tendency towards more highly clustered objects at higher redshifts.
It is now clear that galaxies are subject to so-called ``cosmic downsizing'', by which both star formation (e.g. Cowie et al 1996; Juneau et al 2005) and AGN activity (e.g. Granato et al 2004; De Lucia et al 2006, Croton et al 2006; Merloni \& Heinz 2008; Martini et al 2009; Shen 2009) become increasingly more prevalent in lower-mass  objects at later cosmic epochs, and in less dense environments. 

The inherent changes in the host galaxy population highlight the essential need to effectively characterise their properties at all redshifts. Without a clear handle on what a radio galaxy {\it is} at any given epoch, it is practically impossible to determine how it may or may not differ from the underlying inactive galaxy population, or from the population of radio-quiet AGN.  What are the characteristic sizes of powerful radio galaxies, and how does this parameter vary with redshift and/or radio power? Is the scatter in the $K-z$ relation solely due to variations in galaxy mass, or can it be accounted for at least in part by variations in the host galaxy stellar populations?  Do the host galaxies show signs of recent merger activity, and how does the proportion of merging/disturbed sources compare with that of normal galaxies? Is there any dependence of galaxy morphology on local environment, or on the radio source/AGN properties? With such physical characteristics of the host galaxies pinned down, deeper questions may then be posed on the nature of {\it how} AGN and radio source activity are triggered, e.g.: whether triggering occurs due to major or minor interactions or the more gradual infall of material; the prevalence of different triggering modes; the details of the timeline of events involved and any timelags between star formation and AGN/radio source activity; and the links between such processes and the evolution and  observed morphological characteristics of the host galaxies.

In order to investigate these issues, we turn not to the 3CR sample, but instead to the similarly well studied 2Jy sample of southern radio sources at $z < 0.5$ (Tadhunter et al 1993, Morganti et al 1993). This sample is unique in the sense that it is the {\it only} major sample of powerful radio galaxies for which complete, high quality optical continuum and emission line spectra, as well
as radio imaging data, exist for all the objects, alongside near- and mid-IR imaging and spectroscopy for most sources.  One particular advantage of this sample is that the combination of the mid-IR data and optical spectra allows us to separately determine the extent of both obscured and unobscured star formation activity within the host galaxies, as well as providing two independent measures of the degree of nuclear activity (via the 24$\mu$m emission and the optical line emission).
Our current dataset consists of ground-based near-IR imaging of the majority of the sample objects.  These data will be used to obtain accurate photometric magnitudes, and to thoroughly characterise the host galaxy structural properties as observed in this wavelength regime (galaxy profile, effective radius and the extent of any nuclear point source contamination from the AGN), and how they compare to other radio galaxy samples.  We will also make a preliminary assessment of the prevalance of visible merger activity within the sample (either ongoing or having occured in the recent past); this issue is expanded on in our analysis of optical imaging data for the 2Jy sample (Ramos Almeida et al in prep, hereafter RA10).

Having characterised the morphological properties of the host galaxies, in paper 2 of this series we will extend our investigation to the more global properties of the sample.  By combining galaxy morphologies with our data on the host galaxy stellar populations, the results of the current paper will be used to investigate the nature of the scatter in the $K-z$ relation, and how well it can be accounted for in terms of galaxy mass and recent star formation activity. We will also use literature data for other radio galaxy surveys at higher redshift to examine how redshift and radio power impact on the observed trends.
Finally, together with the results of RA10, we will assess in paper 2 how closely morphological signatures of merger activity correlate with the degree of recent star formation, the host galaxy morphologies, and the size/age of the radio source, helping us to build a clearer picture of the variety of different physical modes of AGN triggering in these sources.
Further to this work, these data will also be used in conjunction with the existing optical data (RA10) to also constrain the local environments of the host galaxies (Ramos Almeida et al, in prep), and to comprehensively assess the impact of environment on the properties of powerful radio sources and their host galaxies.

In section 2, we provide details of our observed sample. Our observations and data reduction processes are described in section 3. In section 4, we present our observational results and our 2-d modelling of the host galaxies, while our discussion and conclusions are presented in sections 5 and 6 respectively.  A more extensive analysis of this dataset is postponed to a second paper in this series.

Throughout this paper, we assume cosmological parameters of $\Omega_0 = 0.27$, $\Omega_{\Lambda} =
0.73$ and $H_0 = 71 \rm{km\,s^{-1}\,Mpc^{-1}}$.

\section{Sample details}

The sample of galaxies studied in this paper has been selected from the 2Jy sample of Wall \& Peacock (1985) which has flux densities $S_{2.7 GHz} > 2$Jy and declinations $\delta < 10^{\circ}$, with the addition of PKS0347+05 which was later found to meet the same criteria (di Serego-Alighieri et al 1994). Optical classifications have been determined for these objects on the basis of previous optical spectroscopic observations (Tadhunter et al 1998) and their optical appearance (Wall \& Peacock 1985). Sources with [O\textsc{iii}]$\lambda$5007 emission line equivalent widths $<10$\AA\ are classified as weak line radio galaxies (WLRGs). The remainder are classified as narrow line radio galaxies (NLRGs), broad line radio galaxies (BLRGs) and quasars. The quasars were classified by Wall \& Peacock (1985) on the basis of their stellar appearance on optical images, while the NLRGs and BLRGs are classified according to whether or not their optical spectra show evidence for broad line emission. Our sample excludes sources optically-classified as quasars, and was primarily selected within the redshift range $0.05 < z< 0.5$ and with right ascensions of $19hr00 < RA < 14hr00$.  Due to the availability of observing time on different instruments, this was later extended to redshifts of $0.03 < z< 0.5$ within the right ascension range of $21hr00 < RA < 14hr00$, and $0.05 < z< 0.5$ within the ranges $14hr00 < RA < 16hr00$ and $17hr30 < RA < 21hr00$. Of the resulting sample of 46 sources, we have obtained observations of 43. PKS0620-52 was dropped from the sample due to its proximity to the bright star Canopus, while PKS1839-48 was lost due to observing time constraints.  PKS1954-55 is known to have a bright foreground star in close proximity to its nucleus, and was also excluded. Despite not reaching 100\% completeness, this sample is a representative, unbiased 93\% complete subset of the whole. 
 
\section{Observations and data reduction}

\subsection{UKIRT K-band imaging}

The UKIRT observations were carried out on 2004 September 13, 2004
September 14 and 2004 September 15 using UFTI (the
UKIRT Fast--Track Imager; Roche et al 2002).  UFTI is a 1--2.5$\mu$m camera with a $1024
\times 1024$ HgCdTe array and a plate scale of 0.091\arcsec\, per
pixel, which gives a field of view of 92\arcsec.   Details of the
observing conditions are provided in Table 1, with further details of
the sources observed and the individual exposure times listed in Table 2.

\begin{table*}
\caption{Details of the observational data used in this paper.
Seeing measurements are given for the wavelength
  of the observations.  Observations were photometric except where noted otherwise.
}
\begin{center} 
\begin{tabular}{cccccccc}
Date & Instrument & Filter & Plate Scale & Typical Field of View &Seeing & Notes \\\hline
20040913 & UFTI & $K$ & 0.0909$^{\prime\prime}/$pixel & $1.5^{\prime}\times 1.5^{\prime}$& 0.4-0.5$^{\prime\prime}$& \\
20040914 & UFTI & $K$ & 0.0909$^{\prime\prime}/$pixel & $1.5^{\prime}\times 1.5^{\prime}$&0.5-1.0$^{\prime\prime}$&\\
20040915 & UFTI & $K$ & 0.0909$^{\prime\prime}/$pixel & $1.5^{\prime}\times 1.5^{\prime}$&0.5-0.6$^{\prime\prime}$&\\
20041114 & SOFI - large field& $K_S$&  0.288$^{\prime\prime}/$pixel&$5.0^{\prime}\times 5.0^{\prime}$ -- $8.0^{\prime}\times 8.0^{\prime}$ &0.9-1.35$^{\prime\prime}$ & 1\\
20041115 & SOFI - large field & $K_S$& 0.288$^{\prime\prime}/$pixel&$5.0^{\prime}\times 5.0^{\prime}$ -- $8.0^{\prime}\times 8.0^{\prime}$ & 0.6-0.95$^{\prime\prime}$& 2\\
20050226 & SOFI - small field & $K_S$& 0.144$^{\prime\prime}/$pixel& $3.0^{\prime}\times 3.0^{\prime}$&   0.55-1.1$^{\prime\prime}$ & 3\\
20050228 & SOFI - small field & $K_S$&  0.144$^{\prime\prime}/$pixel&  $3.0^{\prime}\times 3.0^{\prime}$& 0.55-1.2$^{\prime\prime}$ \\
20050301 & SOFI - small field & $K_S$&  0.144$^{\prime\prime}/$pixel&  $3.0^{\prime}\times 3.0^{\prime}$&  0.65-1.1$^{\prime\prime}$\\
20050315 & SOFI - small field & $K_S$&   0.144$^{\prime\prime}/$pixel&  $3.0^{\prime}\times 3.0^{\prime}$& 0.65-1.4$^{\prime\prime}$\\
20050712 & ISAAC & $K_S$ & 0.072$^{\prime\prime}/$pixel& $1.5^{\prime}\times 1.5^{\prime}$ -- $2.0^{\prime}\times 2.0^{\prime}$& 0.6-2.5$^{\prime\prime}$ \\
20050713 & ISAAC & $K_S$ & 0.072$^{\prime\prime}/$pixel& $1.5^{\prime}\times 1.5^{\prime}$ -- $2.0^{\prime}\times 2.0^{\prime}$&0.6-1.5$^{\prime\prime}$& 4 \\
20061113 & SOFI - small field & $K_S$ & 0.144$^{\prime\prime}/$pixel&  $3.0^{\prime}\times 3.0^{\prime}$ &0.6-0.9$^{\prime\prime}$ &\\

\hline\multicolumn{7}{l}{Notes:}\\
\multicolumn{7}{l}{[1] Initially photometric; observations later abandoned due to cloud}\\
\multicolumn{7}{l}{[2] patchy thin cirrus at times, and partially photometric. }\\
\multicolumn{7}{l}{[3] non photometric ($\sim 0.1-0.6$mag extinction) }\\
\multicolumn{7}{l}{[4] Observations cut short due to cloud}\\

\end{tabular}
\end{center} 
\end{table*}

\begin{table*}
\caption{Sample details.  Columns 2, 3 and 4 list any alternative name, the optical classification and redshift of each of the sources in our sample.  The instrument 
  used, observation date and exposure time for our infrared
  observations are listed in columns 5, 6 and 7 respectively. Column 8 lists the point spread function full width at half maximum (FWHM) for the observations of each source, in arcsec} 
\begin{center} 
\scriptsize{}
\begin{tabular} {lccccccc}

{Source}& {Alt. Name}&{Optical Classification} &{z}&{Instrument}& {Date}&{Exposure} & {PSF FWHM}\\
{[1]}&{[2]}&{[3]}&{[4]}&{[5]}&{[6]}&{[7]}&{[8]}\\\hline
PKS0023-26 &      & NLRG & 0.32 & SOFI & 20041115 & 94 & 0.91 \\
PKS0034-01 & 3C15 & WLRG & 0.07 & SOFI & 20041115 & 20 & 0.86  \\
PKS0035-02 & 3C17 & BLRG & 0.22 & UFTI & 20040913, 20040914 & 36, 18 & 0.54 \\
PKS0038+09 & 3C18 & BLRG & 0.19 & UFTI & 20040915 & 45 &0.52  \\
PKS0039-44 &      & NLRG & 0.346 & SOFI-archive & 20050815 & 1& -- \\
PKS0043-42 &      & WLRG & 0.12 & SOFI & 20041115 & 25 & 0.93 \\
PKS0055-01 & 3C29& WLRG  & 0.05 & SOFI & 20041115 & 5 & 0.89 \\
PKS0105-16 & 3C32 & NLRG & 0.40 & SOFI & 20061113 & 75 & 0.86 \\
PKS0131-36 & NGC0612 & WLRG  & 0.03 & SOFI & 20041114 & 20 & 1.15  \\
PKS0213-132& 3C62 & NLRG & 0.15 & SOFI & 20041115 & 30  & 0.95\\
PKS0305+03 & 3C78&  WLRG  & 0.03 & SOFI & 20041114 & 4 & 1.35 \\
PKS0325+02 & 3C88& WLRG  & 0.03 & SOFI & 20041114 & 5 & 1.24 \\
PKS0347+05 & 4C+05.16&WLRG&0.34 & UFTI & 20040913, 20040914 & 36, 18 & 0.45 \\
PKS0349-27 &      & NLRG & 0.07 & SOFI & 20041115 & 30 & 0.84 \\
PKS0404+03 &3C105 & NLRG & 0.09 & SOFI & 20050226, 20050228 & 95, 95 & 0.73 \\
PKS0427-53 &IC2082 & WLRG    & 0.04 & SOFI & 20041115 & 5 & 0.88 \\
PKS0430+05 & 3C120 & BLRG    & 0.03 & SOFI & 20061113 & 10 & 0.76 \\
PKS0442-28 & & NLRG & 0.15 & SOFI & 20041115 & 30 & 0.86 \\
PKS0453-20 &NGC1692 & WLRG & 0.04 & SOFI & 20041115 & 5  & 0.79\\
PKS0518-45 &Pictor A & BLRG& 0.04 & SOFI & 20041115 & 25 & 0.85 \\
PKS0521-36 &ESO362-G021 & BLRG & 0.06 & SOFI & 20041115 & 20& 0.82  \\
PKS0625-53 &ESO161-IG007 & WLRG & 0.05 & SOFI & 20041115 & 5 & 0.79 \\
PKS0625-35 & & WLRG & 0.06 & SOFI & 20041115, 20061113 & 20, 20& 0.59  \\
PKS0806-10 &3C195 & NLRG & 0.11 & SOFI & 20041115, 20061113 & 20, 20 & 0.66 \\
PKS0859-25 & & NLRG& 0.31 & SOFI & 20050228 & 20&0.55  \\
PKS0915-11 & Hydra A &  WLRG& 0.05 & SOFI, SOFI-archive & 20041115, 20031213 & 10, 38 & 0.86 \\
PKS0945+07 & 3C227& BLRG& 0.09 & SOFI & 20050228 & 22 &0.68 \\
PKS1306-09 & & NLRG & 0.46 & SOFI & 20050301 & 95 & 0.66 \\
PKS1547-79 & & BLRG & 0.48 & SOFI & 20050315 & 95& 0.67  \\
PKS1549-79 & & BLRG?& 0.15 & ISAAC & 20050712 & 27 & 0.78 \\
PKS1559+02 &3C327 & NLRG& 0.11 & ISAAC & 20050713 & 9 & 0.66\\
PKS1733-56 & &BLRG & 0.10 & ISAAC & 20050714 & 20 & 1.05  \\
PKS1814-63 & &NLRG & 0.06 & ISAAC & 20050712 & 18 & 0.93\\
PKS1932-46 & & BLRG & 0.23 & SOFI & 20061113 & 50 & 0.78  \\
PKS1934-63 & & NLRG & 0.18 & SOFI & 20061113 & 45  &0.89 \\
PKS1949+02 &3C403 &NLRG & 0.06 & ISAAC & 20050712 & 36 & 0.93  \\
PKS2104-25 &NGC7018 &   & 0.039 & SOFI-archive & 19990414 & 60 &-- \\ 
PKS2153-69 &ESO075-G041 & BLRG& 0.03 & ISAAC & 20050712 & 9 &0.76  \\
PKS2211-17 &3C444 & WLRG
& 0.15 & ISAAC & 20050712 & 18 &0.86 \\
PKS2221-02 & 3C445& BLRG& 0.06 & SOFI & 20061113 & 5 &0.77 \\ 
PKS2250-41 & &NLRG & 0.31 & SOFI & 20061113 & 50  &0.67 \\
PKS2313+03 &3C459 &NLRG & 0.22 & UFTI, UFTI, SOFI & 20040914, 20040915, 20061113 &
27, 27, 45 & 0.53\\
PKS2356-61 & &NLRG & 0.10 & ISAAC & 20050712 & 27 &0.68 \\
\end{tabular}                       
\end{center} 
\end{table*}                       
\normalsize
 
All observations used a nine point jitter pattern, with offsets of
roughly 10\arcsec\ between each 1 minute exposure.  The observational data were dark subtracted, and 
masked for bad pixels.  The data for each source (or several
consecutive sources for the lower-redshift objects for which only a handful of individual exposures were obtained)  were
combined and median filtered to create a first-pass sky flat-field
image, which accounted for the
majority of the pixel-to-pixel variations of the chip.  However,
large-scale illumination gradients remained, due to the changing position of
the telescope over the night.  To correct for this, smaller groups of
9 -- 18 first pass flat-fielded images were similarly combined to create
residual sky flat-field images.  Applying these residual flat fields to the first-pass flat fielded data successfully 
accounted for this effect.  Bright objects on the fully flat-fielded images
were then masked out, and the whole process repeated, allowing the data
to be cleanly flat fielded without any contamination from stars or
galaxies. 
The flat-fielded data for each source were sky-subtracted and combined
using the IRAF package DIMSUM, creating a final mosaiced image of
approximately 115arcsec $\times 115$arcsec, with the highest
signal-to-noise level restricted to the central $70 \times 70$
arcsec$^2$.  The data were flux calibrated using observations of standard stars
selected from the UKIRT faint  
standards catalogue.

 Photometry was carried using
the IRAF package \textsc{apphot} and a single sky annulus; $4^{\prime\prime}$, $5^{\prime\prime}$ and
$9^{\prime\prime}$ diameter circular apertures were used, and also a
64kpc diameter circular aperture in the rest-frame of the source (for consistency with earlier $K-$band studies of radio galaxy hosts: Eales et al 1997; Jarvis et al 2001; Inskip et al 2002; Willott et al 2003).
These data are presented in Table 3.  We also include the 64kpc
aperture magnitudes for each source after the modelling and removal of any
bright contaminating objects within the aperture, with the exception
of any object which appears to be physically interacting with the
target galaxy.   The error on our zero point measurements are combined in quadrature
with the standard  \textsc{apphot} uncertainties: the percentage error due to the
Poisson noise from the detected counts, the percentage random error due to the sky counts in the
aperture, and the percentage systematic error due to the accuracy to which the
mean sky value can be calculated.
The resulting magnitudes have been corrected
for Galactic extinction using the $E(B-V)$ for the Milky Way from the
NASA Extragalactic Database (NED) and the parametrized Galactic
extinction law of Howarth (1983).

\subsection{SOFI Ks-band imaging}

\begin{table*}
\caption{Results of our infrared observations. In columns 2-6 we
  present the results of aperture photometry carried out within four
  different aperture sizes: 4, 5, and 9\arcsec\ diameter aperture 
$K$ or $K_S$ magnitudes are given in columns 2, 3 and 4 respectively, along
  with the associated errors.  Column 5 lists the aperture magnitudes
  and errors obtained using a fixed metric aperture 64kpc in
  diameter in the rest-frame of each source.  In column 6, we present
  the same aperture magnitude after removal of the flux from any
  contaminating object (see text for full details). Column 7 presents the same data after K-correcting the $K_S-$band observations to the $K$-band filter. All magnitudes have been 
corrected for galactic extinction using E(B-V) values taken from the Nasa Extragalactic Database and the parametrized galactic extinction law of Howarth (1983).} 
\begin{center} 
\footnotesize{}
\begin{tabular} {lcccccc}
{Source}&\multicolumn{4}{c}{magnitudes in different apertures}\\
 & {4$^{\prime\prime}$}& {5$^{\prime\prime}$}& {9$^{\prime\prime}$}&{64kpc}&{64kpc (clean)} & {64kpc (K-corrected)}\\
{[1]}&{[2]}&{[3]}&{[4]}&{[5]}&{[6]}&{[7]}\\\hline
PKS0023-26 & $15.632 \pm 0.057$ & $15.501 \pm 0.055$ & $15.176 \pm
 0.051$ & $14.651 \pm 0.045$ & $15.036 \pm 0.054$ & $15.017 \pm 0.056$\\
PKS0034-01 & $13.571 \pm 0.036$ & $13.403 \pm 0.035$ & $13.050 \pm 0.034$ & $12.569 \pm 0.037$& $12.569 \pm 0.037$& $12.571 \pm 0.037$\\
PKS0035-02 & $14.459 \pm 0.056$ & $14.384 \pm 0.056$ & $14.205\pm  0.055$ &
 $14.100 \pm 0.055$ & $14.104 \pm 0.055$& $14.104 \pm 0.055$\\
PKS0038+09 & $14.402 \pm 0.035$ & $14.368 \pm 0.034$ & $14.305 \pm 0.033$ & $14.216 \pm 0.036$&$14.296 \pm 0.037$& $14.296 \pm 0.037$\\
PKS0039-44 & $15.771 \pm 0.053$ & $15.644 \pm 0.057$& $15.537 \pm 0.086$ & $15.411 \pm 0.122$ & $15.411 \pm 0.122$& $15.388 \pm 0.122$\\
PKS0043-42 & $13.853 \pm 0.037$ & $13.703 \pm 0.036$ & $13.369 \pm
 0.034$ & $12.977 \pm 0.036$ & $12.999 \pm 0.036$& $12.988 \pm 0.036$\\
PKS0055-01 & $13.560 \pm 0.262$ & $13.237 \pm 0.259$ & $12.496 \pm
 0.255$ & $11.554 \pm 0.300$ & $11.596 \pm 0.282$& $11.599 \pm 0.282$\\
PKS0105-16 & $15.801 \pm 0.055$ & $15.686 \pm 0.052$ & $15.475 \pm 0.048$ & $15.419 \pm 0.049$ & $15.419 \pm 0.049$& $15.380 \pm 0.049$\\
PKS0131-36 & $11.760 \pm 0.031$ & $11.465 \pm 0.031$ & $10.783 \pm
 0.030$ & $ 9.553 \pm 0.030$ & $9.563 \pm 0.030$& $ 9.566 \pm 0.030$\\
PKS0213-132 & $14.515 \pm 0.365$ & $14.385 \pm 0.364$ & $14.029 \pm 0.364$ & $13.502 \pm 0.366$& $13.502 \pm 0.366$& $13.492 \pm 0.366$\\
PKS0305+03 & $11.491 \pm 0.031$ & $11.207 \pm 0.031$ & $10.576 \pm
 0.030$ & $9.155 \pm 0.030$ & $9.156 \pm 0.031$& $ 9.159 \pm 0.031$\\
PKS0325+02 & $12.938 \pm 0.033$ & $12.659 \pm 0.033$ & $12.027 \pm
 0.032$ & $10.366 \pm 0.034$ & $10.388 \pm 0.034$& $10.391 \pm 0.034$\\
PKS0347+05 & $14.685 \pm 0.058$ & $14.576 \pm 0.057$ & $14.308 \pm 0.056$ & $13.963 \pm 0.054$ & $14.283 \pm 0.056$& $14.283 \pm 0.056$\\
PKS0349-27 & $13.882 \pm 0.037$ & $13.750 \pm 0.037$ & $13.433 \pm
 0.035$ & $12.675 \pm 0.034$ & $12.853 \pm 0.037$& $12.855 \pm 0.037$\\
PKS0404+03 & $14.131 \pm 0.051$ & $14.000 \pm 0.051$ & $13.713 \pm 0.050$ & $13.343 \pm 0.050$ & $13.417 \pm 0.055$& $13.411 \pm 0.055$\\
PKS0427-53 & $12.320 \pm 0.032$ & $12.086 \pm 0.032$ & $11.536 \pm
 0.031$ & $ 9.984 \pm 0.031$ & $10.067 \pm 0.031$& $10.071 \pm 0.031$\\
PKS0430+05 & $10.778 \pm 0.011$ & $10.727 \pm 0.011$ & $10.587 \pm 0.011$ & $10.018 \pm 0.100$& $10.018 \pm 0.100$& $10.021 \pm 0.100$\\
PKS0442-28 & $13.742 \pm 0.036$ & $13.644 \pm 0.036$ & $13.410 \pm
 0.035$ & $13.134 \pm 0.034$ & $13.160 \pm 0.035$& $13.150 \pm 0.035$\\
PKS0453-20 & $12.578 \pm 0.032$ & $12.312 \pm 0.032$ & $11.698 \pm
 0.031$ & $10.218 \pm 0.032$ & $10.252 \pm 0.032$& $10.256 \pm 0.032$\\
PKS0518-45 & $12.542 \pm 0.032$ & $12.483 \pm 0.032$ & $12.316 \pm
 0.032$ & $11.820 \pm 0.034$ & $12.015 \pm 0.037$& $12.019 \pm 0.037$\\
PKS0521-36 & $11.342 \pm 0.031$ & $11.283 \pm 0.031$ & $11.151 \pm
 0.031$ & $10.901 \pm 0.031$ & $10.913 \pm 0.031$& $10.916 \pm 0.031$\\
PKS0625-53 & $12.537 \pm 0.192$ & $12.269 \pm 0.192$ & $11.641 \pm
 0.191$ & $10.022 \pm 0.191$ & $10.042 \pm 0.191$& $10.045 \pm 0.191$\\
PKS0625-35 & $12.052 \pm 0.014$ & $11.889 \pm 0.013$ & $11.506 \pm
 0.012$ &  $9.817 \pm 0.011$ &  $10.724 \pm 0.047$& $10.727 \pm 0.047$\\
PKS0806-10 & $12.820 \pm 0.017$ & $12.719 \pm 0.016$ & $12.467 \pm 0.016$ & $12.137 \pm 0.016$ & $12.137 \pm 0.016$& $12.126 \pm 0.016$  \\
PKS0859-25 & $15.185 \pm 0.060$ & $15.048 \pm 0.058$ & $14.732 \pm
 0.055$ & $14.370 \pm 0.054$ & $14.758 \pm 0.069$& $14.742 \pm 0.069$\\
PKS0915-11 & $13.246 \pm 0.108$ &
 $12.946 \pm 0.108$ & $12.269 \pm 0.108$ & $10.817 \pm 0.108$ &
 $10.868 \pm 0.108$& $10.871 \pm 0.108$\\
PKS0945+07 & $12.678 \pm 0.047$ & $12.635 \pm 0.047$ & $12.535 \pm
 0.046$ & $12.367 \pm 0.047$ & $12.376 \pm 0.048$& $12.370 \pm 0.048$\\
PKS1306-09 & $15.300 \pm 0.051$ & $15.234 \pm 0.050$ & $14.906 \pm 0.046$ & $14.757 \pm 0.045$ & $15.120 \pm 0.062$& $15.080 \pm 0.062$\\
PKS1547-79 &$15.296 \pm 0.044$ & $15.259 \pm 0.044$ & $15.090 \pm
 0.044$ & $15.030 \pm 0.044$ & $15.185 \pm 0.044$& $15.142 \pm 0.044$\\
PKS1549-79 & $12.466 \pm 0.043$ & $12.429 \pm 0.043$ & $12.362 \pm
 0.043$ & $12.280 \pm 0.043$ & $12.319 \pm 0.043$& $12.309 \pm 0.043$\\
PKS1559+02 & $13.388 \pm 0.044$ & $13.189 \pm 0.044$ & $12.777 \pm 0.043$ & $12.205 \pm 0.044$ & $12.205 \pm 0.044$& $12.194 \pm 0.044$\\
PKS1733-56 & $13.396 \pm 0.044$ & $13.222 \pm 0.044$ & $12.892 \pm 0.043$ & $12.058 \pm 0.043$ & $12.485 \pm 0.043$& $12.474 \pm 0.043$\\
PKS1814-63 & $12.775 \pm 0.043$ & $12.596 \pm 0.043$ & $12.317 \pm
 0.043$ & $11.851 \pm 0.043$ &$11.896 \pm 0.043$& $11.899 \pm 0.043$	\\
PKS1932-46 & $15.413 \pm 0.046$ & $15.301 \pm 0.044$ & $15.083 \pm 0.040$ & $14.971 \pm 0.044$& $14.971 \pm 0.044$& $14.962 \pm 0.044$\\
PKS1934-63 & $14.845 \pm 0.036$ & $14.665 \pm 0.034$ & $14.220 \pm 0.028$ & $13.931 \pm 0.026$&$14.023 \pm 0.030$& $14.016 \pm 0.030$\\
PKS1949+02 & $12.662 \pm 0.043$ & $12.453 \pm 0.043$ & $12.001 \pm
 0.043$ & $10.121 \pm 0.043$ & $11.333 \pm 0.043$& $11.336 \pm 0.043$ \\
PKS2153-69 & $11.998 \pm 0.043$ & $11.808 \pm 0.043$ & $11.362 \pm 0.043$ & $10.036 \pm 0.043$	 & $10.180 \pm 0.044$& $10.183 \pm 0.044$	\\
PKS2211-17 & $14.975 \pm 0.046$ & $14.671 \pm 0.045$ & $14.008 \pm 0.044$ & $13.111 \pm 0.044$ & $13.422 \pm 0.048$& $13.412 \pm 0.048$	\\
PKS2221-02 & $11.602 \pm 0.013$ & $11.569 \pm 0.013$ & $11.503 \pm 0.013$ & $11.448 \pm 0.013$ & $11.448 \pm 0.013$& $11.451 \pm 0.013$\\
PKS2250-41 & $15.813 \pm 0.055$ & $15.739 \pm 0.053$ & $15.573 \pm 0.051$ & $15.508 \pm 0.057$& $15.508\pm 0.057$& $15.492 \pm 0.057$\\
PKS2313+03 (K) & $13.961 \pm 0.043$ & $13.890 \pm 0.042$ & $13.752 \pm 0.042$ & $13.639 \pm 0.042$ & $13.639 \pm 0.042$& $13.639 \pm 0.042$\\
PKS2313+03 (KS)& $14.005 \pm 0.026$ & $13.932 \pm 0.025$ & $13.795 \pm 0.024$ & $13.672 \pm 0.024$& $13.672 \pm 0.024$& $13.664 \pm 0.024$\\
PKS2356-61 & $13.704 \pm 0.044$ & $13.515 \pm 0.044$ & $13.115 \pm 0.044$ & $12.559 \pm 0.044$ & $12.559 \pm 0.044$& $12.548 \pm 0.044$\\
\end{tabular}                       
\end{center}                        
\end{table*}                       
\normalsize

$K_S-$band  observations were obtained on the nights of 2004 November
14, 2004 November
15, 2005 February 26, 2005 February 28, 2005 March 1, 2005 March 15
and 2006 November 13, using the Son of ISAAC (SOFI; Moorwood, Cuby \&
Lidman 1998) instrument on the ESO 3.5-m New Technology Telescope
(NTT),  as part of the
European Southern Observatory (ESO) observing programmes
074.B-0296(A), 074.B-0296(B) and 078.B-0500(A).
On most nights the instrument was
used in the small field mode, which results in a plate scale of 0.144
arcsec per pixel.  The large field mode, with a plate scale of 0.288
arcsec per pixel, was used for the observations obtained on 20041115.
Details of the
observing conditions are provided in Table 1, with further details of
the sources observed and the individual exposure times listed in Table 2.
Generally, the observations of each source consisted of a number of
1 min exposures,  with each
observation subject to a random offset within a 40 arcsec diameter box.
For the lower redshift objects in our sample (typically those with
$z<0.1$) we switched to five 1 min integrations at fixed offsets.

The data were corrected for SOFI's interquadrant row cross talk effect
using an adapted version of the SOFI crosstalk.cl IRAF script.
Flat-fielding of the data was carried out using the following process:
all target frames were combined, median filtered and normalised to a
mean pixel value of 1.0 to create a first-pass flat-field image, which
was then applied to each frame.  Bright objects on the flat-fielded
images were then masked out, and the process repeated with the masked
frames, allowing the data to be cleanly flat fielded without any
contamination from stars or galaxies. The subsequent data reduction
uses well-established techniques: the flat-fielded data were
sky-subtracted and combined using the IRAF package DIMSUM.

Flux calibration used observations of NICMOS Photometric Standard
stars (Persson et al 1998).   Aperture photometry, galactic
extinction corrections and uncertainties for the resulting 
magnitudes were determined in the same way as for the UFTI observations.

The observations carried out on the nights of 2004 November 15 and 2005 February 26 were unfortunately subject to the presence of variable patchy cirrus at the start and the end of the night.  Regular observations of standard stars and good knowledge of the instrumental zero point based on the previous night's observations allowed us to keep track of which sources were observed in partially non-photometric conditions, and the changing extent of the extinction due to thin clouds.

As an example, the bulk of our observations of PKS0023-26 were obtained in non-photometric conditions. (Note that the use of repeated standard star observations confirmed that some of the observations made of this object were obtained in clear transparancy conditions.) This presents several challenges. Firstly, due to the strongly varying sky background level, the data are no longer best reduced in time-sequence order, such that sky subtraction is carried out for each frame based on the mean sky level of the frames immediately preceding and following the observation. Instead, the data are re-stacked in order of background sky brightness, and sky subtraction is carried out based on the mean sky background of the frames closest in sky luminosity.  This procedure results in a high-quality final data product indistinguishable from one obtained in better observing conditions, which would not be the case for the more standard mode of sky subtraction.  Secondly, while the non-photometric data are still useful for morphological studies, simple combination of frames with equal weighting does not result in the best signal-to-noise level for the final image.  We therefore monitor the flux of bright objects and the background noise level in each frame, and our final image is produced from a weighted combination of individual frames according to the signal-to-noise level applicable for each.  Monitoring the flux level of bright objects in each frame is also of great importance for selecting any frames which may be free of extinction due to cirrus, particularly for the stars with 2MASS magnitudes available within the instrumental field of view. In the case of PKS0023-26, the final five frames obtained were identified as being free of extinction, and combined together to make a 'photometric-equivalent' final product separate from that used for our analaysis of the galaxy morphologies.  For PKS0213-132, even the best frames were not obtained in clear transparency conditions, but from examination of 2MASS object fluxes we have confirmed the extinction level at 1.2 magnitudes for these frames, and applied the appropriate correction to the derived magnitudes.

For the data obtained on the night of 2004 November 15, the bulk of the sources were observed under clear-transparency conditions. For most of the remainder, either reliable photometric-equivalent images were produced, or further data were obtained.  Photometric observations of PKS0404+03 were obtained on 2005 February 28, while photometric observations of PKS0806-10 and PKS0625-35 were obtained on 2006 November 13.
For the two outstanding sources observed only under non-photometric conditions, standard star fluxes and the variation of 2MASS object fluxes within the observational fields of view confirm that the observations of PKS0055-01 and PKS0043-42 were subject to a stable extinction of $0.9$magnitudes and $0.2$magnitudes respectively.  This, along with an associated estimate of the error on the extinction due to thin cirrus, has been accounted for in our photometry of these sources.

\subsection{ISAAC Ks-band imaging}

Infrared Ks-band imaging observations of nine sources in our sample were carried out
on 2005 July 12 and 2005 July 13 using the Infrared Spectrometer And
Array Camera (ISAAC; Moorwood et al 1998)  as part of the
European Southern Observatory (ESO) observing programme
075.B-0674(A).  ISAAC is mounted on the Nasmyth B focus of the UT1 Antu unit of the
Very Large Telescope (VLT).  The plate scale for our observations is
0.148 arcsec per pixel.  Details of the
observing conditions are provided in Table 1, with further details of
the sources observed and the individual exposure times listed in Table
2.  We used a nine-point jitter pattern for these observations, and
the data were reduced and flux-calibrated using the same process as for our SOFI
observations.   Aperture photometry, galactic
extinction corrections and uncertainties for the resulting 
magnitudes were determined in the same way as for our other observations.

\subsection{Archival data}

Limited observations of three sources were obtained from the ESO archive.

For PKS0039-44, four photometric 10s spectroscopic acquisition observations (from program 075.B-0777(A)) taken on the night of 2005 August 18 using the SOFI instrument on the NTT are available.  While the total exposure time is too brief for the observations to be of use for morphological studies, they are still useful for determining a galaxy magnitude.

Although we observed PKS0915-11 on the night of 15 November 2004, none of our
data for this source is photometric.  We have obtained additional archival
observations of this source (from program 072.A-0549(A)), consisting of 38 10$\times$6s photometric exposures using the SOFI instrument in the large field mode on the NTT on the night of 2003 December 13.

PKS2104-25 had been previously observed on the night of 1999 April 14 using the SOFI instrument in the large field mode on the NTT. These archive observations (from program 63.O-0448(A), studying the cluster Abell 3744 of which PKS2104-25 is a member) consist of 36 20$\times$5s photometric exposures. However, due to the relatively small size of the random offsets used in the observational sequence for PKS2104-25 relative to the size of the galaxies in the cluster observed, the outer regions of the radio source host galaxy are unavoidably oversubtracted as part of the data reduction process, and these data are not useable for our purposes.

\subsection{Photometric transformations}

\begin{figure*}
\vspace{4.75 in}
\begin{center}
\includegraphics{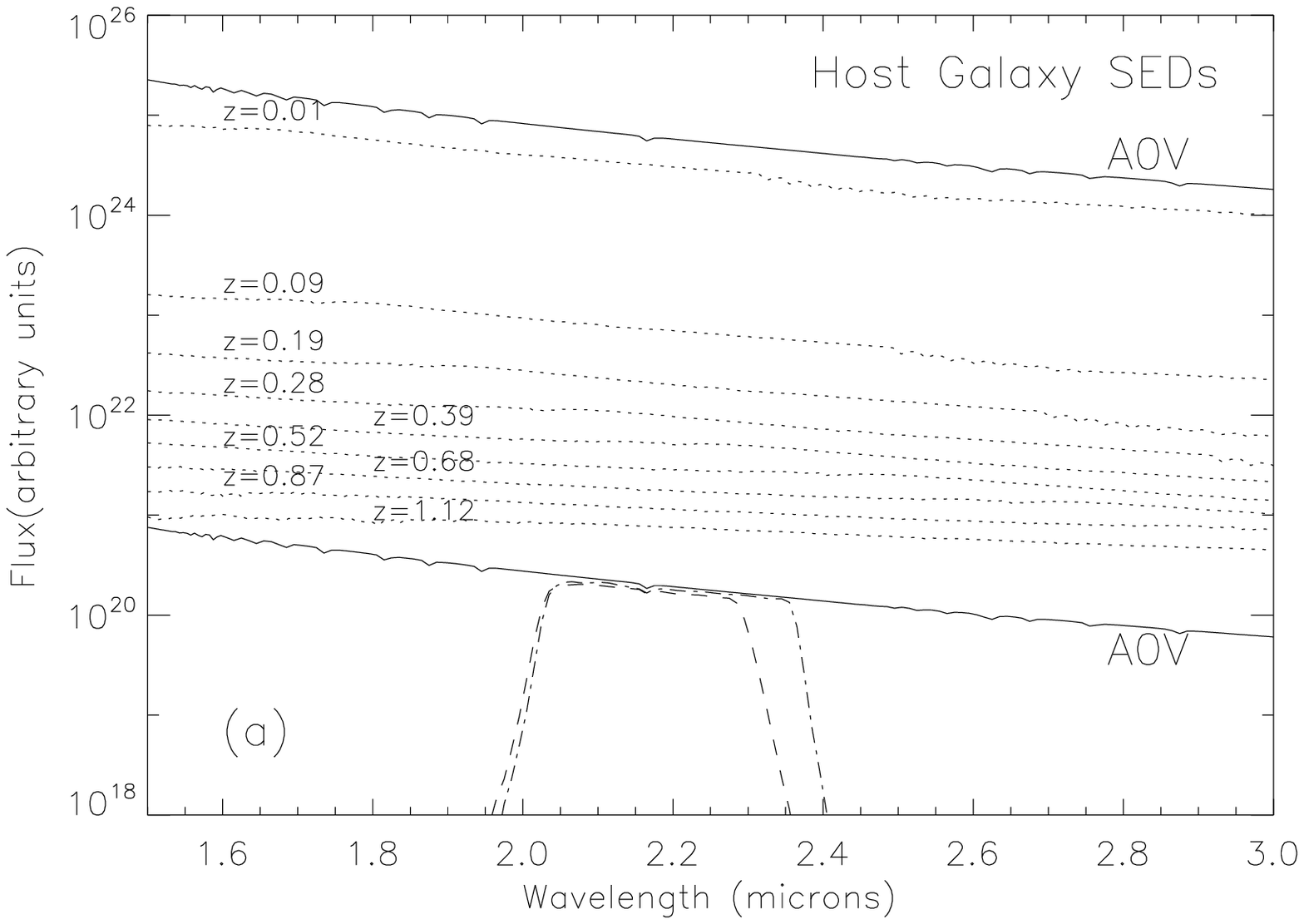}
\includegraphics{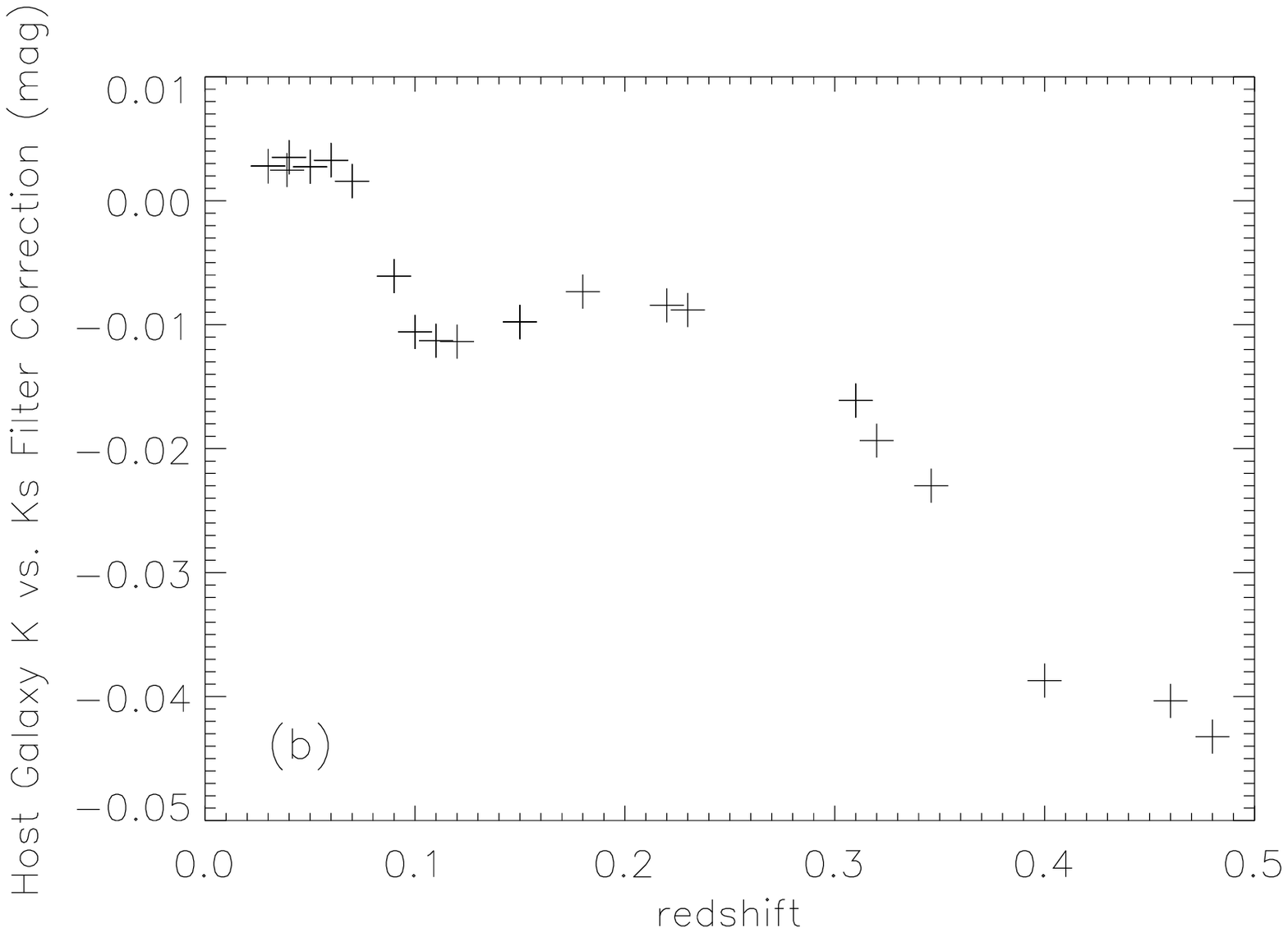}
\includegraphics{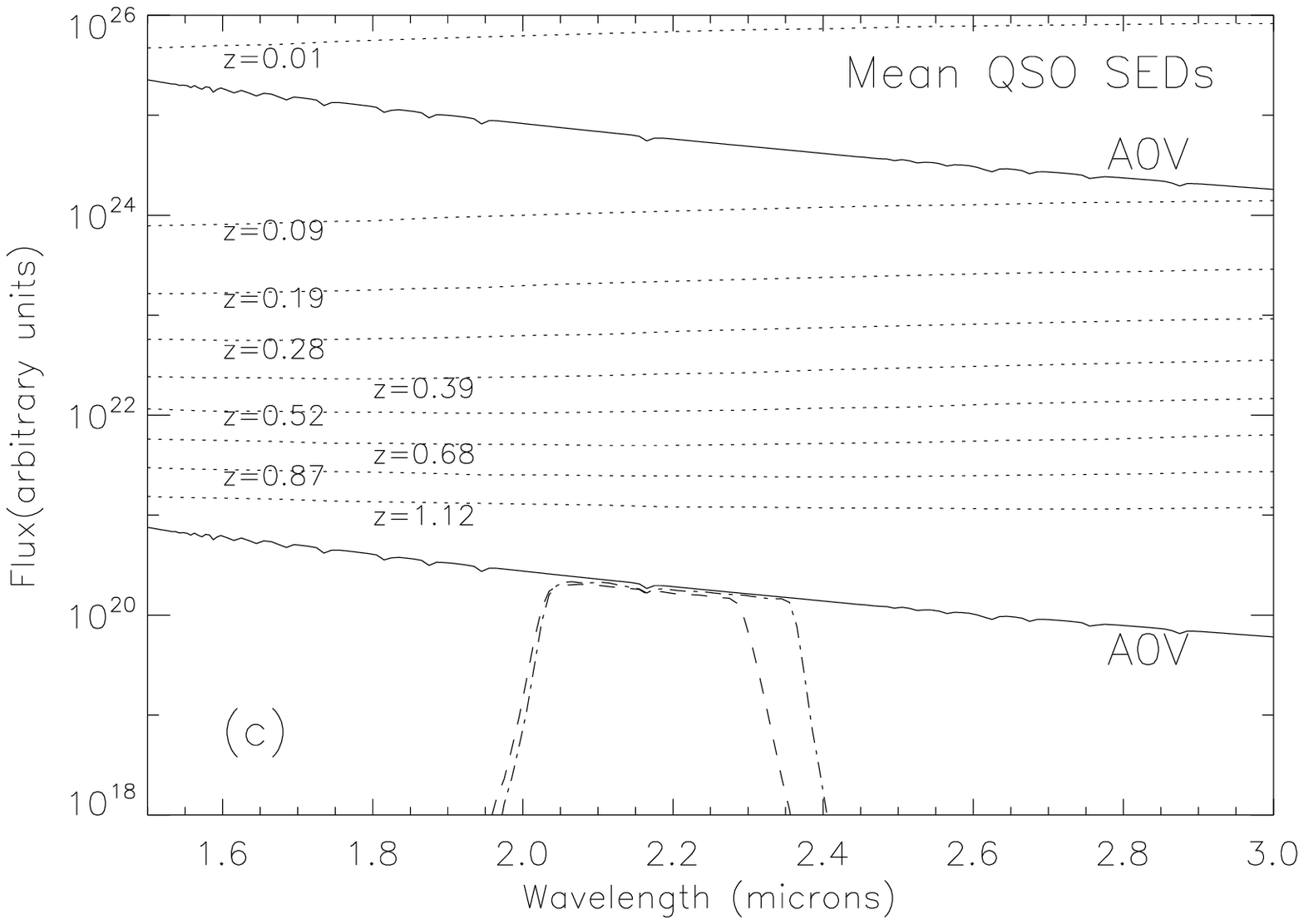}
\includegraphics{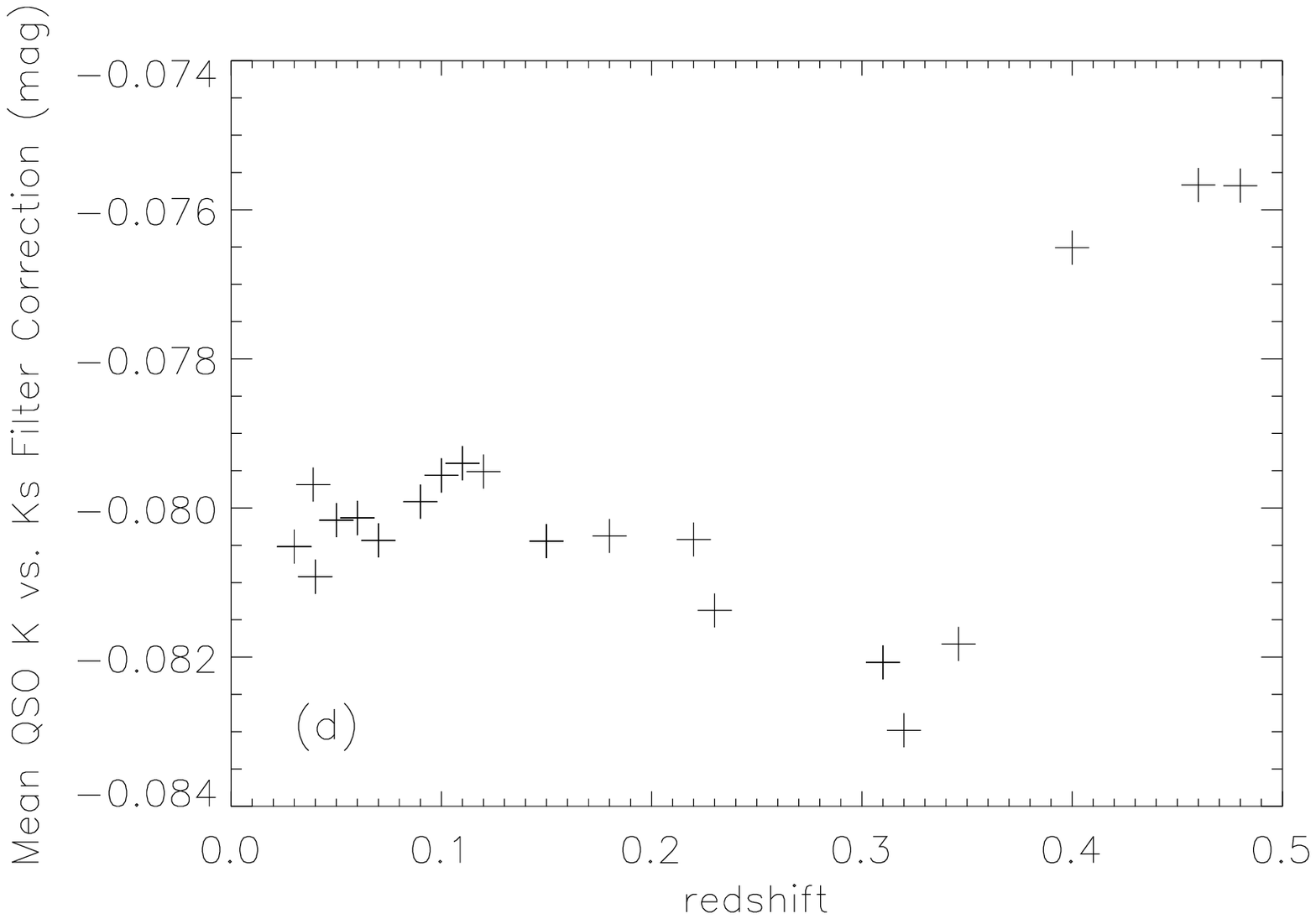}
\end{center}
\caption{(a - top left) Filter profiles vs. various spectral energy distributions, using the salpeter IMF solar metallicity models of Maraston (2009), and assuming a red horizontal branch morphology. Dotted lines represent the SED of an elliptical galaxy formed at a redshift of 10 and observed at redshifts ranging from 0.01 to 1.12, with appropriate relative flux scaling. The solid lines display the spectral shape of the reference A0V star Vega. Dashed and dot-dashed lines display the $K_S-$ and $K-$band filter profiles respectively, after convolution with the lower A0V SED. At low redshifts, the spectral shapes of star and galaxy are comparable, while at higher redshifts the elliptical galaxy is significantly redder. This corresponds to a magnitude difference of effectively zero between $K_S-$ and $K-$band observations of such a galaxy at $z=0$, while at higher redshifts the omission of flux in the shorter $K_S$ filter leads to a numerically larger value for magnitudes obtained in the $K_S-$band. (b - bottom left) Plot of the difference between observed  $K_S-$ and $K-$band magnitudes for galaxies at the redshifts of our sample objects which were observed in the $K_S$ filter; objects are generally brighter in the $K-$band, except at the lowest redshifts.  (c - top right) Filter profiles vs. the redshifted mean QSO SED of Richards et al (2006), with appropriate relative flux scaling. The reference A0V SED and filter profiles are as in frame (a).  The spectral shape of the mean QSO SED is redder than that of the the A0V reference star, and remains roughly constant over the redshift range of interest. (d - bottom right) Plot of the difference between observed  $K_S-$ and $K-$band magnitudes for a QSO at the redshifts of our sample objects which were observed in the $K_S$ filter; a typical QSO at these redshifts would be brighter in the $K-$band by roughly 0.08 magnitudes.
\label{Fig: kks}}
\end{figure*}

Our observations make use of two different filters: the Mauna Kea broad-band $K-$band filter in use with UFTI on UKIRT, and the shorter $K_S-$band filter in use with SOFI on the NTT and ISAAC on the VLT.  Broadly speaking, at low redshifts the numerical difference in magnitudes obtained in either filter is negligible, due to the comparable shapes of the SEDs of both galaxies and the zero-magnitude A0V reference star Vega.  However, for observations of higher redshift galaxies the situation becomes marginally more complex. Due to the flattening of the galaxy SED as redshift increases, the redder emission which would be included in the full $K-$band filter but not in the shorter $K_S-$band filter increases in strength relative to the emission included in both filters. Thus, the magnitude of a higher-redshift galaxy observed in the $K_S-$band is numerically larger than the value obtained in the $K-$band.  We illustrate this effect in figure \ref{Fig: kks}a, which displays the SED of an elliptical galaxy formed at a redshift of 10 and observed at redshifts ranging from 0.01 to 1.12, alongside the two filter profiles and two scaled SEDs representative of Vega, which (in the Vega-based magnitude system) has a magnitude of zero in both bands.  We have calculated the necessary numerical corrections which would need to be applied in order to convert a $K_S-$band magnitude to a $K-$band magnitude at the redshifts of all sources observed in the $K_S-$filter; these values are displayed in Fig.~\ref{Fig: kks}b.  We also apply these K-corrections to the cleaned 64kpc aperture photometry (Table 3, column 6), producing a consistent set of $K-$band magnitudes for the whole sample (Table 3, column 7).  Also displayed in  \ref{Fig: kks}c and  and  \ref{Fig: kks}d are similar plots illustrating the K-corrections which would be applicable to the mean SED of a quasar (from Richards et al 2006) at the same redshifts of interest; these allow us to account for the minimal K-corrections of any modelled QSO point source components found in the $K_S-$band data.

\section{Results}

The results of our aperture photometry, carried out as described in section 3.1, are tabulated in Table 3. The errors on our final K-corrected magnitudes range from 1.5\% to 12\%, and for over 70\% of our sample the photometric error is below 5\%.  In addition to photometry we have also  modelled the morphological properties of the sample galaxies, as detailed in the following sections.

\subsection{GALFIT modelling}
\subsubsection{Initial considerations}
The aim of our modelling is not solely to derive an accurate model for each galaxy, but also one which is useful in a wider sense, allowing us to compare sources both within the sample and in the wider literature in a simple, consistent and transparent manner. In brief, our modelling process asks the following questions: how well can the host galaxy be described as (A) a pure de Vaucouleurs elliptical (i.e. a galaxy with an $n=4$ S\'ersic (1968) profile\footnote{The S\'ersic surface brightness profile has the form $\Sigma(r)=\Sigma_{\rm{eff}} e^{-\kappa[(r/r_{\rm{eff}})^{1/n})-1]}$, where $\Sigma(r)$ is the surface brightness at a radius of $r$, $n$ is the S\'ersic index, $\kappa$ a numerical parameter which varies with $n$, $r_{\rm{eff}}$ is the effective radius of the galaxy (within which is contained half the light), and $\Sigma_{\rm{eff}}$ is the surface brightness at the effective radius.}), (B) a combination of de Vaucouleurs elliptical and an unresolved nuclear point source, and (C) a combination of an alternative S\'ersic index profile and a nuclear point source, or more complex models as required.  Our modelling utilises \textsc{galfit}\footnote{http://users.obs.carnegiescience.edu/peng/work/galfit/galfit.html} (Peng et al 2002, version 3.0 Peng et al 2010), a well-documented two-dimensional fitting algorithm which allows the user to simultaneously fit a galaxy image with an arbitrary number of different model components, and thus to extract structural parameters of the galaxy.  The model galaxy is convolved with a point spread function (PSF) and, using the downhill-gradient Levenberg-Marquardt algorithm, is matched to the observational data via the minimisation of the $\chi ^2$ statistic.

As the data for some objects have been obtained over the course of several nights and with different observing conditions, the point spread function for the combined data can be rather complex.  We therefore derive point spread profiles for each object by extracting 2-d images of stars in the field of view of each source, normalizing to unit flux and taking an average profile weighted by the signal-to-noise of the component extracted stellar profiles.  Where the resulting PSF data is still relatively noisy (i.e. for approximately half of our sources), we then use \textsc{galfit} to model the PSF as a combination of gaussian functions, resulting in a less noisy PSF image with the same characteristic shape and FWHM. This process results in an appropriate, high-quality point spread function for each of our sample objects, regardless of the observing conditions or availability of bright stars in the field of view.

The host galaxies of our sample objects are modelled over an area equivalent to a 100kpc $\times$ 100kpc box on the plane of the sky at the redshift of the source in question. Alongside the morphological parameters of the host galaxy, our \textsc{galfit} modelling is also intended to determine the percentage contamination by any unresolved nuclear point source emission from the AGN. As demonstrated by Pierce et al (2010), even a low level of AGN contamination can substantially affect the derived morphological parameters if it is left unaccounted for.    As our data span a wide range of magnitudes and spatial resolutions, it is also important to consider how well we can disentangle nuclear point source emission from that of the host galaxy, particularly in cases of poor seeing. Is the fitting biased towards identifying higher or lower nuclear point source contributions, under the conditions of different host galaxy sizes, different magnitudes and spatial resolutions?  

In order to address this question, it is important to understand how reliably \textsc{galfit} can extract morphological information from an image. This issue has been tested both by comparing \textsc{galfit} with other software and via the analysis of simulated data. 
For example, Floyd et al (2008) tested the results produced by \textsc{galfit} in comparison with those of 2DM (McLure et al 1999, Floyd et al 2004), which is optimised for the analysis of quasars. Both routines produced very consistent results, though the error treatment of the latter was deemed better for sources dominated by their nuclear emission, and overall \textsc{galfit}'s separation of the extended host galaxy emission from genuine unresolved emission from the galaxy nucleus was deemed to be very reliable.  This result is backed up by our prior modelling of several sources in the sample, which used our own least-squares fitting routines, also based on the  Levenberg-Marquardt algorithm.  For PKS2250-41 (Inskip et al 2008) and PKS1932-46 (Inskip et al 2007), the structural parameters derived with  \textsc{galfit} are in very close agreement with those determined for our previous work.  

A number of different groups (primarily working with HST data) have used extensive sets of simulated images to test the performance of \textsc{galfit} (e.g. Sanchez et al 2004; Jahnke et al 2004; Simmons \& Urry 2008; Kim et al 2008; Gabor et al 2009). In general, the magnitudes of bright point sources are very accurately recovered, although the morphological parameters of the host galaxies of such high-contrast systems were found to be less accurately constrained. However, spurious point-source contributions were identified in a small number of bulge-dominated systems (Simmons \& Urry 2008). Such issues are in part due to PSF undersampling and/or mismatches (e.g. Kim et al 2008), and the fact that the ACS PSF follows roughly an $r^{1/4}$ profile (Jahnke et al 2004).  
In cases where the PSF and galaxy data are well sampled, \textsc{galfit} accounts for observational point spread functions very robustly  (Peng et al 2002; Peng et al 2010; \textsc{galfit} website; Kim et al 2008), and the size of the observational PSF has no significant impact on the further separation of host galaxy from nuclear point source.  In the rest frame of our target objects, the physical resolution of our observations ranges from $\sim 0.04$kpc per pixel at best down to a minimum of $1.3$kpc per pixel for the highest redshift object in our sample. For our worst-resolution object (PKS0023-26), the seeing FWHM is equivalent to a spatial scale of $\sim 3.1$kpc, and the host galaxy is clearly well resolved beyong this level.  For our sample as a whole, undersampling of the host galaxy model components is not an issue, nor is undersampling of our PSF images, even for our large-field SOFI observations.

\begin{table*}
\vbox to220mm{\vfil Landscape table to go here in journal article. See the astro-ph only Appendix A for the same data.

\caption{}
\vfil}
\label{landtable}
\end{table*}

\begin{table*}
\vbox to220mm{\vfil Landscape table to go here in journal article. See the astro-ph only Appendix A for the same data.

\contcaption{}
\vfil}
\label{landtable}
\end{table*}

\begin{table*}
\setcounter{table}{3}
\caption{continued.} 
\begin{center} 
\scriptsize{}
\begin{tabular} {lcccccc}
{Source}& S\'ersic n & M$_{gal}$ & Point source \% & $R_{eff}$ (arcsec)& $R_{eff}$ (kpc) & reduced$\chi^2$ \\ \hline

PKS1949+02 & 4 & 11.36 & --    & $4.51 \pm 0.23$ & $5.07 \pm 0.26$ & 2.156 \\\vspace{0.05cm}
           & 4 & 11.29 & $5.0 \pm 0.1$\% & $6.29 \pm 0.31$  & $7.08 \pm 0.35$ & \textbf{1.955} \\
PKS2153-69 & 4 & 10.29 & --    & $8.45 \pm 0.42$ & $4.69 \pm 0.23$ & 1.254\\
           & 4 & 10.07 & $4.4 \pm 0.1$\% & $14.39 \pm 0.72$ & $7.97 \pm 0.40$ & 1.116\\\vspace{0.05cm}
           & 4,1 (bulge+disk)$^{2}$&9.94,13.23 & $3.6 \pm 0.1$\%$^2$ &$21.84 \pm 1.10$, $1.51 \pm 0.08$  &$12.09 \pm 0.61$, $0.84 \pm 0.04$ & \textbf{1.102}\\\vspace{0.05cm}
PKS2211-17 & 4 & 12.24 & --    & $19.03 \pm 0.95$ & $50.03 \pm 2.51$ & \textbf{0.5733}\\
          & \multicolumn{6}{c}{No adequate fit for n=4 + point source} \\
PKS2221-02 & 4 & 12.17 & --    & $0.003 ^{+0.06}_{-0.003}$& $0.004 ^{+0.08}_{-0.004}$& 1.830\\\vspace{0.05cm}
           & 4 & 12.40 & $64.7 \pm 0.7$\%& $7.02 \pm 0.41$ & $7.53 \pm 0.44$ & \textbf{1.505}\\
PKS2250-41 & 4 & 15.60 & --    & $0.44 \pm 0.02$ & $1.98 \pm 0.11$ & 1.045\\\vspace{0.05cm}
           & 4 & 15.63 & $16.6 \pm 0.7$\%& $1.09 \pm 0.07$ & $4.88 \pm 0.33$ & \textbf{1.040}\\
PKS2314+03 & 4 & 13.90 & --    & $0.35 \pm 0.02$ & $1.23 \pm 0.06$ & 1.612\\\vspace{0.05cm}
           & 4 & 13.74 & $35.9 \pm 0.4$\%& $5.74 \pm 0.31$ &$20.18 \pm 1.09$ &\textbf{1.068}\\
PKS2356-61 & 4 & 12.40 & --    & $4.66 \pm 0.23$ &  $8.19 \pm 0.41$ & 0.9418\\ 
           & 4 & 12.36 & $2.4 \pm 0.1$\% & $5.56 \pm 0.28$ & $9.77 \pm 0.49$ &\textbf{0.9037}\\

\hline\multicolumn{7}{l}{Notes:}\\
\multicolumn{7}{l}{[1] Best fit model for PKS0131-36 includes an edge on disk with a central surface brightness of }\\
&\multicolumn{6}{l}{14.81 mag/arcsec$^2$, a scale height of $1.2^{\prime\prime}$ and a scale length of $3.4^{\prime\prime}$.}\\
\multicolumn{7}{l}{[2] Point source is given as percentage of total (bulge+disk+point source) galaxy flux.}\\
\multicolumn{7}{l}{[3] The eastern galaxy of this pair is the radio source host.}\\
\multicolumn{7}{l}{[4] The eastern galaxy of this pair is the radio source host.}\\
\multicolumn{7}{l}{[5] Point source centroid offset from galaxy centroid by $\sim 0.5^{\prime\prime}$}\\
\multicolumn{7}{l}{[6] Companion modelled as $K=16.15$ point source}\\
\multicolumn{7}{l}{[7] The eastern galaxy of this interacting pair is the radio source host.}\\
\end{tabular} 
\end{center}                        
\end{table*}                       
\normalsize

\subsubsection{Details of the modelling strategy}

The free parameters for our most basic model (case A) are therefore the host galaxy centroid, its luminosity, scale length (effective radius $r_{eff}$), the position angle of its major axis ($\theta$) and its ellipticity ($b/a$).  For case B, we add a further three free parameters, two for the centroid of the nuclear point source emission (which is not always precisely aligned with that of the host galaxy) and another for its luminosity. At this stage the observations have already been background-subtracted, and no background emission pedestal value was used initially.  Once a good fit was obtained, we also allowed the residual background level to vary as an additional free parameter (within the constraints imposed by the residual background flux measured across the full field of view of our data), and in general this rarely led to any variation in the results obtained.  For objects in crowded fields, we also iteratively model any neighbouring stars or galaxies which interfere with the host galaxy model fit; once a good model for these adjacent objects has been obtained, their parameters are held fixed, effectively removing them from consideration.  Final reduced-$\chi^2$ values are determined by holding the fit values fixed and then modelling all other objects in the field of view; as the reduced-$\chi^2$ value is sensitive to the total residual flux as well as to the estimated value of the noise level in the data, the presence of companion objects would skew the resultant reduced-$\chi^2$ away from an ideal value of 1.0.  For several sources (PKS0043-42, PKS0453-20, PKS1934-63) the large values of the reduced-$\chi^2$ are due to the presence of unmodelled faint close companion objects, rather than denoting a poor fit to the host galaxy itself.

For our more complex model fits (case C), we consider a range of different S\'ersic indices, and identify the best model resulting from the possible values of $n=1$, 2, 4 or 6.  While it would be possible to allow the S\'ersic index $n$ to vary freely, in general there is some level of degeneracy between the  nuclear point source contribution and S\'ersic index (and to a lesser extent the background level and galaxy scale length).  By holding $n$ fixed over an appropriate range of values, although we may overlook the 'numerical best-fit', we can more reliably obtain solid values for the other model parameters and a measure of the goodness of fit in each case. This also gives us a better understanding of how the source deviates from a more standard model which can be easily contrasted with other sources and previous models of the same galaxy within the literature.  Where the galaxy is obviously more complex than a single S\'ersic profile plus a point source, we also add further components; these are noted separately on a case-by-case basis.  Finally, following the work of Donzelli et al (2007) we also attempt to fit each source with a combination of an $n=4$ S\'ersic bulge with an $n=1$ S\'ersic disk component; unlike Donzelli et al, we fix the S\'ersic parameter of the bulge at $n=4$ and also explicitly account for a nuclear point source component as an additional optional parameter of our modelling, as opposed to excluding the innermost point source contaminated regions from the model fits.

The results of our modelling are tabulated in Table 4, where we include the relevant values for the pure $n=4$ models (with and without a nuclear point source) and also the best fit models with alternative values of $n$ and/or the Donzelli-style bulge/disk combination where appropriate. The modelling results are also presented pictorially in Figures 2-8; machine readable images can be obtained from the main 2Jy website\footnote{http://2jy.extragalactic.info}. 
We display contour plots of only the central 50kpc $\times$ 50kpc part of this modelled region in Figures 2-42a, together with contours of the best-fit model (as denoted by the bold reduced-$\chi^2$ values in Table 4) overlaid on a greyscale of the model-subtracted residual images in the same region.  The choice of displaying only the central area of our modelled regions was made as it provides a far clearer illustration of the match between observation and model, and any residual features observed.  In the following section, we provide more detailed notes on the results obtained for each individual object.

\subsection{Notes on individual objects}

In this section, we present notes on each individual object in turn. Where comparisons are made with other morphological modelling studies in the literature, these all account for the presence of a nuclear point cource component, except where otherwise noted.

\subsubsection{PKS0023-26}
PKS0023-26, at a redshift of $z=0.322$, is one of the higher redshift objects in our sample.  The source appears to lie within a dense cluster environment: VLT spectroscopy has shown that the two neighbouring galaxies to either side (at the edges of the frame in fig.~\ref{Fig: 2}a) lie at the same redshift (RA10; Tadhunter et al 2010).  

Our \textsc{galfit} modelling of this source is consistent with a de Vaucouleurs elliptical, though there is considerable degeneracy between effective radius and S\'ersic index for this source owing to the lower signal-to-noise and smaller spatial scale. For fits with $n=4$, we find $r_{eff}=1.0^{\prime\prime}$ ($\sim$5kpc) for fits without a point source, and $r_{eff}=4.4^{\prime\prime}$ ($\sim$20kpc) with a best-fit unresolved nuclear point source contribution of 12.2\%. The large effective radius in this case is not unrealistic; deeper $r^{\prime}-$band imaging observations of this source (Tadhunter et al 2010) reveal that it lies at the centre of a cluster and is surrounded by a large diffuse halo which extends out to projected distances at least as far as the two neighbouring objects in the field.

\subsubsection{PKS0034-01 (3C15)}

For this source, our best-fit model is a de Vaucouleurs elliptical with an effective radius of $3.7^{\prime\prime}$ ($\sim 5$kpc), and a point source contriubtion of $\sim 4\%$. 
By comparison, $R$-band observations and $r^{1/4}$ modelling of this source by Govoni et al (2000) found a larger effective radius of $6.3^{\prime\prime}$. Zirbel (1996) records an even large effective radius of $6.9^{\prime\prime}$/9.1kpc (after conversion to our assumed cosmological model, and with no nuclear point source component) at even bluer (observed V-band) wavelengths. However, Fasano, Falomo \& Scarpa (1996) find a more comparable effective radius ($5.24^{\prime\prime}$), also in the R-band. At infrared wavelengths, $H-$band NICMOS observations (Floyd et al 2008) also find very similar values to our own ($r_e = 4.96 \pm 0.45$kpc with $n=4.17$).

\begin{figure*}
\vspace{8.45 in}
\begin{center}
\includegraphics{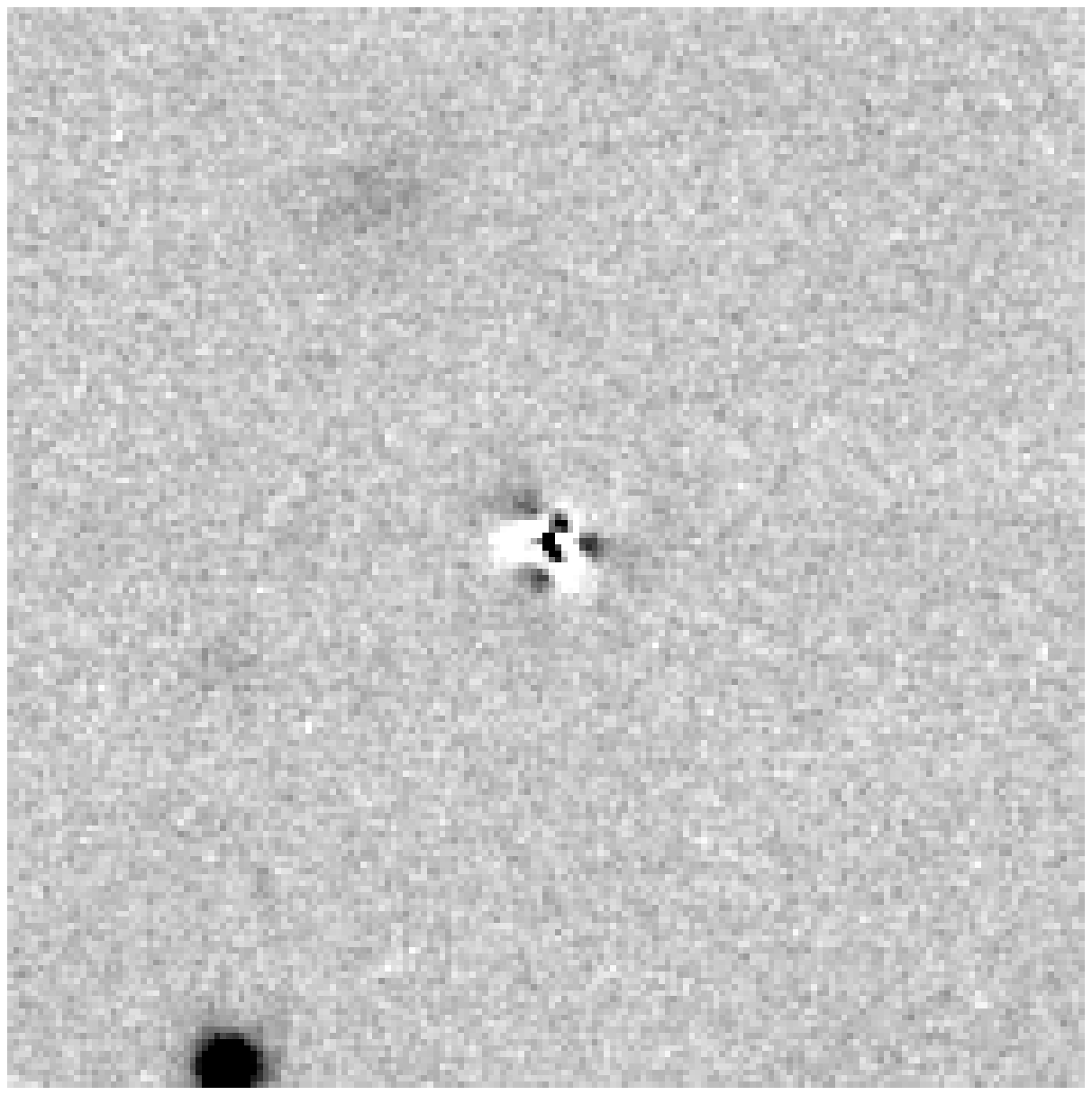}
\includegraphics{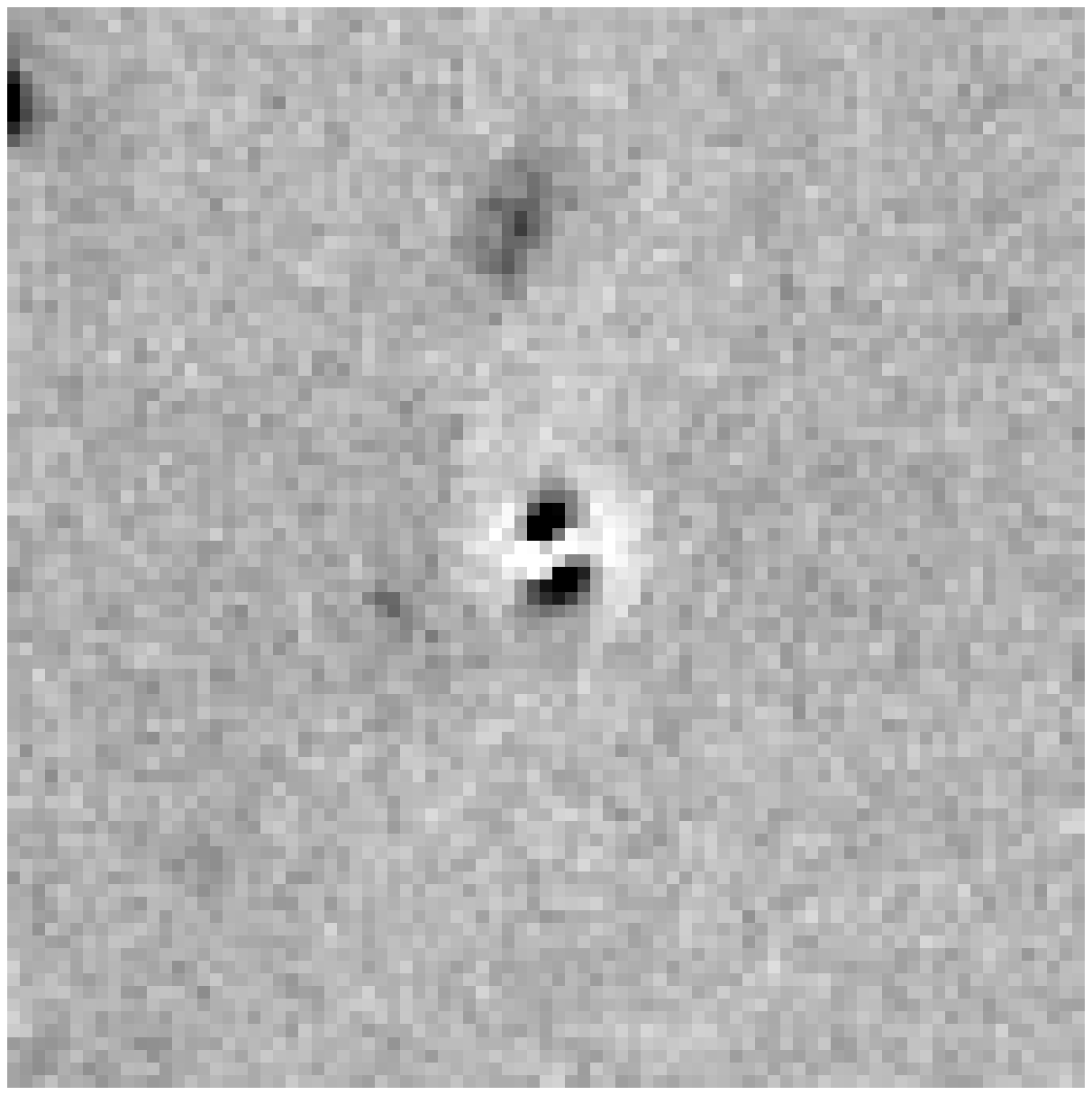}
\includegraphics{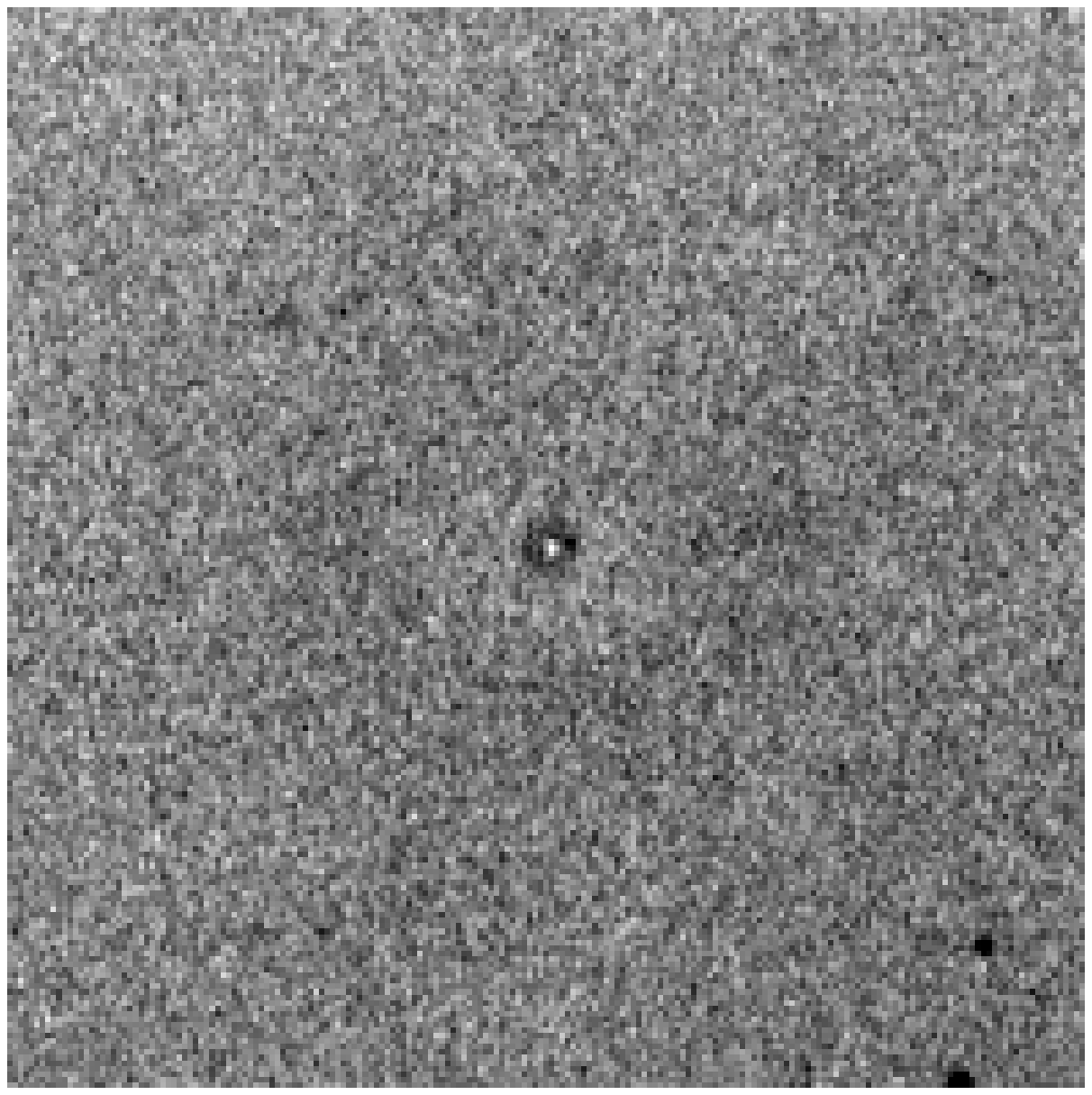}
\includegraphics{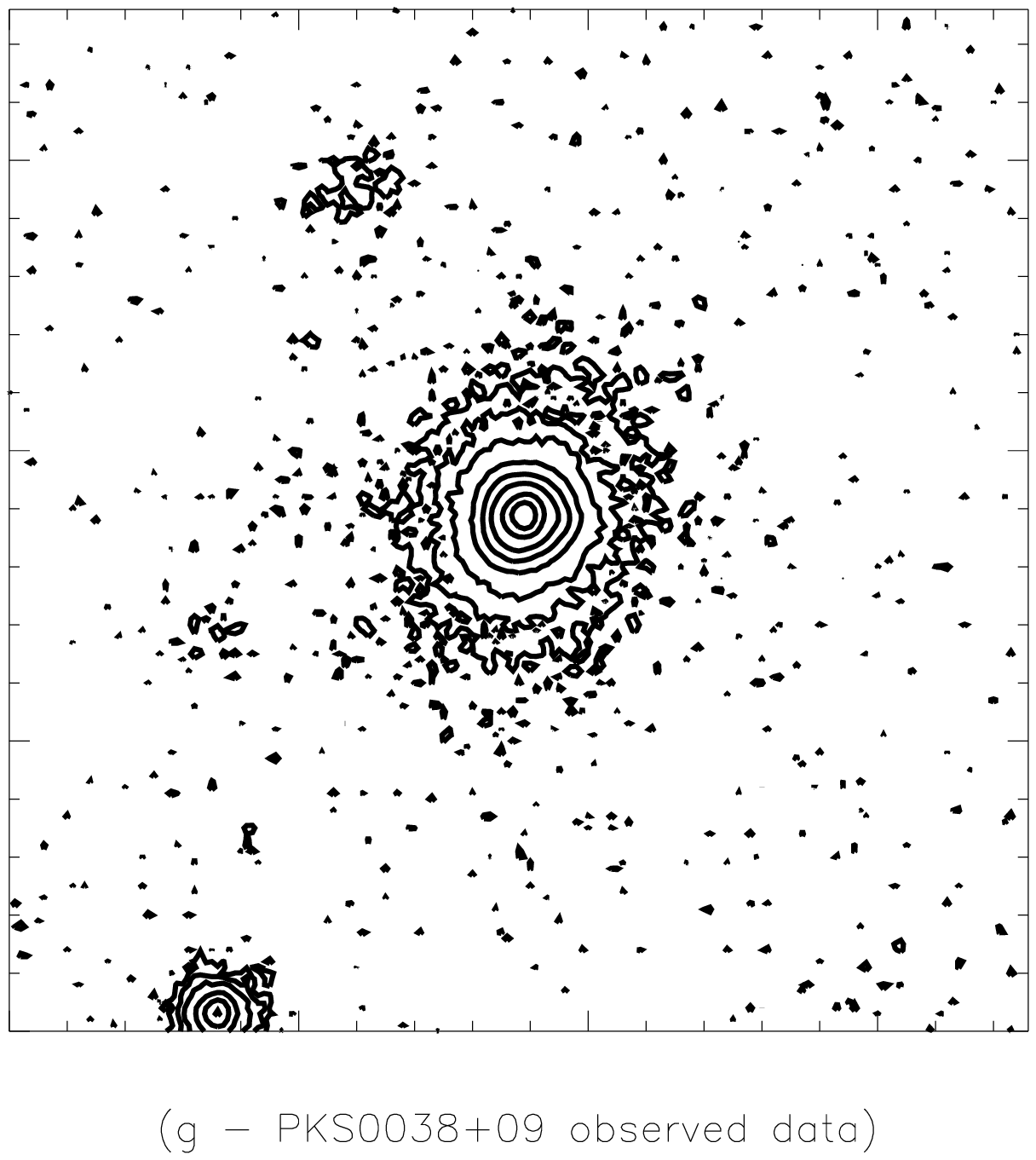}
\includegraphics{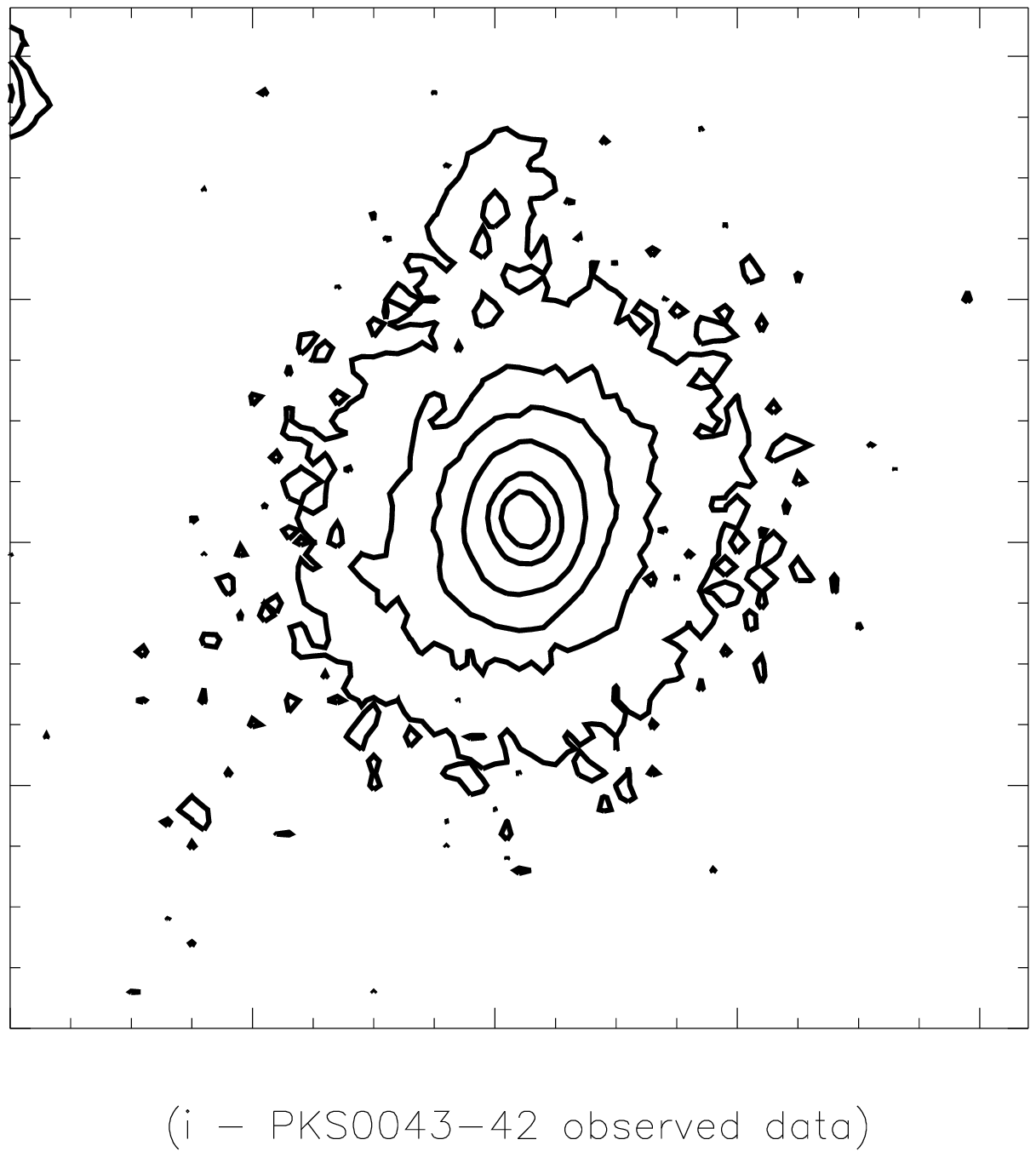}
\includegraphics{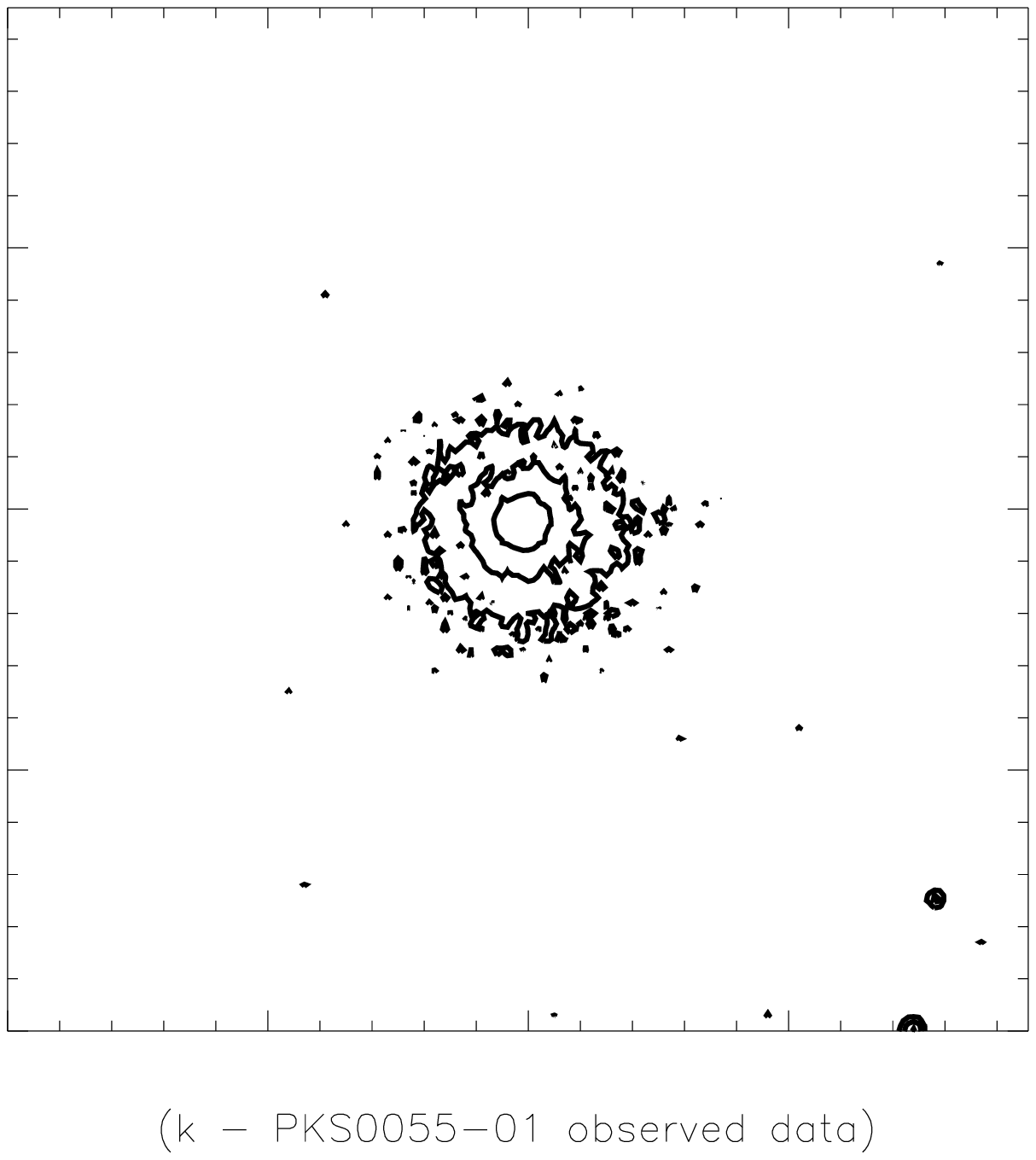}
\includegraphics{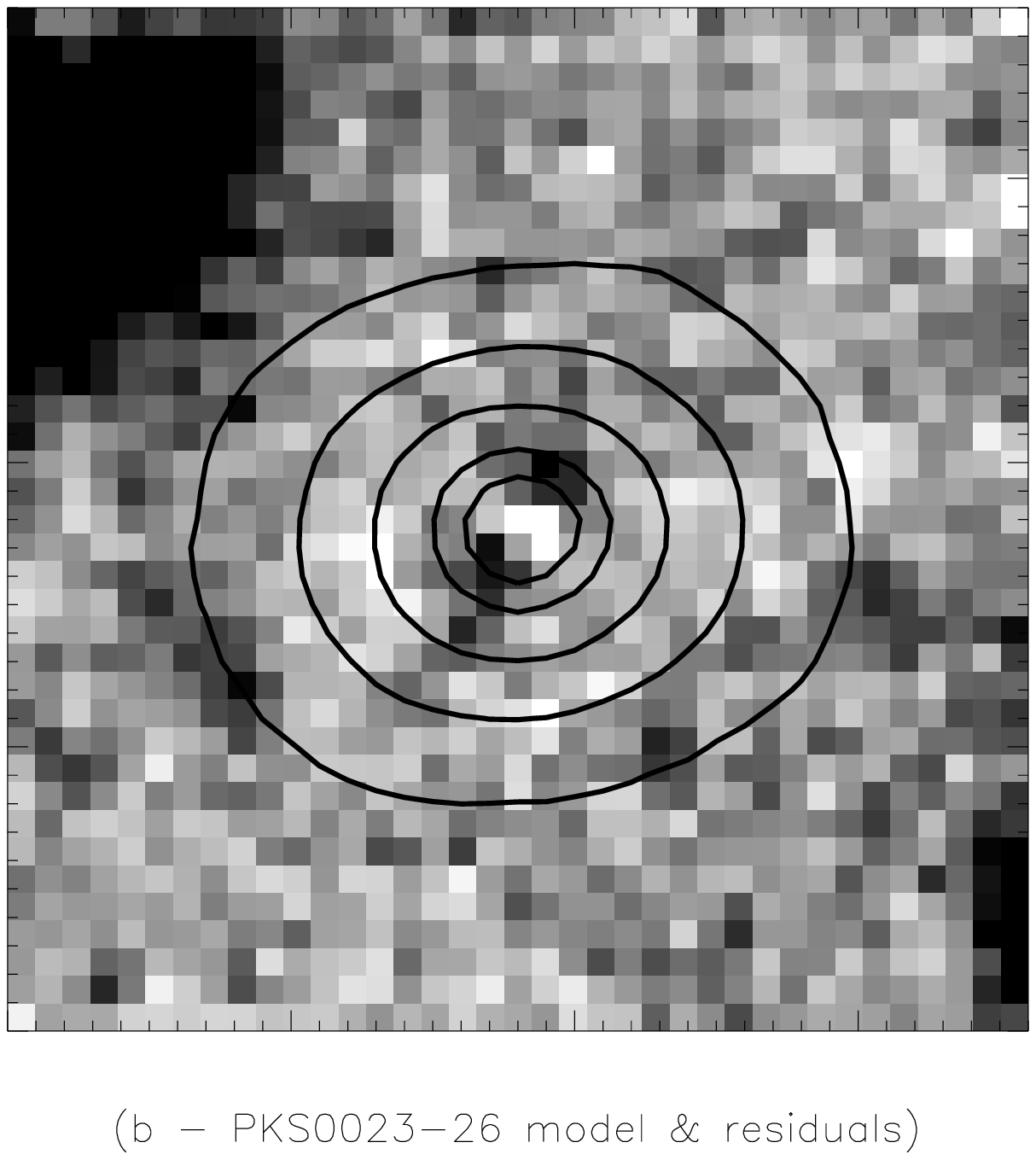}
\includegraphics{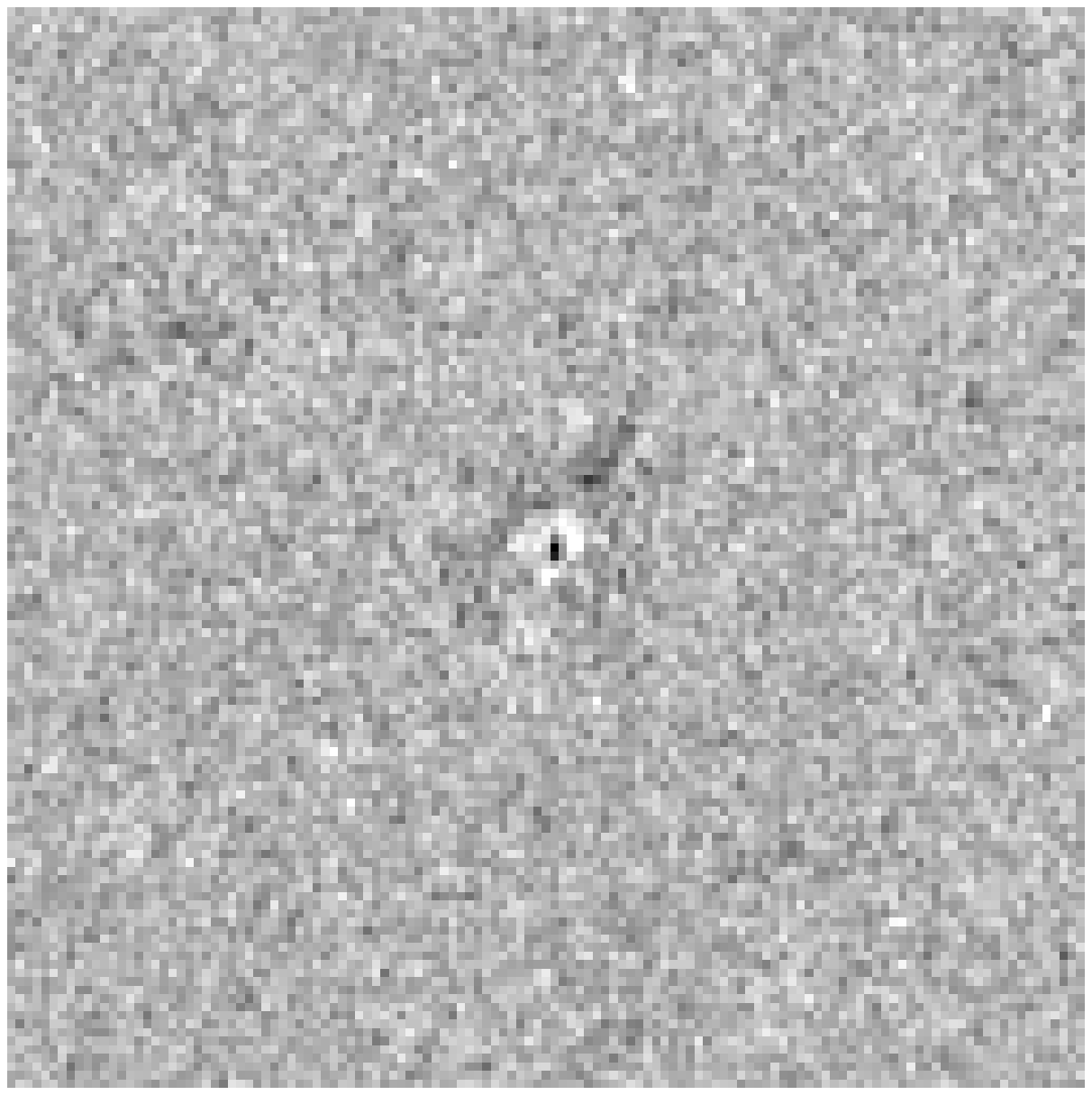}
\includegraphics{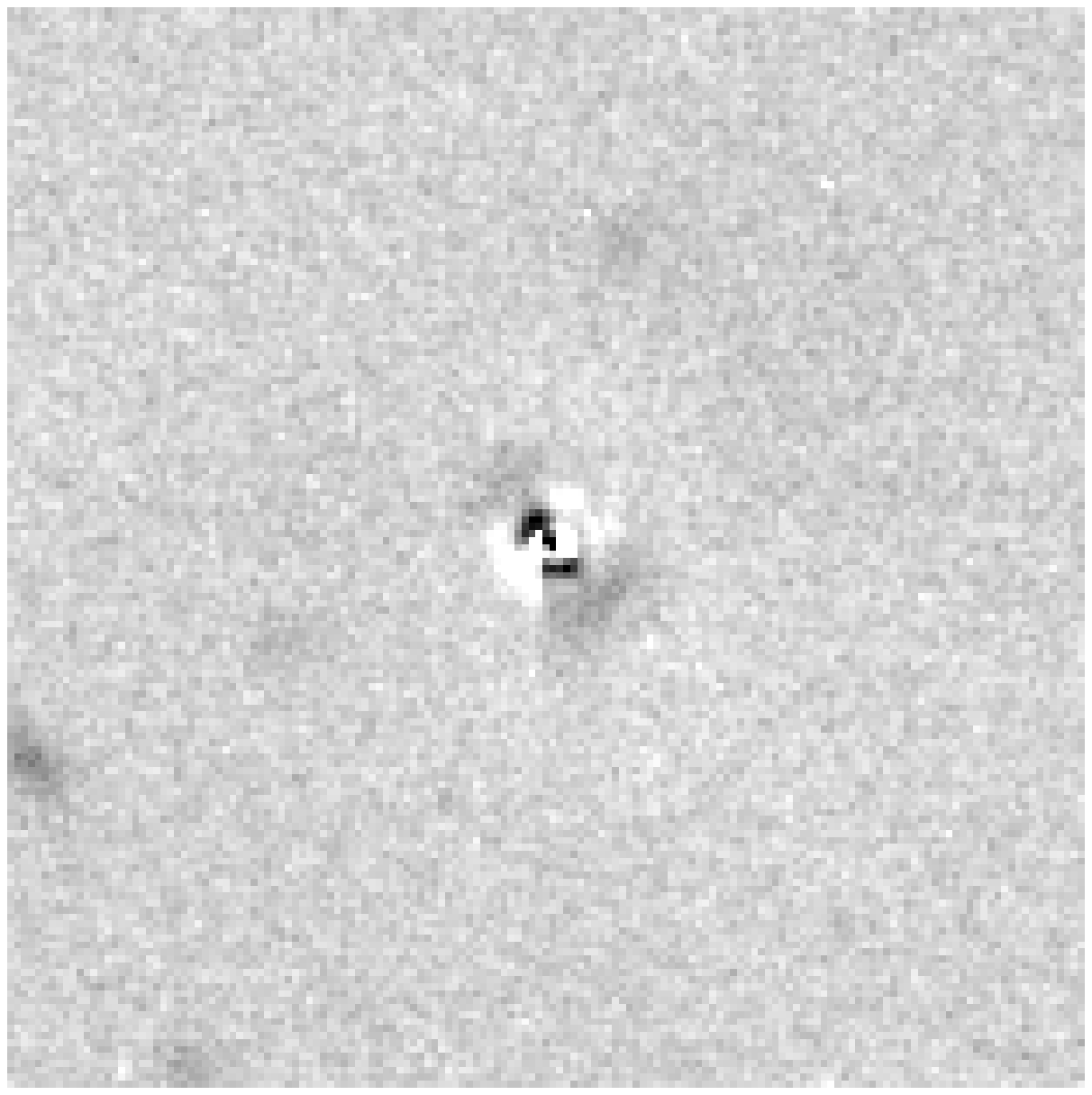}
\includegraphics{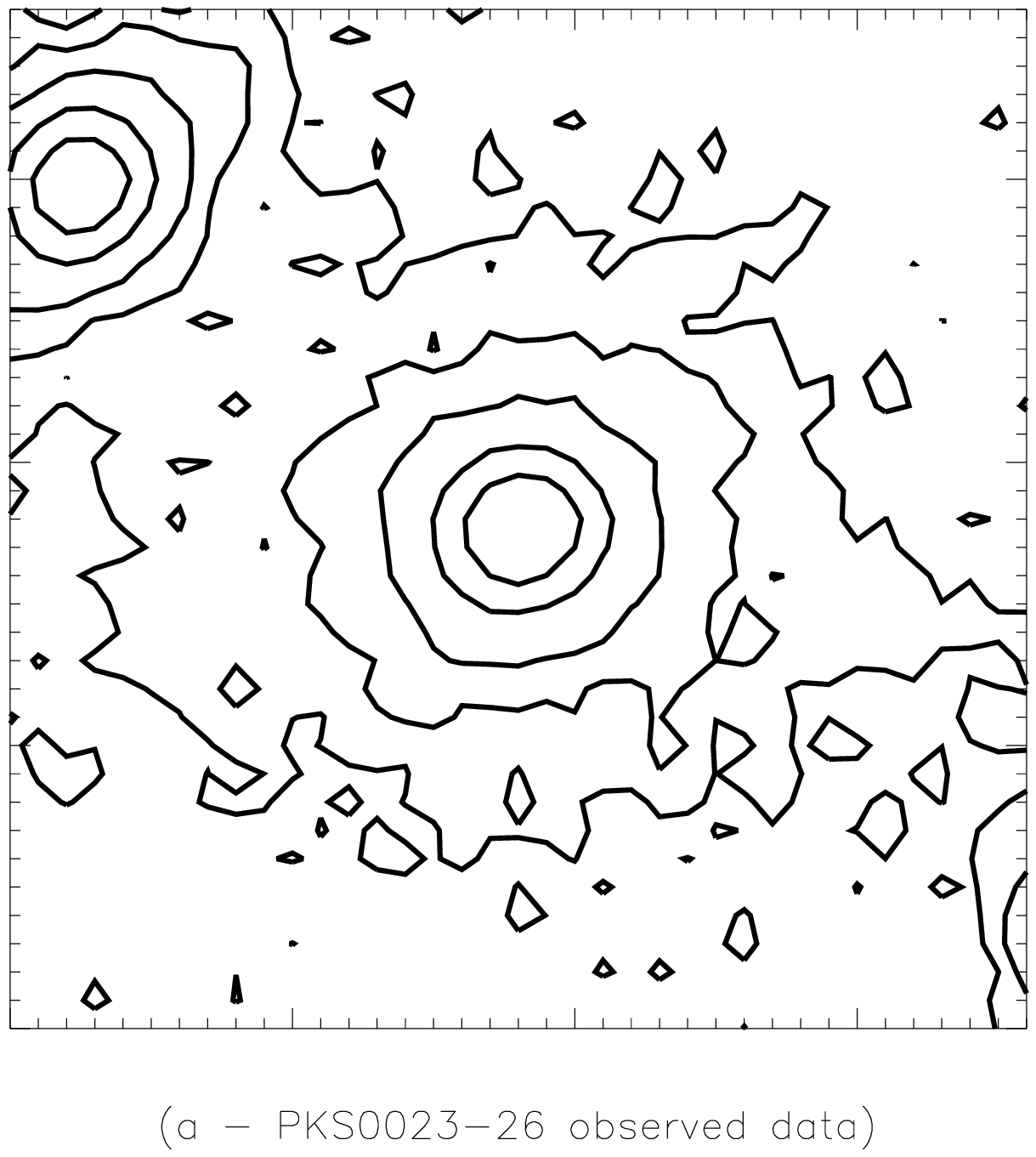}
\includegraphics{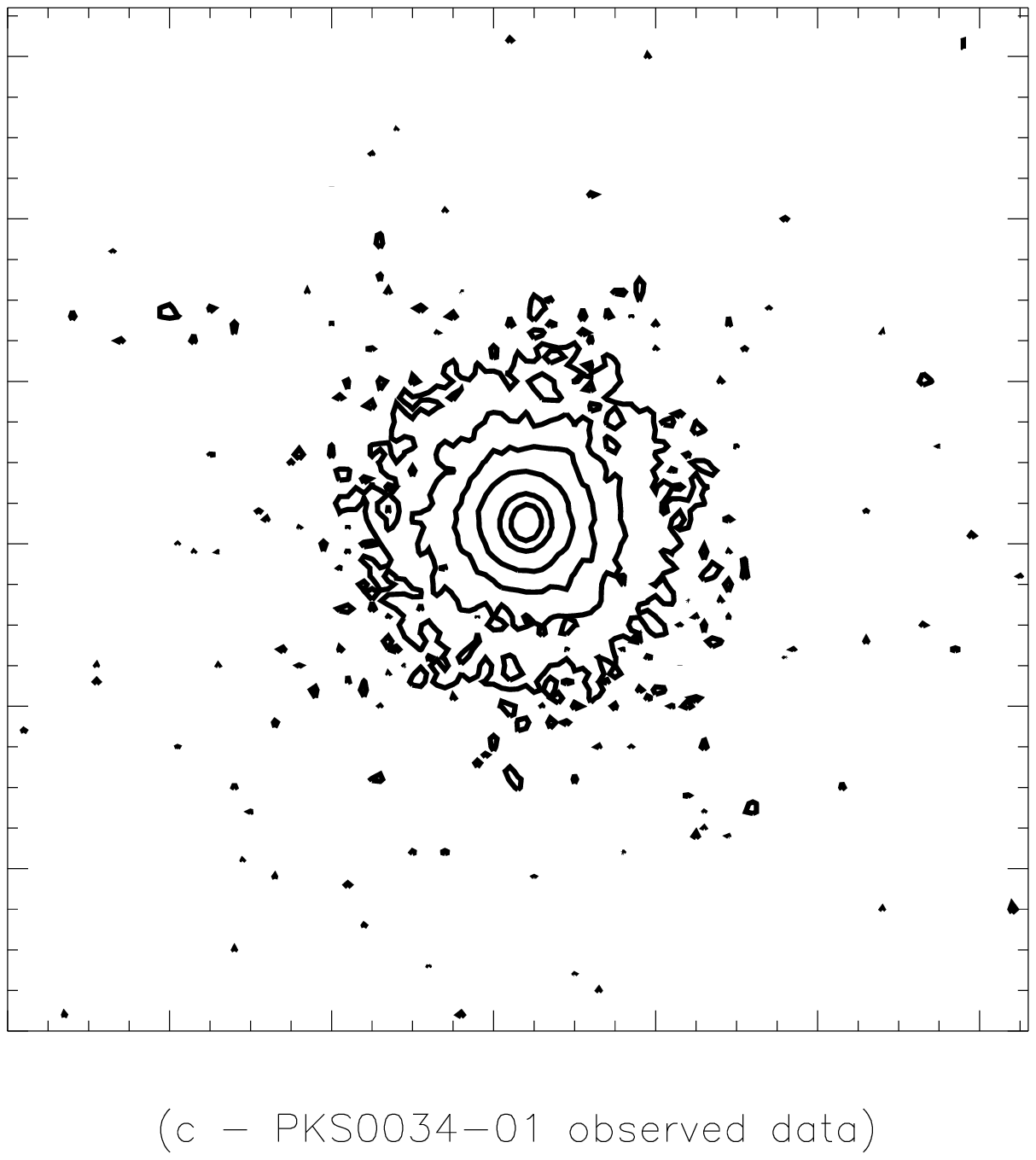}
\includegraphics{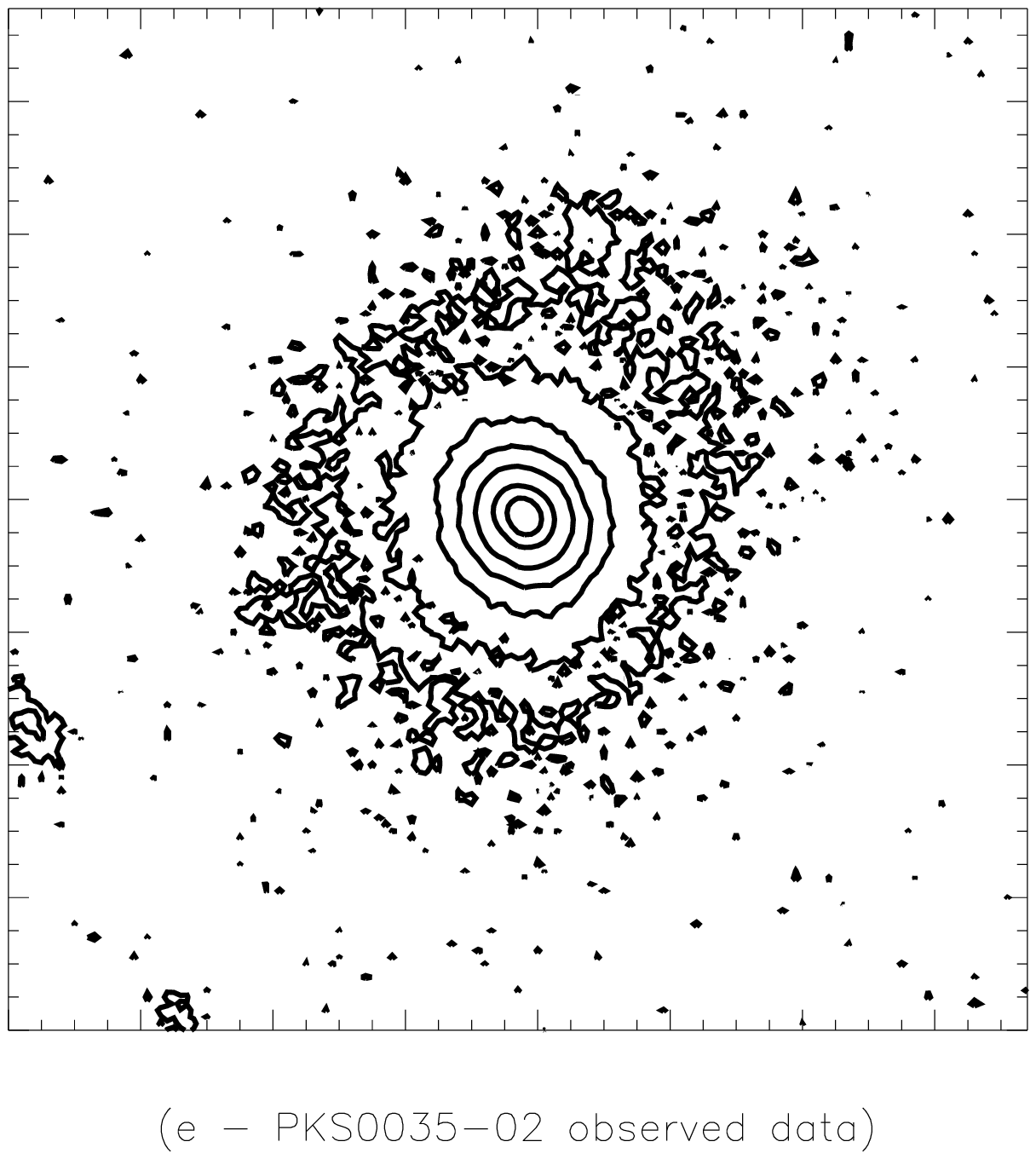}
\end{center}
\caption{50kpc by 50kpc images of PKS0023-26, PKS0034-01, PKS0035-02, PKS0038+09, PKS0043-42 and PKS0055-01. The observed data contours are displayed in frames (a), (c), (e), (g), (i) and (k), while frames (b), (d), (f), (h), (j) and (l) show the best-fit model contours on greyscale images of the model-subtracted residuals.  The maximum contour level is 50\% of the peak flux for that source in all cases, with subsequent contours at 25\%, 10\%, 5\%, 2.5\%, 1\%, 0.5\% and 0.25\% (latter flux levels not shown in all cases). The minimum contours displayed are at 0.25\% for PKS0038+09, 0.5\% for PKS0035-02, 1\% for PKS0034-01 and PKS0043-42, 2.5\% for PKS0023-26, and 10\% for PKS0055-01. 
\label{Fig: 2}}
\end{figure*}

Interestingly, an optical synchrotron jet has been observed in this source (Martel et al 1998) at a position angle of $-30^{\circ}$.  We also observe this feature in the K-band, visible both in the K-band contour plot (fig.~\ref{Fig: 2}a) and our model-subtracted residual image (fig.~\ref{Fig: 2}c). Martel et al (1998) and Sparks et al (2000) also observe complex emission and dust features in the nuclear region of PKS0034-01. The complex  nuclear residuals present in our data could potentially be the same features, although we cannot rule out small-scale PSF variations as the cause in this case.

Our modelling of this source gives a best-fit $r_{eff} = 7.64$kpc ($2.17^{\prime\prime}$) with a S\'ersic index of $n=4$, and a nuclear point source contribution of 23.9\%, consistent with its status as a BLRG.  Our derived galaxy effective radius is a little larger than that found by Floyd et al (2008) in their study of NICMOS observations of this source, and roughly twice that determined at the same wavelength range by Tremblay et al (2007) and Donzelli et al (2007), though this is likely due to the smaller S\'ersic index of $n \sim 2.7$ used by the latter. Our derived radius is smaller than the 10kpc value found in the optical de Vaucouelurs-only modelling of Zirbel (1996).  Our model residuals display an excess of flux extending linearly to the NE-SW (aligned with the galaxy major axis in optical WFPC2 data; de Koff et al 1996), with a slight oversubtraction in the region immediately beyond this excess flux. Beyond this, we observe subtle variations in the background emission surrounding a number of fainter features/companion objects lying close to the host galaxy. It is possible that these features have biased our model towards a larger effective radius, and it should be noted that a secondary minimum in our modelling is observed at a similar effective radius to that found for the zero point source model.  

PKS0035-02 has a very interesting radio morphology (Morganti et al 1999). The south-eastern radio jet displays an extreme bend and a number of bright hotspots. Intriguingly, these structures appear to be almost perfectly aligned with the features observed in our K-band images.

\subsubsection{PKS0038+09 (3C18)}
Approximately two thirds of the $K-$band flux from this source is modelled as an unresolved nuclear point source, consistent with PKS0038+09's status as a BLRG. We find a de Vaucouleurs effective radius of $\sim 5.5$kpc ($1.77^{\prime\prime}$), slightly larger than the value measured in the $V-$band for this source in the de Vaucouelurs-only modelling of Zirbel (1996).  Considering alternative S\'ersic indices, the model with $n=2$ provides a comparably good fit to the data, with an effective radius of $4.85$kpc ($1.56^{\prime\prime}$) and a nuclear point source contribution of 70\%.

\subsubsection{PKS0043-42}
PKS0043-42 is a WLRG, which appears to be interacting with a companion object to the north. Three further objects lie to the south and north east at projected distances of $\sim 17^{\prime\prime}$. Our best fit de Vaucouleurs model includes a nuclear point source contribution of order $\sim 10\%$, and an effective radius of 4.35$^{\prime\prime}$/9kpc. 
It should be noted that the inner regions of this galaxy are rather complex, and there are other possible causes for the excess flux currently fitted by the point source. We observe a central isophotal twist and an E-W dimming/N-S excess flux, possibly due to a dust lane and/or excess flux associated with the apparent interaction.

\subsubsection{PKS0055-01 (3C29)}

Assuming a S\'ersic index of $n=4$, our measured effective radius for this source is $11.46^{\prime\prime}$ (10.00kpc), comparable to those found in the previous $R-$band studies by Govoni et al (2000), Fasano, Falomo \& Scarpa (1996) and Smith \& Heckman(1989) of $9.7^{\prime\prime}$, $10.3^{\prime\prime}$ and  $12.7^{\prime\prime}$ respectively.

However, our best fit model includes a very small nuclear point source contribution and a smaller (4.06kpc; $4.65^{\prime\prime}$), disk-type S\'ersic index of $n=1$, with an excess of flux in the inner regions of the galaxy. This ties in with Fasano et al's comments on the shape of the luminosity profile of PKS0055-01, which they find to be too bright in the centre to be explained by an exponential law, and too faint for a de Vaucouleurs law, and explain in terms of strong nuclear absorption. We confirm that the same luminosity profile is observed in our $K-$band data.

\begin{figure*}
\vspace{8.45 in}
\begin{center}
\includegraphics{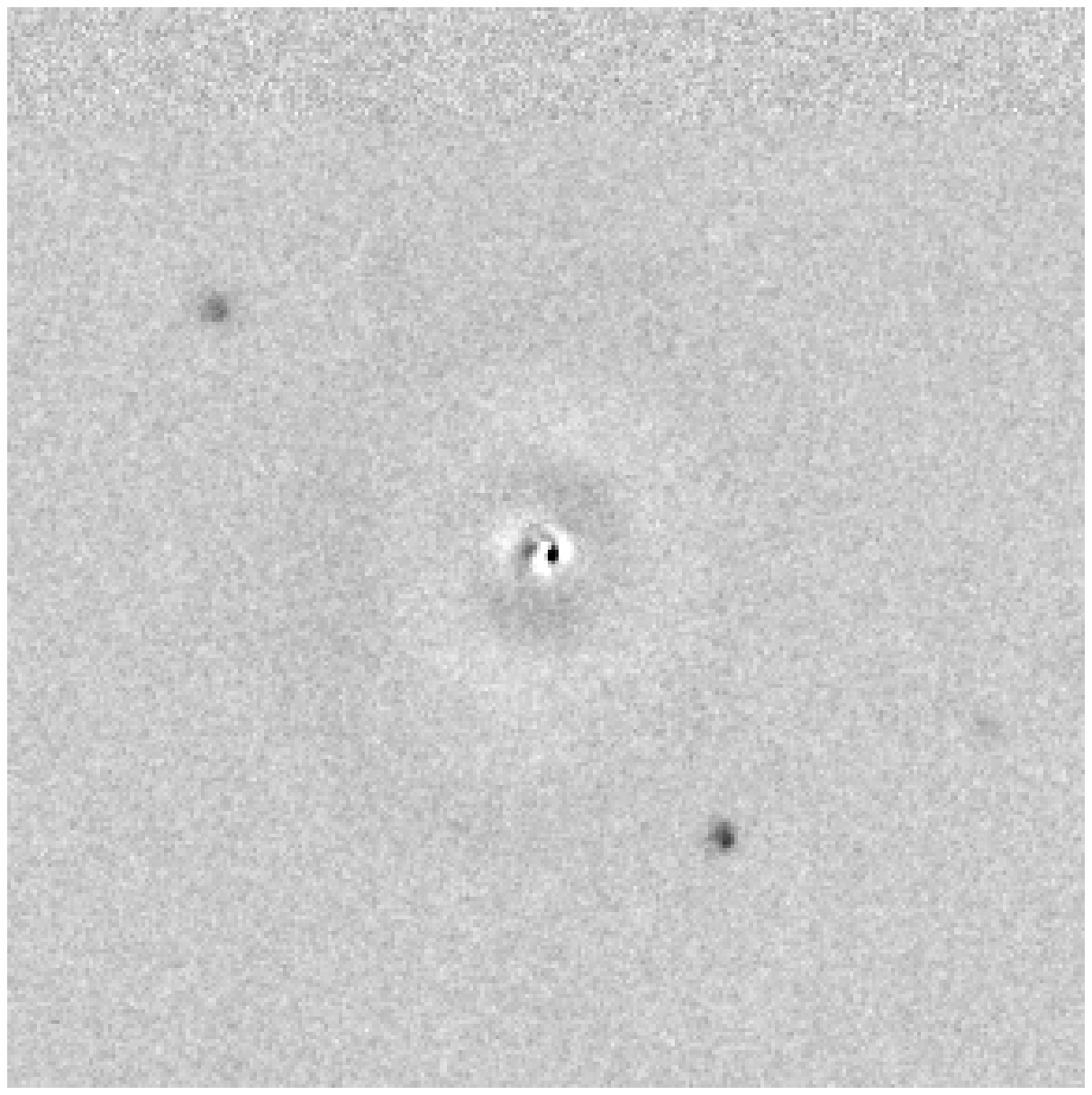}
\includegraphics{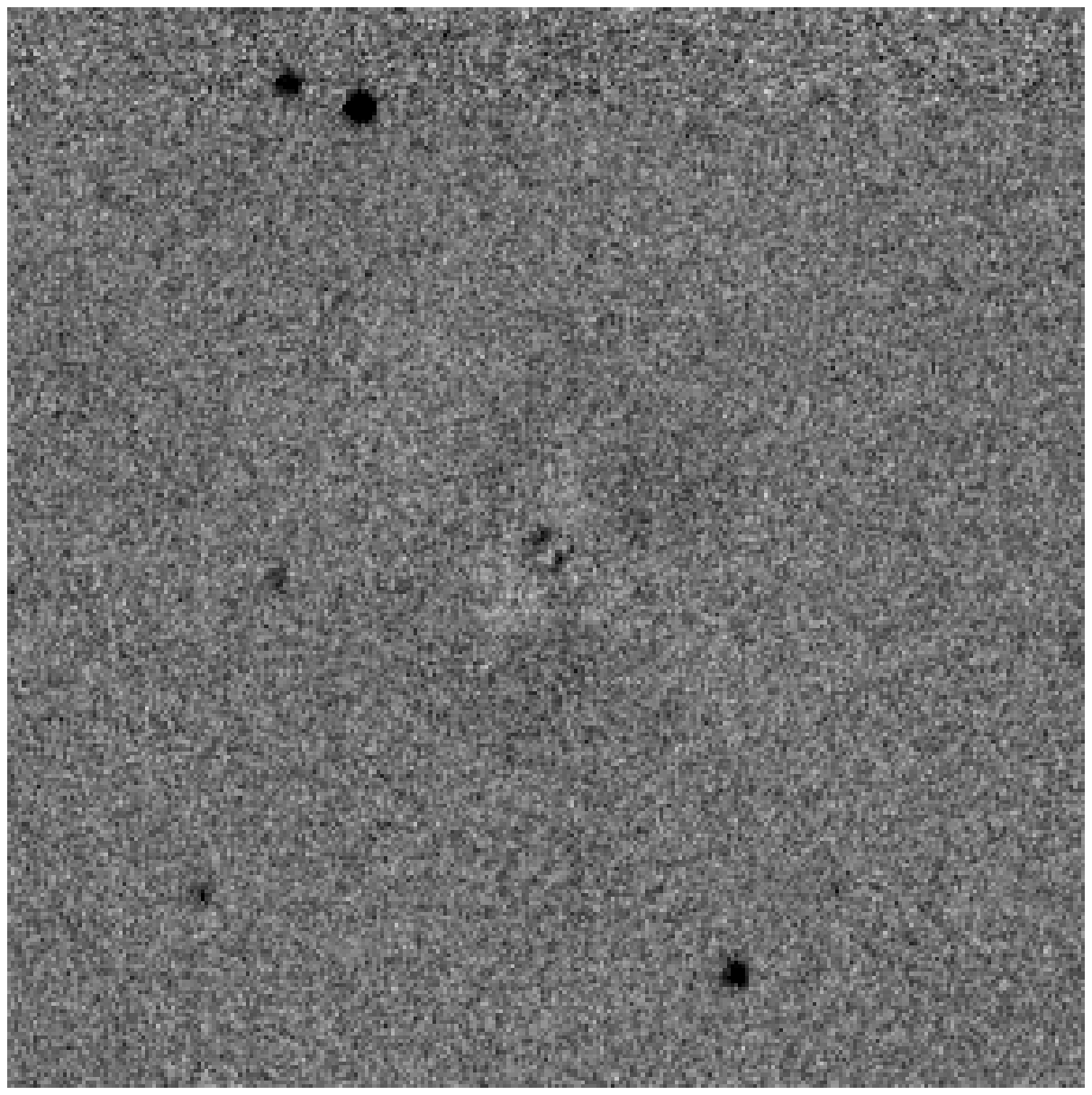}
\includegraphics{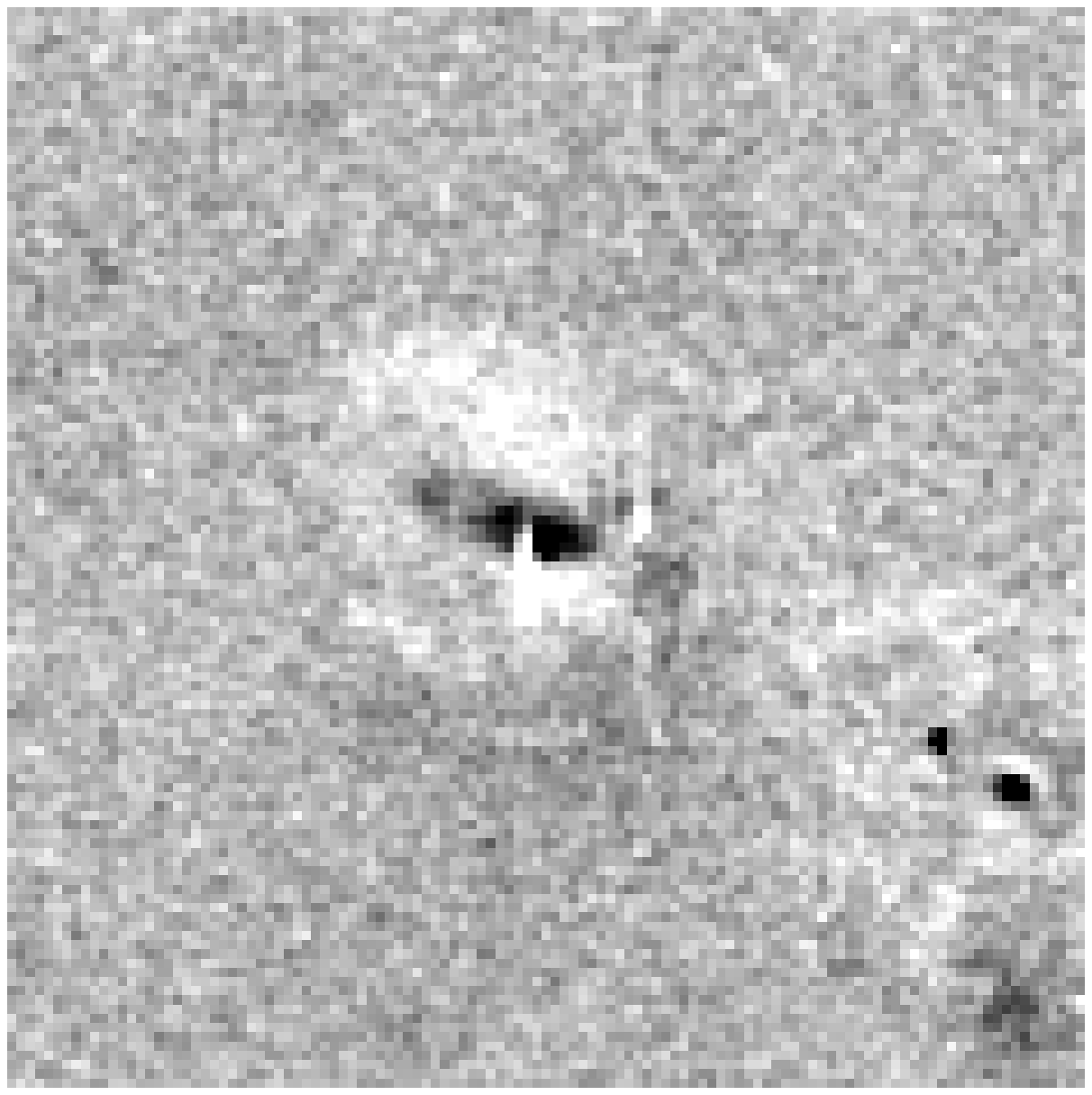}
\includegraphics{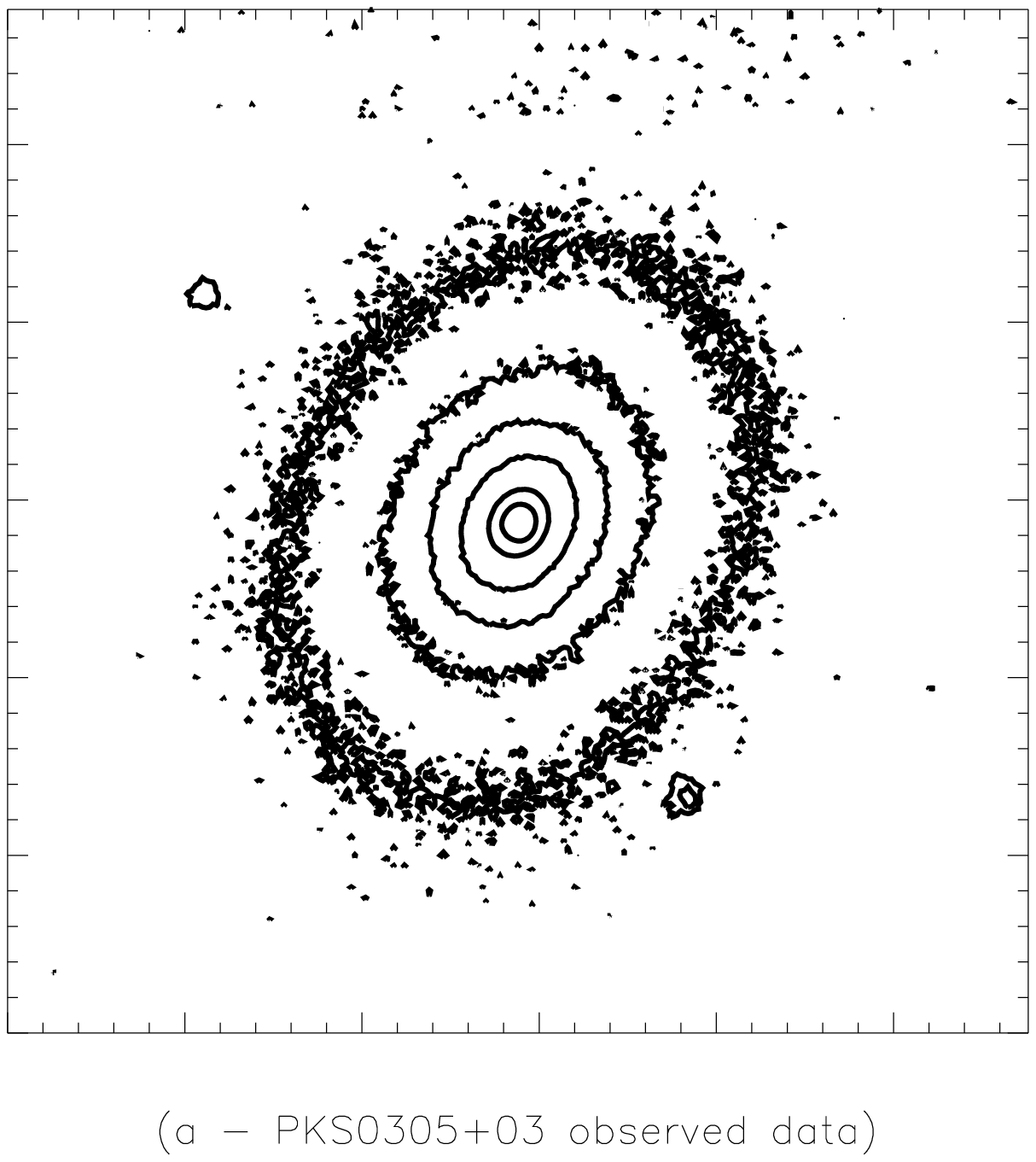}
\includegraphics{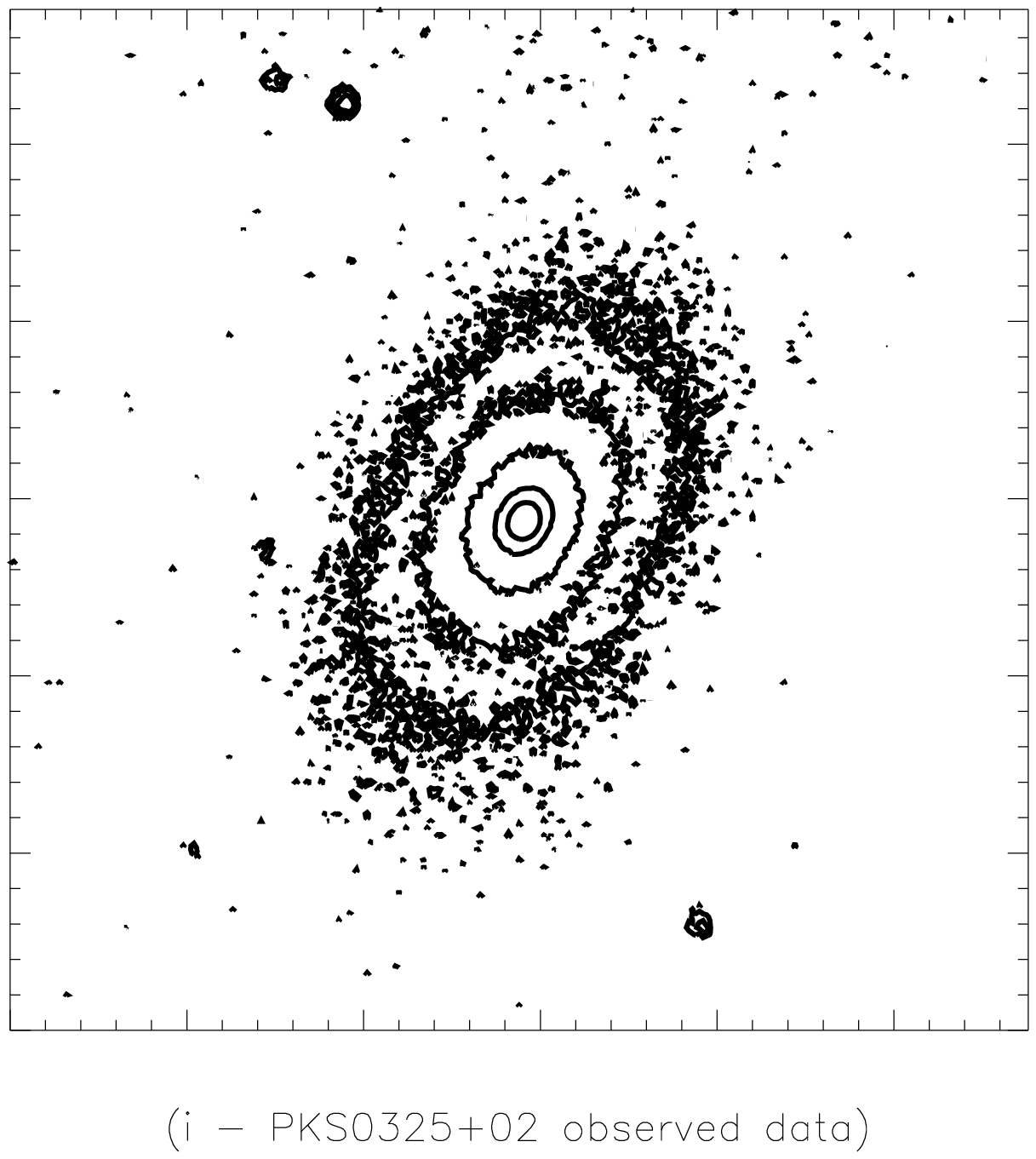}
\includegraphics{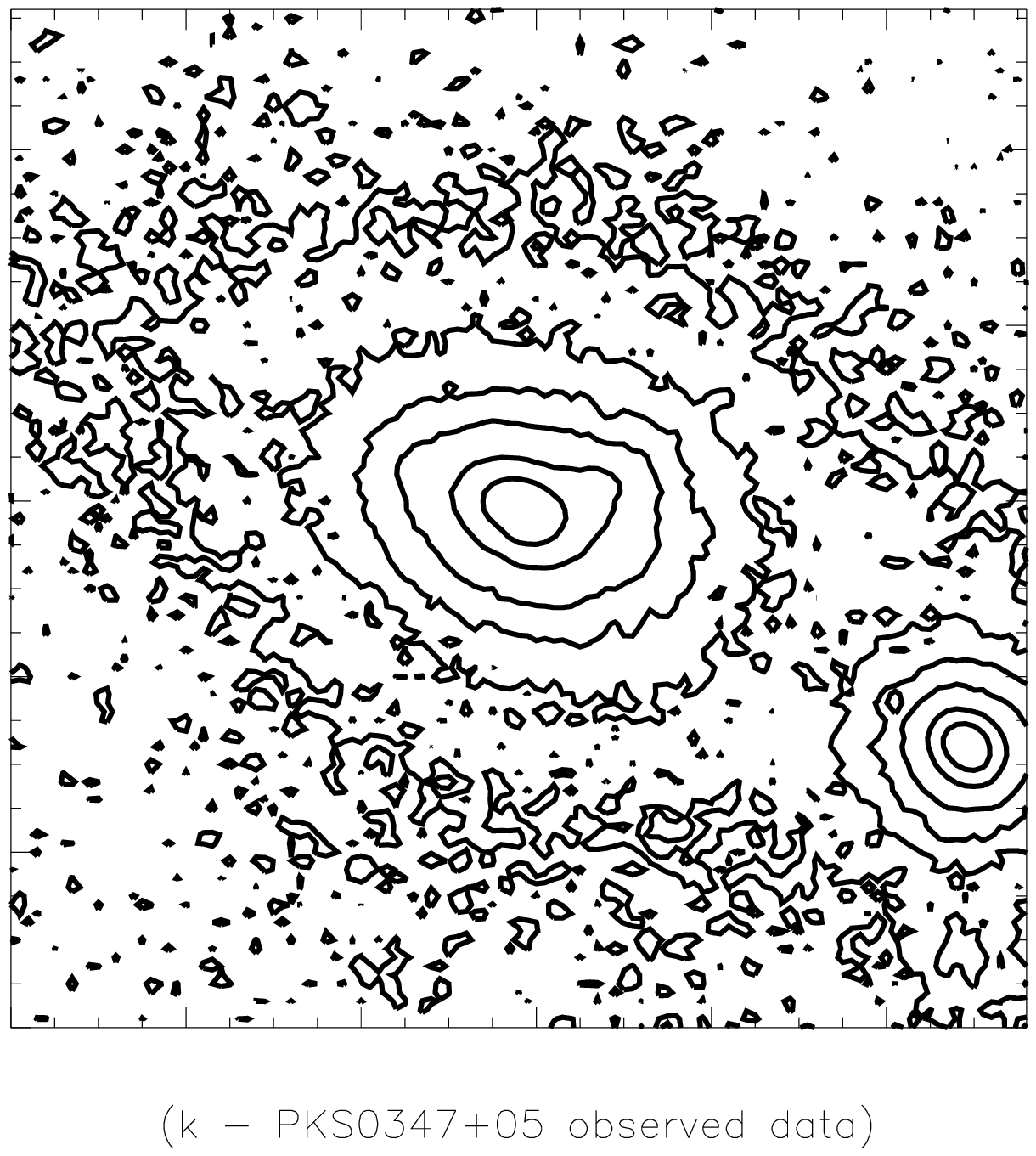}
\includegraphics{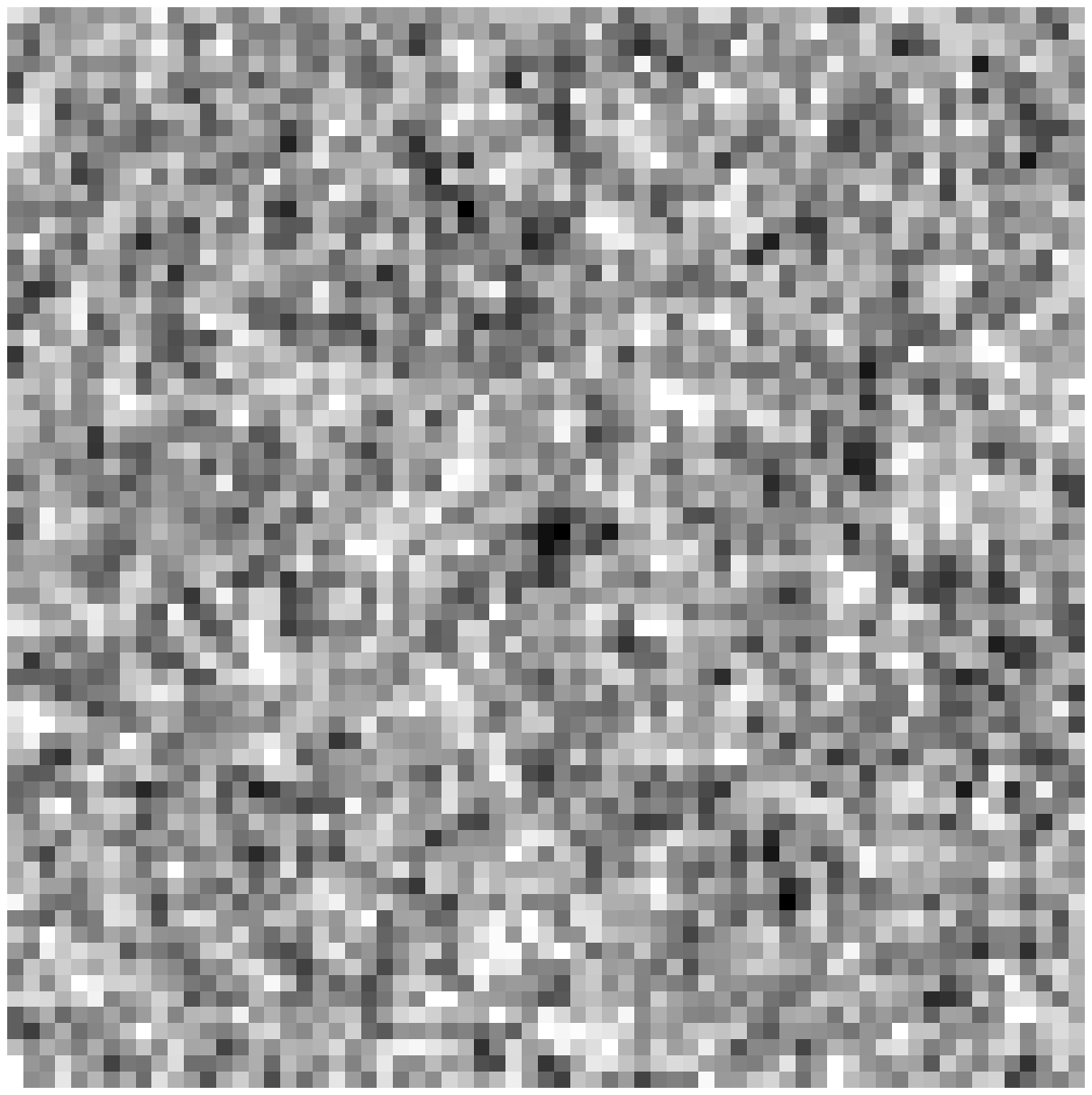}
\includegraphics{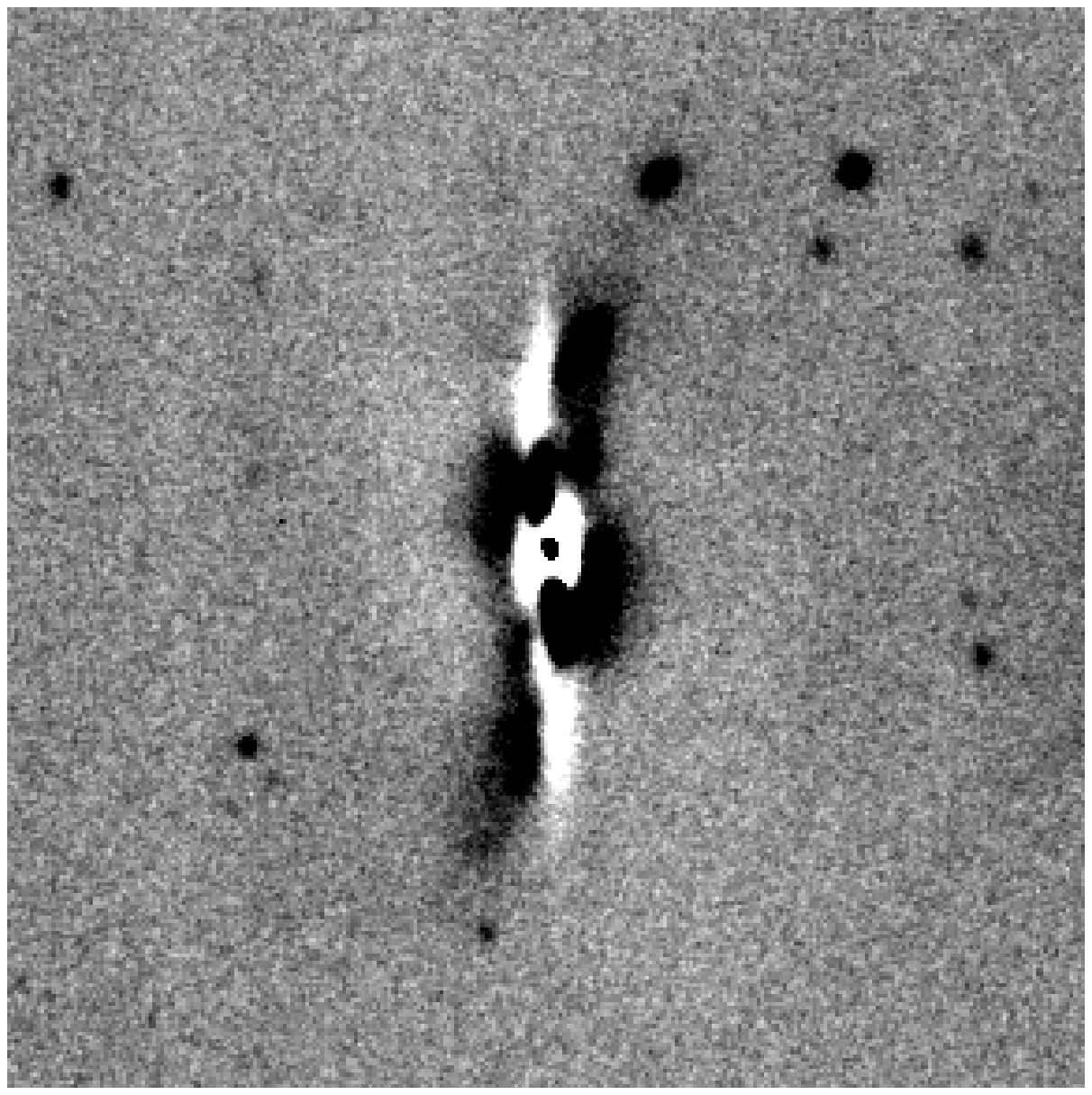}
\includegraphics{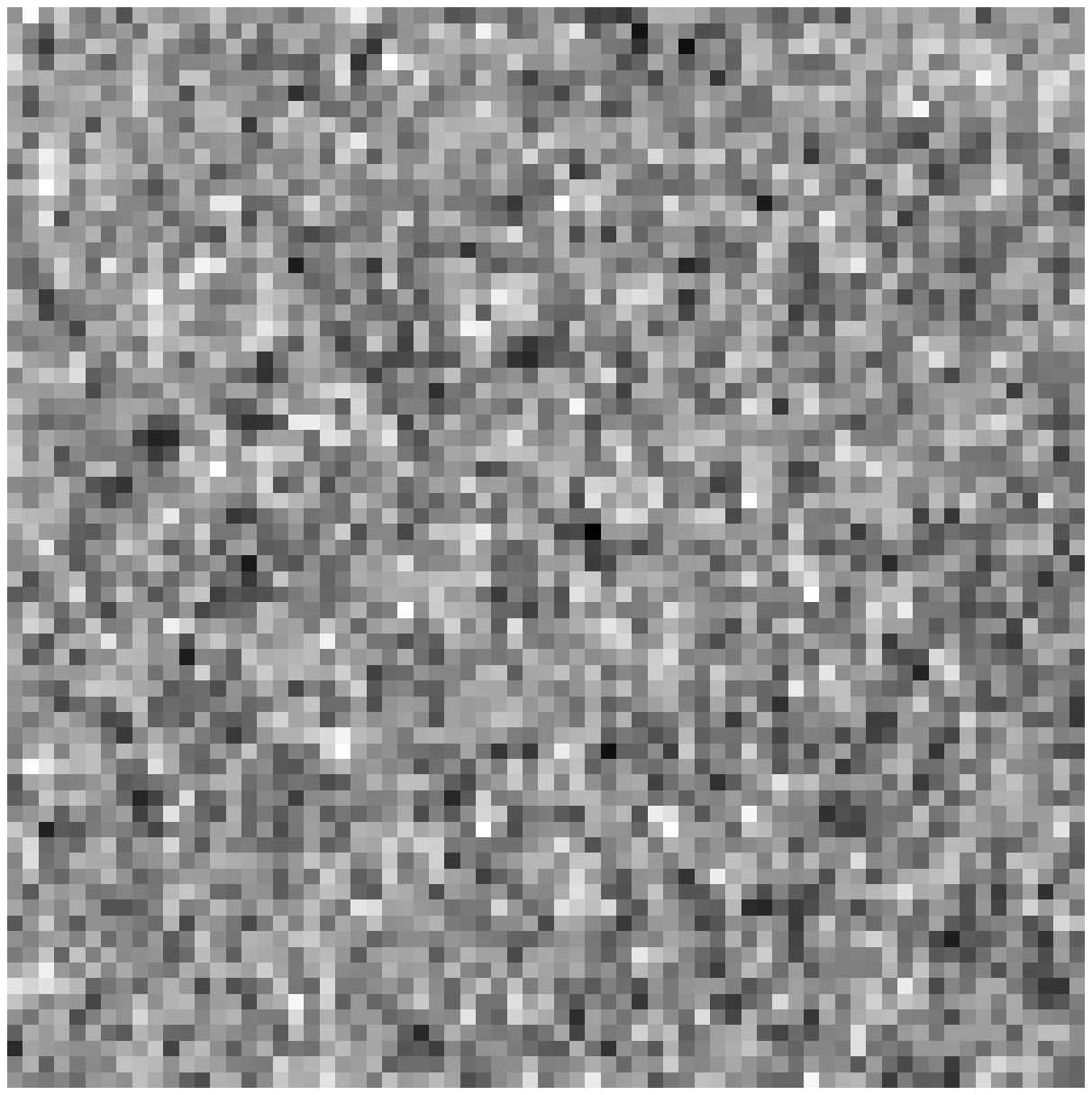}
\includegraphics{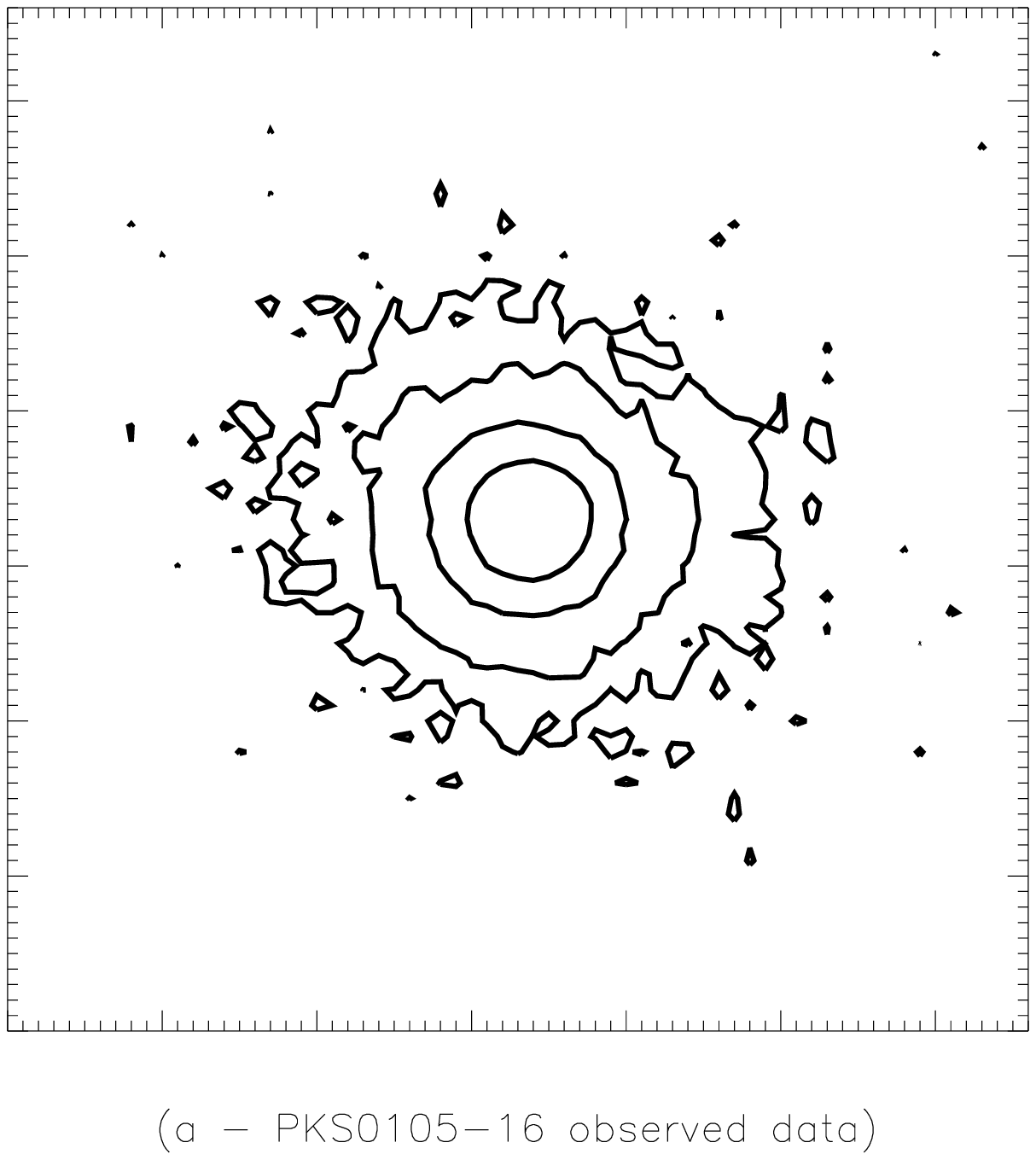}
\includegraphics{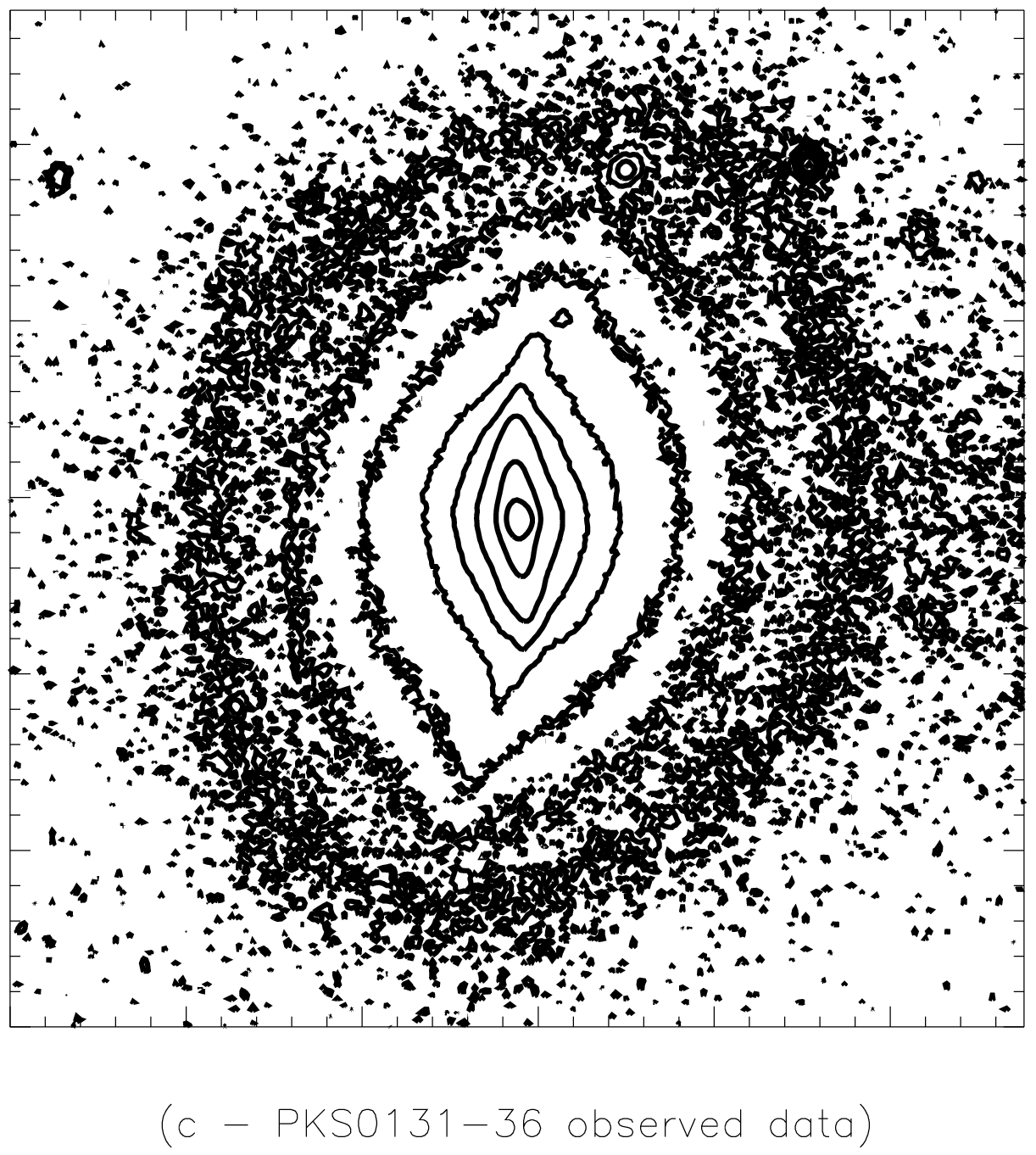}
\includegraphics{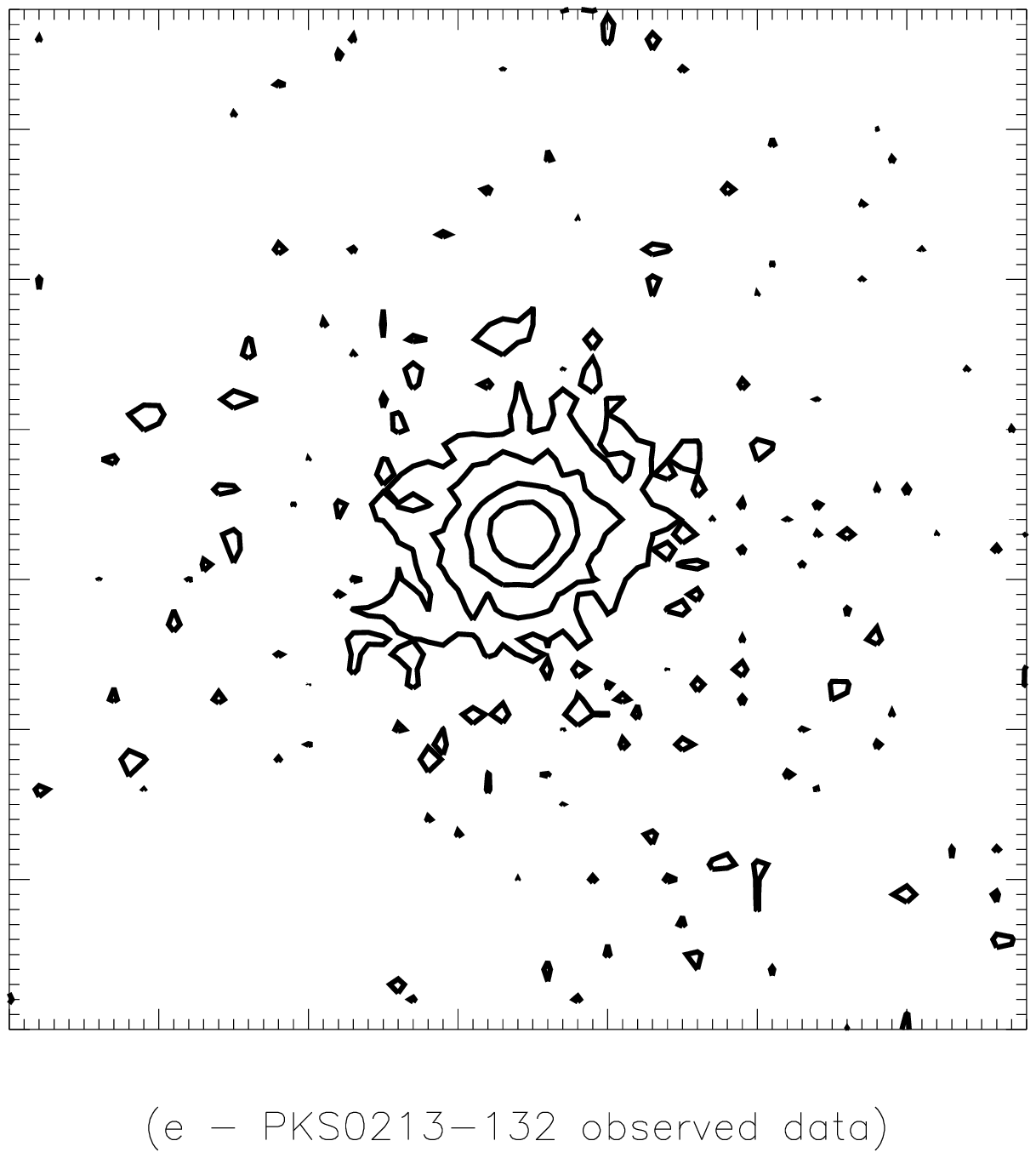}
\end{center}
\caption{50kpc by 50kpc images of PKS0105-16, PKS0131-36, PKS0213-132, PKS0305+03, PKS0325+02 and PKS0347+05. The observed data contours are displayed in frames (a), (c), (e), (g), (i) and (k), while frames (b), (d), (f), (h), (j) and (l) show the best-fit model contours on greyscale images of the model-subtracted residuals.  The maximum contour level is 50\% of the peak flux for that source in all cases, with subsequent contours at 25\%, 10\%, 5\%, 2.5\%, 1\%, 0.5\% and 0.25\% (latter flux levels not shown in all cases). The minimum contours displayed are at 0.25\% for PKS0131-36, 1\% for PKS0305+03 and PKS0347+05, 2.5\% for PKS0325+02, and 5\% for PKS0105-16 and PKS0213-132. 
\label{Fig: 3}}
\end{figure*}

\subsubsection{PKS0105-16 (3C32)}
PKS0105-16 is well fit by a de Vaucouleurs elliptical with $r_{eff} = 6.5$kpc and a $K-$band nuclear point source contribution of 32.6\%.

\subsubsection{PKS0131-36 (NGC0612)}
PKS0131-36 has been previously classified as an S0 galaxy (Kotanyi \& Ekers 1979, Sansom et al 1987), and more recently as being well described by a de Vaucouleurs elliptical model (Fasano et al 1996, V\'eron-Cetty \& V\'eron, 2001).  It has a luminous near edge-on disk, which has a gaseous counterpart (Emonts et al 2008), and a strong dust ring or lane.  Our modelling agrees with the latter elliptical classification, and our derived effective radius after inclusion of an edge-on disk component in our fit is $r_{eff}=20.2^{\prime\prime}$/12kpc, broadly comparable to the values obtained in the $i-$band ($17.7^{\prime\prime}$, V\'eron-Cetty \& V\'eron) and also in the $R-$band ($16.72^{\prime\prime}$ and $16.7^{\prime\prime}$ by Fasano et al (1996) and Govoni et al (2000) respectively), although our disk-free model provides a closer match to the literature values with a value of  $r_{eff}=16.4^{\prime\prime}$/9.7kpc.

\subsubsection{PKS0213-132 (3C62)}
For this source, our best fit model is a de Vaucouleurs host galaxy with an effective radius of $\sim 11$kpc ($4.3^{\prime\prime}$) and a nuclear point source contributing 10\% of the total flux. This compares favourably with the value of 13.7kpc obtained by Smith \& Heckmann (1989) in the $V$-band.

\subsubsection{PKS0305+03 (3C78)}

Considering first the de Vaucouleurs elliptical model, we find a value of $R_{eff} \sim 11.7$kpc (19.7$^{\prime\prime}$) for this galaxy with a nuclear point source contribution of 1\% of the total flux. 
After conversion to the same cosmological model, Smith \& Heckman (1989) derive a best-fit effective radius (assuming a de Vaucouleurs $r^{1/4}$ profile) of $r_{eff} = 9.0$kpc in the V-band, smaller than our K-band value.  Colina \& de Juan (1995) find a similar result in the $r-band$ for a de Vaucoueurs-only model, but also note the presence of light excesses above the $r^{1/4}$ profile fit. It is of note that our model residuals clearly display ring-type features at large radii, and a slight over-subtraction of light just within this radius, which most likely explain the differences in our derived values. However, a better fit is obtained with a S\'ersic index of $n=6$ and a resulting effective radius of $R_{eff} \sim 22$kpc, and no additional point source emission. The quality of the fit is improved on further with the Donzelli-style bulge+disk model, which has a bulge effective radius of $R_{eff} \sim 13$kpc and a small central disk with $R_{eff} \sim 1$kpc accounting for $\sim 2$\% of the total galaxy flux.

Sparks et al (2000) observed an optical synchrotron jet and dust disk for this source; we see no clear sign of these features in the K-band.

\subsubsection{PKS0325+02 (3C88)}
De Vaucouleurs modelling of PKS0325+02 in the optical derived effective radii of $\sim 30^{\prime\prime}$ (Smith \& Heckman 1989, Govoni et al 2000), comparable to our measured values of $\sim 28-34^{\prime\prime}$ without and with a point source contribution respectively. However, we find that a higher-order S\'ersic index of $n=6$ with a minimal point source contribution and a very large $r_{eff} \sim 82^{\prime\prime}$ provides a better fit to the data.
The residuals of our modelling of PKS0325+02 suggest the presence of a major-axis dust lane. This feature lies along the same position angle as the dust-feature observed by de Koff et al (2000), and could potentially have skewed our fitting towards the large galaxy sizes obtained.

In a different study, Donzelli et al (2007) model this source as part of their analysis of the NICMOS snapshot survey, fitting both a bulge and a disk-like component to this galaxy, and obtaining a significantly smaller bulge effective radius of only $\sim 5$kpc. As our single S\'ersic component model residuals also display excess flux at large radii, we also carry out a fit with two S\'ersic components, one with $n=4$ and another with $n=1$.  For this fit, the bulge component has a value of $r_{eff} \sim 12.5^{\prime\prime}$ or $\sim 7$kpc, smaller than that of our previous models but still slightly larger than that of Donzelli et al.  The magnitude ratio of the two components of our model is close to that of Donzelli et al, despite the disk-component radius being larger.

\subsubsection{PKS0347+05 (4C+05.16)}

PKS0347+05 is hosted by an interacting galaxy, which is neighboured by a QSO/Seyfert 1 to the southwest.  Both the host galaxy and the secondary object with which it is interacting can be well modelled with S\'ersic profiles with $n=4$, while we determine a good fit for the quasar with $n=6$ and a $\sim$20\% nuclear point source. Our derived effective radius for the radio source host galaxy is $\sim 8.5kpc$ ($\sim 1.8^{\prime\prime}$).

\subsubsection{PKS0349-27}

De Vaucouleurs modelling of this source gives a best-fit effective radius of $\sim 6.7$kpc (5.3$^{\prime\prime}$) and a point source contribution of $\sim 11$\% of the total flux.
Previous studies at optical wavelengths have derived larger effective radii: Fasano et al (1996) and Govoni et al (2000) find $r_{eff} = 10-11^{\prime\prime}$ in the $R-$band, while the de Vaucouleurs-only modelling of Zirbel (1996) finds $r_{eff} \sim 12^{\prime\prime}$ in the $V$-band. However, this is perhaps due to the known presence of extensive diffuse ionized emitting gas surrounding this system (Danziger et al 1984). We note that we also observe faint extended emission in our model-subtracted residual image, which may be tracing the same emitting regions.

Our best-fit model for PKS0349-27 has a S\'ersic index of $n=6$ and an effective radius of $\sim 10$kpc, with a nuclear point source contribution of 7.0\%.  As well as having a smaller reduced $\chi^2$ value, this model also provides a significantly better fit to the host galaxy with minimal residuals at both large and small radii.

\subsubsection{PKS0404+03 (3C105)}
PKS0404+03 has been previously studied in the optical by Zirbel (1996), and in the infrared by Donzelli et al (2007) and Tremblay et al (2007). We derive a best fit model of a de Vaucouleurs host galaxy with an effective radius of $r_{eff} = 4.4$kpc ($2.7^{\prime\prime}$) and a nuclear point source accounting for $\sim6$\% of the total flux, cf. the values of $r_{eff} = 2kpc$ (Zirbel) and $r_{eff} = 1.2kpc$ for the infrared bulge+disk model (Donzelli et al).   Our bulge+disk+point source model finds $r_{eff} \sim 2$kpc for the bulge component, and is both numerically and visually a better fit than a pure bulge host galaxy. While the bulge radius is comparable, our disk-component radius is larger than that of Donzelli et al (though this is at least in part a consequence of their slightly smaller S\'ersic index of $n=3.17$), and we find that, while close in flux, the bulge-component is the brighter of the two.

\begin{figure*}
\vspace{8.45 in}
\begin{center}
\includegraphics{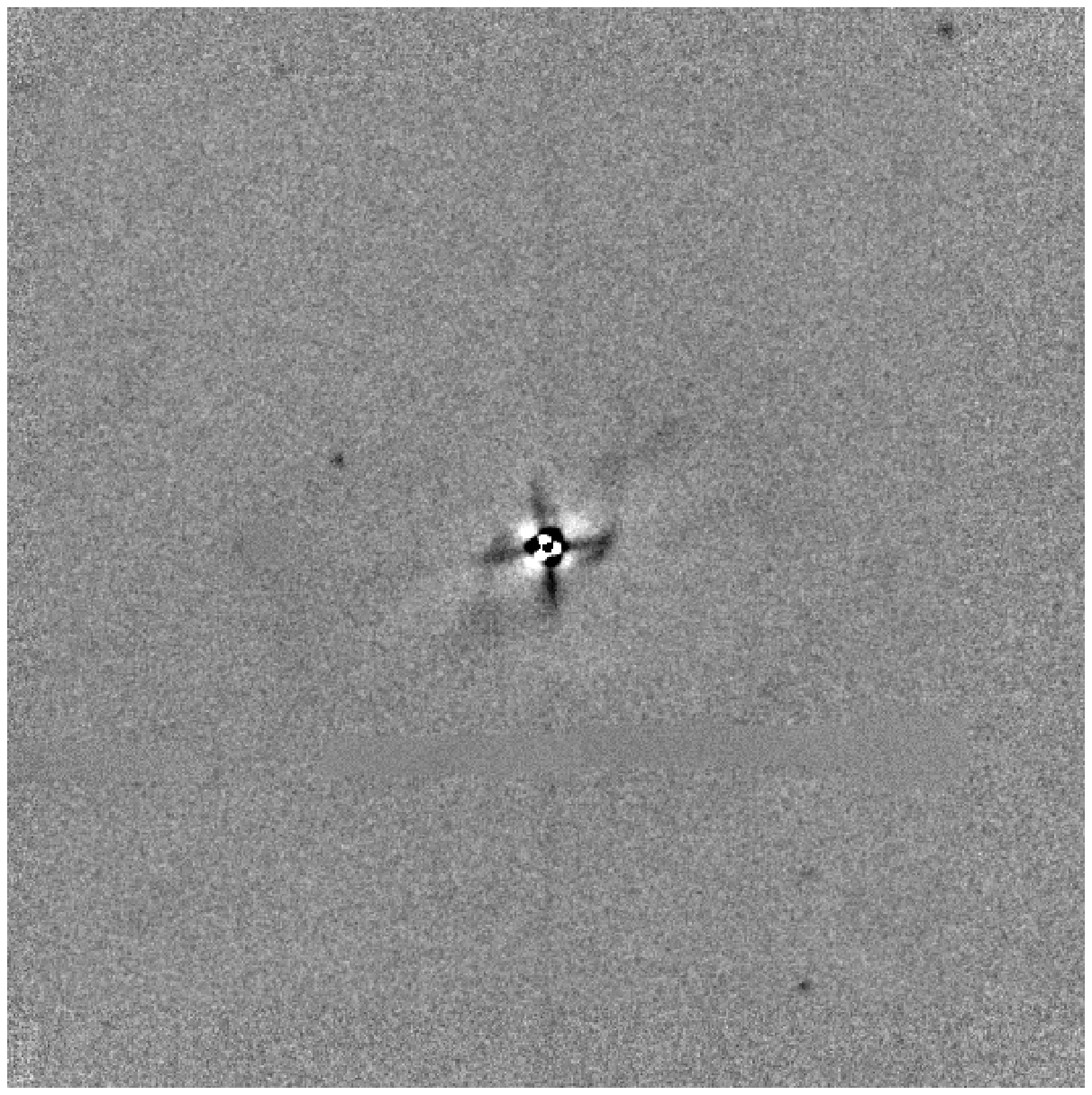}
\includegraphics{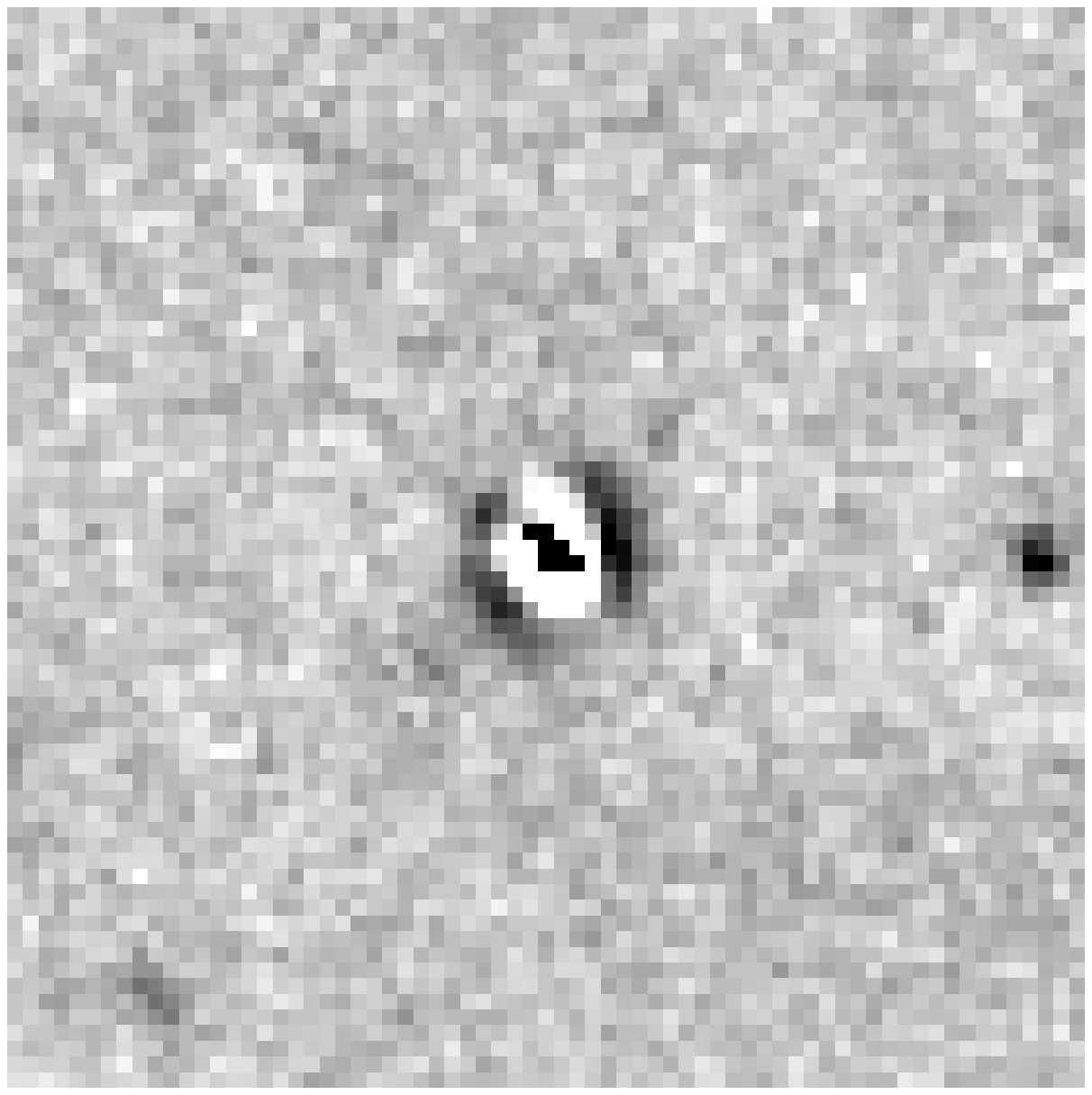}
\includegraphics{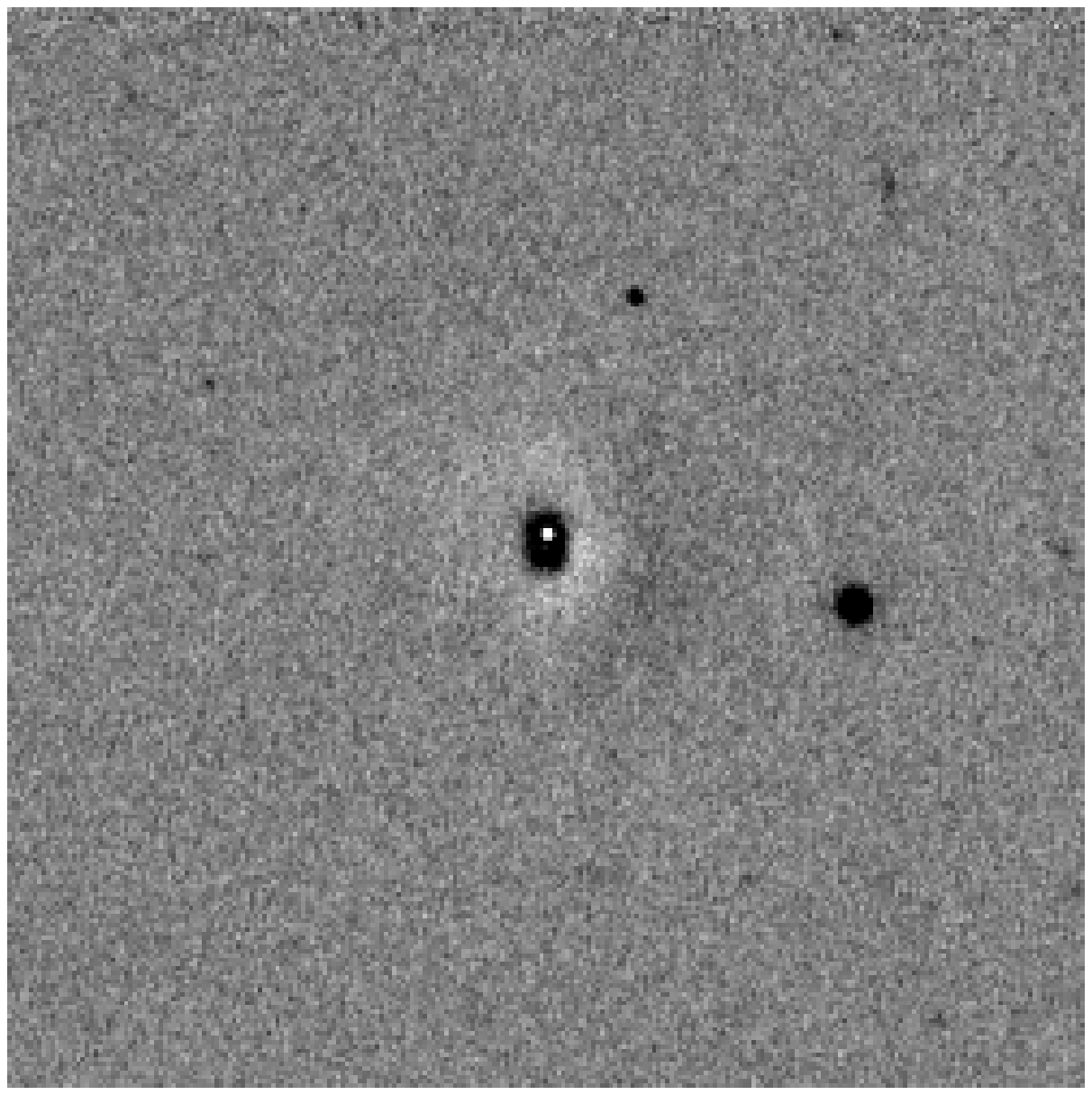}
\includegraphics{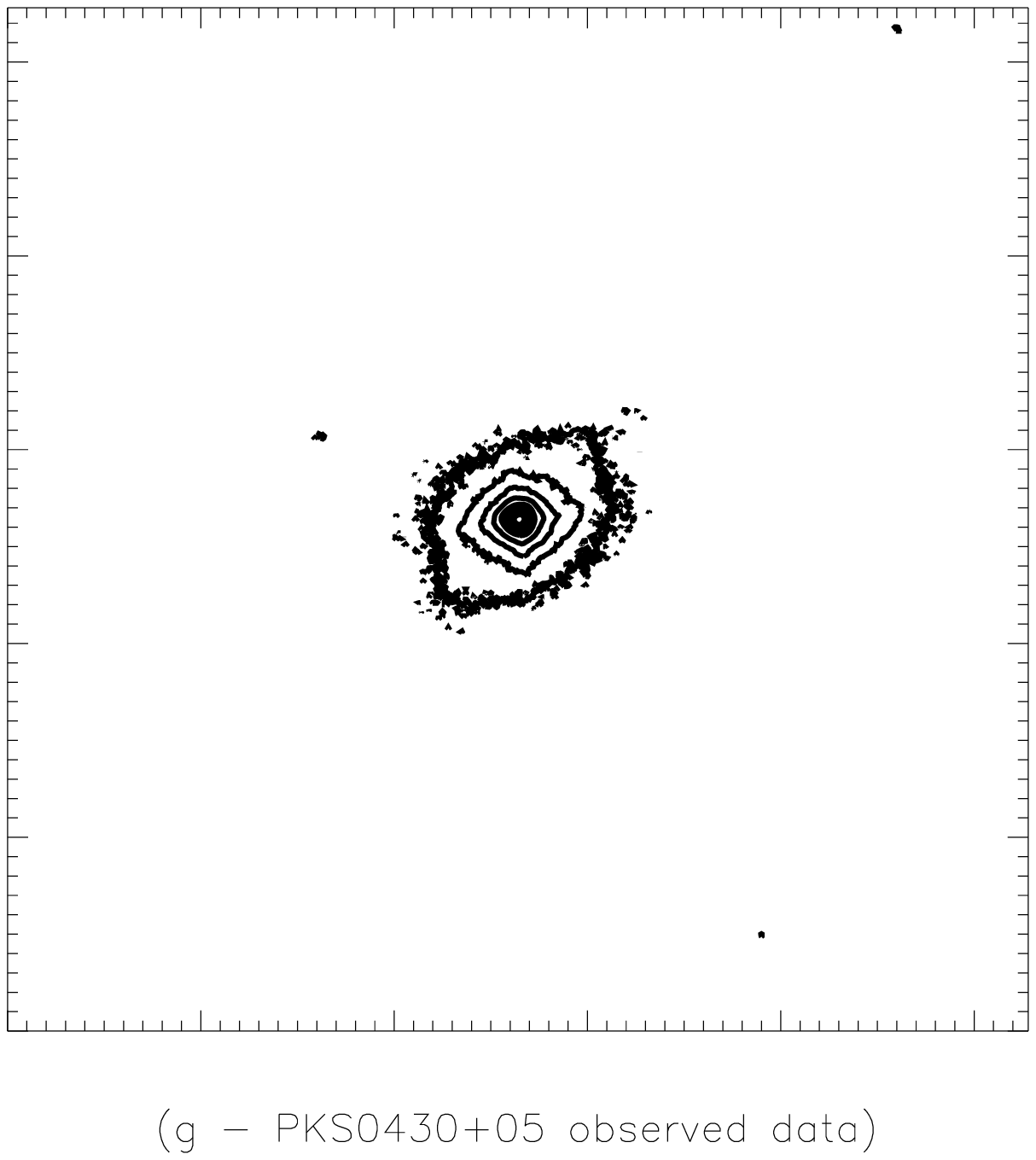}
\includegraphics{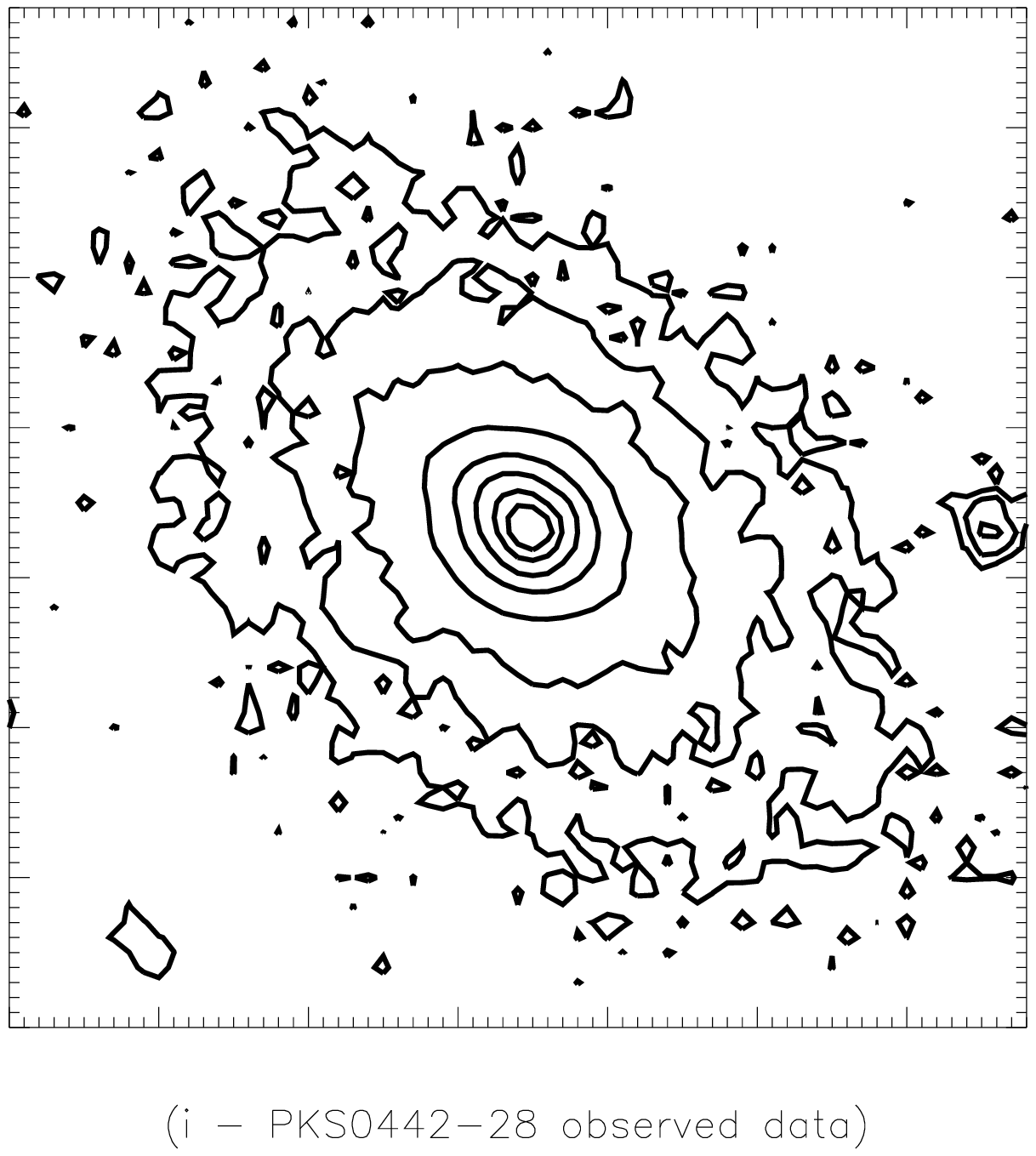}
\includegraphics{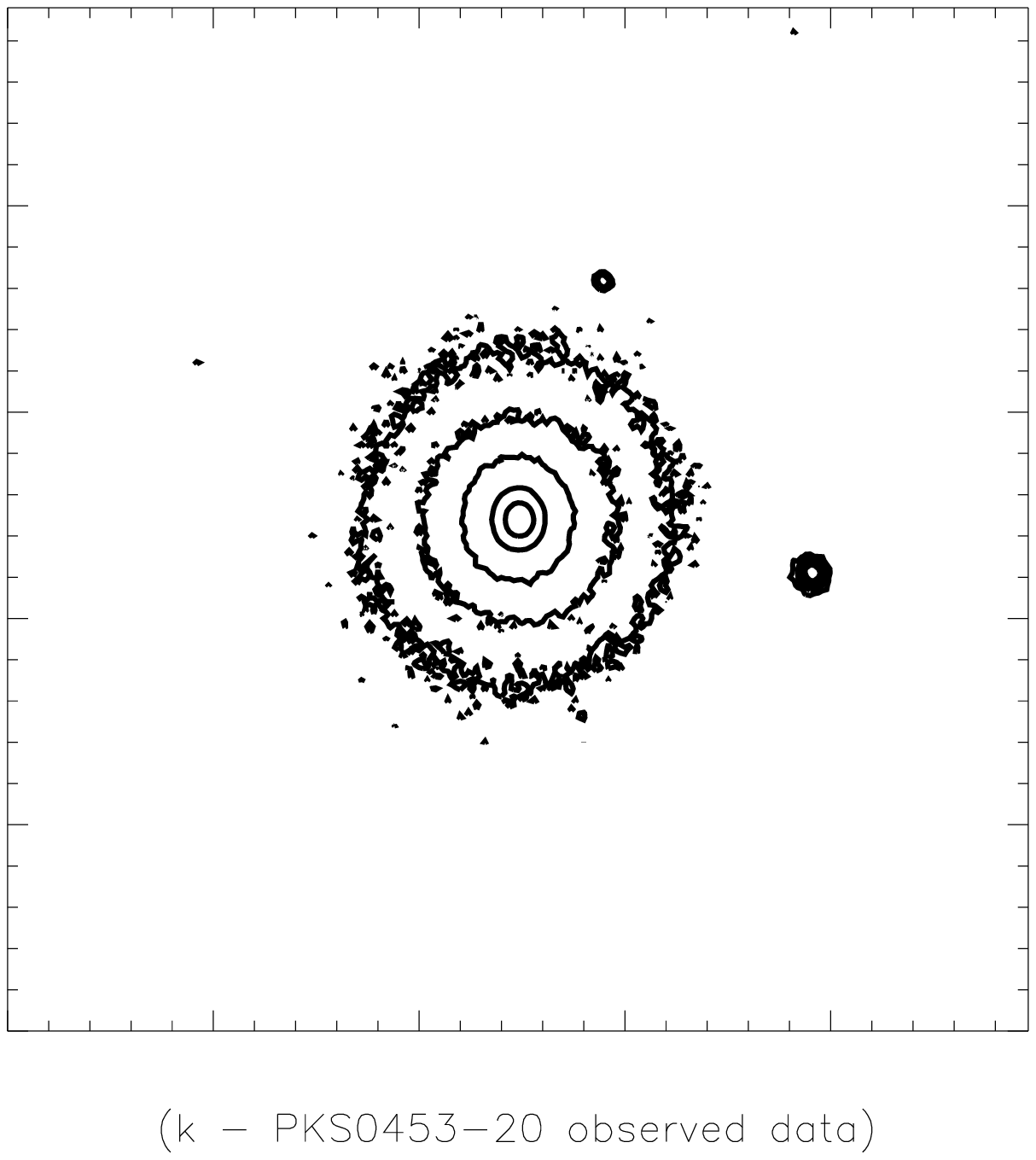}
\includegraphics{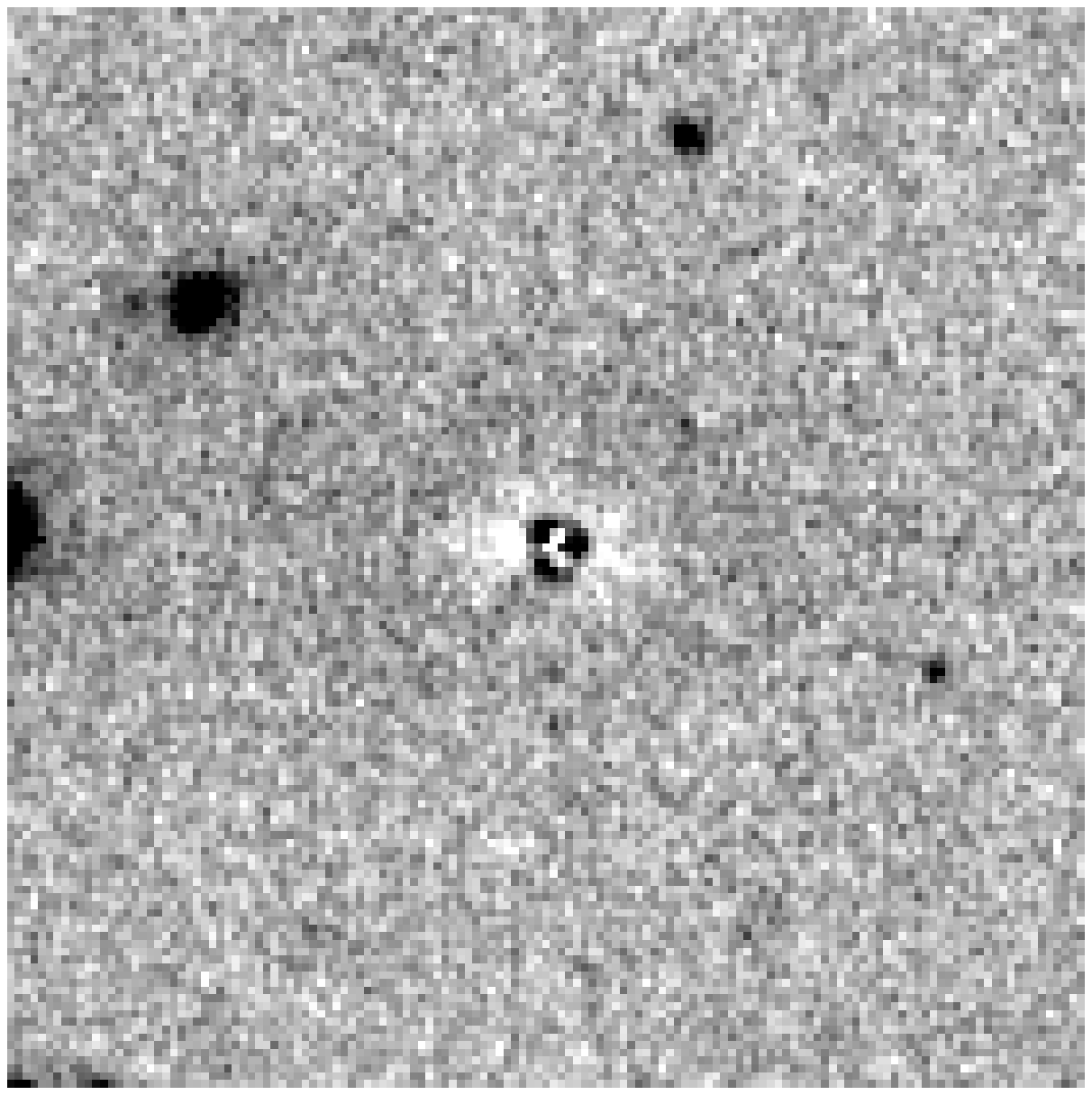}
\includegraphics{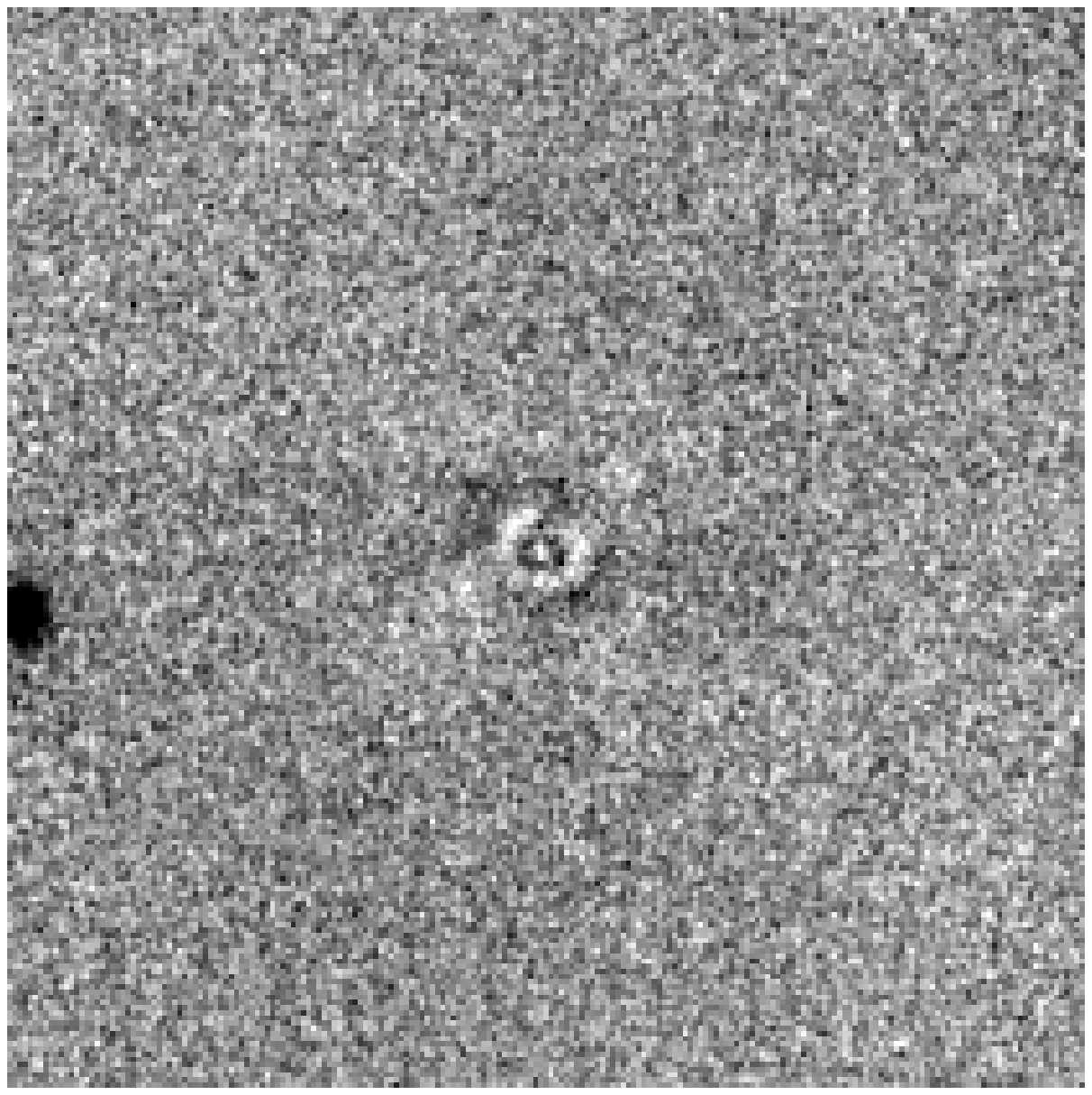}
\includegraphics{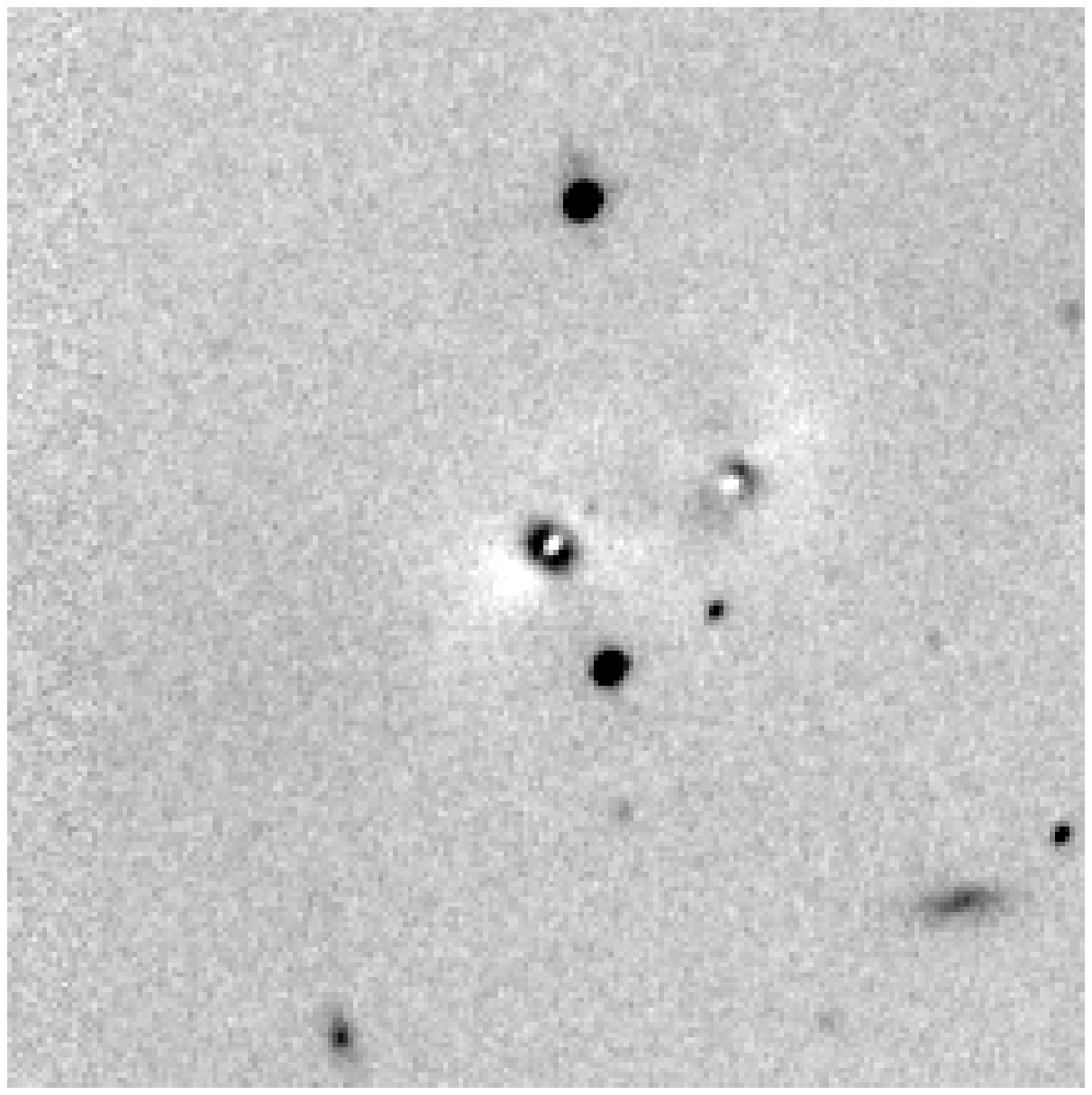}
\includegraphics{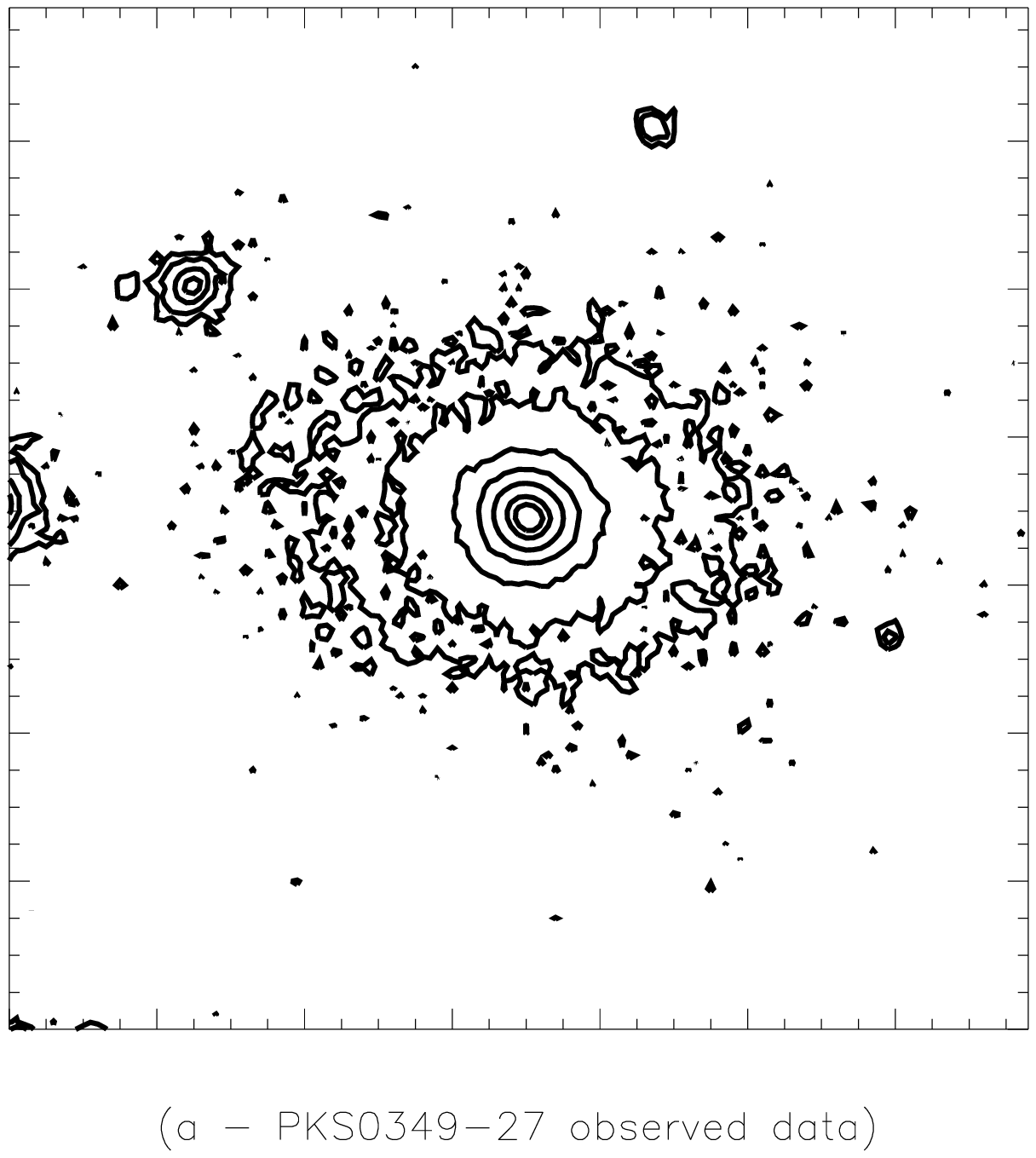}
\includegraphics{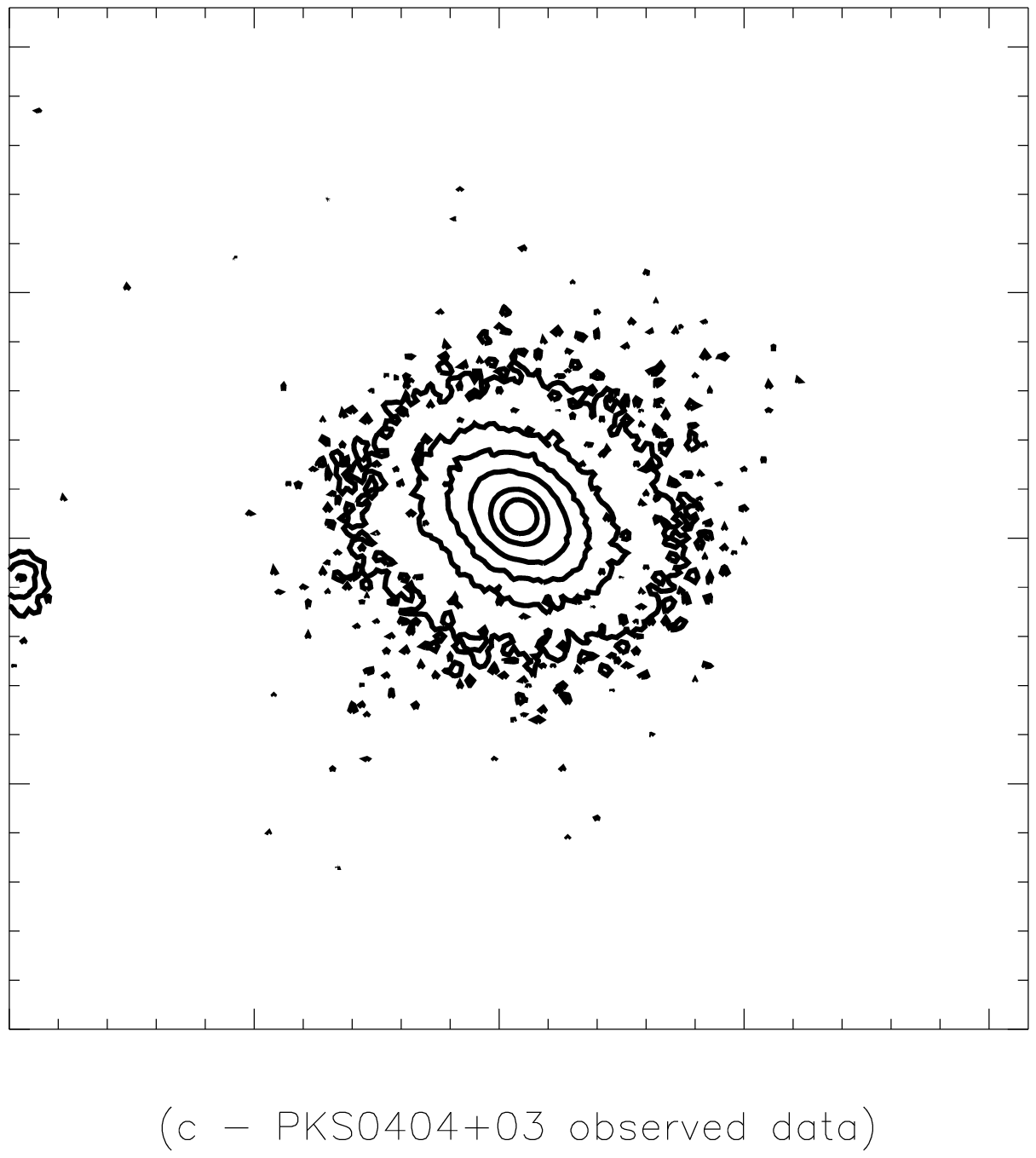}
\includegraphics{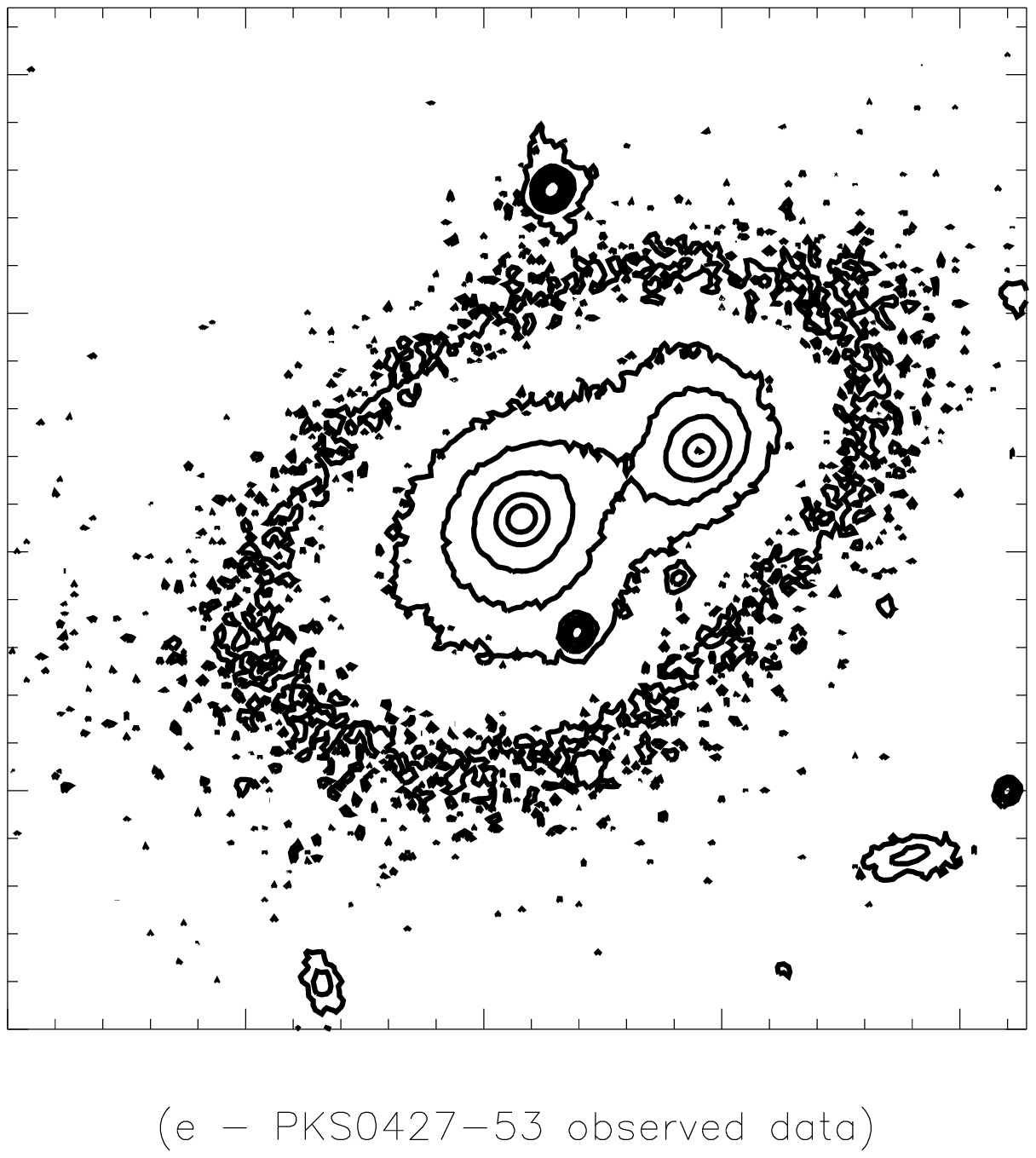}
\end{center}
\caption{50kpc by 50kpc images of PKS0349-27, PKS0404+03, PKS0427-53, PKS0430+05, PKS0442-28 and PKS0453-20. The observed data contours are displayed in frames (a), (c), (e), (g), (i) and (k), while frames (b), (d), (f), (h), (j) and (l) show the best-fit model contours on greyscale images of the model-subtracted residuals.  The maximum contour level is 50\% of the peak flux for that source in all cases, with subsequent contours at 25\%, 10\%, 5\%, 2.5\%, 1\%, 0.5\%, 0.25\% and 0.1\% (latter flux levels not shown in all cases). The minimum contours displayed are at 0.1\% for PKS0430+05, 0.25\% for PKS0442-28, 0.5\% for PKS0349-27, 1\% for PKS0404+03 and PKS0427-53, and 2.5\% for PKS0453-20. 
\label{Fig: 4}}
\end{figure*}

\subsubsection{PKS0427-53 (IC2082)}

The radio source PKS0427-53 is hosted by the eastern component of a dumbbell galaxy system (Carter et al 1989). We model both sources with the inclusion of a fourier mode to account for the shear in the flux distribution caused by the strong interaction between these two galaxies.  Our best fit models do not include any nuclear point source contributions. For the radio source host galaxy, with a S\'ersic index of $4$ we find a best-fit $r_{eff} \sim 9$kpc ($11.3^{\prime\prime}$).
These values are larger than those found by Govoni et al (2000) in the optical ($r_{eff} \sim 6$kpc for $n=4$).  Our model residuals show clear tidal distortions to both galaxies, and the amplitude of the fourier distortion (which accounts for the lopsidedness of the galaxy isophotes, without altering the derived effective radius) is roughly twice as large for the western component ($a=0.29$ cf. 0.16).

\subsubsection{PKS0430+05 (3C120)}

For PKS0430+05 the best de Vaucouleurs model has $r_{eff} = 4.2$kpc (6.5$^{\prime\prime}$) and a nuclear point source contribution of $\sim$39\%.
However, we find a better fit both numerically and visually with a Donzelli-style bulge+disk model (Donzelli et al 2007), which has a nuclear point source contribution of $\sim$33\%, a bulge component with an effective radius $R_{eff} \sim 9$kpc and a central disk component with $R_{eff} \sim 1$kpc accounting for $\sim 8$\% of the {\it total} galaxy flux.
The strong nuclear point source in PKS0430+05 is consistent with its status as a BLRG.  The R-band modelling of Govoni et al (2000) combined an $r^{1/4}$ profile with a nuclear point source and also a disk component, as does that in the infrared by Kotilainen et al (1992) and Kotilainen \& Ward (1994).  Modelled effective radii vary considerably between the different studies, but the K-band bulge fit of Kotilainen \& Ward (1994) is in good agreement with our own results.

Our residual flux image of this galaxy displays a narrow excess of flux along the major axis, which then extends into an s-shape to the NW and SE.  Previous observations of this source (Sargent 1967; Baldwin et al 1980; Garc\'{i}a-Lorenzo et al 2005) highlight the complexities of its light distribution which seem to match our own observed model residuals, and suggest that it has been involved in merger activity in the relativelty recent past.

\subsubsection{PKS0442-28}

This NLRG is well modelled as a de Vaucouleurs elliptical with an effective radius of 6$^{\prime\prime}$/15kpc and an unresolved nuclear point source accounting for 23.5\% of the total flux.

\subsubsection{PKS0453-20 (NGC1692)}

Unusually for a WLRG, Wills et al. (2004) find evidence for young stellar populations in this source. On the western side of this galaxy, the residual flux image shows excess flux beyond the slightly oversubtracted region surrounding the nucleus, suggestive of some underlying asymmetry in the host galaxy.  Our de Vaucouleurs model does not require a strong nuclear point source contribution, and our best-fit effective radius of $\sim 16$kpc ($\sim 24^{\prime\prime}$) is comparable to that found by Govoni et al (2000) in the $R-$band.

\subsubsection{PKS0518-45 (Pictor A)}

PKS0518-45 is a well known southern radio galaxy classifed as a BLRG.
Zirbel (1996) measured a de Vaucouleurs profile effective radius of only 0.84kpc for this source (after correction to our assumed cosmological model, and without including a nuclear point source component); however we were unable to produce a reliable fit for PKS0518-45 using a pure de Vaucouleurs model. Adding a point source contribution, we obtain $r_{eff} = 7.3$kpc ($10.6^{\prime\prime}$) with a S\'ersic index of $n=4$, and a strong point source contribution of 41.3\%. However, our best fit model is obtained with a more disky $n=2$, $r_{eff} = 4.5$kpc ($6.6^{\prime\prime}$) and a nuclear point source accounting for 50\% of the total observed flux.  This is consistent with its status as a BLRG.

\subsubsection{PKS0521-36 (ESO362-G021)}

PKS0521-36 is a well-known BL Lac/BLRG, and our modelling requires roughly equal flux contributions from the host galaxy and the unresolved nuclear point source. For a de Vaucouleurs model, our derived effective radius (4.3$^{\prime\prime}$/4.5kpc) is slightly larger that measured in the R-band (see Scarpa et al 2000, Urry et al 2000, Falomo et al 2000) and $V-$band (Zirbel 1996).  In the infrared, Cheung et al (2003) also find a smaller effective radius, at the expense of a weaker nuclear point source contribution, though their observations are at a lower signal-to-noise than our own.  However, we obtain a better fit to the observational data with a S\'ersic index of $n=2$, a point source contribution of 56\% and an effective radius of 3.5$^{\prime\prime}$/3.7kpc.  A Donzelli-style (Donzelli et al 2007) bulge+disk combination imrpoves on this further, and has a nuclear point source contribution of $49$\% of the total flux, a bulge effective radius $R_{eff} \sim 12$kpc and a disk component with $R_{eff} = 2.4$kpc contributing 14\% of the {\it total} galaxy flux.

The radio jet in this source has previously been detected at optical wavelengths (Danziger et al 1979, Cayatte \& Sol 1987, Scarpa et al 1999); it is clear from our data that this jet is visible in the K-band, as was also noted by Cheung et al.

\subsubsection{PKS0625-53 (ESO161-IG007)}

The radio source PKS0625-53 is hosted by the eastern component of this dumbbell galaxy system.  Our modelled effective radius ($r_{eff} \sim 24^{\prime\prime}$, or $\sim$24.5kpc) is not dissimilar to that found by Govoni et al (2000) in the optical ($r_{eff} \sim 33^{\prime\prime}$); the discrepancy is most likely due to the excess flux in the tidal feature linking the two galaxies.  We find that both galaxy components are well described by $r^{1/4}$ law profiles with the addition of a fourier-component to account for the lopsided flux distribution caused by the interaction (note that the addition of this fourier component to the modelling does not alter the mean effective radius derived), with no strong nuclear point source component in either object.

\subsubsection{PKS0625-35}

Wills et al. (2004) argue that PKS0625-35 is a BL Lac, so the apparent presence of a one-sided jet in our residual image confirms this classification. Our best fit de Vaucouleurs model includes a nuclear point source, and while the derived value of 7.3\% might seem rather small for such an object, this value is relative to the {\it total} flux of the galaxy, which is itself relatively bright and extended.  Indeed, the point source component is clearly dominant in the nuclear regions, where featureless continuum emission is a major component in the optical spectrum of this object (Wills et al 2004). Our modelled  effective radius (9.5$^{\prime\prime}$/10kpc) is consistant with that of the previous analysis of this source in the optical (Govoni et al 2000), and the observed jet feature is aligned with that observed in the radio by Venturi et al (2000).  We also model this source with a bulge+disk combination, finding a comparable $\sim 7$\% nuclear point source contribution, a larger bulge effective radius $R_{eff} \sim 20$kpc and a central disk-component with $R_{eff} \sim 2$kpc contributing $\sim 9$\% of the {\it total} galaxy flux.

\subsubsection{PKS0806-10 (3C195)}

PKS0806-10 can be well modelled as a de Vaucouleurs elliptical with a S\'ersic index of $n=4$, an effective radius of $9.5$kpc ($4.8^{\prime\prime}$) and a nuclear point source contribution of 20.5\%.

The host galaxy has a clear north-south aligned extension, and a broad arc extending from the northern tip of this extension around to the west. It is possible that the host galaxy is interacting with a smaller extended companion object to the east.  Previous modelling of this source in the optical (Govoni et al 2000) has found a larger effective radius (6.8$^{\prime\prime}$) than the value we derive.

\subsubsection{PKS0859-25}
This source is well modelled by a de Vaucouleurs elliptical with an effective radius of $r_{eff} = 6$kpc ($1.4^{\prime\prime}$). Fits also including a point source contribution are numerically slightly better, but have a larger effective radii (12kpc, $2.7^{\prime\prime}$), most likely due to other objects and noise in the relatively crowded surrounding field.

\begin{figure*}
\vspace{8.45 in}
\begin{center}
\includegraphics{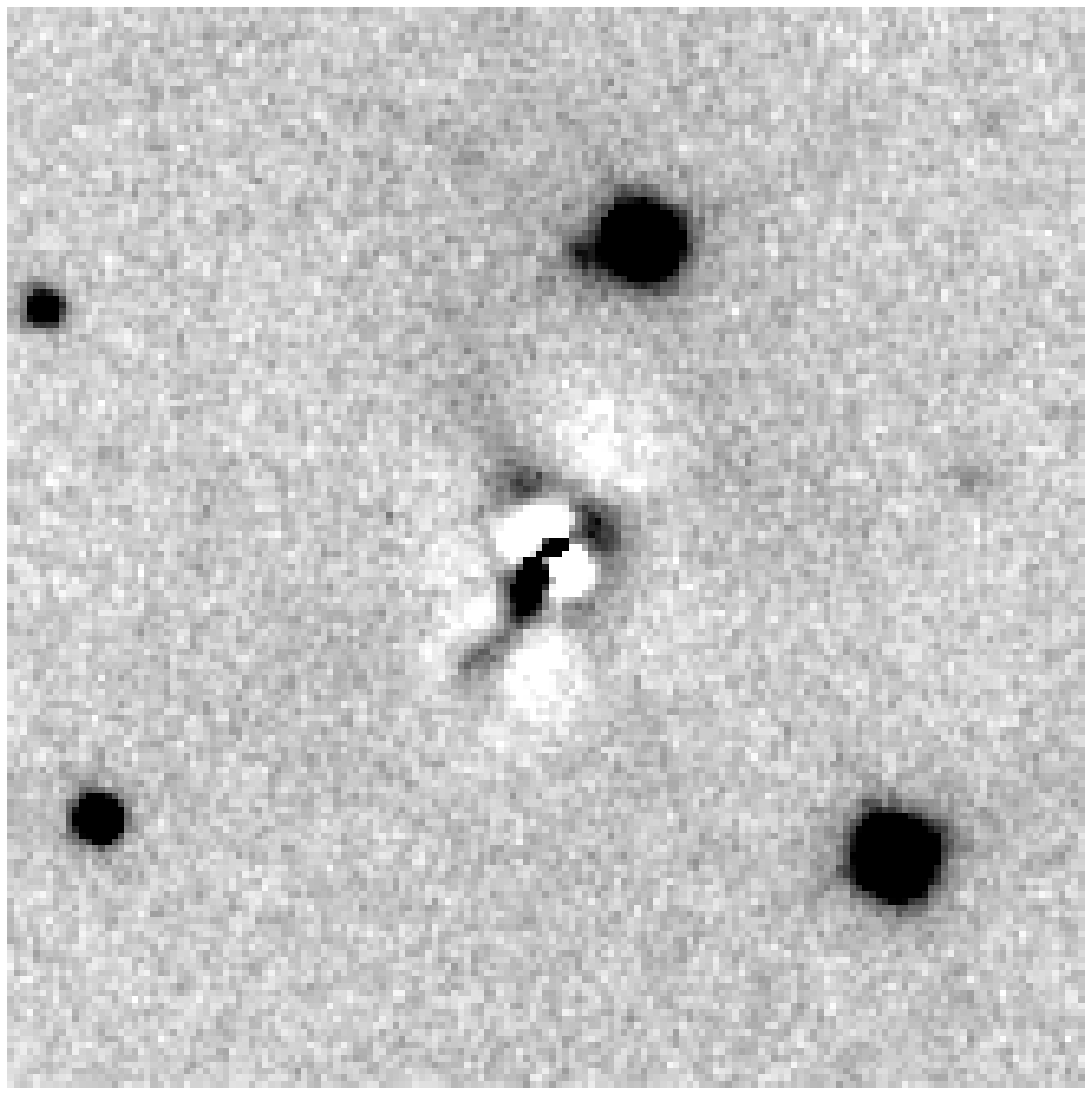}
\includegraphics{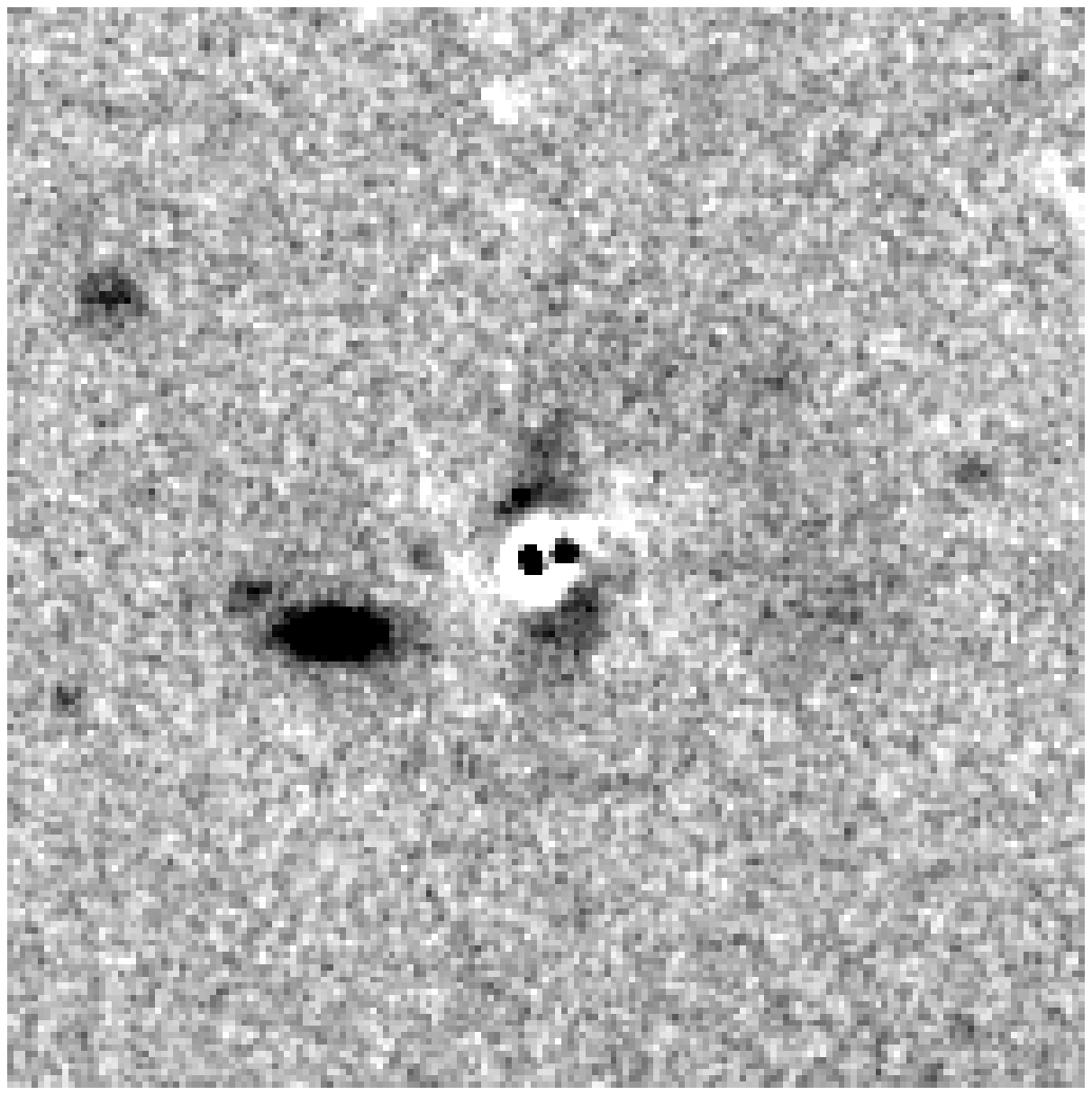}
\includegraphics{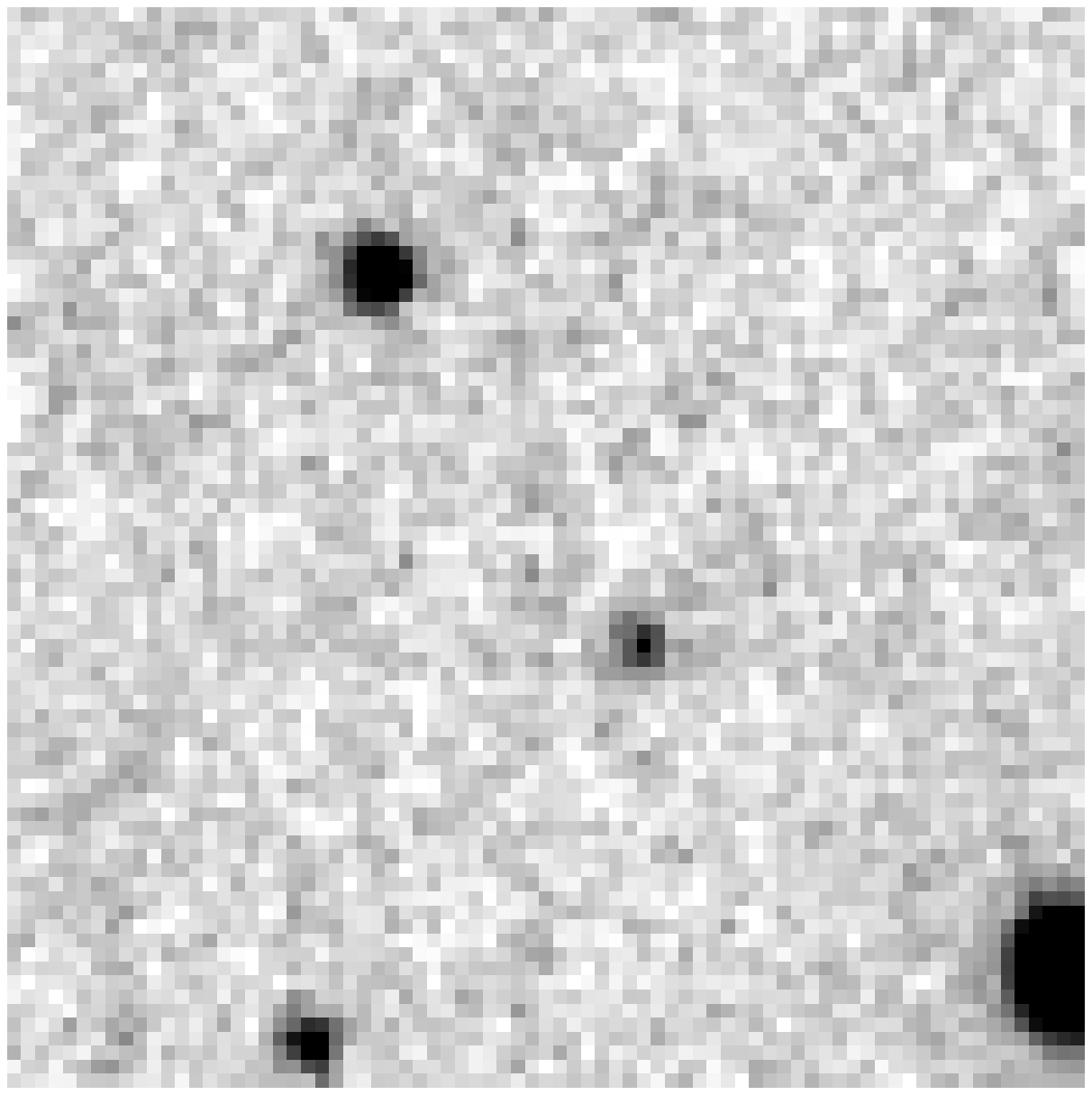}
\includegraphics{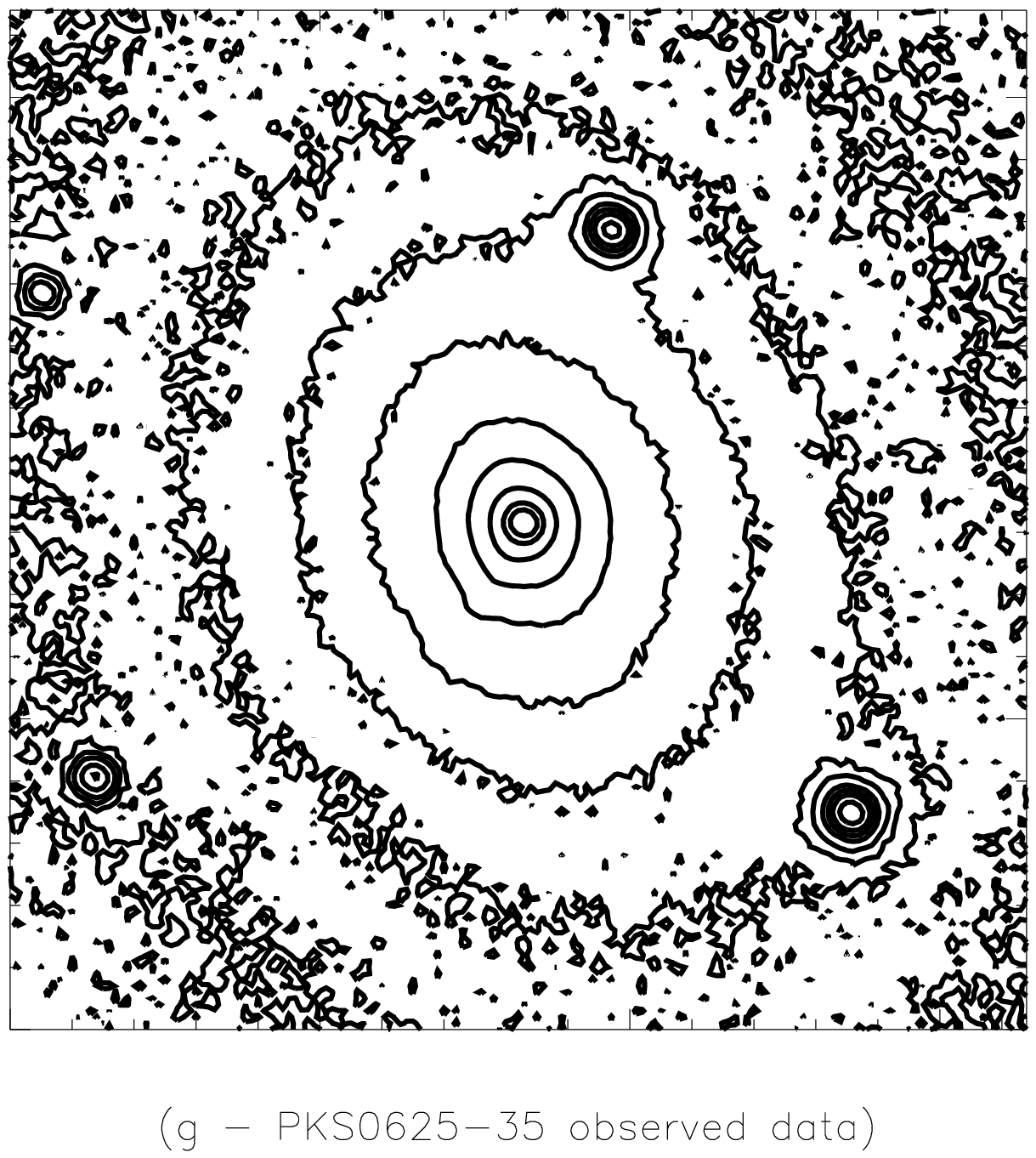}
\includegraphics{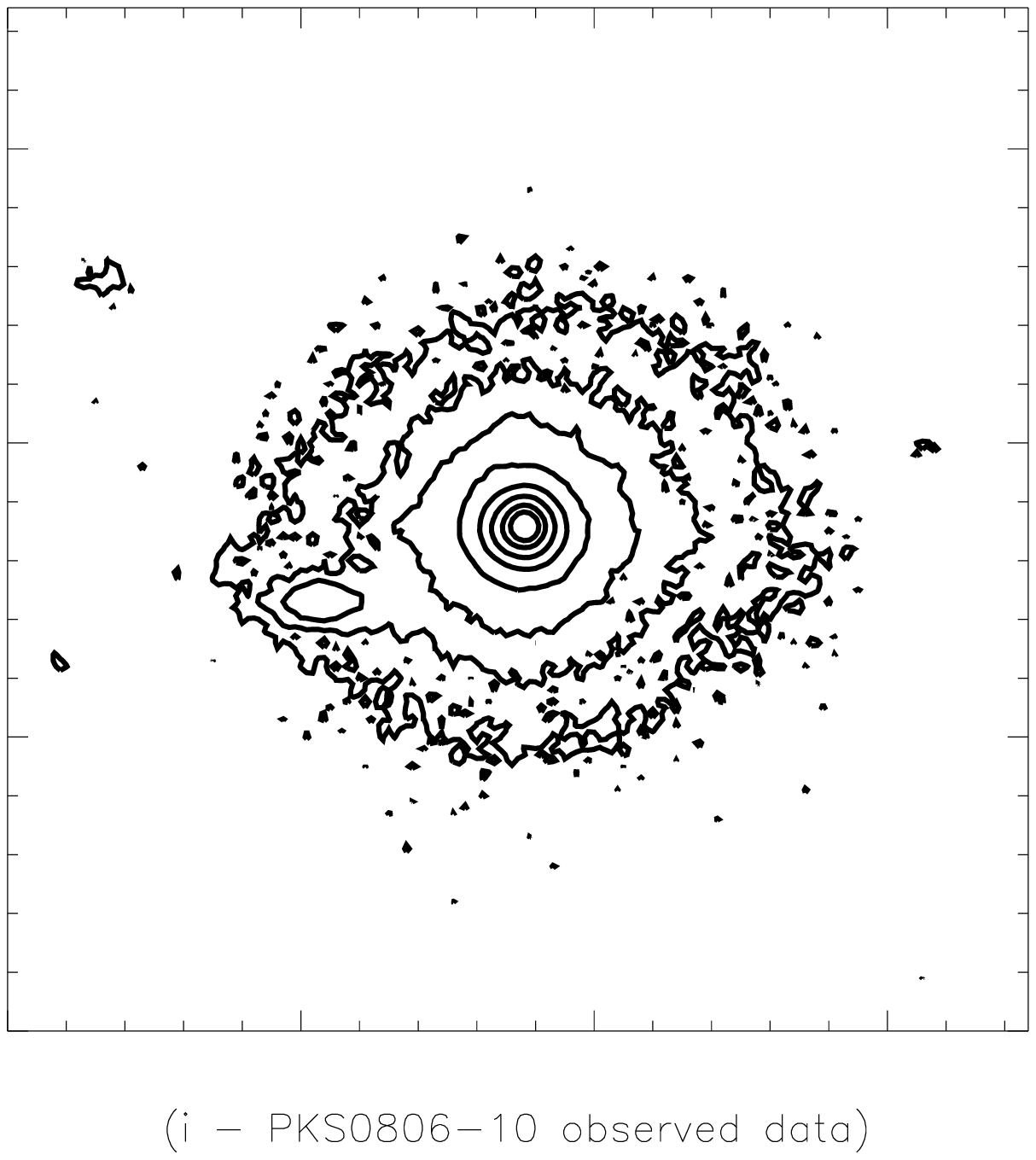}
\includegraphics{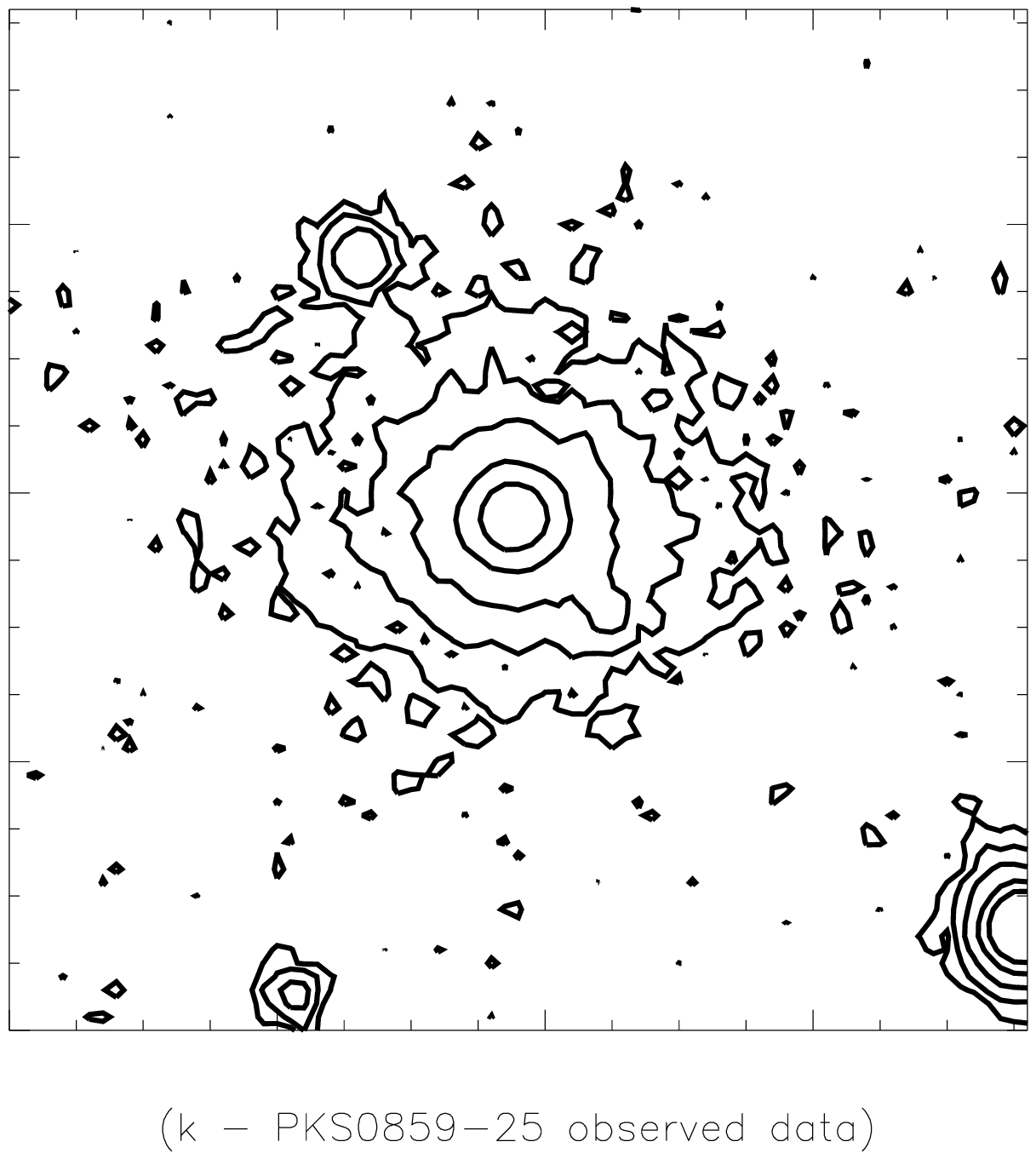}
\includegraphics{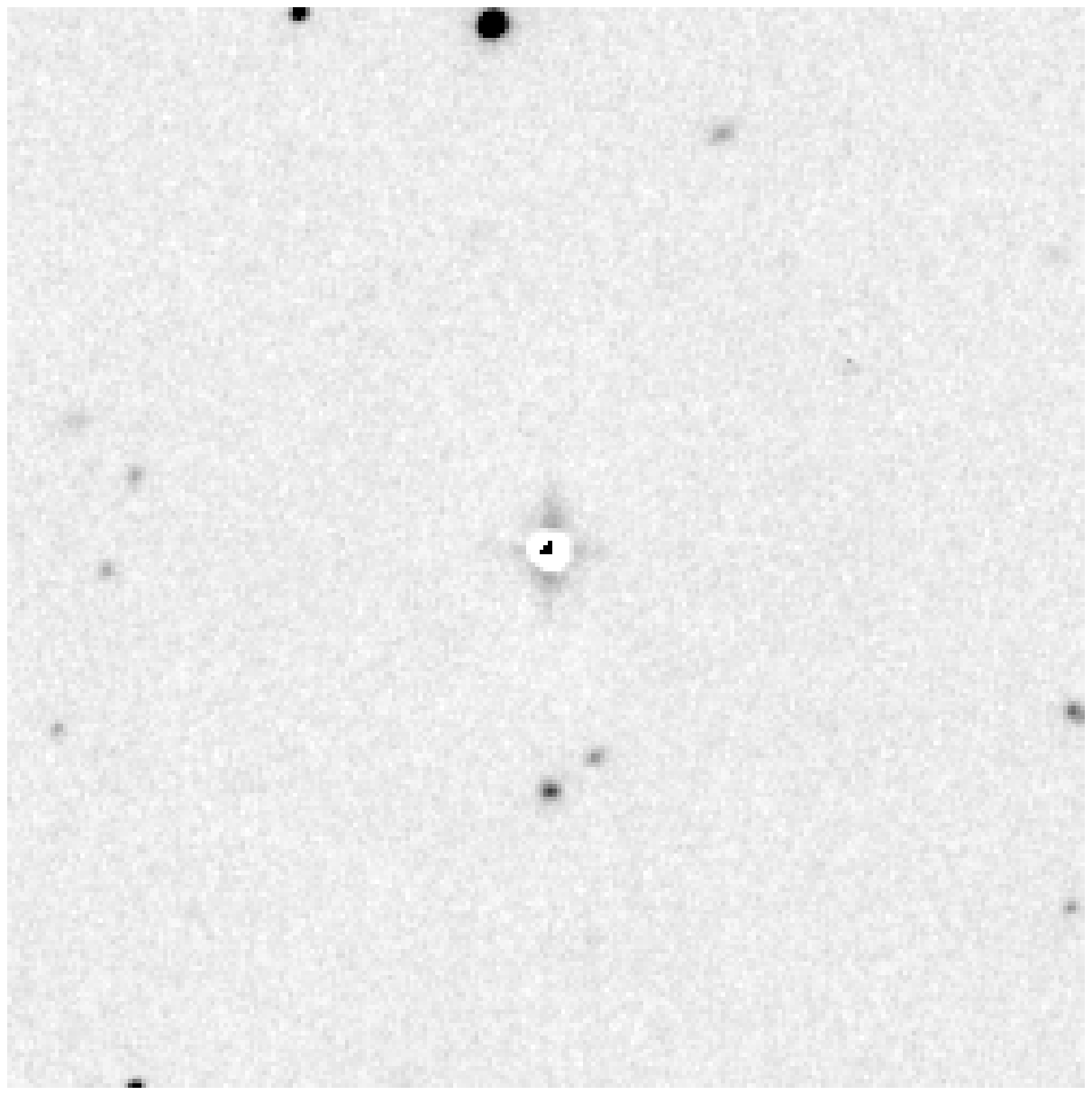}
\includegraphics{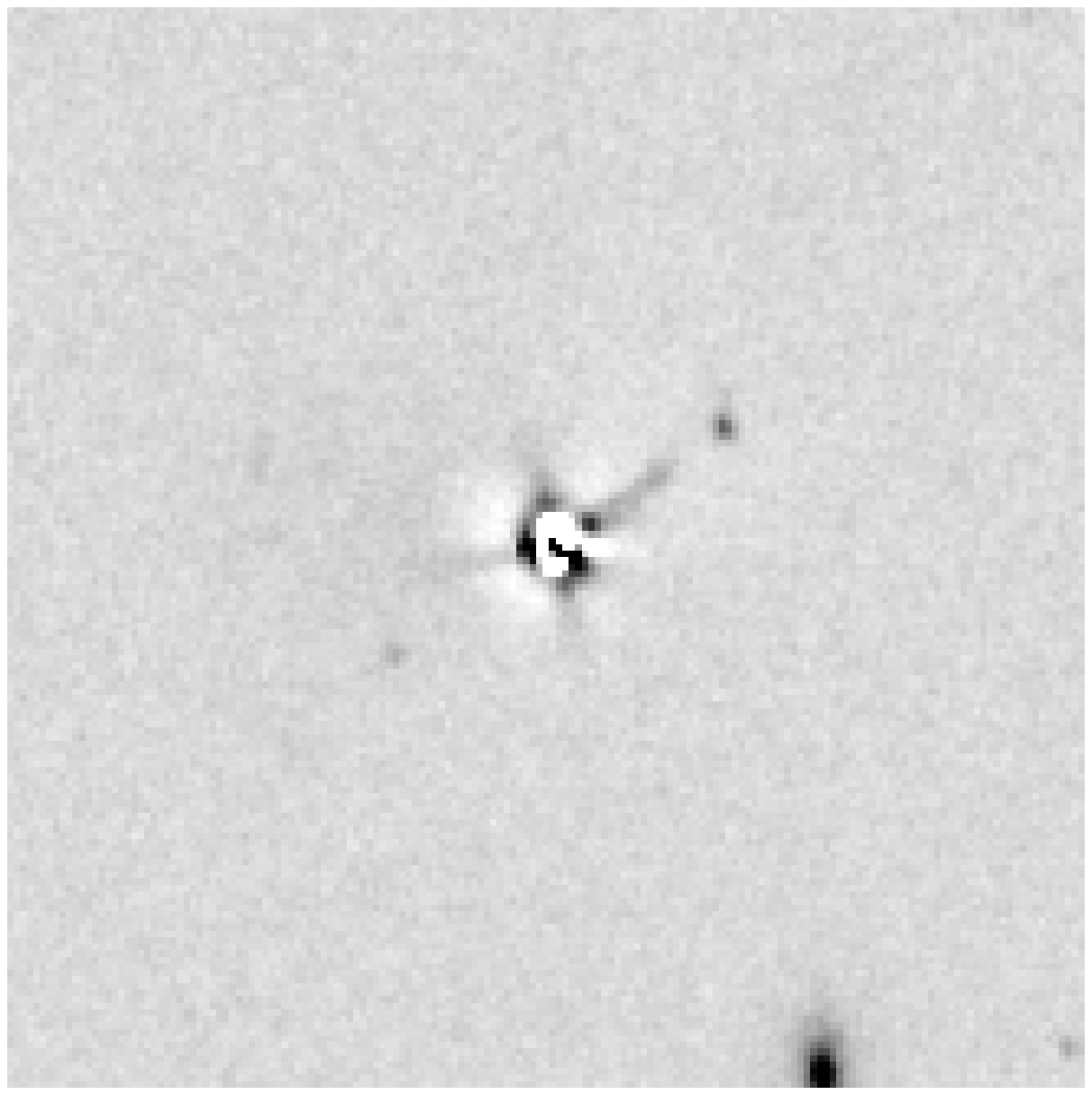}
\includegraphics{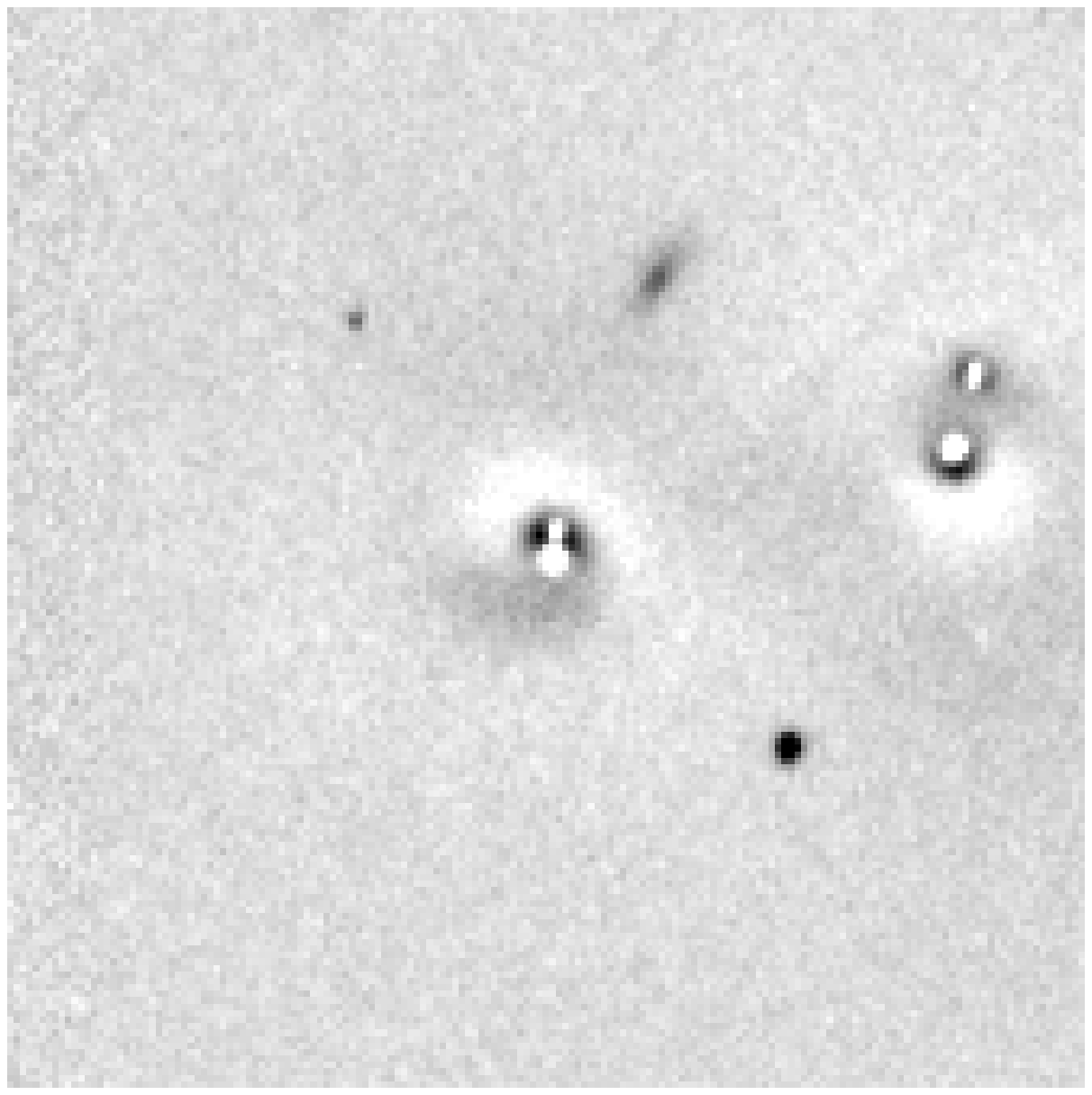}
\includegraphics{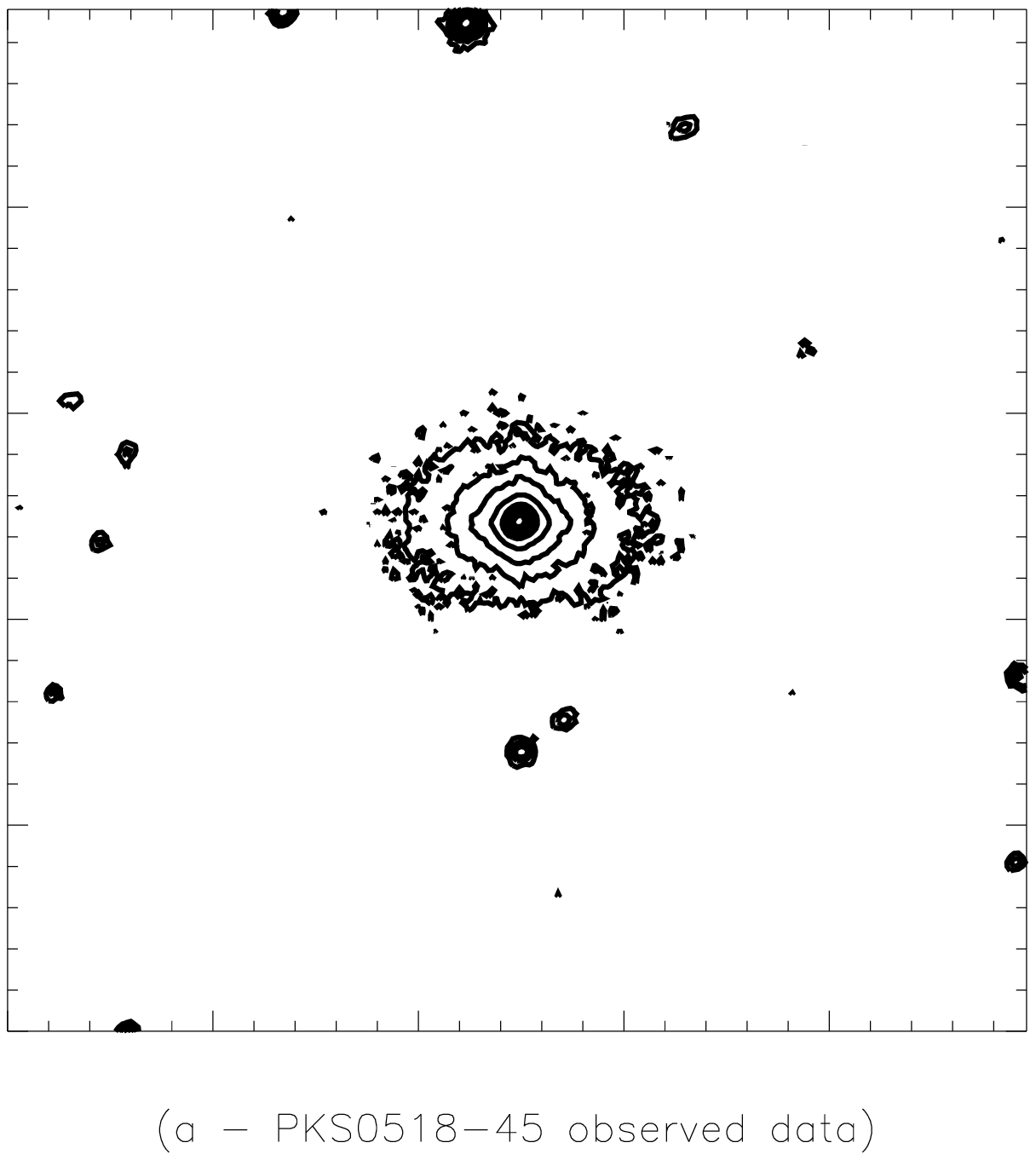}
\includegraphics{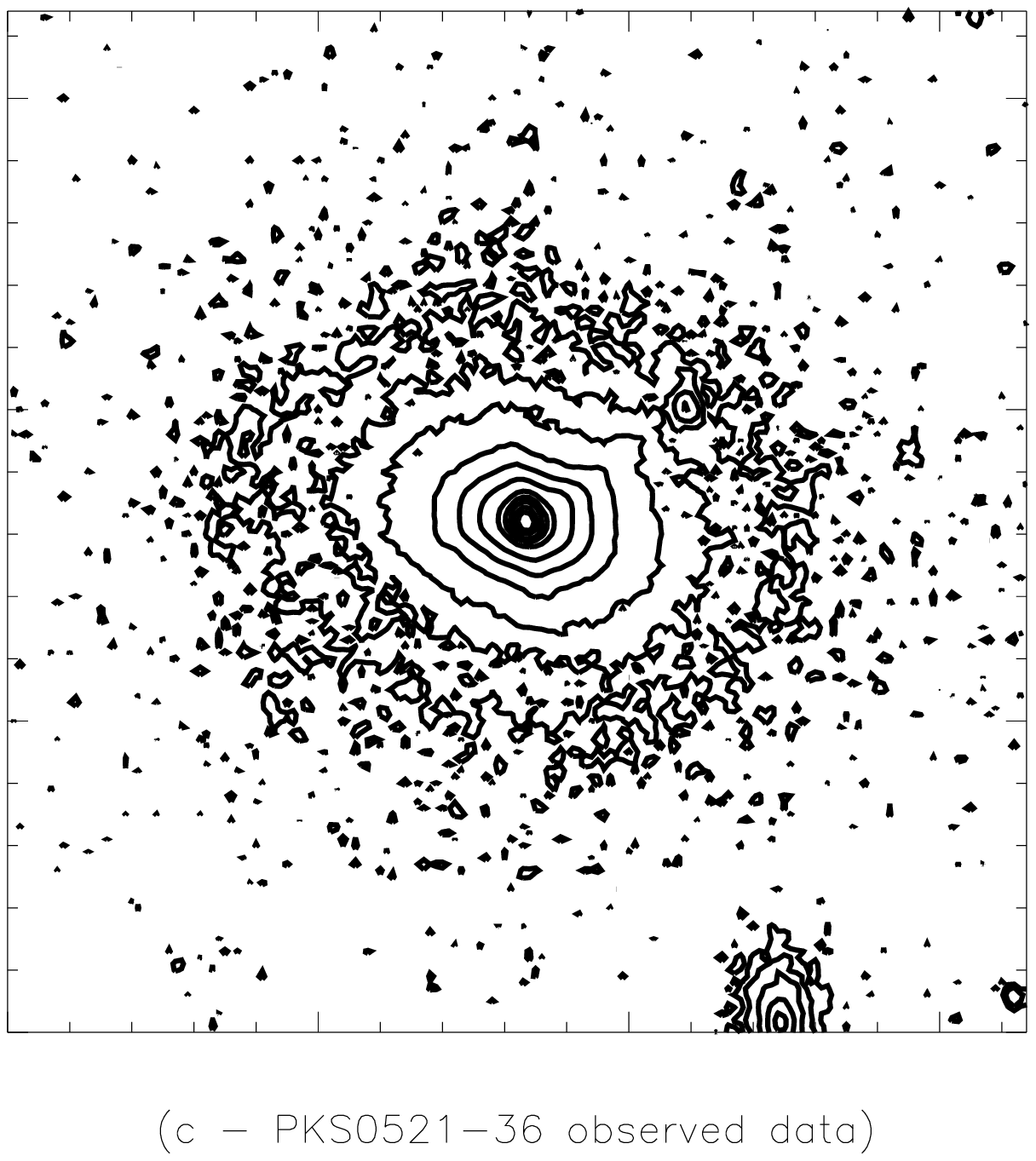}
\includegraphics{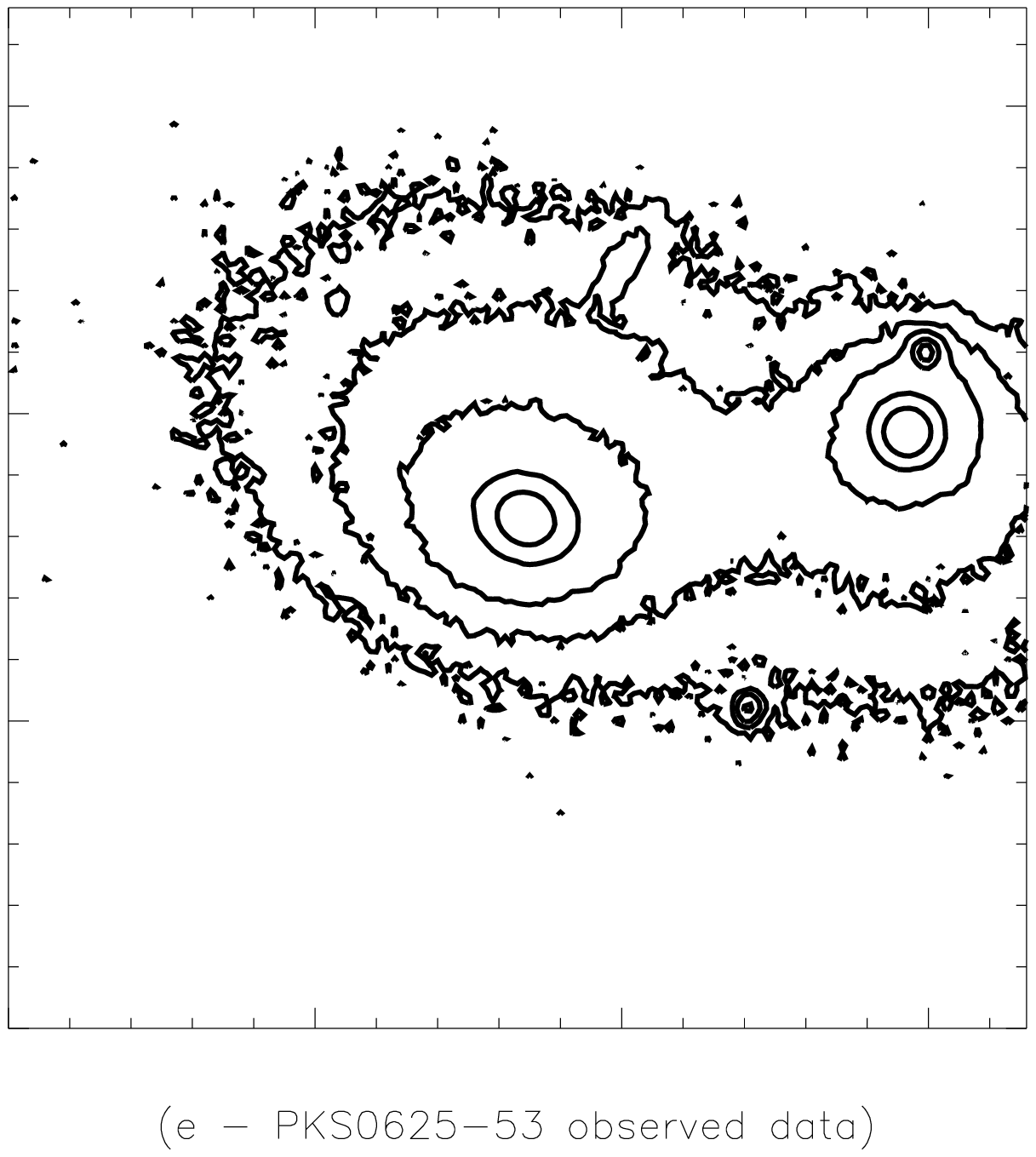}
\end{center}
\caption{50kpc by 50kpc images of PKS0518-45, PKS0521-36, PKS0625-53, PKS0625-35, PKS0806-10 and PKS0859-25. The observed data contours are displayed in frames (a), (c), (e), (g), (i) and (k), while frames (b), (d), (f), (h), (j) and (l) show the best-fit model contours on greyscale images of the model-subtracted residuals.  The maximum contour level is 50\% of the peak flux for that source in all cases, with subsequent contours at 25\%, 10\%, 5\%, 2.5\%, 1\%, 0.5\%, 0.25\%, 0.1\%, 0.05\% and 0.025\% (latter flux levels not shown in all cases). The minimum contours displayed are at 0.025\% for PKS0521-36, 0.1\% for PKS0518-45 and PKS0625-35, 0.25\% for PKS0806-10, and 2.5\% for PKS0625-53 and PKS0859-25. 
\label{Fig: 5}}
\end{figure*}

\subsubsection{PKS0915-11 (Hydra A)}
PKS0915-11 (Hydra A, 3C218) is a well known central cluster galaxy. We find that only models {\it without} a nuclear point source provide good fits to the data. Our best-fit de Vaucouleurs model has an effective radius of approximately 27 arcsec ($\sim 28$kpc) and oversubtraction in the central regions, in good agreement with the 33 arcsec effective radius derived by Govoni et al (2000) in the optical. However, we find that a better fit is obtained with a S\'ersic profile with a shallower S\'ersic index of $n=2$ and an effective radius of $\sim 16^{\prime\prime}$/16kpc, combined with a small $n=0.64$ central component. A dust lane detected in optical imaging of this source (RA10) may have contributed to the shallowing of the profile in the central regions of this object.
There is also a small companion object which can be well modelled with a S\'ersic profile with $n=4$ and $r_{eff} = 0.7^{\prime\prime}$/0.7kpc.

\subsubsection{PKS0945+07 (3C227)}

PKS0945+07 is another well-known BLRG, with spectacular extended emission line
structures (Prieto et al. 1993).  The nuclear point source accounts for $\sim 50-60\%$ of the $K-$band flux in our modelling, and our best fit effective radius is $\sim 7^{\prime\prime}$ (11.9kpc).  This is slightly larger than the value obtained by Govoni et al (2000) in the optical ($\sim 5^{\prime\prime}$), and considerably larger than that obtained in the infrared by Donzelli et al using higher-resolution NICMOS data ($r_{eff} \sim 3$kpc), due in the main to the difficulty in defining the point spread function accurately - it is clear from our model residuals that strong PSF diffraction spikes have most likely skewed our fit towards a larger effective radius than the galaxy truly has.

\subsubsection{PKS1306-09}

In addition to the host galaxy itself, our modelling of PKS1306-09 also includes three neighboring objects to the SW.  The residuals are highly disturbed, and it seems likely that this source is undergoing some form of interaction with other objects in the field. Our best fit model is a de Vaucouleurs elliptical with an effective radius of $\sim$13.6kpc ($2.3^{\prime\prime}$) and a nuclear point source contribution of $\sim 21$\%, offset from the galaxy centroid. The secondary nucleus is also observed in $r^{\prime}-$band observations of this source (RA10).

\subsubsection{PKS1547-79}

PKS1547-79 is one of the higher redshift objects in our sample, and is clearly an interacting system. Fits without a nuclear point source contribution do not adequately explain the data, as expected given its optical BLRG/Q classification.  Wth a S\'ersic index of $n=4$ for both the radio galaxy (north) and its faint companion (south), the host galaxy is modelled as having an effective radius $r_{eff} \sim 0.9^{\prime\prime}$ (6 kpc) and a point source contribution of 29\% (note that the PSF-deconvolved derived effective radius is 50\% larger than the PSF FWHM for the observations of this source). However, a numerically better fit is obtained by modelling the host galaxy as a disk with $n=1$, $r_{eff} \sim 1.3^{\prime\prime}$ (7.5 kpc) and  a point source contribution of 3\%.  Even so, both options underestimate the flux in the centre of the galaxy, and oversubtract it immediately outside the central regions.  Interestingly, this source is surrounded by a large number of faint blobby features which are not duplicated elsewhere in the field, and are plausbily satellite systems.

\begin{figure*}
\vspace{8.45 in}
\begin{center}
\includegraphics{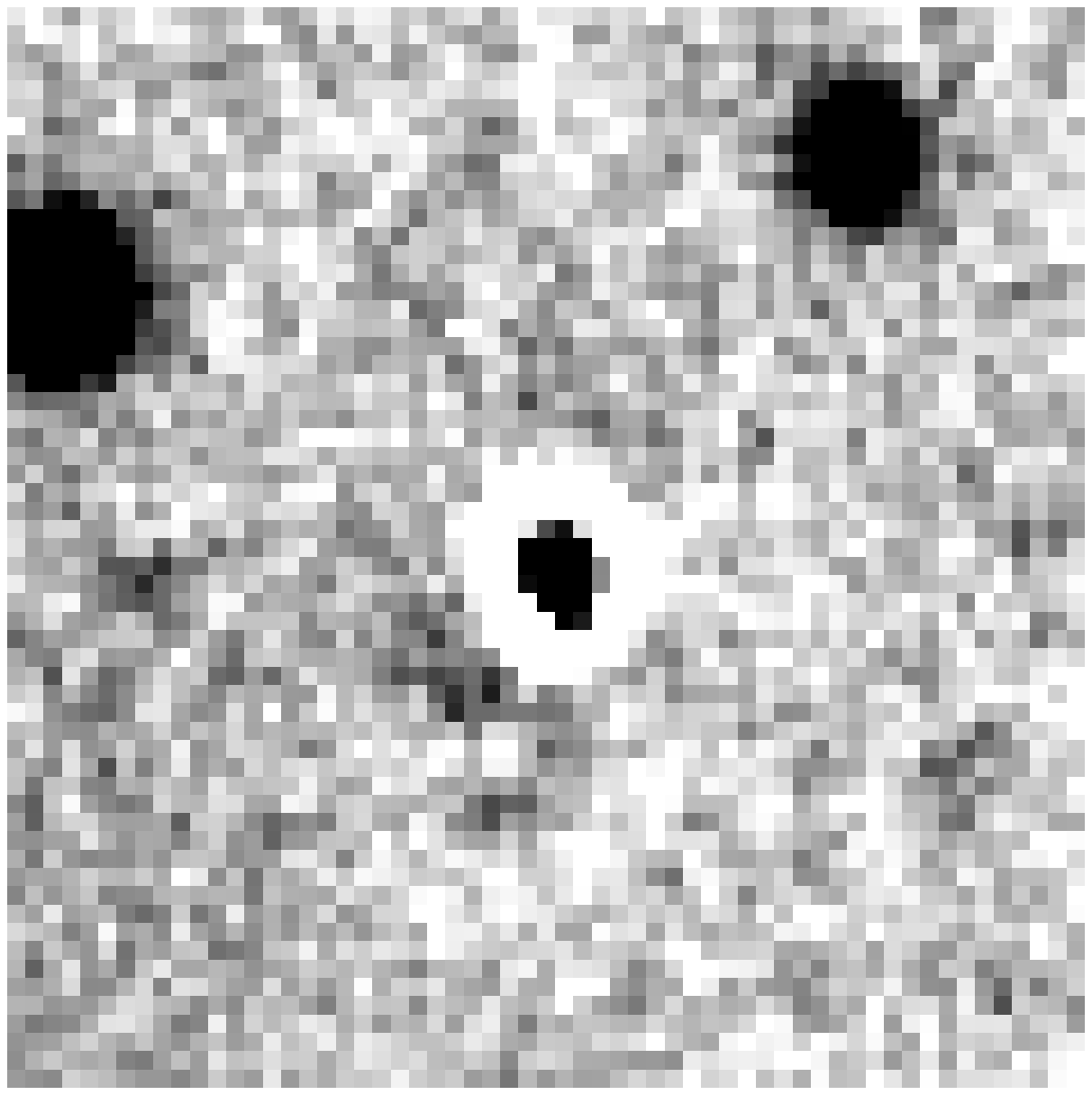}
\includegraphics{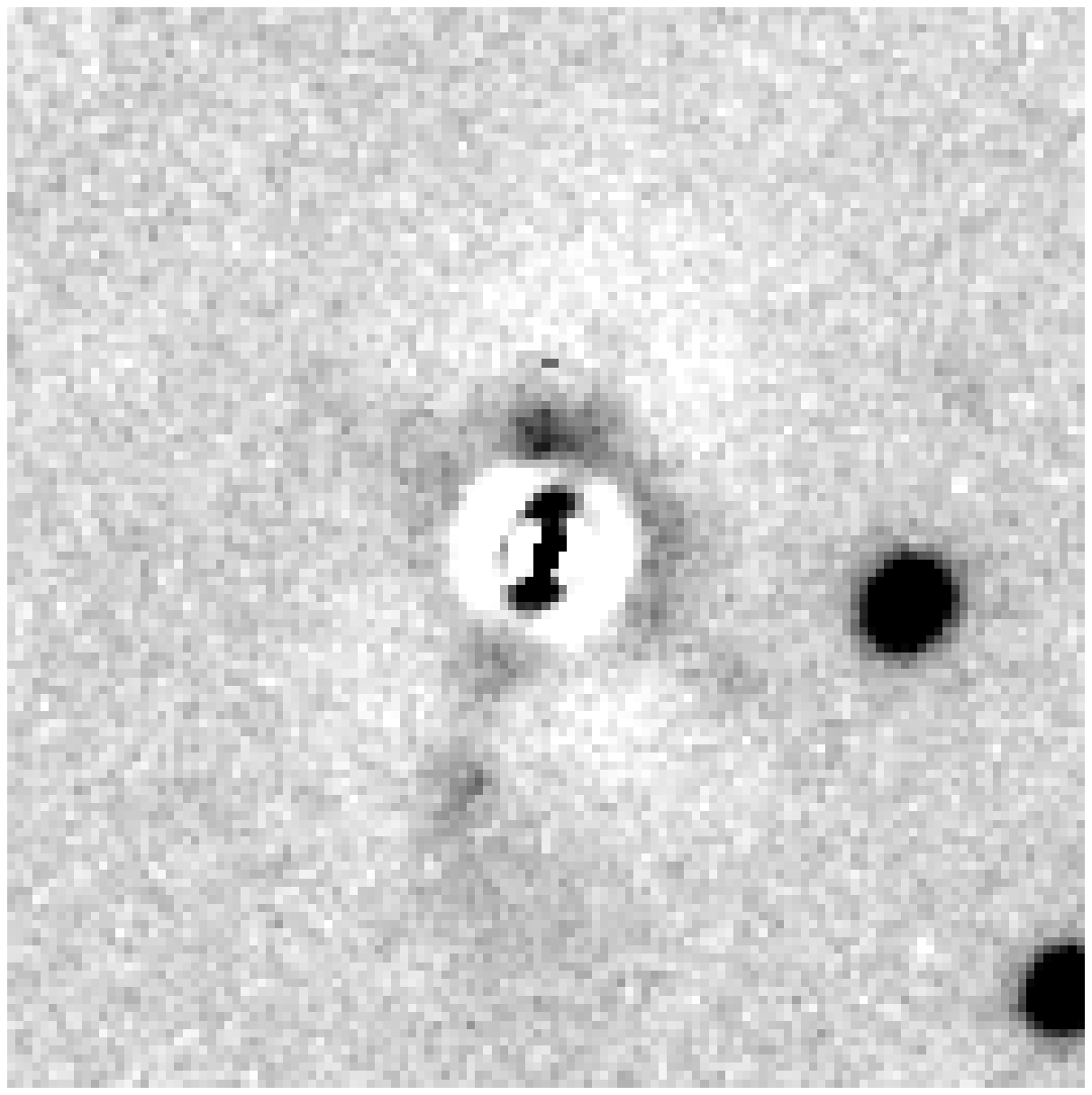}
\includegraphics{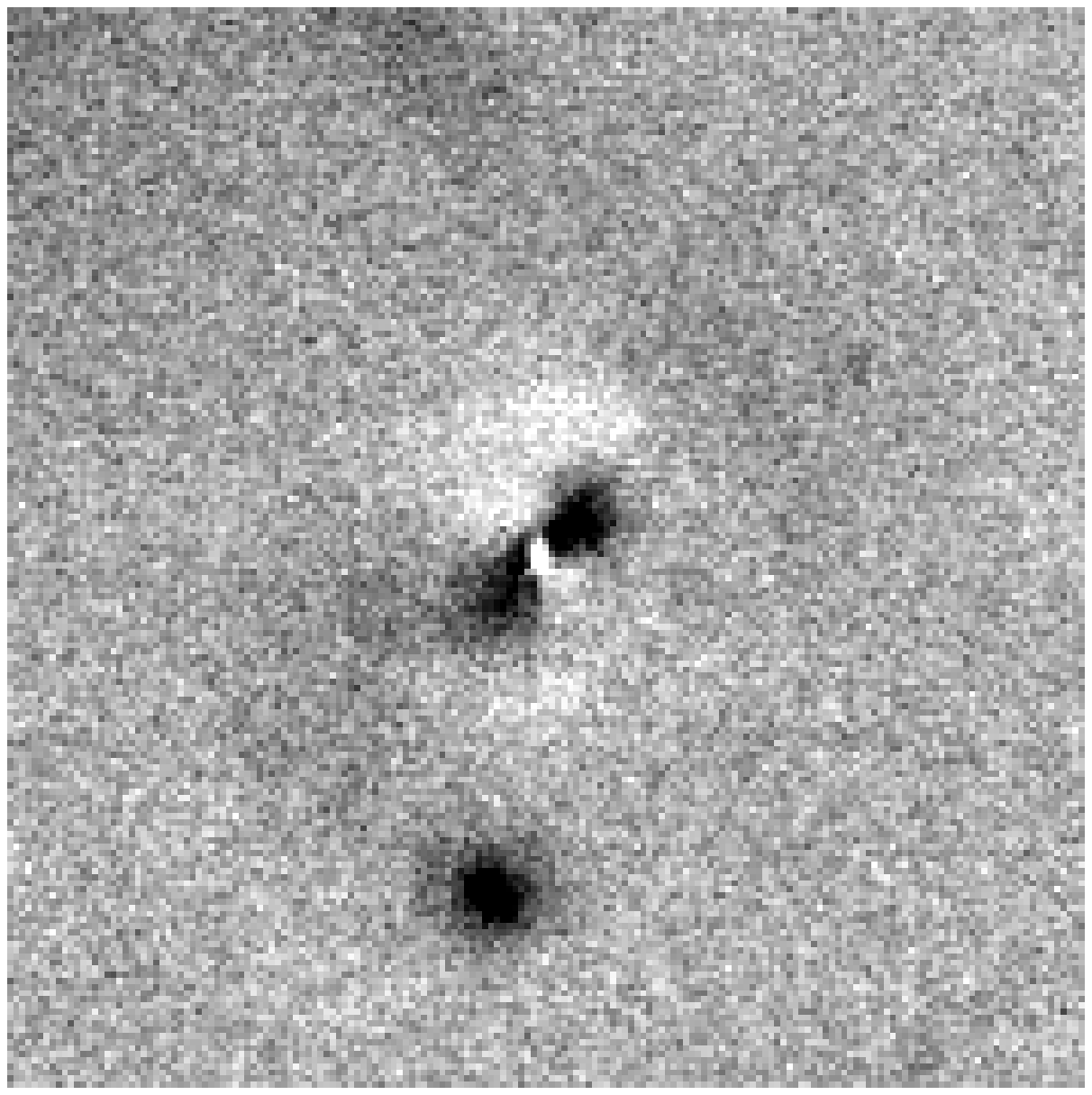}
\includegraphics{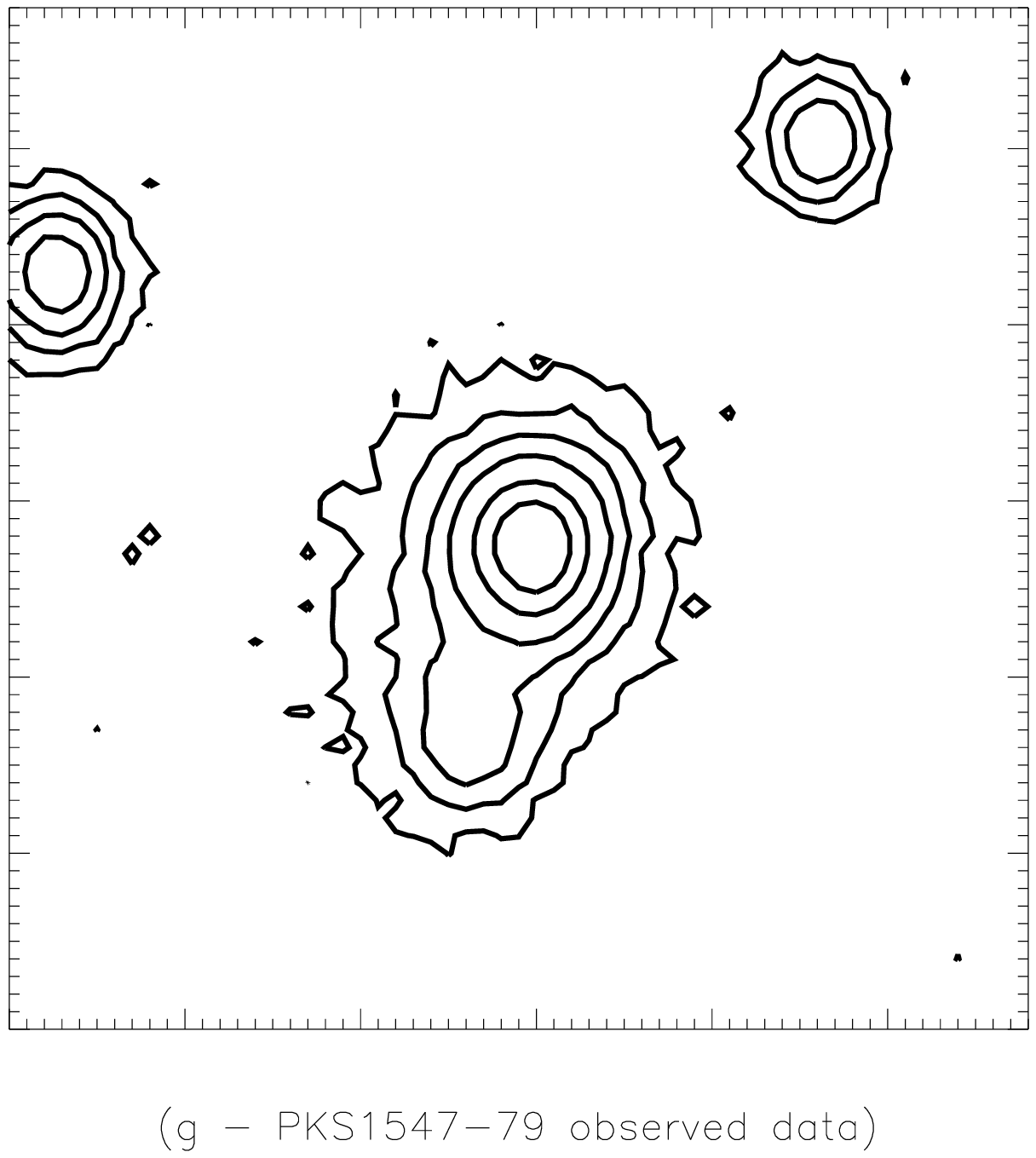}
\includegraphics{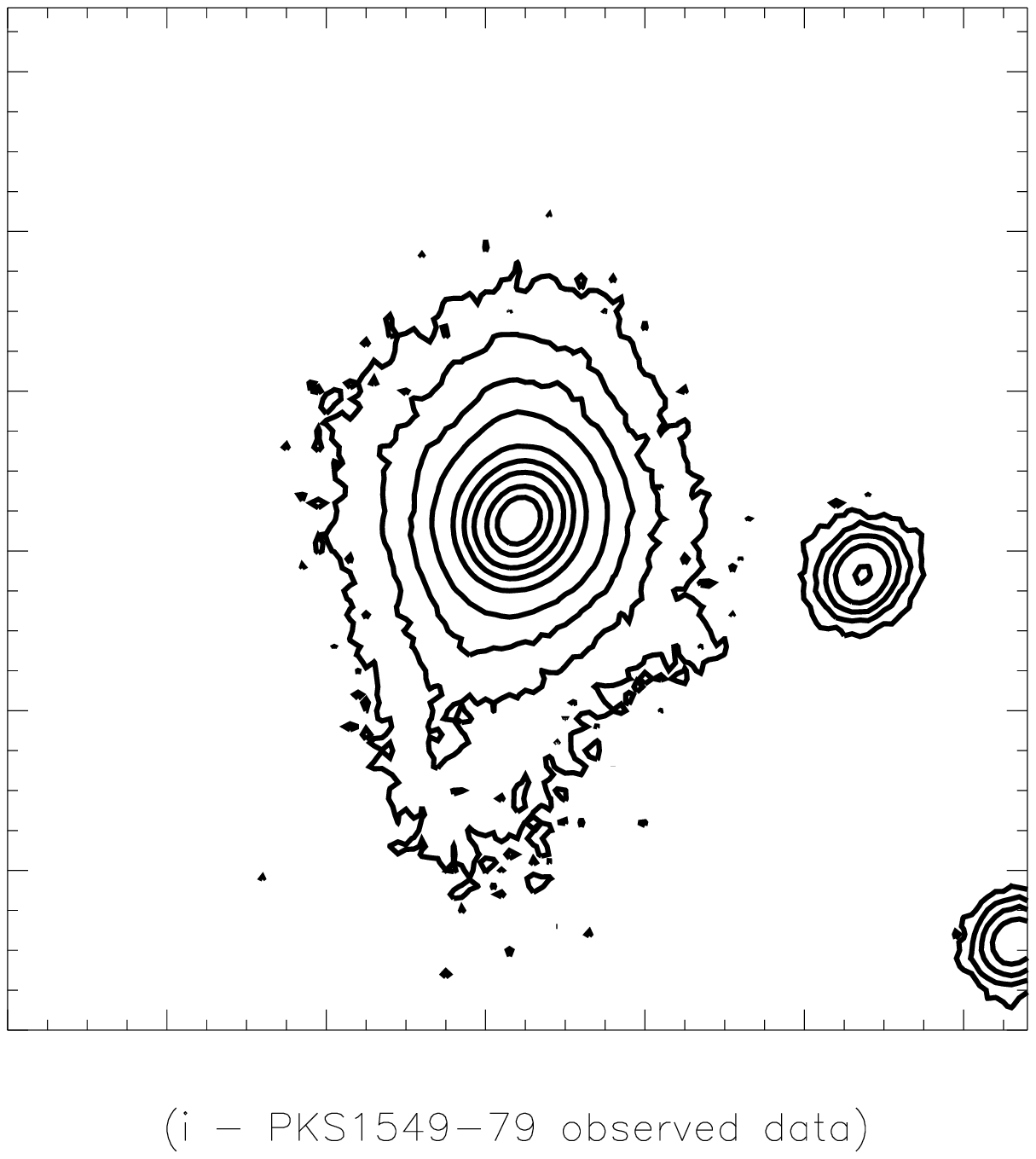}
\includegraphics{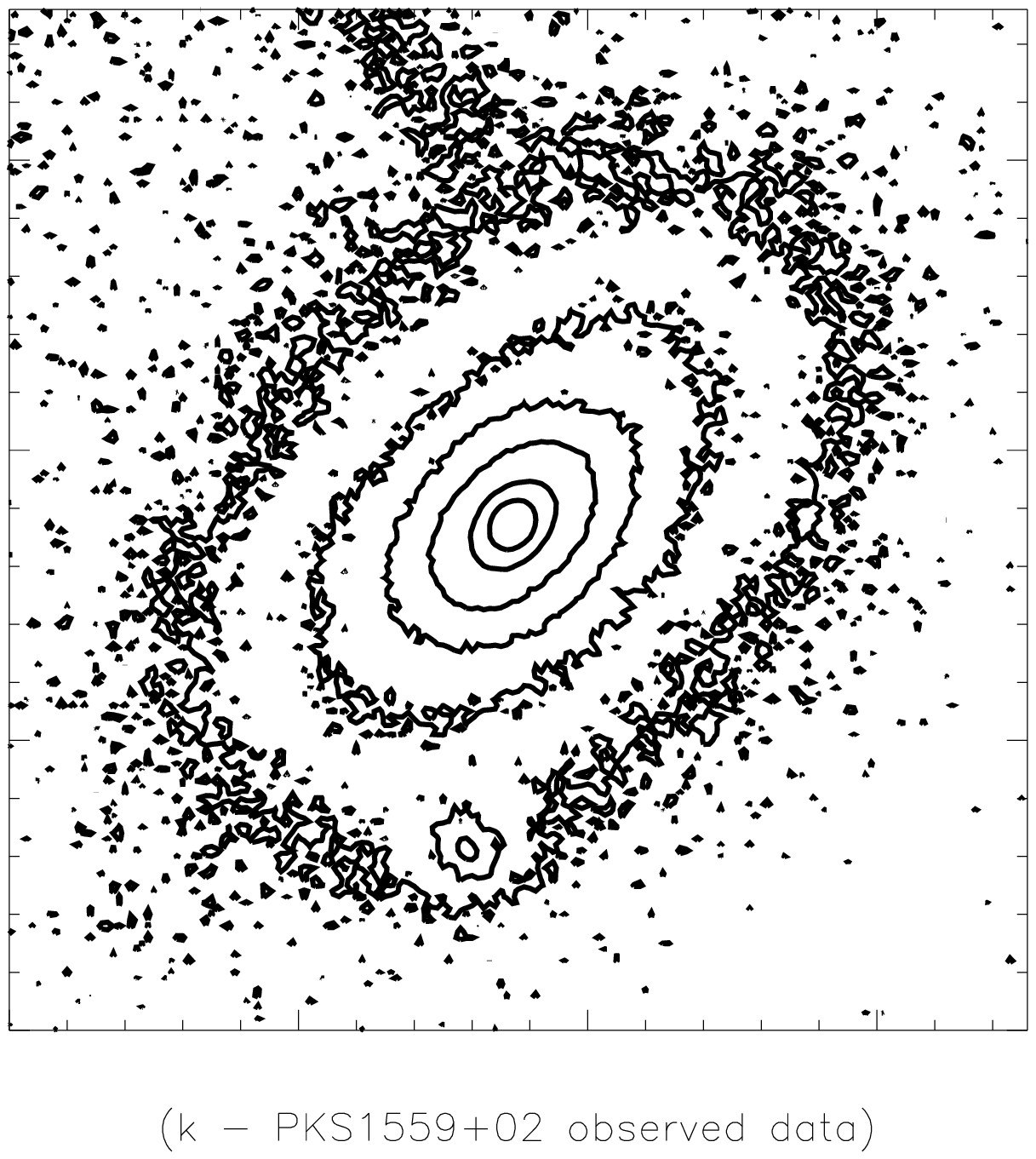}
\includegraphics{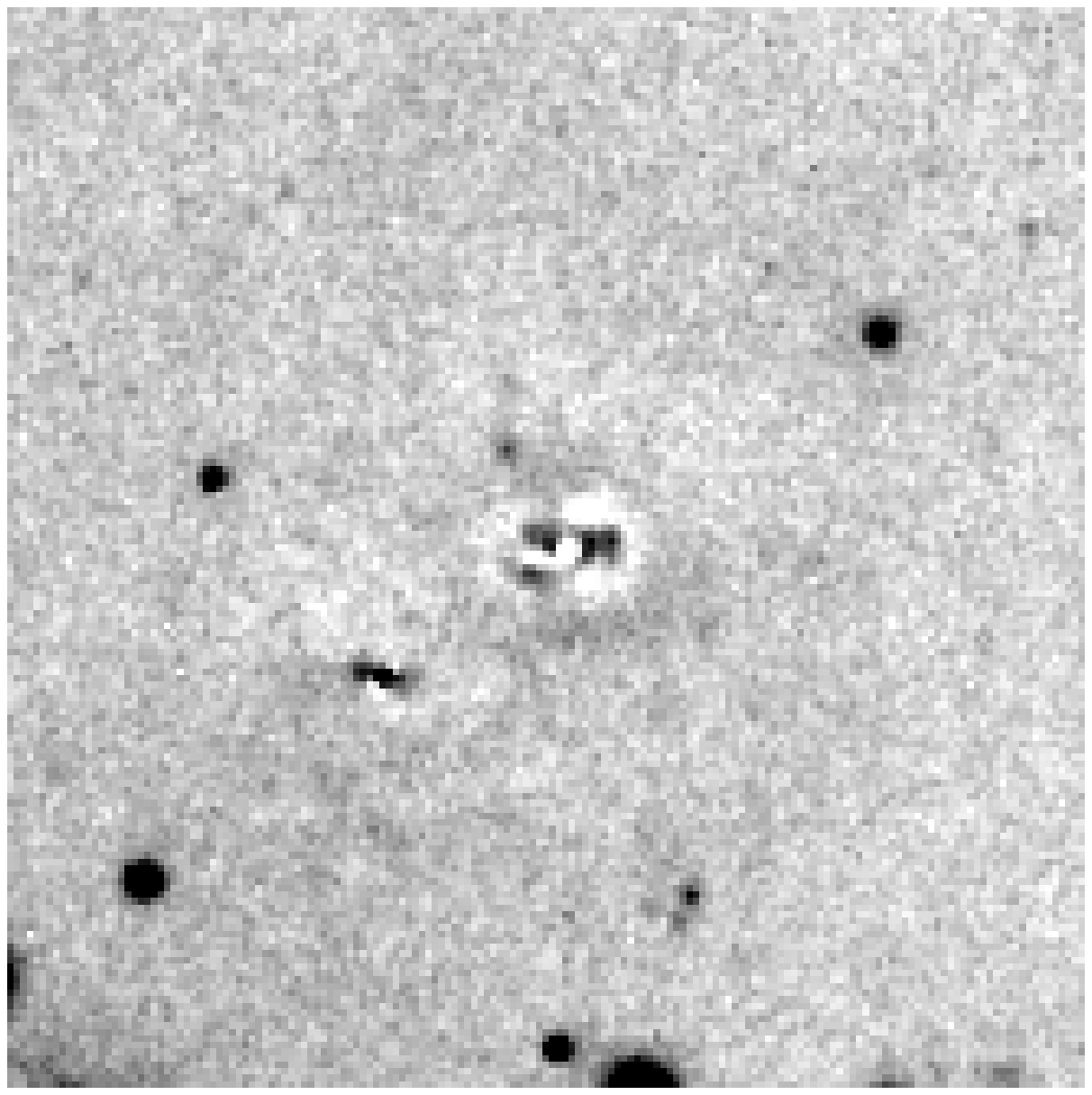}
\includegraphics{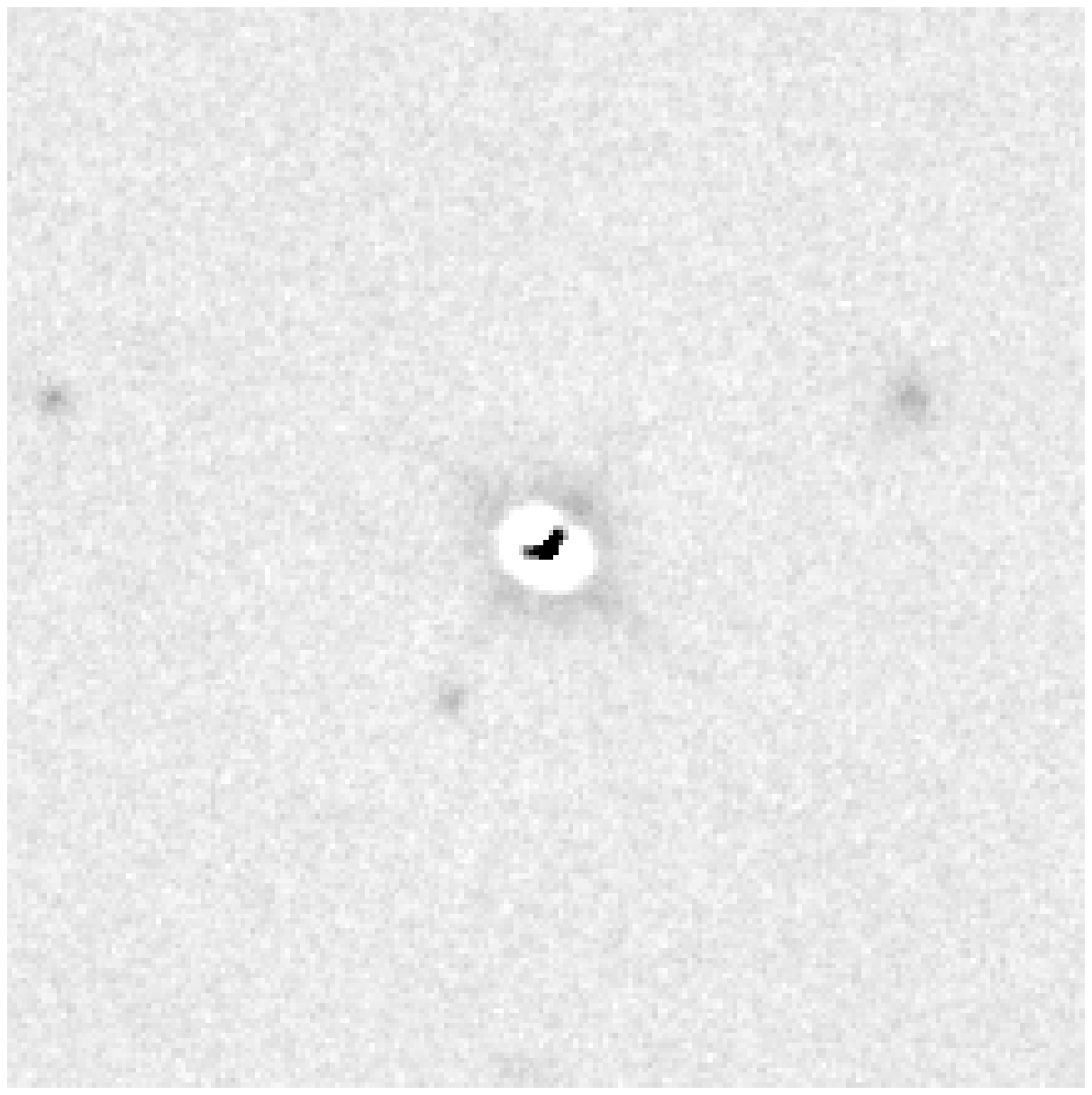}
\includegraphics{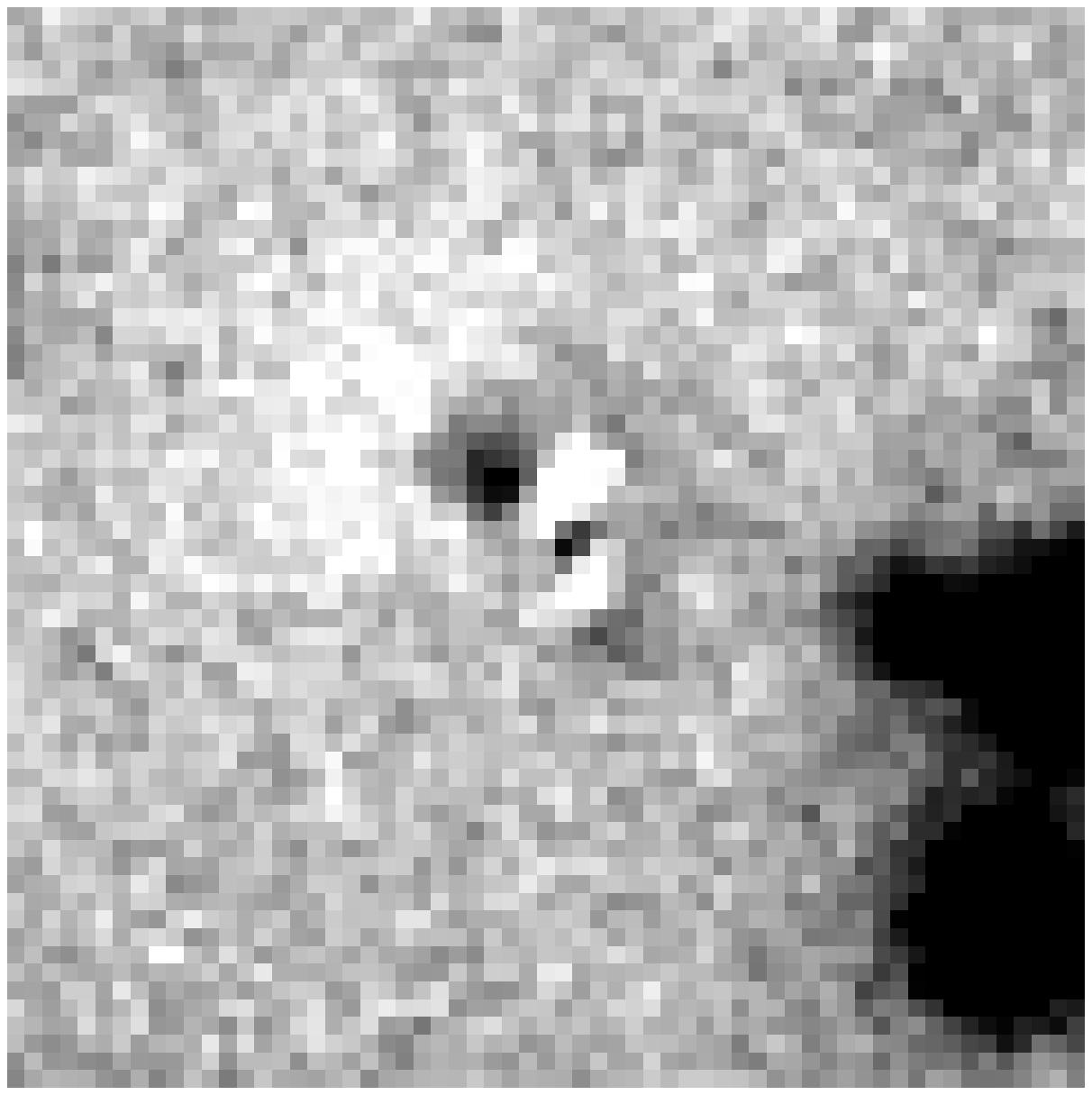}
\includegraphics{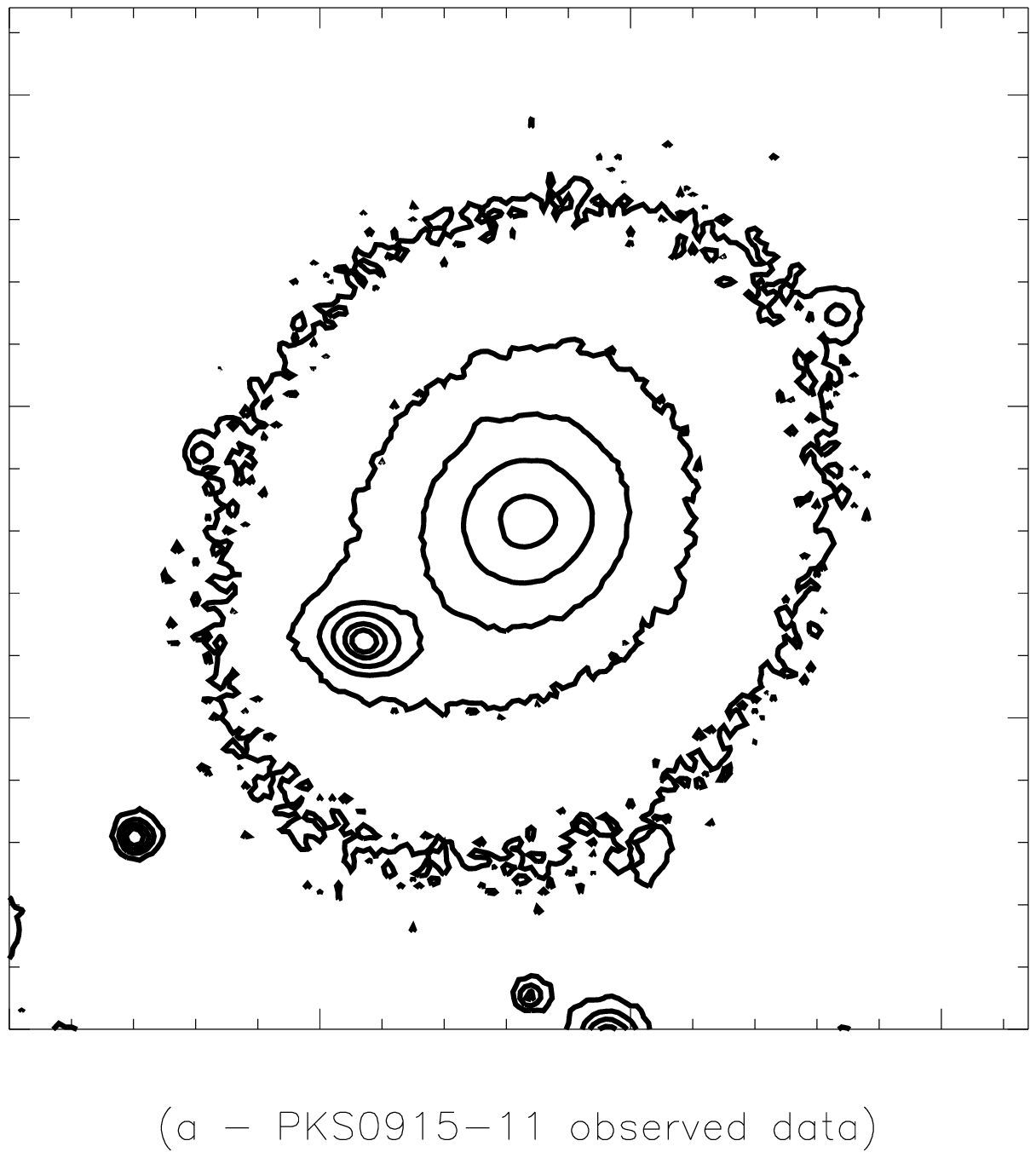}
\includegraphics{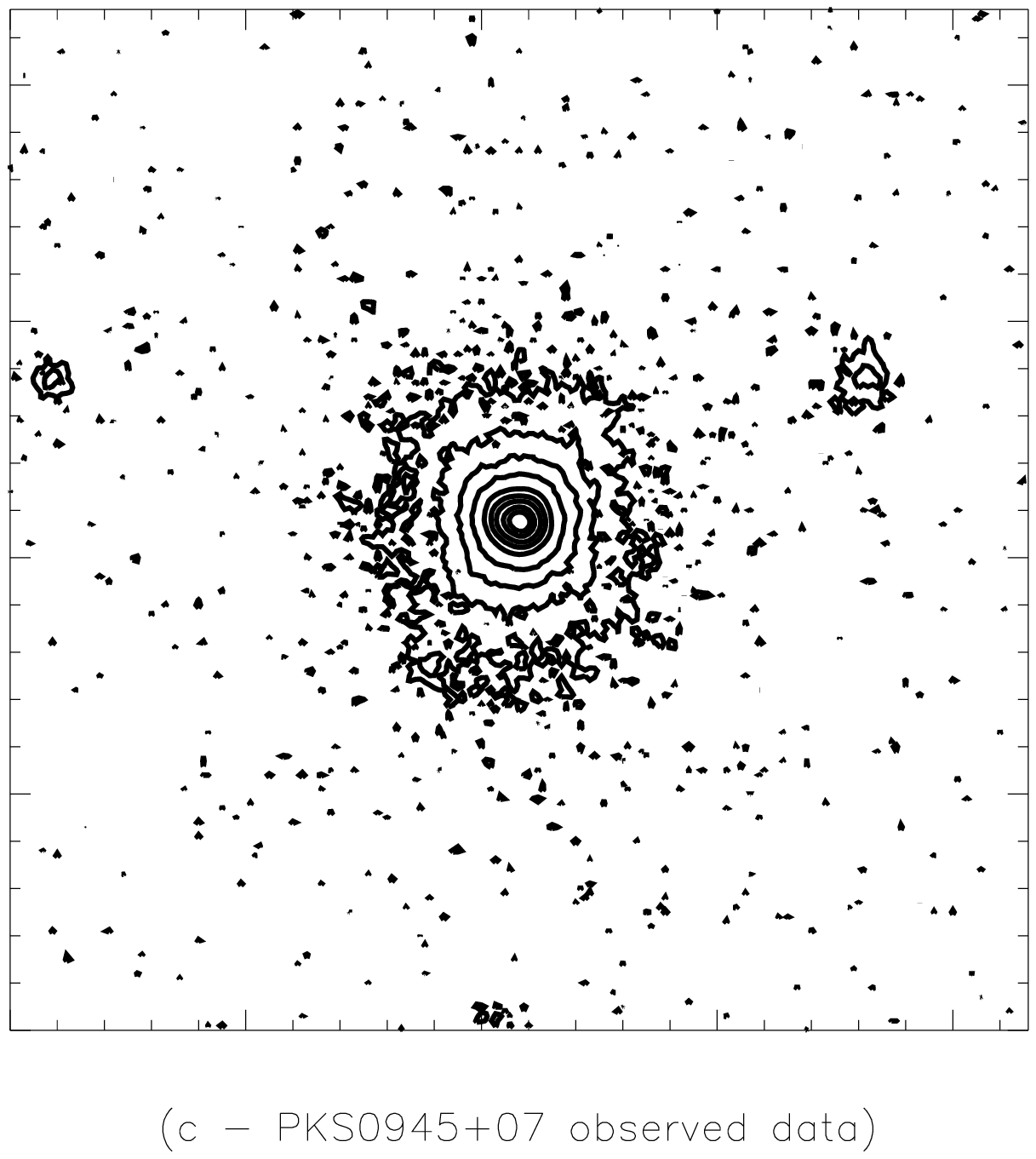}
\includegraphics{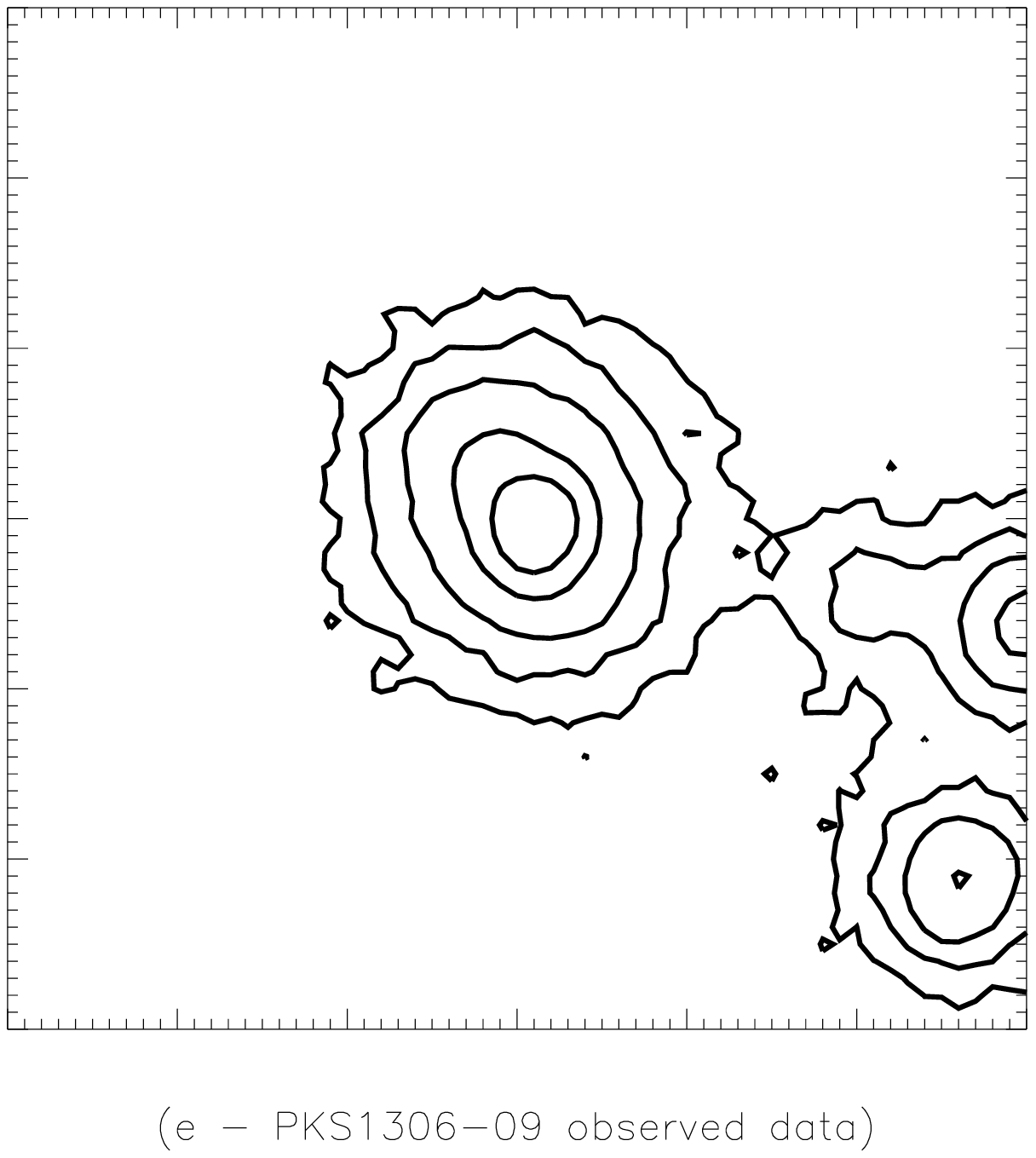}
\end{center}
\caption{50kpc by 50kpc images of PKS0915-11, PKS0945+07, PKS1306-09, PKS1547-79, PKS1549-79 and PKS1559+02. The observed data contours are displayed in frames (a), (c), (e), (g), (i) and (k), while frames (b), (d), (f), (h), (j) and (l) show the best-fit model contours on greyscale images of the model-subtracted residuals.  The maximum contour level is 50\% of the peak flux for that source in all cases, with subsequent contours at 25\%, 10\%, 5\%, 2.5\%, 1\%, 0.5\%, 0.25\% and 0.1\%,  (latter flux levels not shown in all cases). The minimum contours displayed are at 0.1\% for PKS0945+07 and PKS1549-79, 1\% for PKS0915-11, PKS1547-79 and PKS1559+02, and 2.5\% for PKS1306-09. 
\label{Fig: 6}}
\end{figure*}

\subsubsection{PKS1549-79}

PKS1549-79 has a very strong nuclear point source contribution -- the largest in our sample  -- and has been well studied in the past (Holt et al 2006 and references therein).  Our best-fit de Vaucouleurs elliptical model has $r_{eff}=10.8$kpc ($4.2^{\prime\prime}$) and a nuclear point source contribution of 71.5\%, but does not adequately explain the data. We find that the host galaxy is better modelled as a disk-type galaxy; our best-fit S\'ersic model has $n=1$ and an effective radius of $r_{eff} \sim 8$kpc ($3.1^{\prime\prime}$) and a nuclear point source contribution of $\sim 80$\% of the total flux. The model residuals are very distorted due to the presence of multiple high surface brightness tidal tails (Holt et al 2006).

\subsubsection{PKS1559+02 (3C327)}

For this massive galaxy, our best fit de Vaucouleurs models have effective radii of 6.4$^{\prime\prime}$/12kpc without a nuclear point source, and 7.1$^{\prime\prime}$/13.6kpc with a nuclear point source accounting for $\sim 1$\% of the total flux.  However, our best fit model has a S\'ersic index $n=6$, $r_{eff} = 24kpc$ ($\sim 14^{\prime\prime}$), and no additional nuclear point source emission.
Our model residuals show dust features close to the nucleus of the galaxy, and an elongated apparently tidal feature connecting the nucleus of PKS1559+02 with what appears to be a satellite object to the south. Previous optical imaging by Smith and Heckman (1989) derived an even larger effective radius for this host galaxy of approximately 50kpc in our assumed cosmology.

\subsubsection{PKS1733-56}
PKS1733-56 is a well-known BLRG lying in a relatively crowded field. Our best fit model is for a de Vaucouleurs elliptical galaxy with $r_{eff} = 6.6$kpc ($3.9^{\prime\prime}$), comparable to that found by Govoni et al (2000) in the optical, and a nuclear point source contribution of $\sim 13$\%. Our modelling residuals display an excess of flux aligned with the galaxy major axis and connecting it with an unresolved companion to the northwest.  Bryant et al (2002) state on the basis of their integral field spectroscopic observations of this source that the gas dynamics of PKS1733-56 are very disrupted, and that it is in the process of undergoing a merger with companion objects, likely those that we observe close to the host galaxy in our infrared observations.

\subsubsection{PKS1814-63}
This object has a star lying close to the nucleus at optical
wavelengths, which we model alongside the host galaxy. We find that this source is well modelled as a de Vaucouleurs elliptical with $r_{eff} \sim 5.2$kpc ($4.35^{\prime\prime}$ and a nuclear point source contribution of 15\%), but that a numerically better fit is obtained with a S\'ersic index of $n=2$, which gives $r_{eff} \sim 5.1$kpc ($4.3^{\prime\prime}$) and a nuclear point source contribution of $\sim 25$\%.   A faint extended disk feature is visible in our residuals, along with a possible dust lane.  Similar features are observed in $r^{\prime}-$band observations of this source (RA10), where the disk feature becomes particularly prominent. Previous observations of this source in the optical (V\'eron-Cetty et al 2000) note that the optical $B-I$ colour of this source is 0.3 magnitudes bluer than expected for an elliptical galaxy.

\subsubsection{PKS1932-46}
This source, most recently the subject of a detailed study by Inskip et al (2007), is an elliptical galaxy which appears to be a member of a small ineracting group.  Extensive continuum and line emitting star-forming structures are present in the surrounding IGM, and most recently $r^{\prime}-$band imaging has revealed the presence of further irregular structures surrounding the nucleus of this galaxy (Tadhunter, priv. comm.). Assuming $n=4$, we obtain a very good fit to the central regions of the host galaxy with an effective radius of $r_{eff} \sim 8$kpc (2.3$^{\prime\prime}$).  
Despite the BLRG nature of this source, the central nuclear point source component is relatively weak, accounting for only 15\% of the total flux in our $n=4$ model.

\subsubsection{PKS1934-63}
PKS1934-63 is hosted by the eastern component of an interacting galaxy pair. Both systems have rather disturbed morphologies; in the optical (Ramon Almeida et al 2010) a tidal tail leads away from the fainter companion galaxy. Our best fit model for the companion uses a S\'ersic profile with $n=2$ and a small nuclear point source contribution. The host galaxy can be well fit with either $n=4$ or $n=4$ plus a 5\% nuclear point source.  Our derived effective radii for the radio galaxy range from 1.2-1.4$^{\prime\prime}$ (3.5-4.4kpc) depending on the model, and are comparable to that found in the optical ($r_{eff} = 4.85$kpc for a de Vaucouelurs-only model after conversion to the same cosmological parameters) by Zirbel (1996).   As well as faint tidal features, our model residuals also show a third faint object lying just to the south of the interacting companion galaxy.

\subsubsection{PKS1949+02 (3C403)}

For PKS1949+02, our best fit de Vaucouleurs model has an effective radius of $\sim 7$kpc ($6.3^{\prime\prime}$) (very close to that found in the optical  by Govoni et al 2000) and a minimal point source contribution of $\sim 5\%$. Madrid et al (2006) observe a companion galaxy close to the SE, also present in our images.
Our modelling of this source is also in reasonable agreement with the modelling of NICMOS data by Donzelli et al (2007) (see also Tremblay et al 2007), who derive a slightly smaller effective radius than our own, consistent with the smaller S\'ersic index of their fitted bulge component ($n \sim 2.5$). 

\begin{figure*}
\vspace{8.45 in}
\begin{center}
\includegraphics{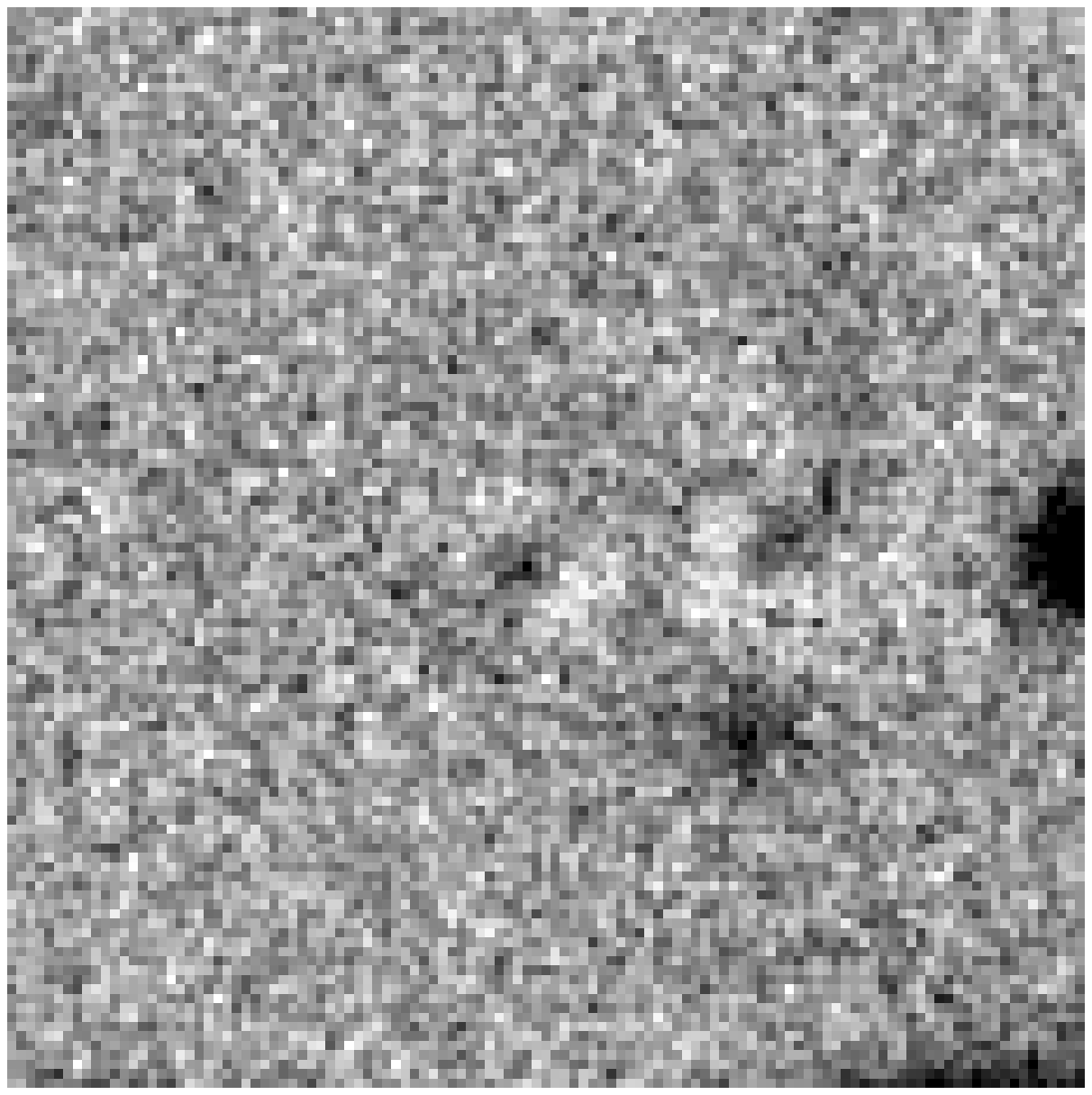}
\includegraphics{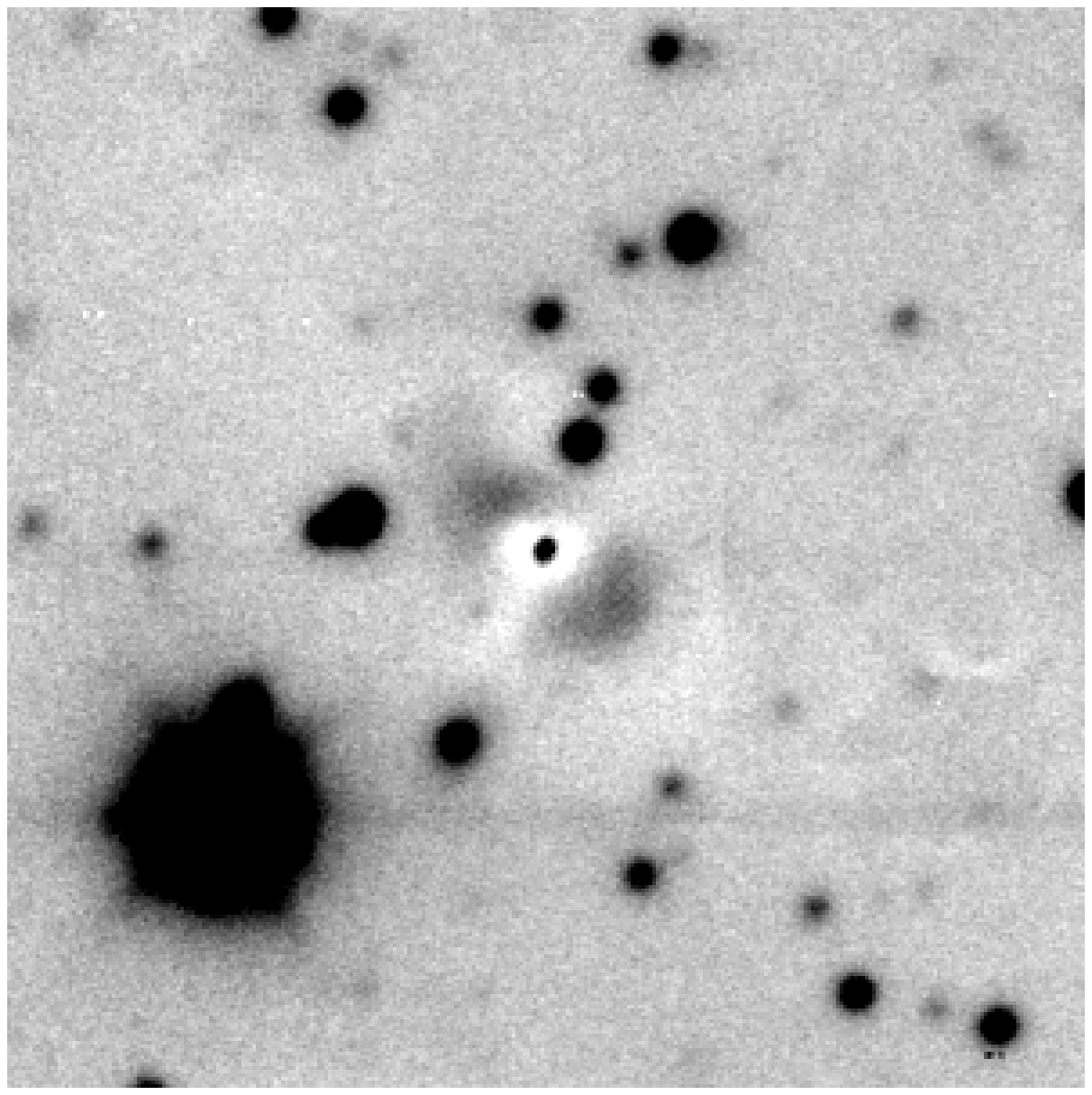}
\includegraphics{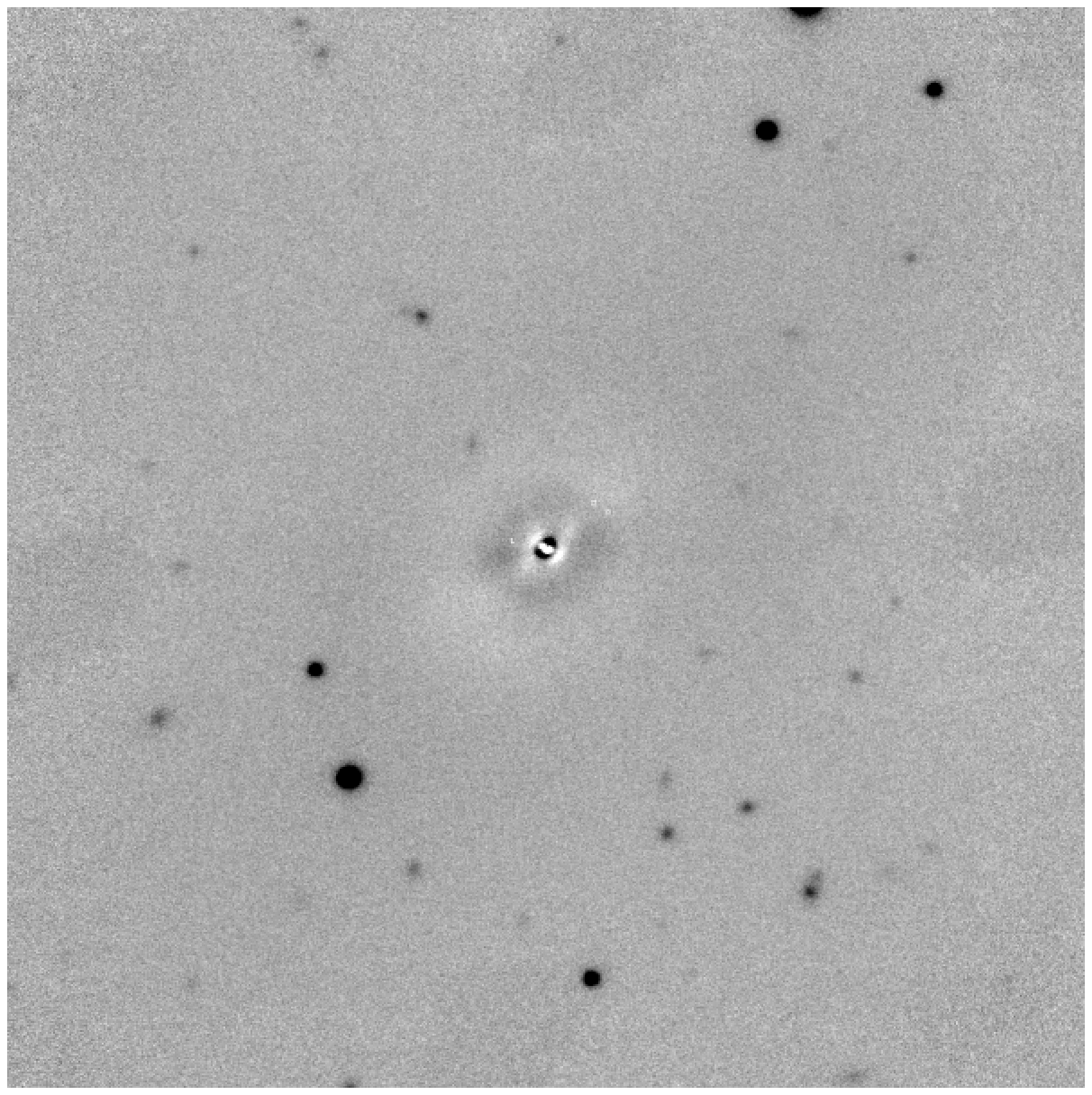}
\includegraphics{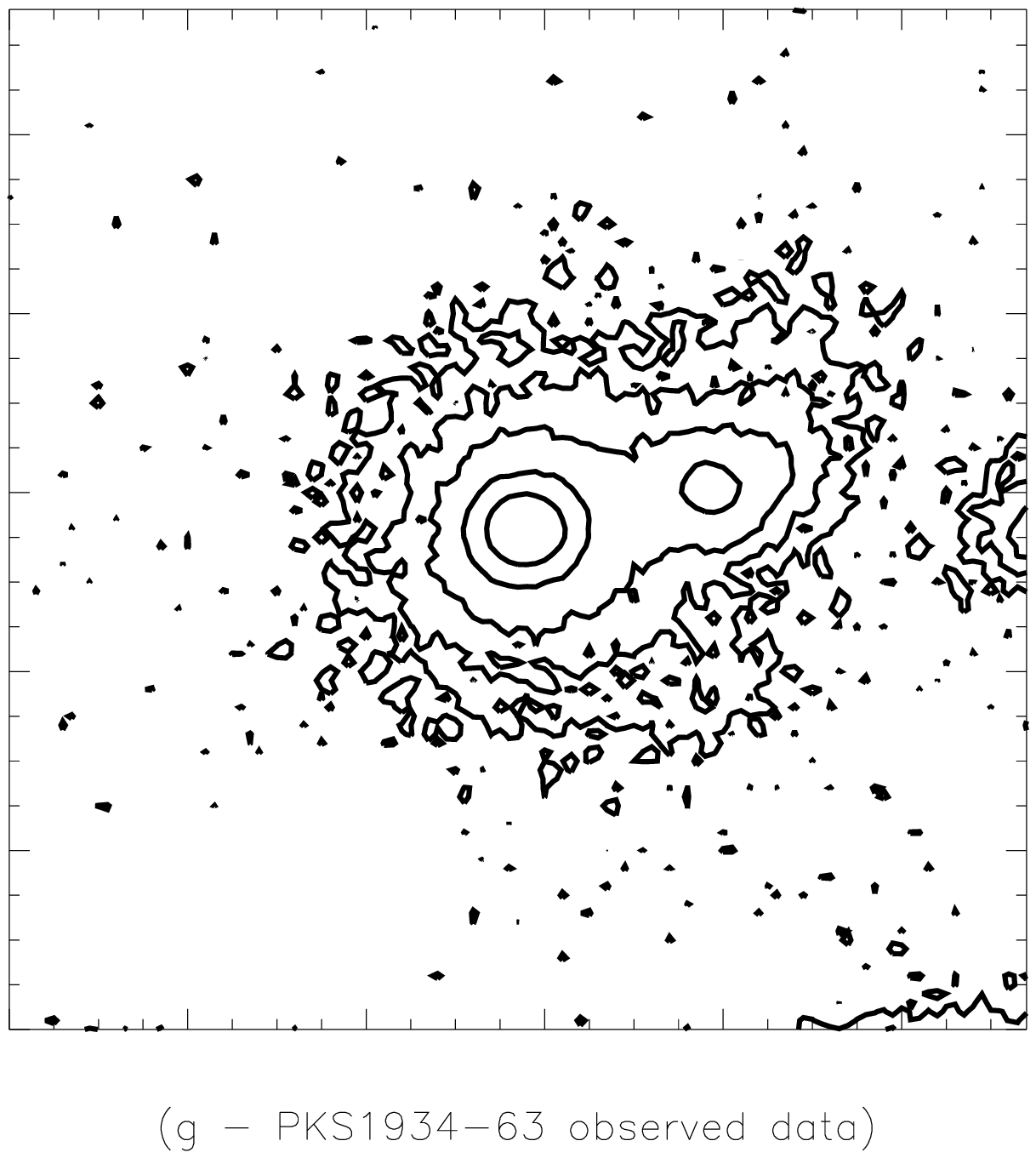}
\includegraphics{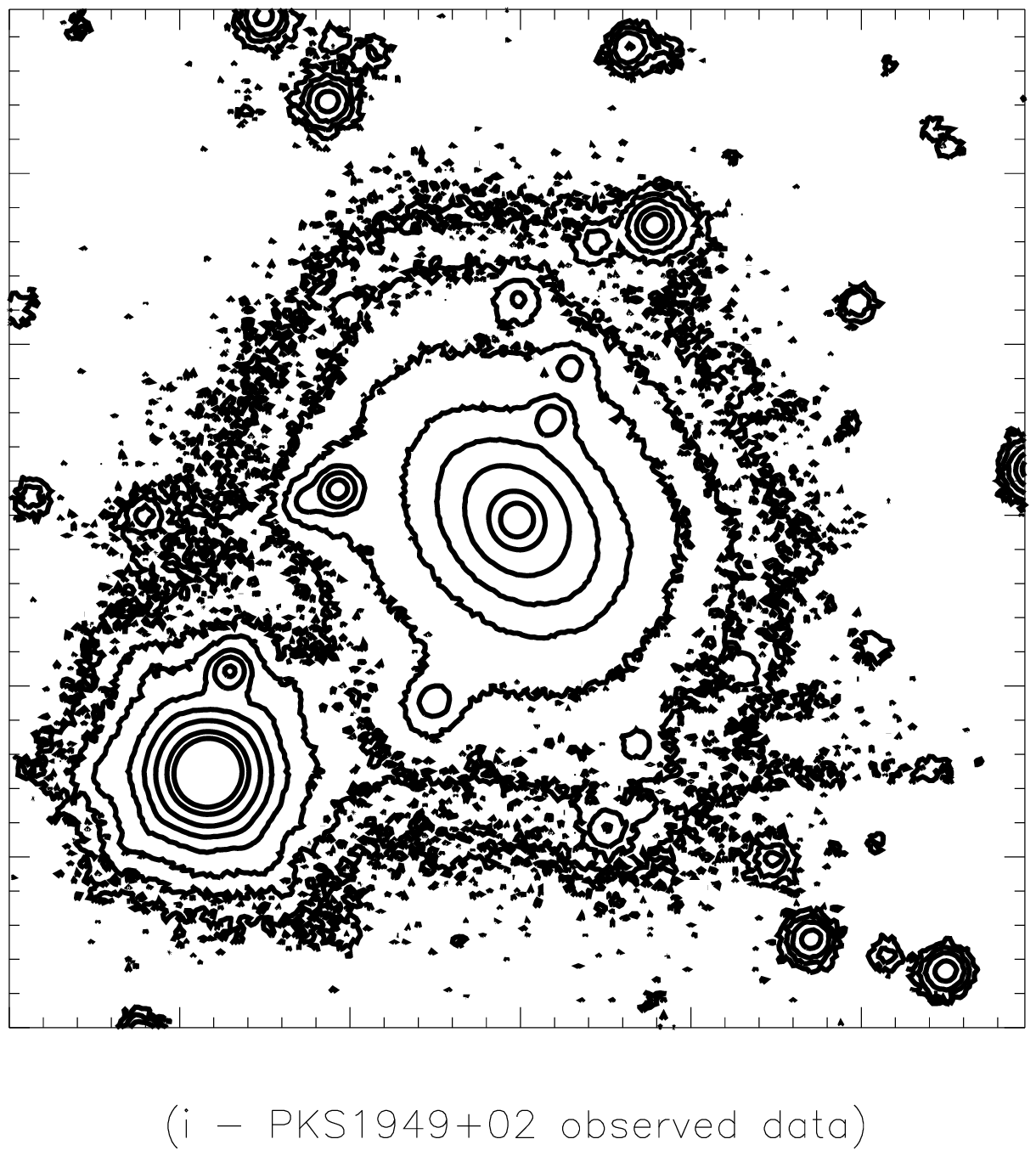}
\includegraphics{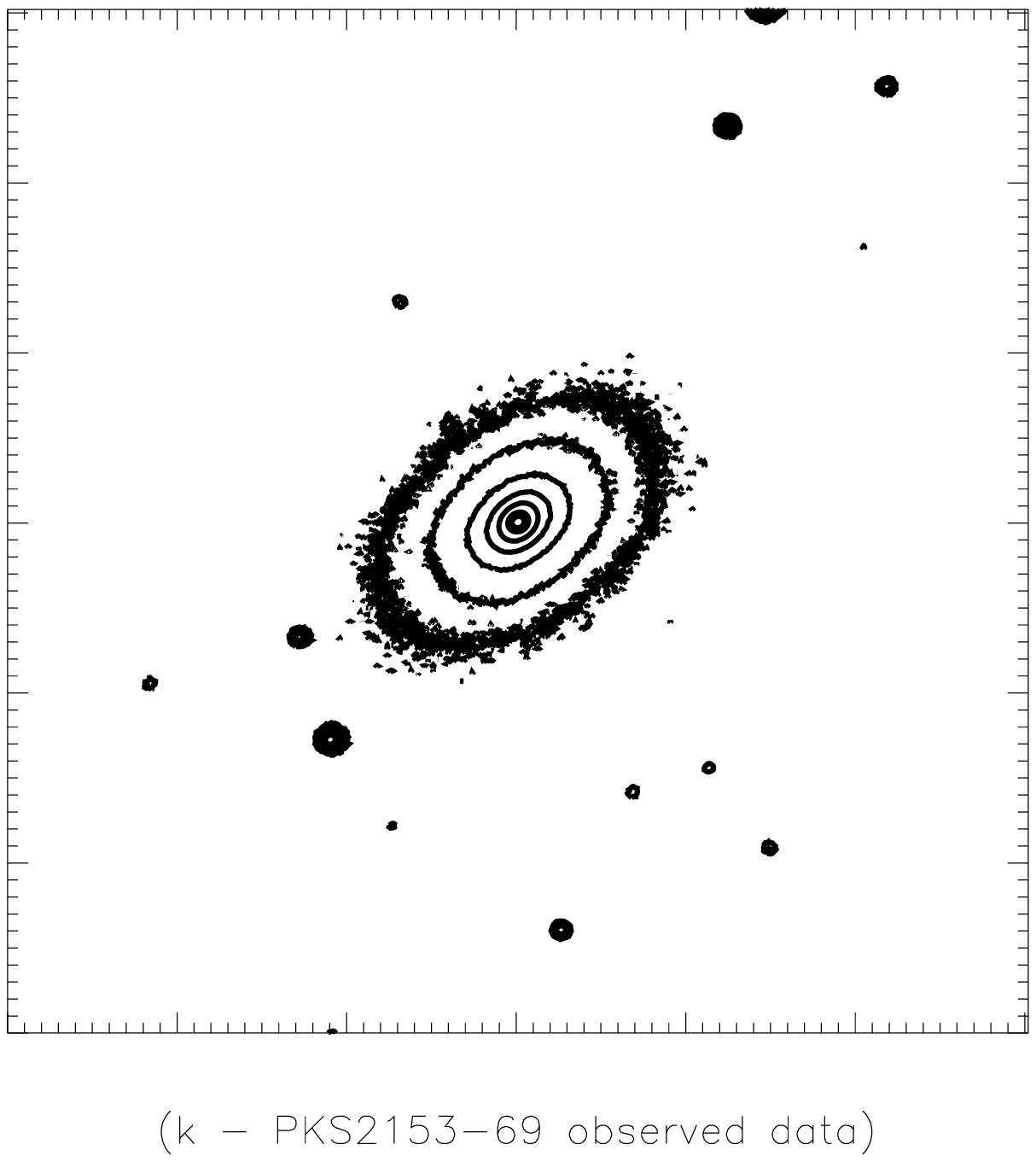}
\includegraphics{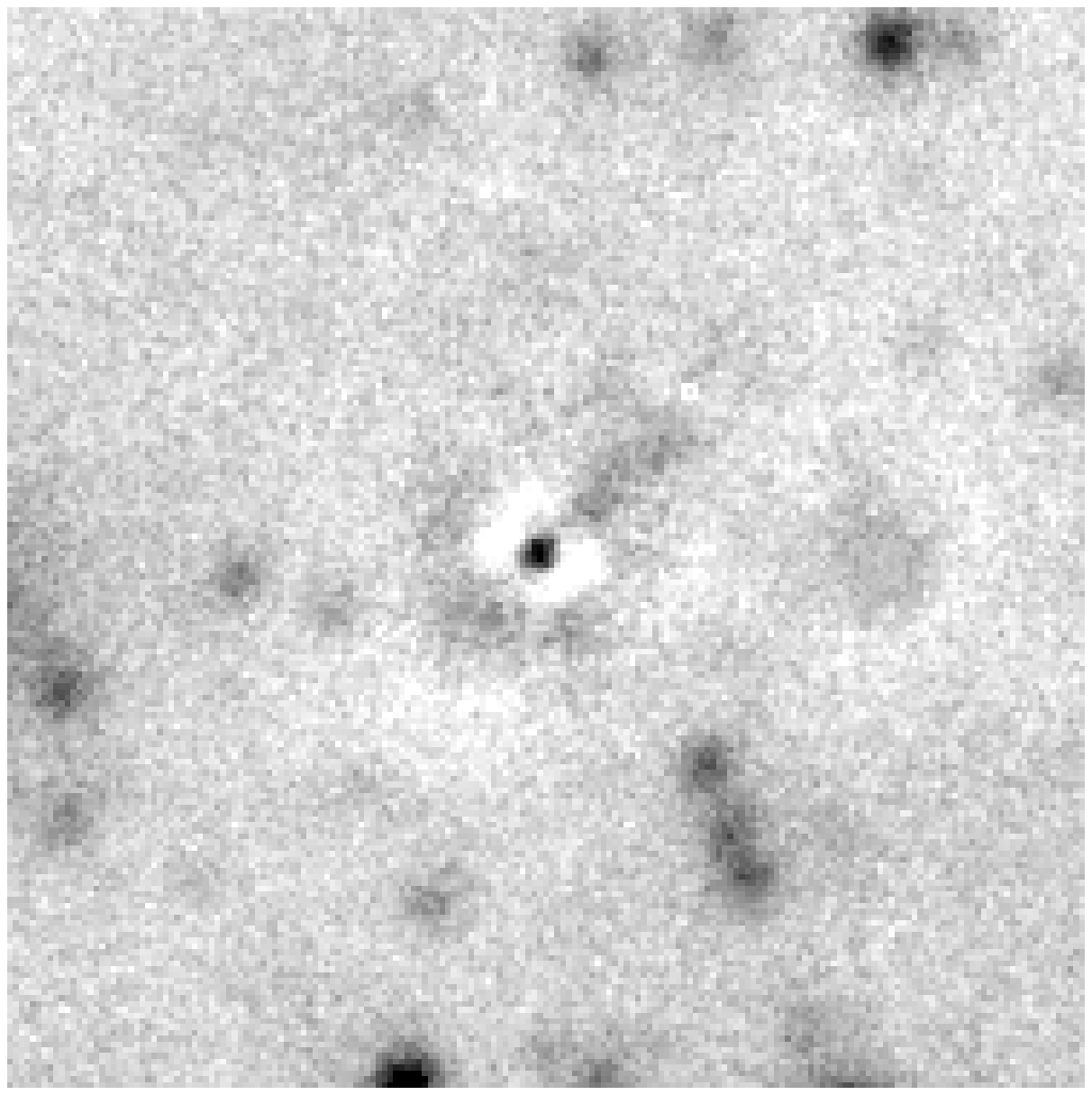}
\includegraphics{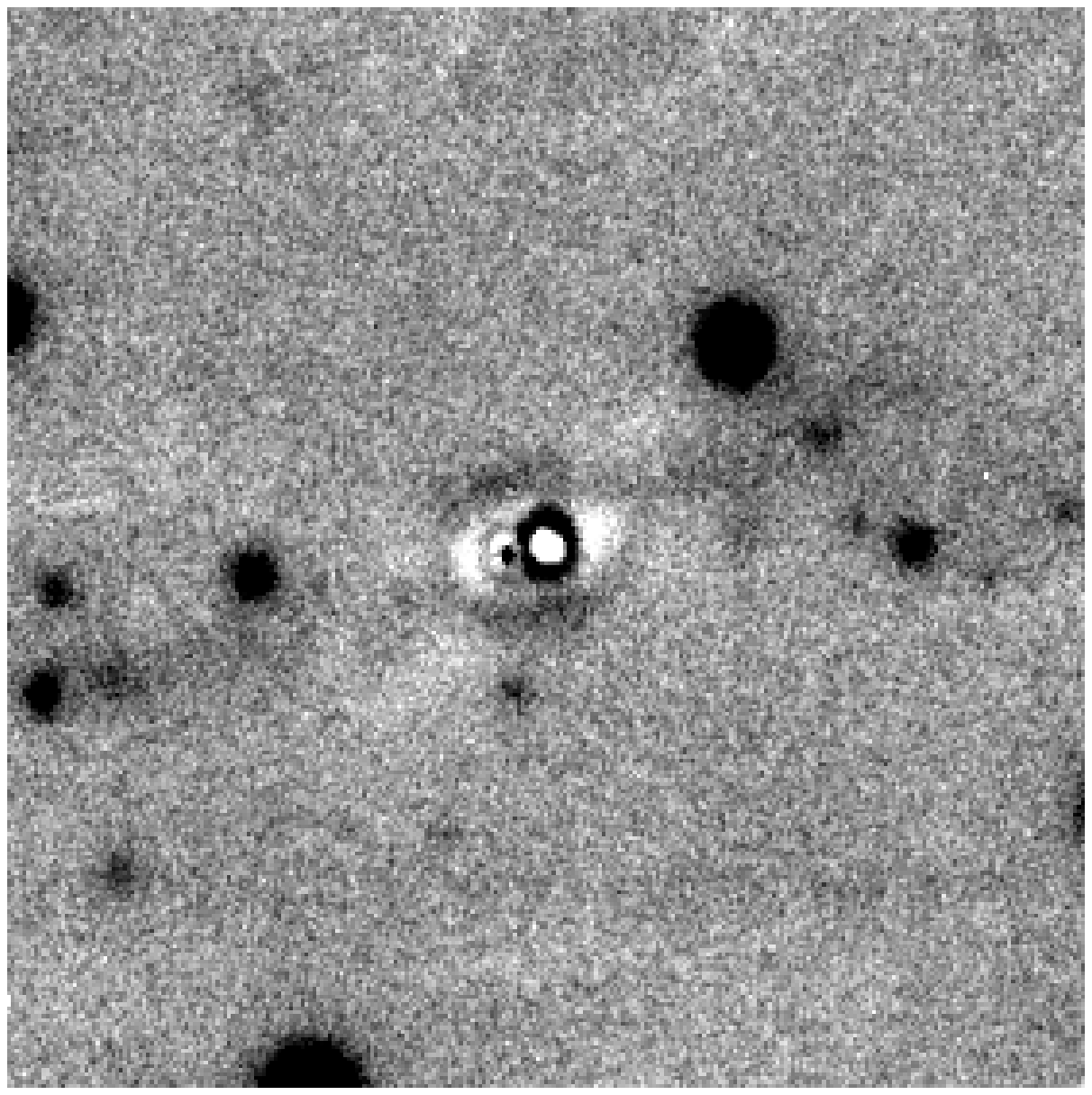}
\includegraphics{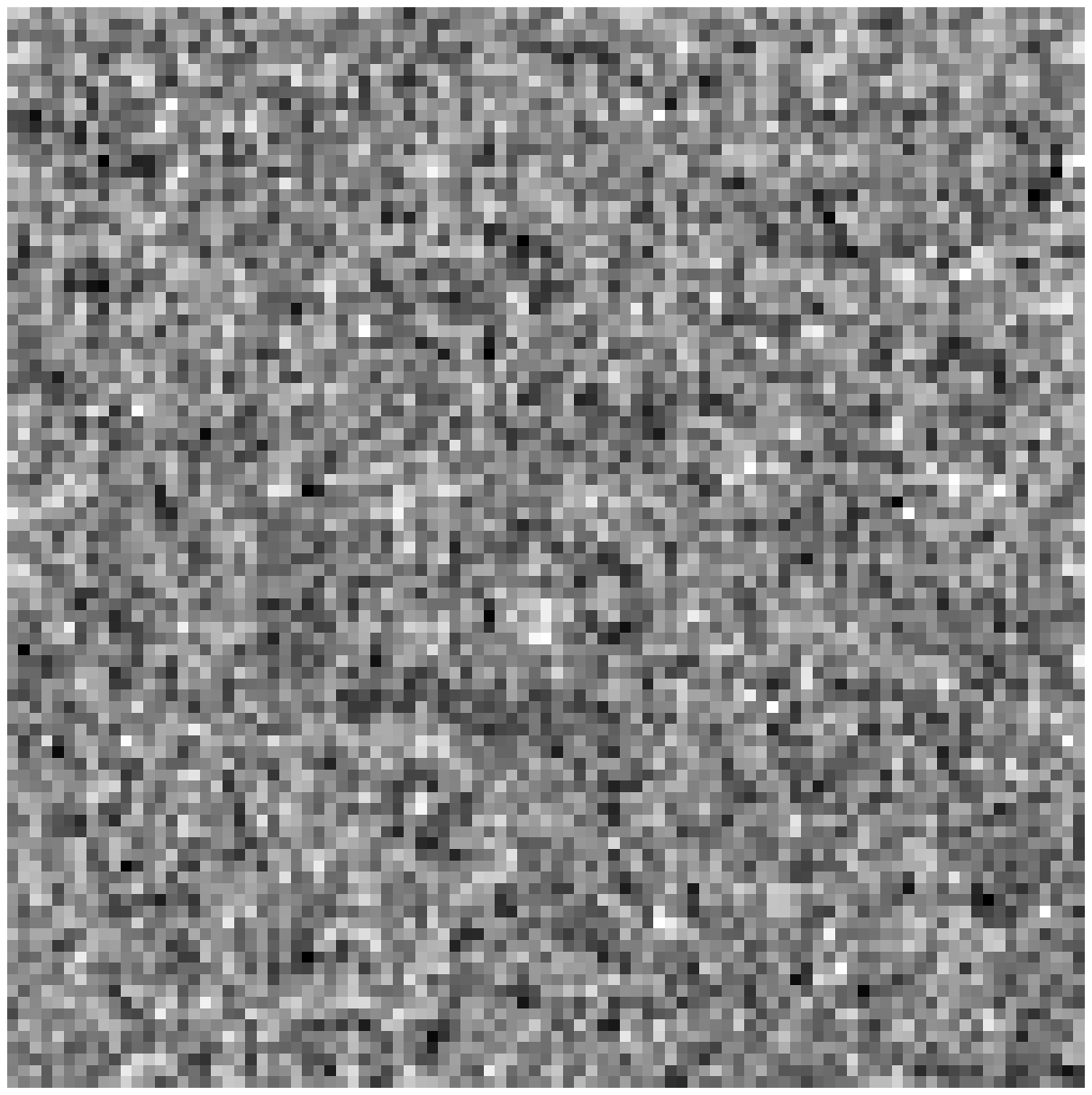}
\includegraphics{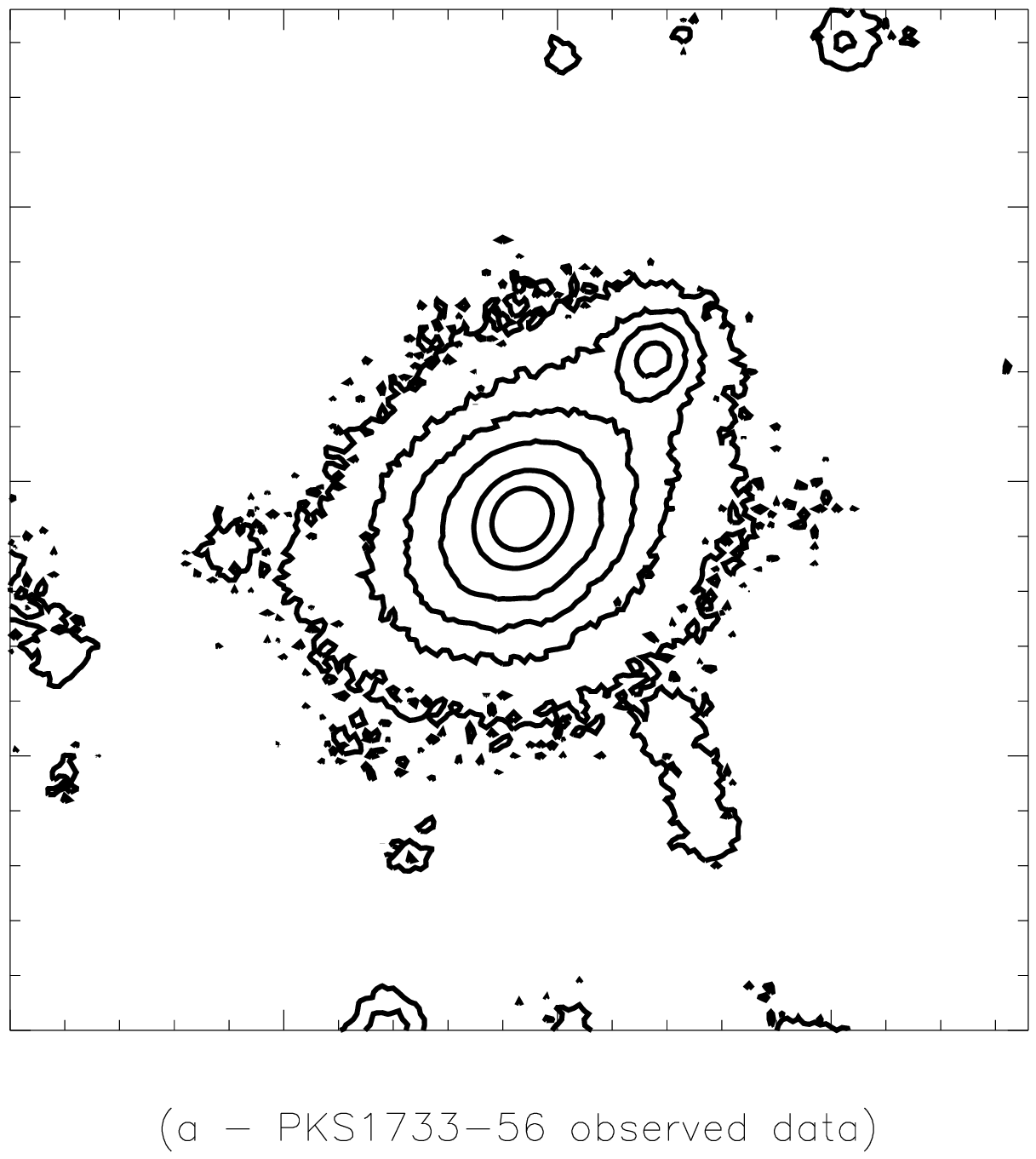}
\includegraphics{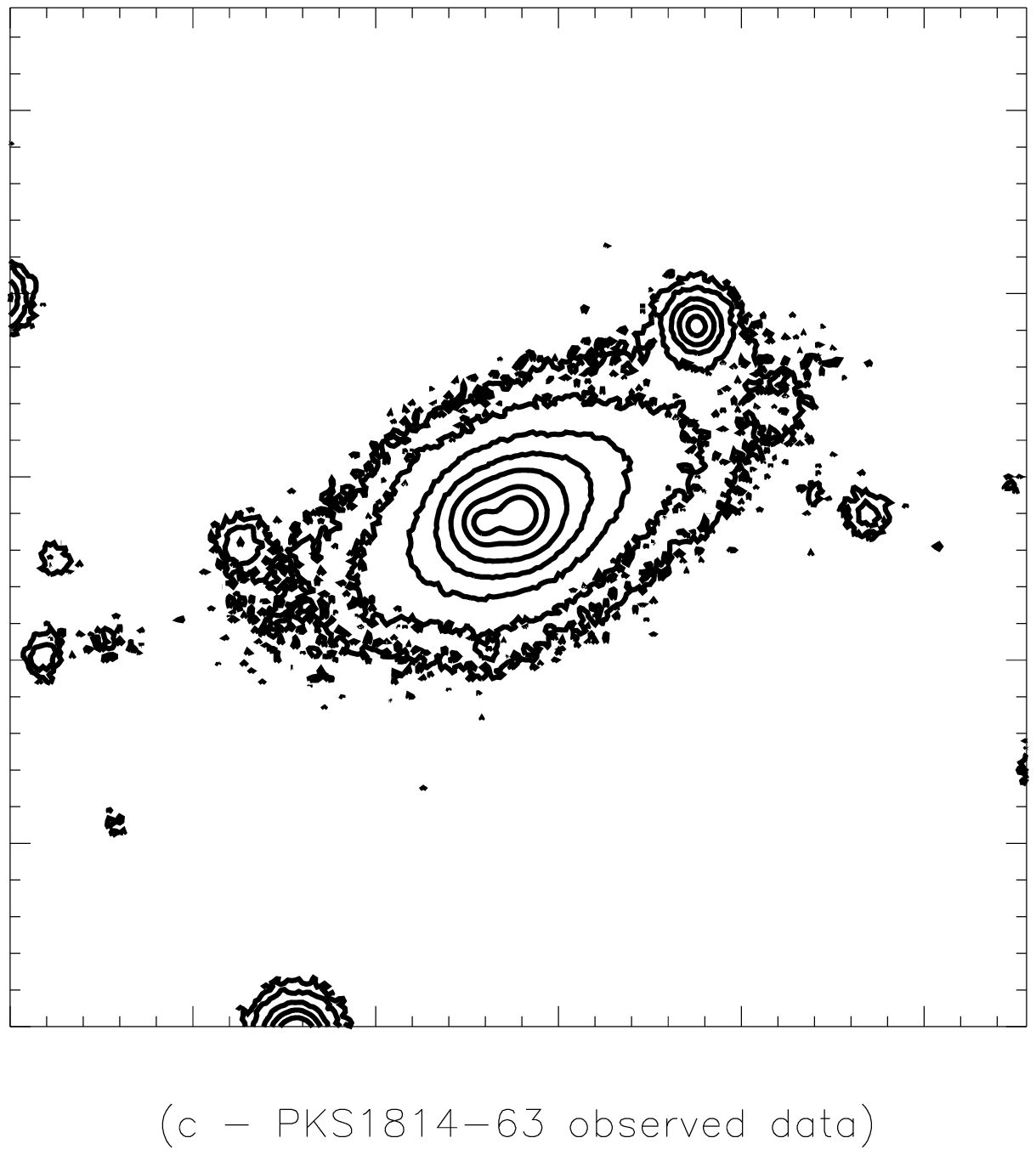}
\includegraphics{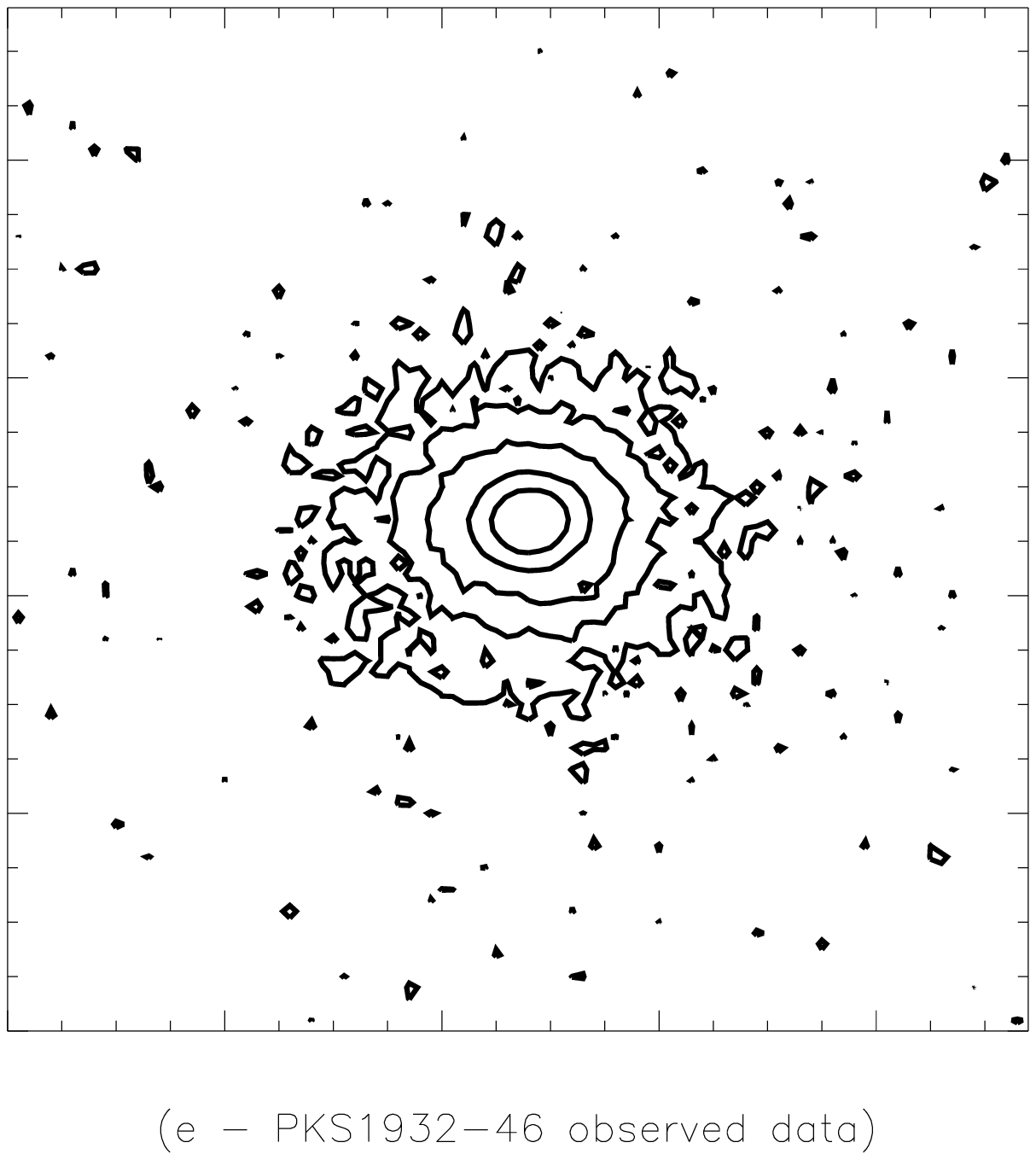}
\end{center}
\caption{50kpc by 50kpc images of PKS1733-56, PKS1814-63, PKS1932-46, PKS1934-63, PKS1949+02 and PKS2153-69. The observed data contours are displayed in frames (a), (c), (e), (g), (i) and (k), while frames (b), (d), (f), (h), (j) and (l) show the best-fit model contours on greyscale images of the model-subtracted residuals.  The maximum contour level is 50\% of the peak flux for that source in all cases, with subsequent contours at 25\%, 10\%, 5\%, 2.5\%, 1\%, 0.5\%, and 0.25\%, (latter flux levels not shown in all cases). The minimum contours displayed are at 0.25\% for PKS1949+02, 0.5\% for PKS1814-63 and PKS2153-69, 1\% for PKS1733-56, and 2.5\% for PKS1932-46 and PKS1934-63. 
\label{Fig: 7}}
\end{figure*}

\begin{figure*}
\vspace{8.45 in}
\begin{center}
\includegraphics{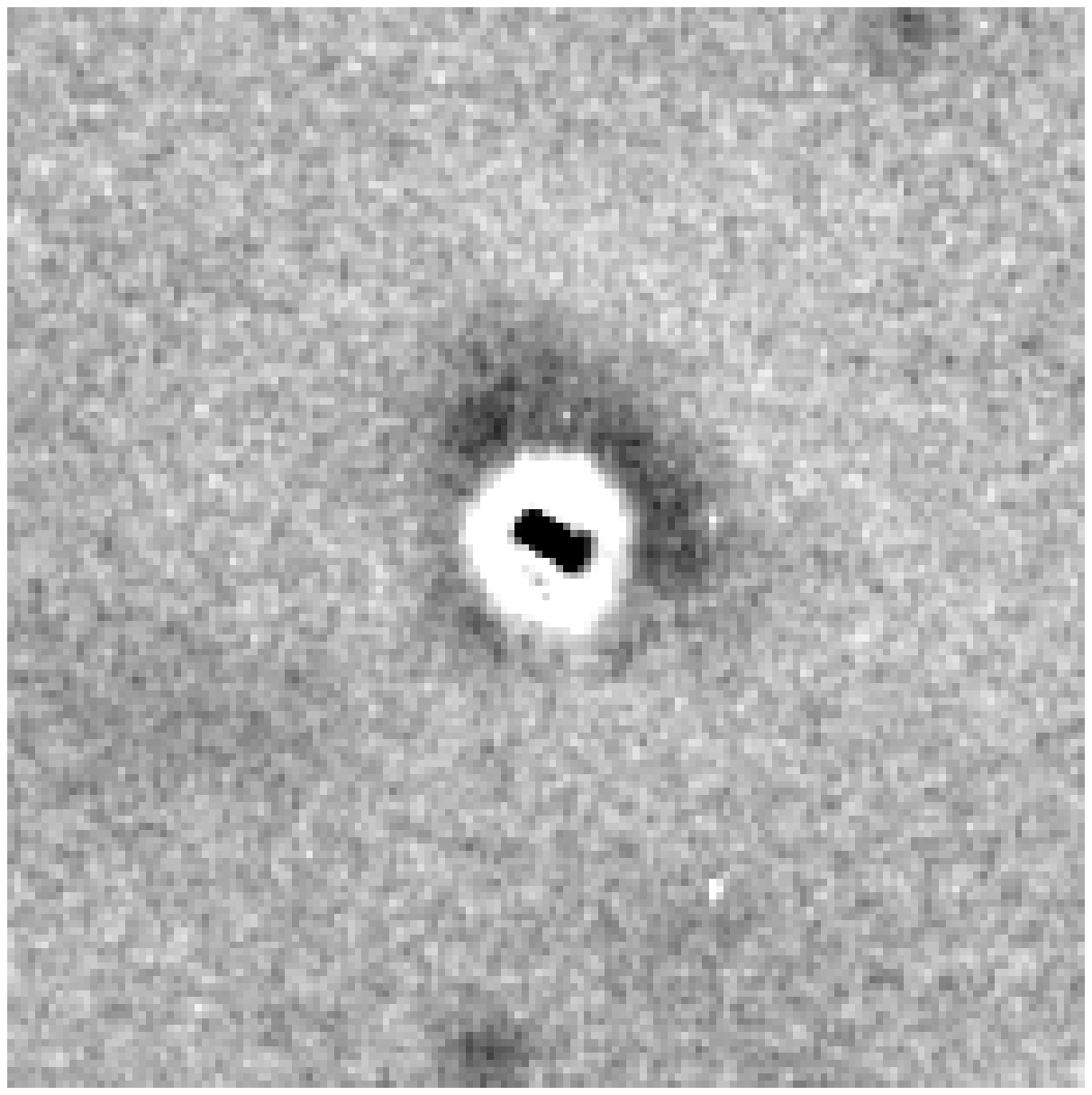}
\includegraphics{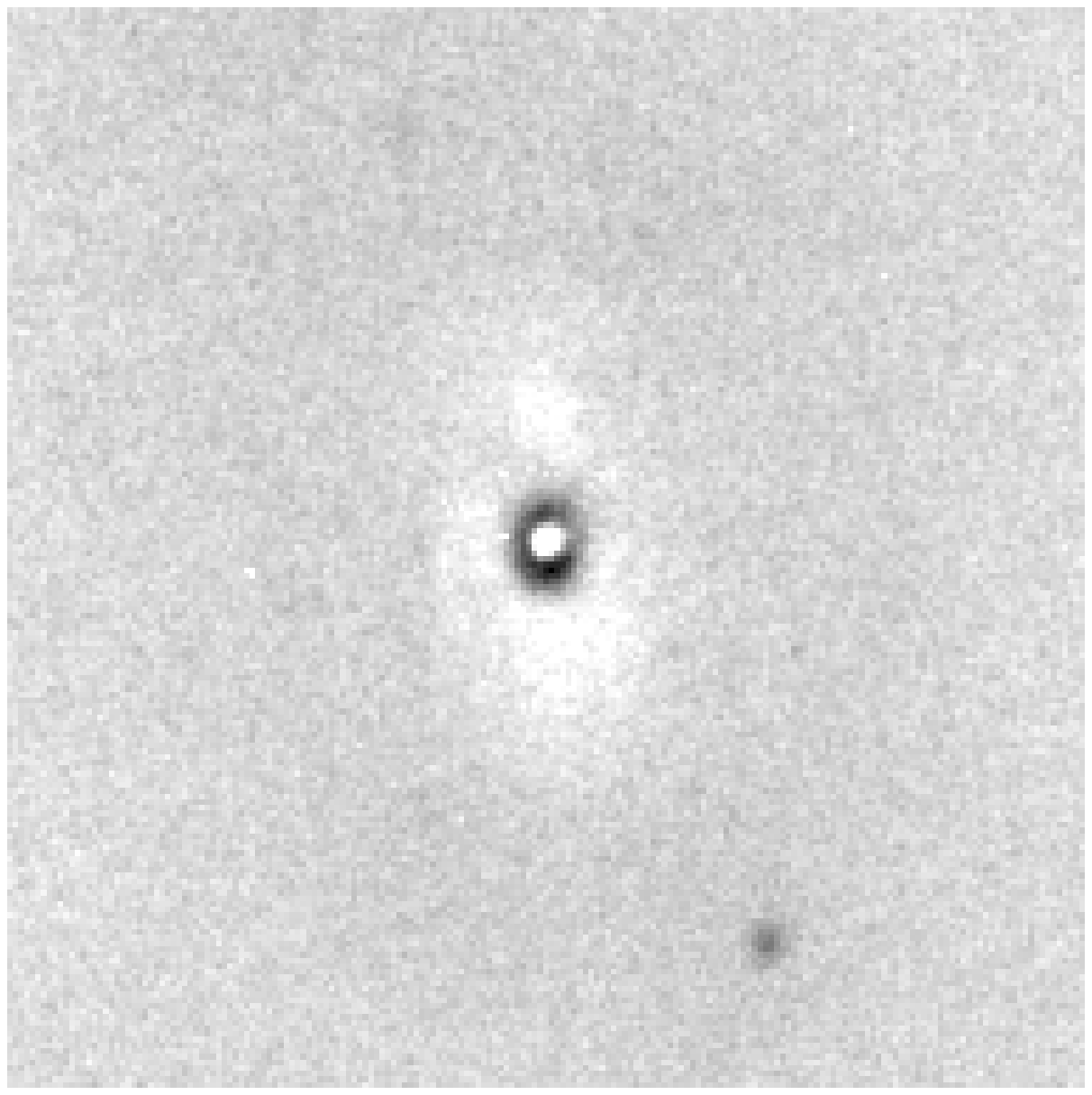}
\includegraphics{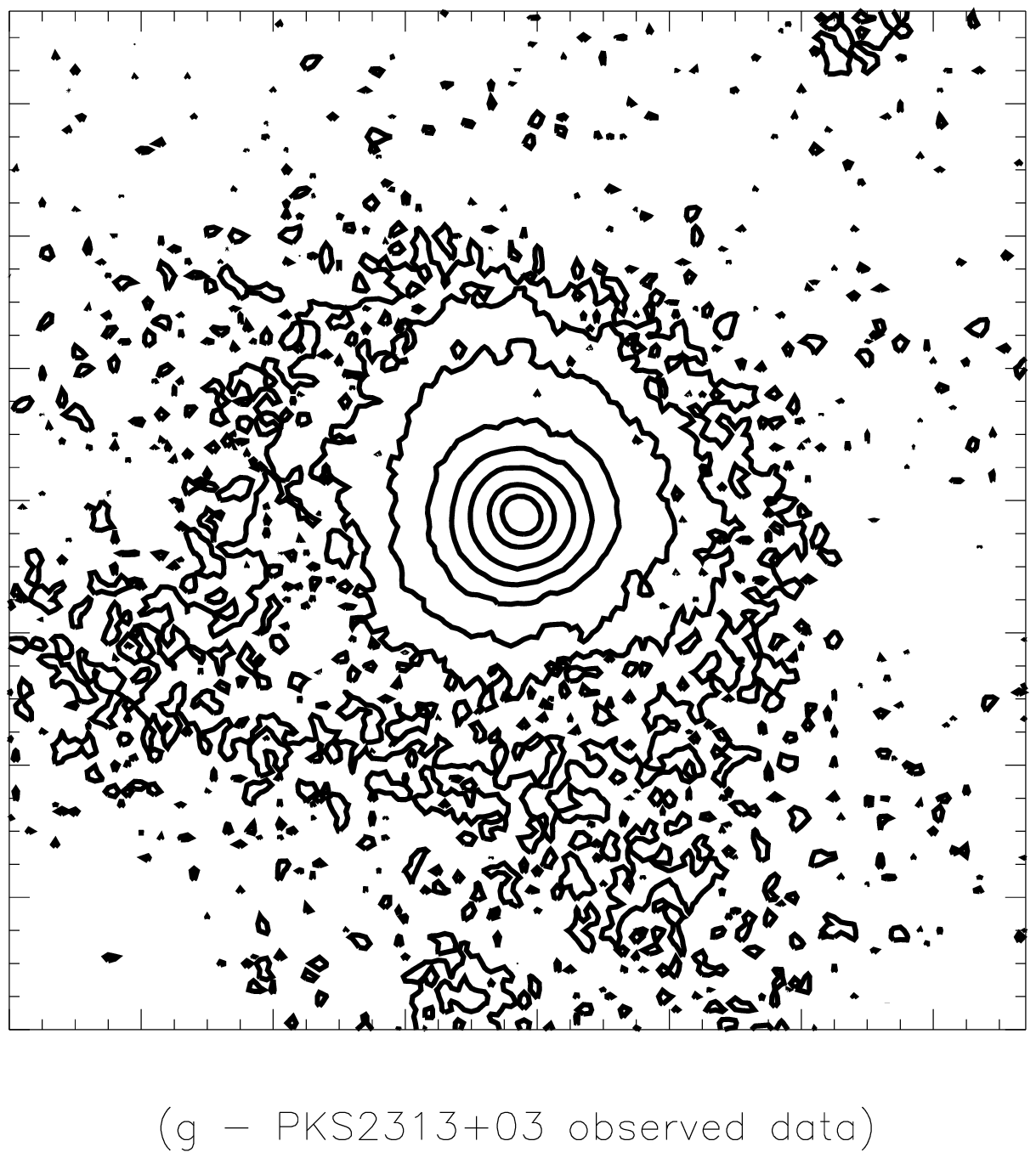}
\includegraphics{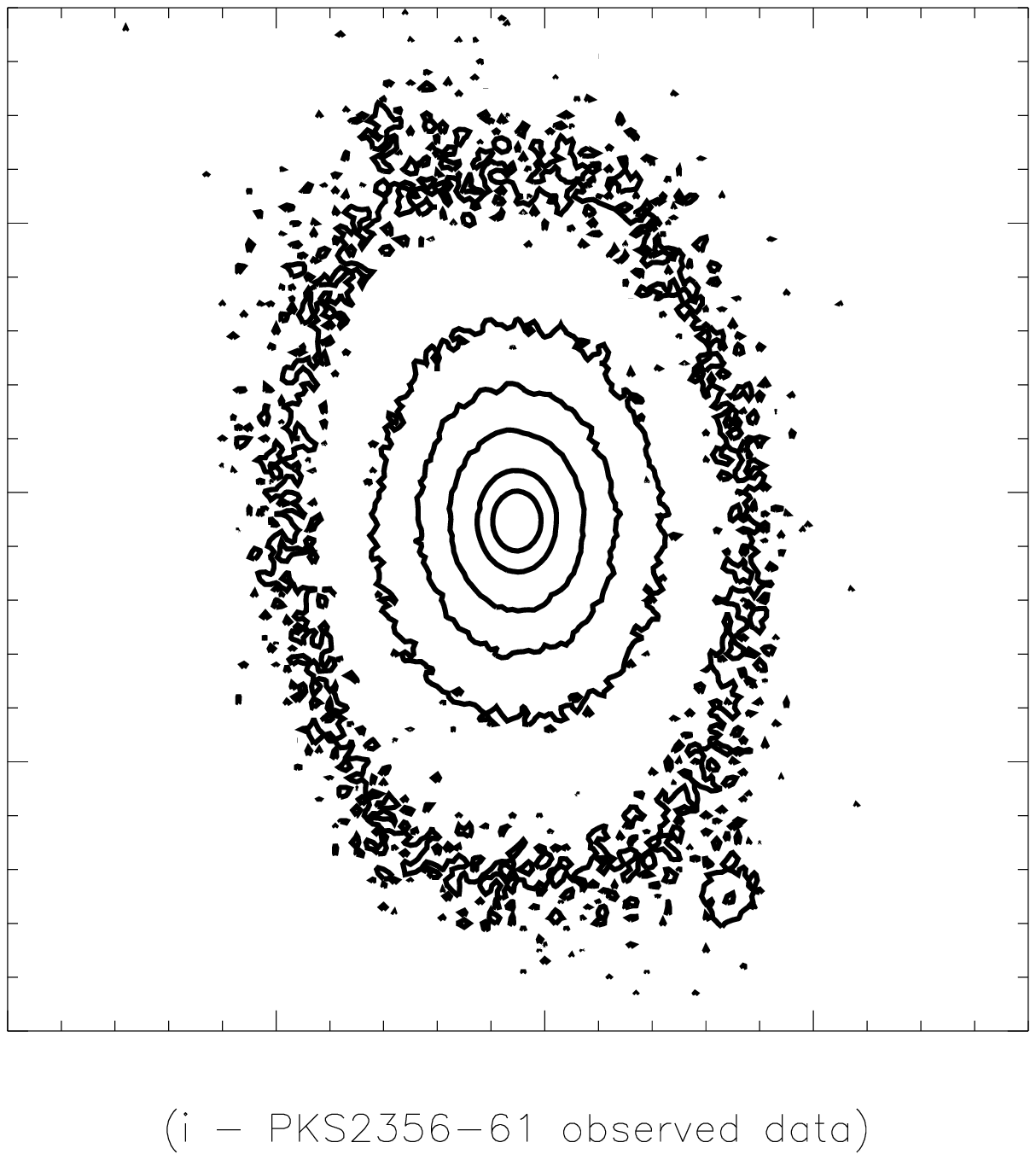}
\includegraphics{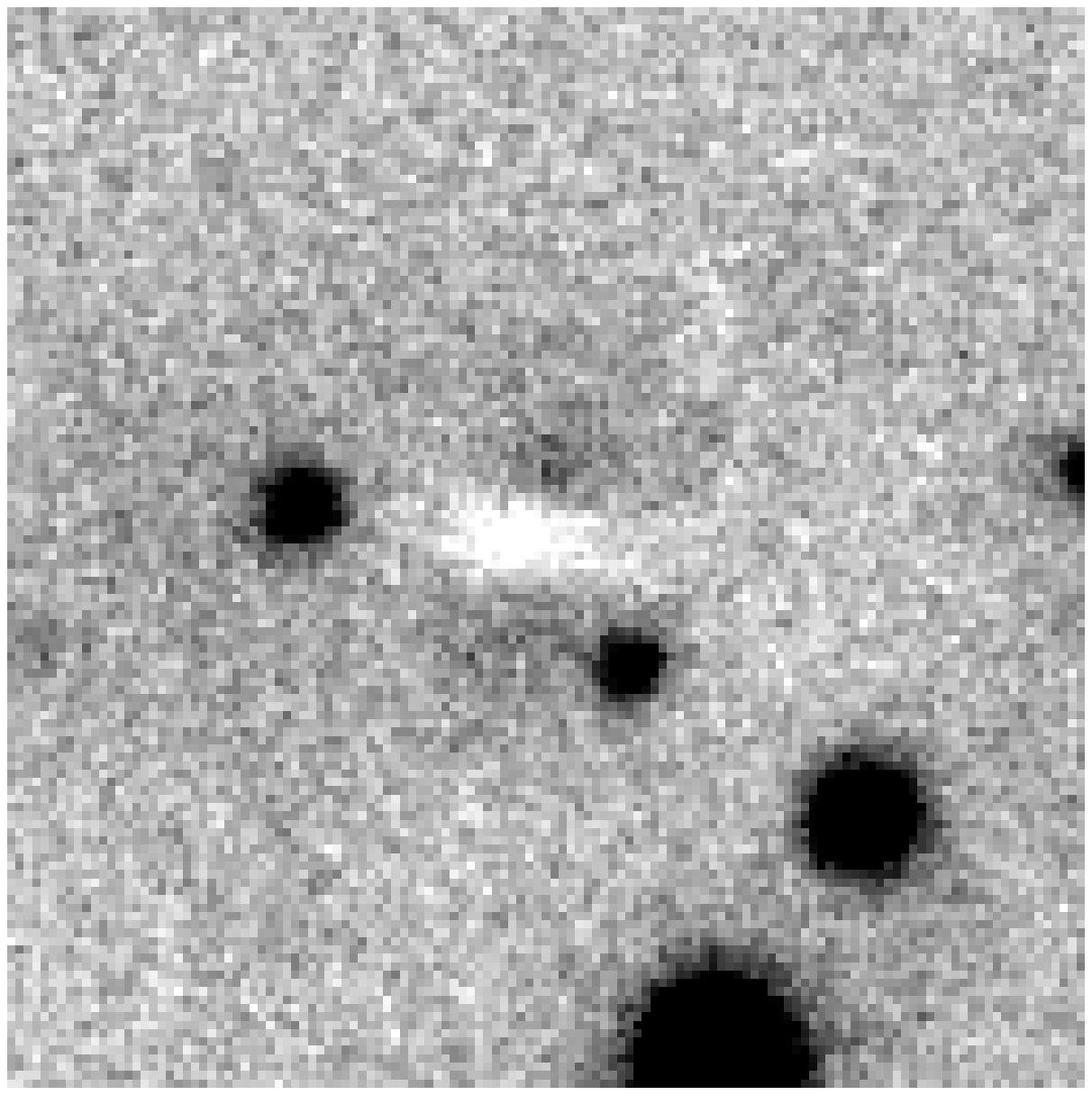}
\includegraphics{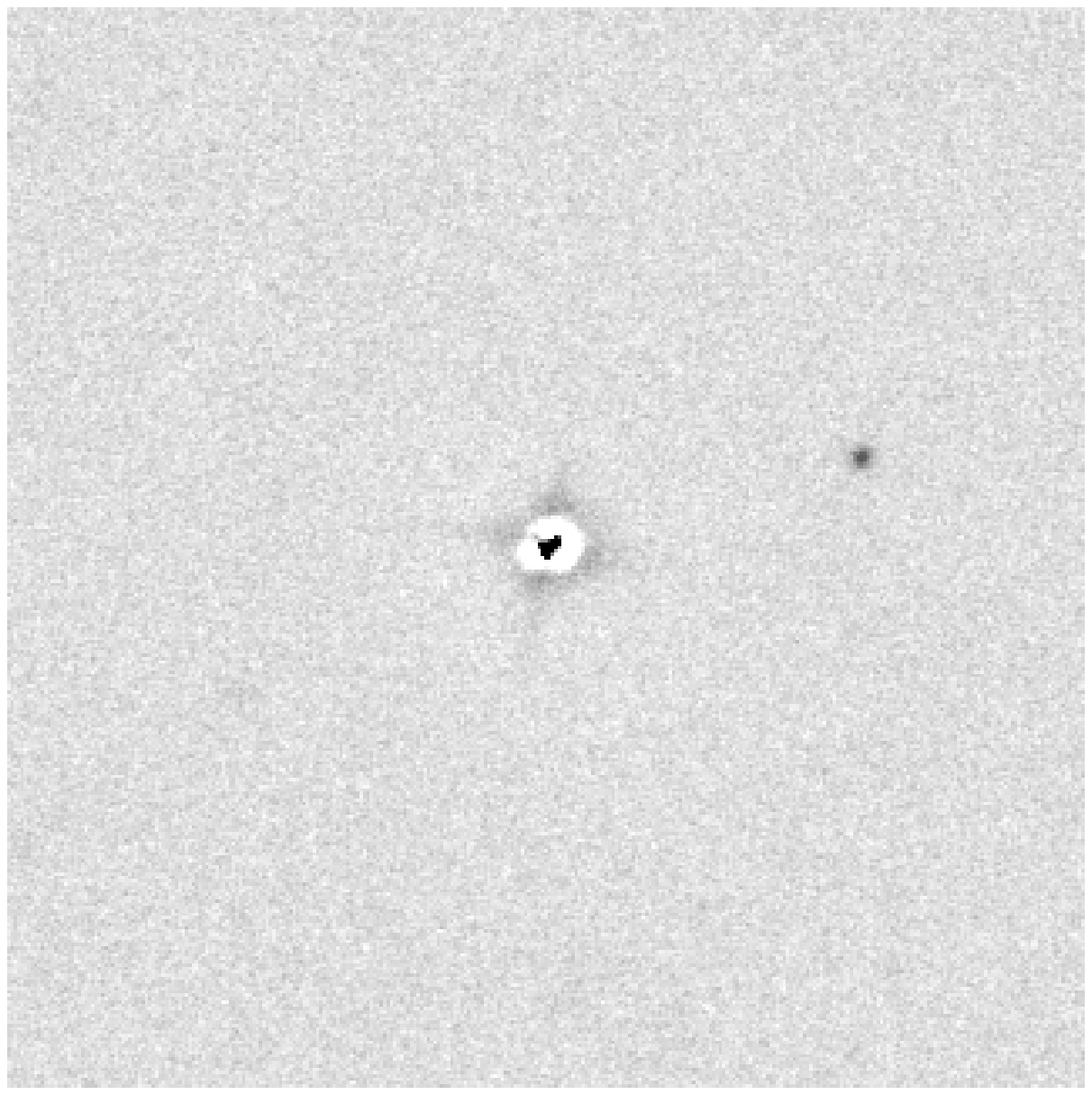}
\includegraphics{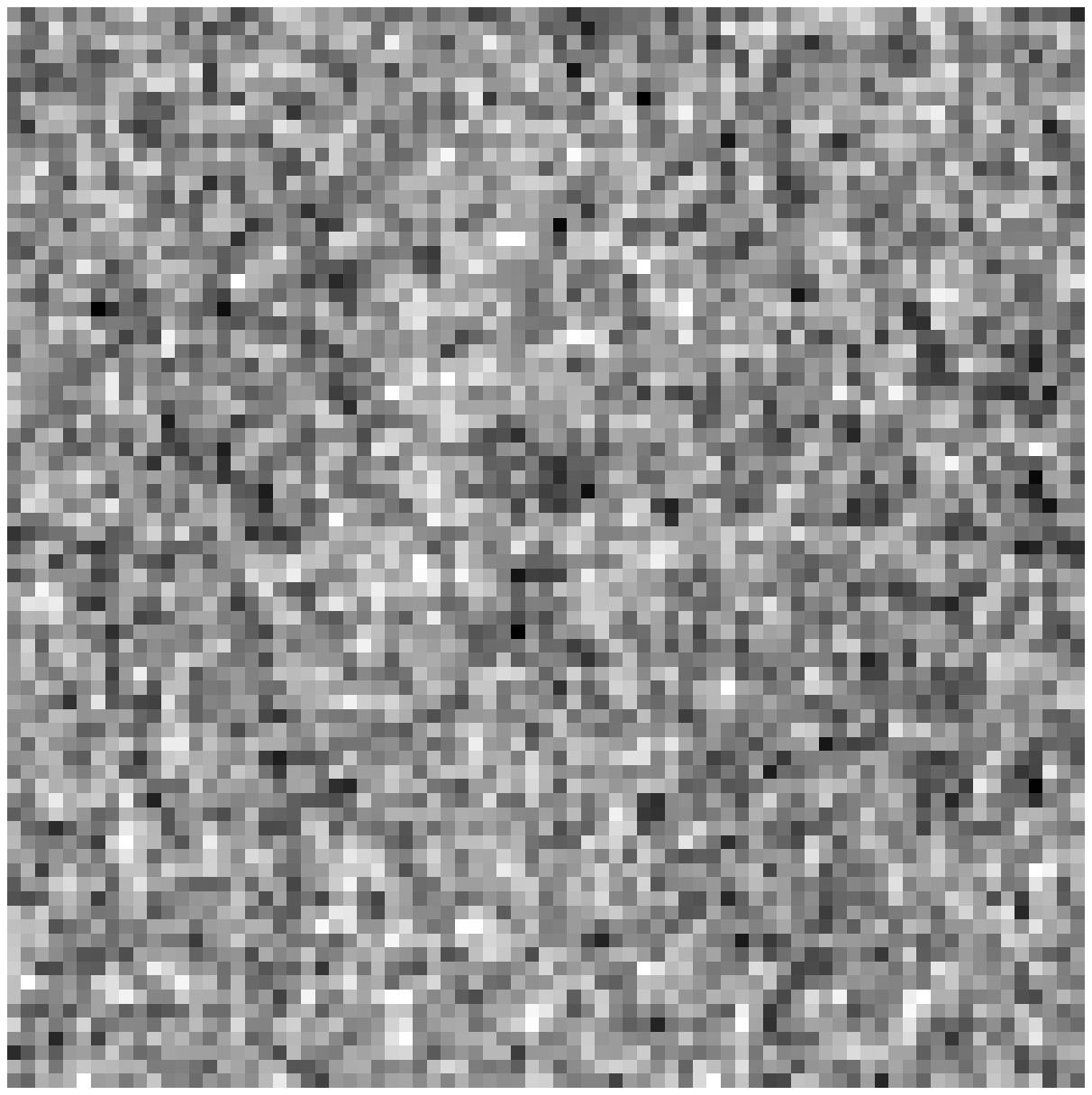}
\includegraphics{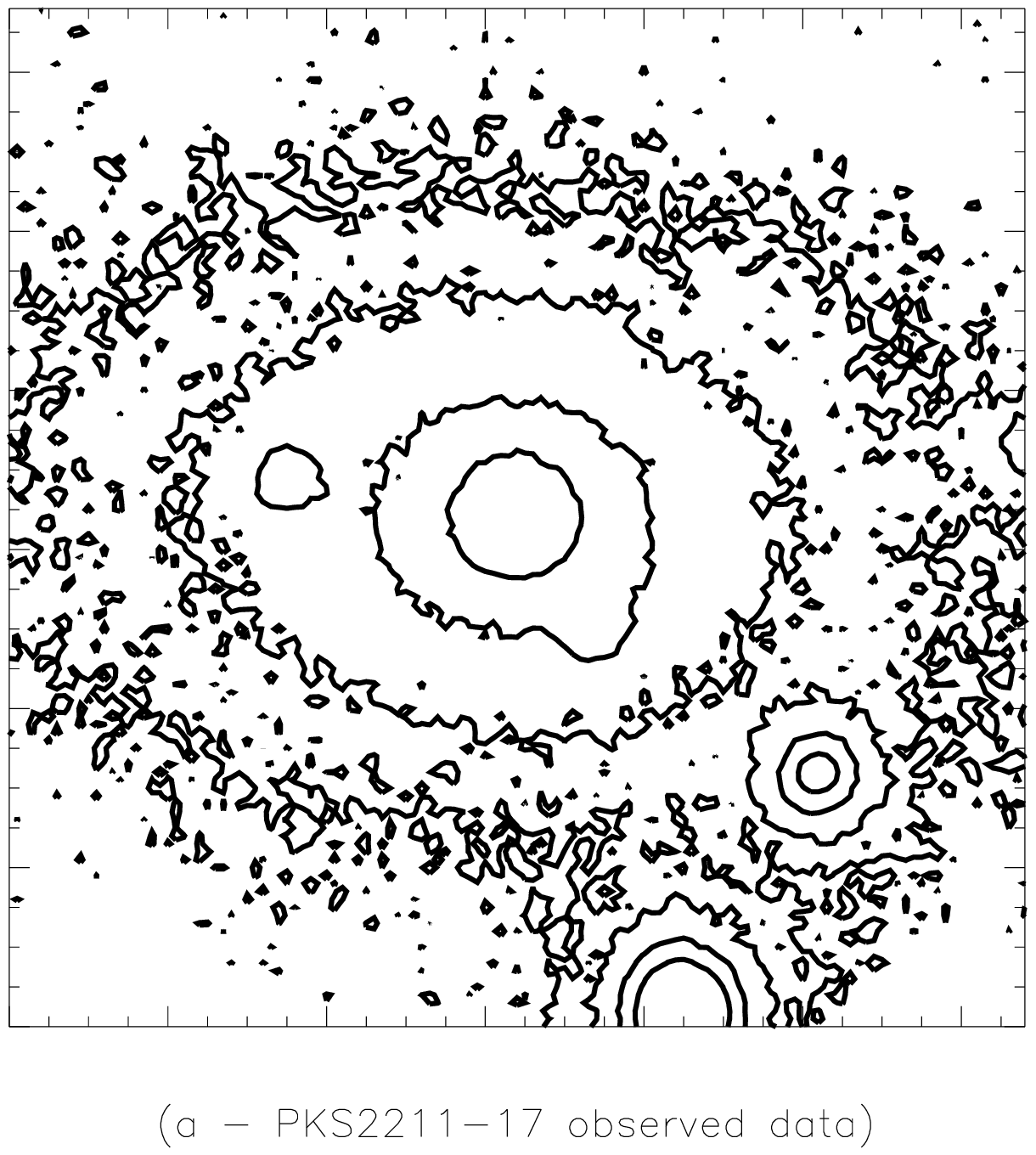}
\includegraphics{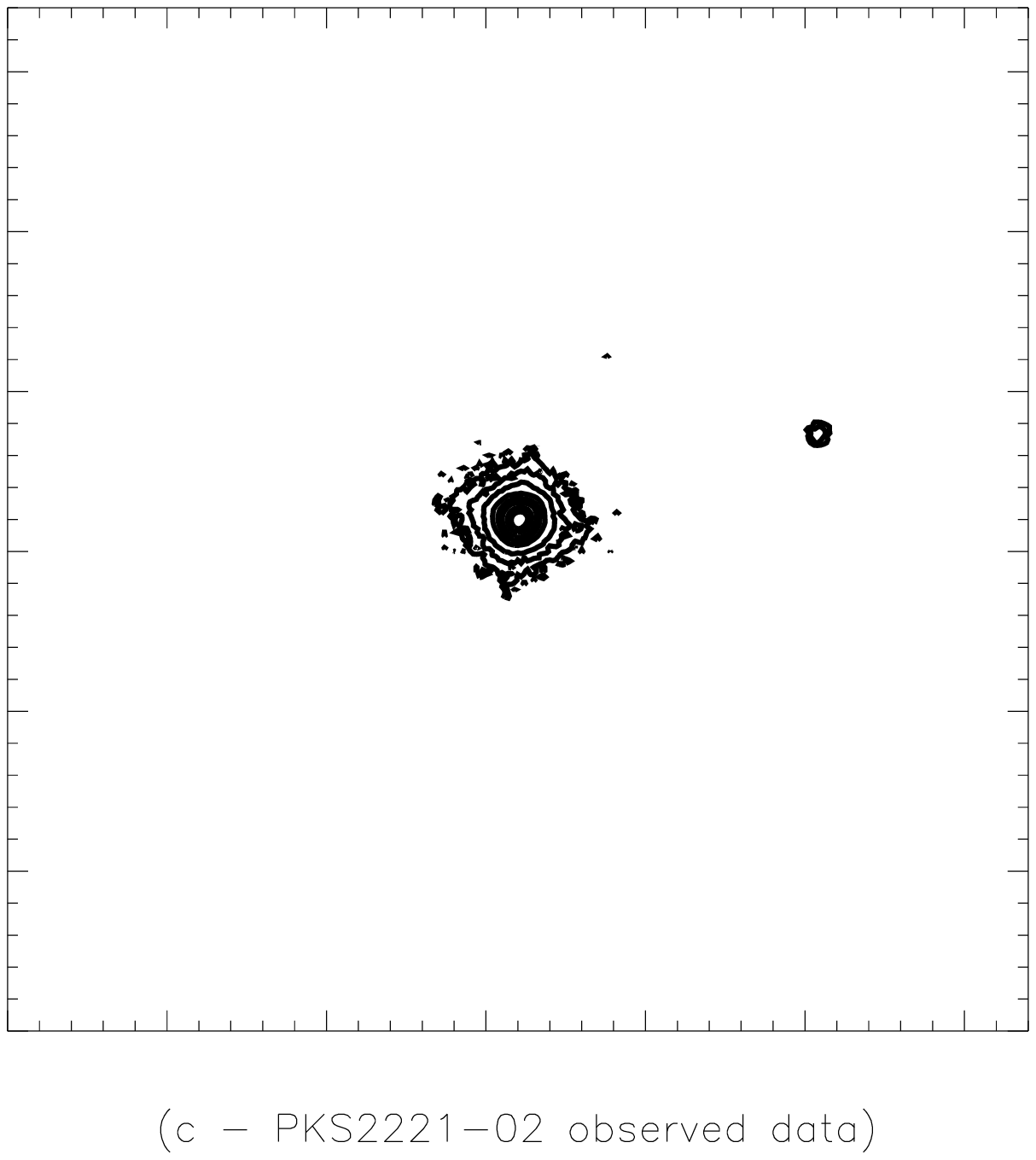}
\includegraphics{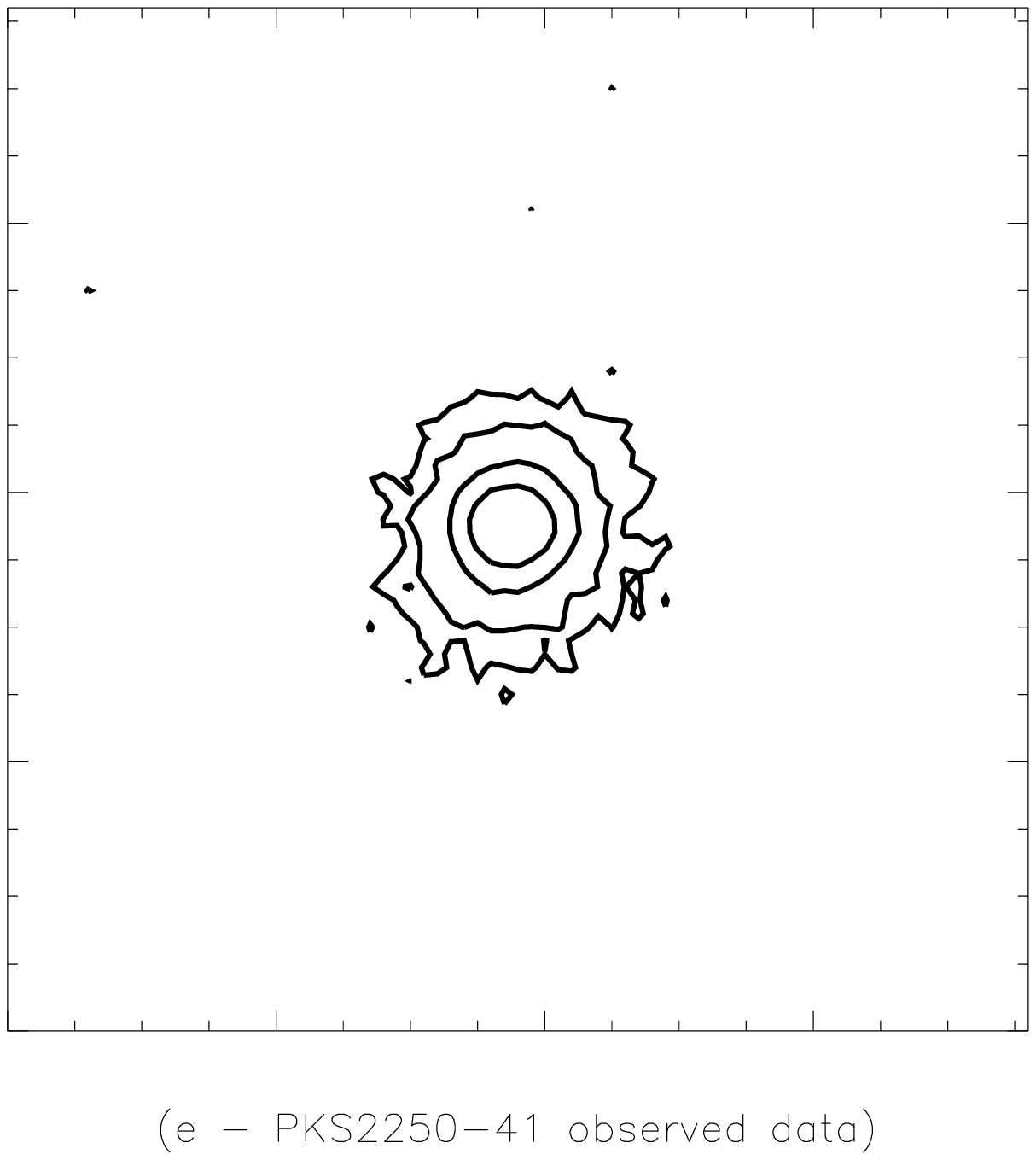}
\end{center}
\caption{50kpc by 50kpc images of PKS2211-17, PKS2221-02, PKS2250-41, PKS2313+03 and PKS2356-61. The observed data contours are displayed in frames (a), (c), (e), (g), and (i), while frames (b), (d), (f), (h), and (j) show the best-fit model contours on greyscale images of the model-subtracted residuals.  The maximum contour level is 50\% of the peak flux for that source in all cases, with subsequent contours at 25\%, 10\%, 5\%, 2.5\%, 1\%, 0.5\%, and 0.25\%, (latter flux levels not shown in all cases). The minimum contours displayed are at 0.25\% for PKS2221-02 and PKS2313+03, 1\% for PKS2356-61, and 5\% for PKS2211-17 and PKS2250-41.
\label{Fig: 8}}
\end{figure*}

It is possible that this source has undergone merger activity in the relatively recent past.  The X-shaped radio source has been linked to rapid jet reorientation $\sim 10^7$ years previously (Dennett-Thorpe  et al 2002).  On the basis of optical HST data, Martel et al (1999) note that this source displays dust lanes, and that the central elliptical region is surrounded by a low-surface brightness halo with a sharp boundary, while de Koff et al (2000) observe extensive dust lane features. Shell structures in the outer envelope due to dust features are also observed in ground-based Gemini imaging of this source (RA10); we see hints of these features in our model-subtracted residual image.

\subsubsection{PKS2153-69 (ESO075-G041)}
PKS2153-69 has been previous studied by Tadhunter et al (1988) and Fosbury et al (1998).  Although Tadhunter et al classify this source as a BLRG on the basis of its optical spectrum, the nuclear point source contribution derived from our modelling is relatively low, with a best-fit value of 4.4\% for an assumed S\'ersic index of $n=4$, and an effective radius of 8kpc/14.4$^{\prime\prime}$. However, while low relative to the {\it total} galaxy flux, the nuclear point source is still a significant contributor to the flux from the nuclear regions of the galaxy. A better fit is obtained using the Donzelli-style bulge+disk model, which results in a nuclear point source contribution of $\sim 4$\% of the total flux, a bulge with $R_{eff} \sim 12$ kpc and a disk component with $R_{eff} \sim 1$kpc contributing $\sim 4$\% of the {\it total} galaxy flux.  The major-axis dust lane  visible in our model residuals has also been detected in other observations of this source.

\subsubsection{PKS2211-17 (3C444)}

PKS2211-17 (3C444) is a well-know central cluster galaxy and weak-line radio galaxy. Our best fit model for this galaxy is a de Vaucouleurs elliptical with an effective radius of $r_{eff} = 50$kpc (19$^{\prime\prime}$) and no detectable nuclear point source component. This is well matched by the value found by Smith \& Heckman (1989) in the optical ($r_{eff} = 42$kpc after correction for the different assumed cosmological parameters).  This source appears to have a number of small satellites (also modelled, although these model components are not included in our model contour plot in Fig.~\ref{Fig: 8}a), each of approximately 18th magnitude or less in the $K-$band, and our residuals suggest the presence of a major-axis dust lane and excess diffuse flux at large radii, particularly on the eastern side of the galaxy.

\subsubsection{PKS2221-02 (3C445)}
This BLRG is dominated by its nuclear point source, which contributes approximately two thirds of the $K$-band flux. Our best-fit host galaxy is a de Vaucouleurs elliptical with $r_{eff} \sim 7.5$kpc ($7^{\prime\prime}$), in good agreement with the value obtained by Govoni et al (2000) in the optical.

\subsubsection{PKS2250-41}
PKS2250-41 is well fit by a de Vaucouleurs elliptical galaxy with a point source contribution of $\sim17$\% of the total flux, and $r_{eff} \sim 5$kpc (1.1$^{\prime\prime}$). The current modelling results, obtained using \textsc{galfit}, are very close to those obtained by Inskip et al (2008) for this source using a different least-squares minimisation modelling process.  Changing the S\'ersic index $n$ does not greatly alter the value of $r_{eff}$ obtained.

\subsubsection{PKS2313+03 (3C459)}

PKS2313+03 is an extreme starburst radio galaxy and ULIRG whose stellar population is dominated by young stars (Tadhunter et al 2002; Wills et al 2008).  Previous studies of this source  have derived a variety of different galaxy sizes: optical HST imaging with the F814W ($\sim I-$band) filter was used by Zheng et al (1999) to derive a (cosmology-corrected) value of $r_{eff}=3.9$kpc (excluding a nuclear point source component), while Donzelli et al (2007) use a combination of bulge (with S\'ersic index $n \sim 2$) and disk components of approximately equal flux for their NICMOS data, and derive $r_{eff,bulge}=0.81$kpc. 
For our de Vaucouleurs model without a nuclear point source, we derive a value of $r_{eff}=1.23$kpc (0.35$^{\prime\prime}$), comparable to that obtained with NICMOS. Although this effective radius is smaller than the FWHM of the PSF with which the model is convolved in the fitting process, even values as low as this can be reliably extracted, provided they are still larger than the pixel scale (as is the case for this source). However, a strong nuclear point source is a preferred component in our modelling, and for our best fit model we obtain a  nuclear point source contributing $\sim 36$\% of the total flux for a galaxy with a de Vaucouleurs effective radius of $r_{eff}=5.7^{\prime\prime}$/20kpc.   Our model tends towards a much larger galaxy size, due to the extensive diffuse emission surrounding the source, the larger $n=4$ S\'ersic index used, and potentially our inability to resolve the innermost regions of the host galaxy to a sufficiently high degree of accuracy and the fact that the surrounding diffuse emission has {\it not} been resolved out in our ground-based data. Fits combining disk, bulge and point source components are very degenerate in the results obtained, and do not provide a better fit to the $K-$band emission in our data. In general,  the fitting difficulties for this object are due to the fact that it is clearly undergoing a major merger, similar to PKS1549-79.

\subsubsection{PKS2356-61}

PKS2356-61 is well fit by a standard de Vaucouleurs law, with a fairly typical effective radius for a massive elliptical of $\sim$10kpc (5.6$^{\prime\prime}$), and a very minor nuclear point source contribution (0-3\%).  The model residuals suggest the possible presence of a major-axis dust lane.

\section{Discussion}
\begin{figure*}
\vspace{4.5 in}
\begin{center}
\includegraphics{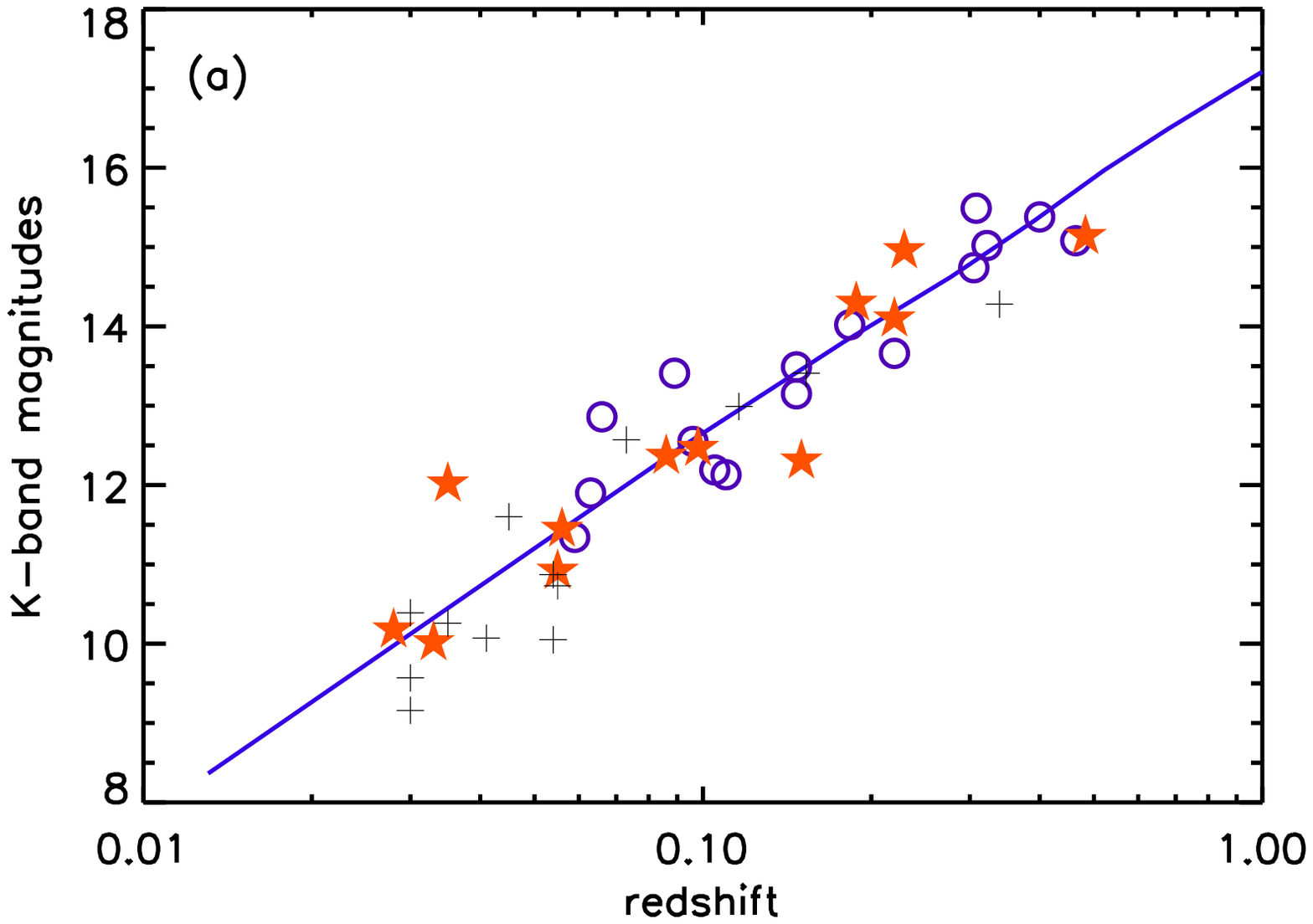}
\includegraphics{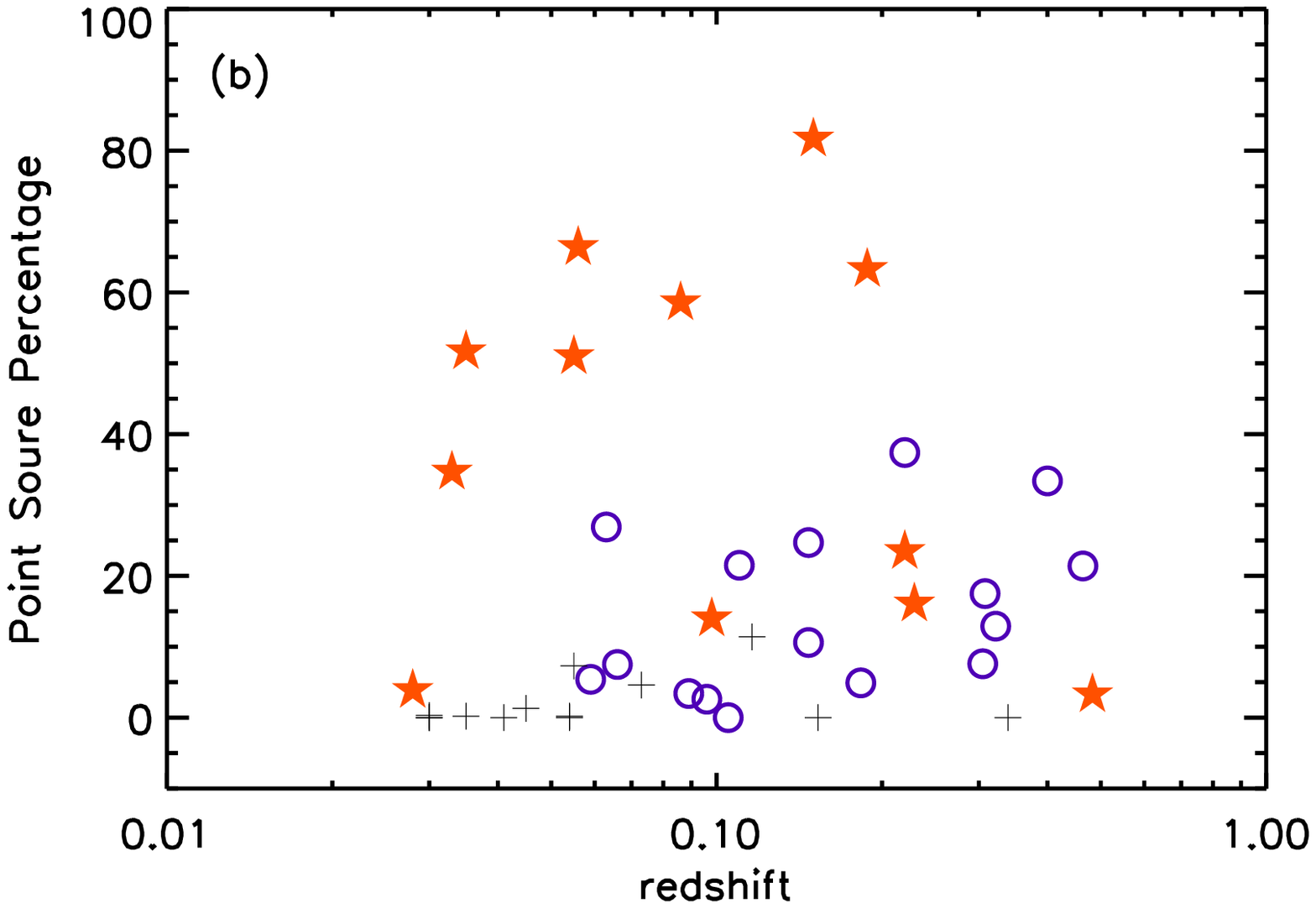}
\includegraphics{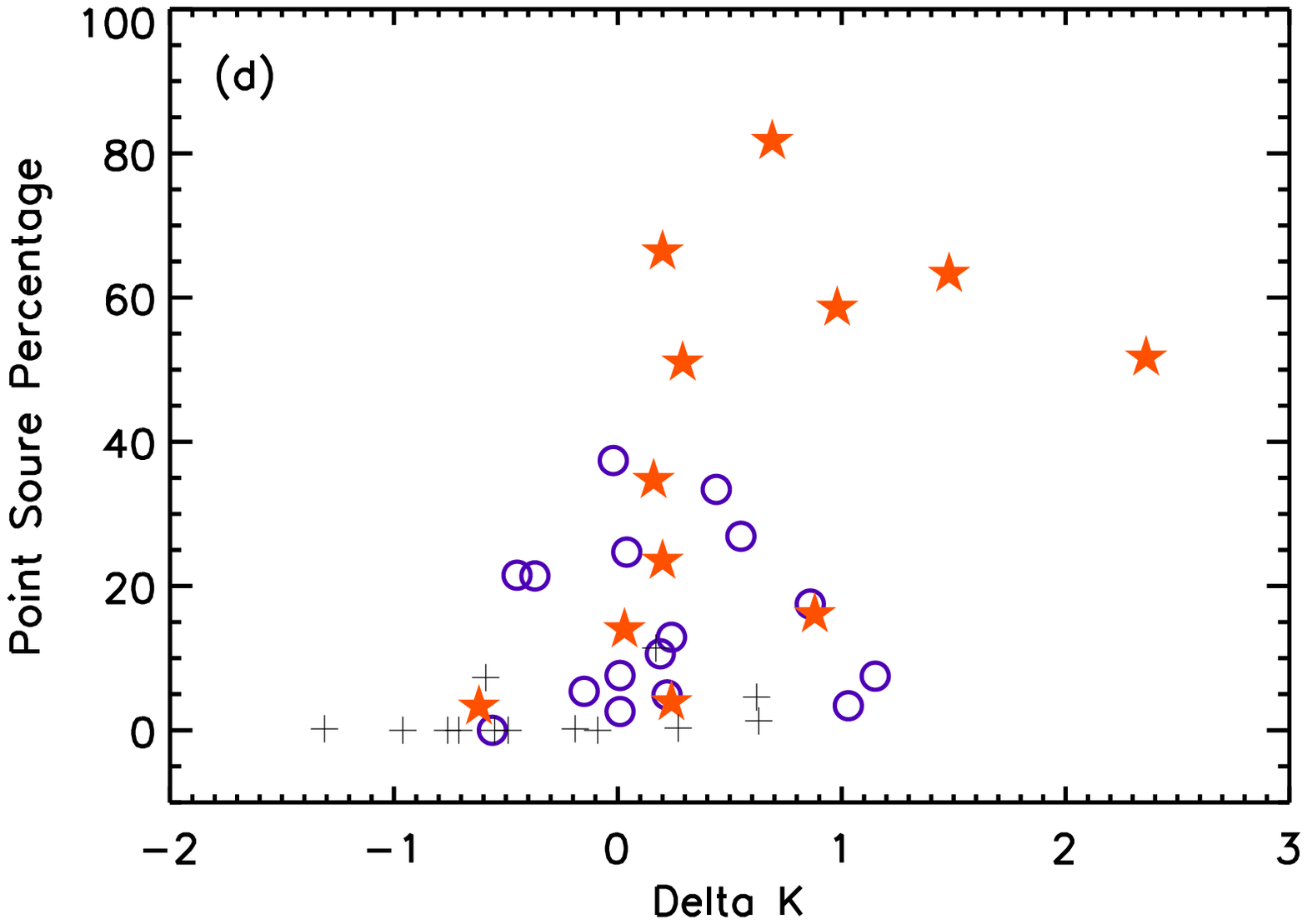}
\includegraphics{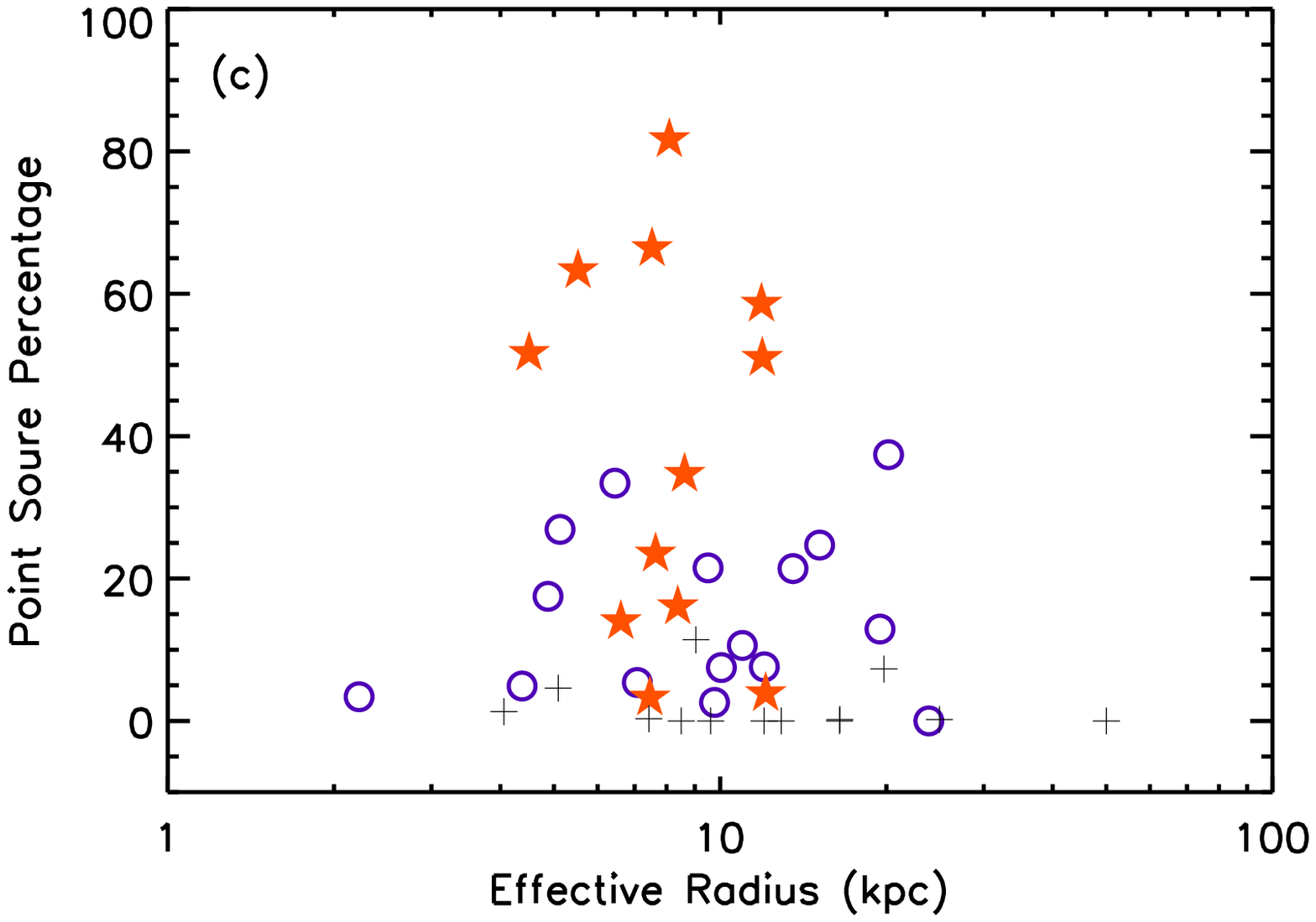}
\end{center}
\caption{(a -- top left)  K-z relation for the 2Jy subsample. The cleaned, K-corrected and extinction corrected 64kpc aperture magnitudes ({\it including} the flux from any nuclear point source) for the sample (see Table 3 column 7) are displayed 
alongside the computed track for passively evolving galaxies of mass $\sim 6.8 \times 10^{11} \rm{M_{\odot}}$ formed at a redshift of 10. The different plot symbols represent the different optical classes of the galaxies: stars = BLRG, circles = NLRG, crosses = WLRG. (b -- top right) Plot of k-corrected percentage nuclear point source contamination in the $K-$band vs. redshit; symbols as in frame (a). (c -- lower left)  K-corrected percentage nuclear point source contribution in the $K-$band vs. modelled host galaxy effective radius; symbols as in frame (a). (d -- lower right)  Plot of k-corrected percentage nuclear point source contamination in the $K-$band vs. the difference in K-band magnitude between the 2Jy sample galaxies and the passively evolving track for a galaxy formed at a redshift of 10; symbols as in frame (a). The difference value $\Delta K$ is determined by subtracting the $K-z$ relation magnitude at a given redshift from the host galaxy magnitude, after also removing the flux contribution from any nuclear point source emission.
\label{Fig: newkz}}
\end{figure*}

In this section, we consider the results for the sample as a whole. Using the results of our photometry (from Table 3), Figure \ref{Fig: newkz}a displays the $K-$band magnitude--redshift relation for the 2Jy subsample at redshifts $0.03 < z < 0.5$.  The different symbols represent the different optical classes of the sample objects (NLRGs, BLRGs and WLRGs) and are displayed alongside the track for passively evolving galaxies of mass $\sim 6.8 \times 10^{11} \rm{M_{\odot}}$ formed at a redshifts of 10 (the mass is chosen so as to normalise the fiducial track to the mean of the low redshift data), produced using the spectral synthesis models of Maraston et al (2009).  Also plotted in Figure \ref{Fig: newkz} are the the nuclear point source contributions for the sample galaxies.  As the spectral energy distribution of the point source emission will not trace that of the host galaxy, the modelled values obtained in the $K_S-$band have been k-corrected to appropriate $K-$band point source percentages using the mean QSO SED of Richards et al (2006); see Fig.~\ref{Fig: kks}c,d.  Typically, the modelled point source percentages vary by very little, increasing by up to a maximum of 1.5\%.  The revised $K-$band point source contributions are plotted as a function of redshift (\ref{Fig: newkz}b), modelled effective radius (\ref{Fig: newkz}c), and the magnitude difference, $\Delta K$, between the host galaxy emission (after subtraction of the unresolved nuclear point source component) and the predictions of the passively evolving galaxy formed at $z=10$ (\ref{Fig: newkz}d).  The average values of these data are also listed in Table 5, separated according to the optical classes of the sample galaxies. 

\begin{table*}
\caption{Average properties of the 41 sample galaxies which have been successfully modelled. In this table we display the typical redshifts, host galaxy effective radius, the percentage nuclear point source contribution to the observed aperture magnitude, and the luminosity relative to a fiducial passively evolving galaxy formed at $z=10$ ($\Delta K$) both prior to and after applying a correction for the nuclear point source emission. Values displayed give the mean value plus its standard error ($\sigma/\sqrt N$) for sources optically classified as NLRGs, BLRGs and WLRGs, plus the combination of NLRGs and WLRGs}
\begin{center} 
\begin{tabular}{cccccc}
Property & NLRGs & BLRGs & WLRGs & NLRGs+WLRGs& NLRGs+BLRGs\\\hline
Redshift                                   & $0.19 \pm 0.03$ &$0.14 \pm 0.04$ &$0.08 \pm 0.02$&$0.14 \pm 0.02$ &$0.17 \pm 0.02$ \\
Effective Radius                           &$10.9\pm 1.6$kpc&$8.4\pm 0.7$kpc&$15.1\pm 3.4$kpc &$12.8\pm 1.7$kpc &$9.8\pm 1.0$kpc\\
Percentage nuclear point source contribution&$14.9 \pm 2.9$\%&$39.0 \pm 7.7$\%&$2.0 \pm 1.0$\%&$9.1 \pm 2.0$\%&$25.2 \pm 4.3$\%\\
$\Delta K$ (uncorrected for point source)  & $0.02\pm 0.13$ &$0.01\pm 0.20$ &$-0.33\pm 0.16$&$-0.14\pm 0.11$ &$0.01\pm 0.11$ \\
$\Delta K$ (corrected for point source)    & $0.20 \pm 0.13$ &$0.57 \pm 0.22$ &$-0.30\pm 0.17$&$-0.03\pm 0.11$ &$0.36\pm 0.12$ \\
\end{tabular}                       
\end{center}                        
\end{table*}

So, how uniform truly is the sample as a whole?  On the $K-z$ relation itself, the NLRGs and BLRGs lie extremely close to the fiducial $K-z$ relation, while the WLRGs (which appear at lower redshifts on average than the other radio galaxy types due to Malmquist bias) are brighter than the fiducial $K-z$ relation by $\sim0.3$mag on average. The observed scatter in the $K-z$ relation is typical of other radio galaxy samples; we further investigate this scatter (and its dependence on the quantifiable properties of our sample objects) in the second paper of this series. However, the most obvious causes of this scatter are variations in the host galaxy total stellar mass, and the contamination of the $K-$band emission by nuclear light from the AGN.  We now consider the nuclear emission and the morphological properties of the galaxies in more depth.

\subsection{Unresolved nuclear emission}
We observe a similar spread of nuclear point source percentages to other studies. As Figure \ref{Fig: newkz} illustrates, it is clear that the measured nuclear point source contribution has no dependence on either  source redshift or on the intrinsic host galaxy luminosity. This confirms that resolution and/or signal-to-noise issues have not led to increasing levels of confusion between genuine AGN point source emission and unresolved emission from the central regions of the host galaxies at higher redshifts/smaller spatial scales. We also observe no bias in the measured point source contributions towards more compact galaxies having brighter or fainter nuclear point source components.  While weaknesses in the modelling of the central regions of the galaxies could lead to strong degeneracies between total point source contribution, effective radius and S\'ersic index, this does not appear to be the case for our sample. Although we cannot rule out a non-AGN contribution to the unresolved emission from the galaxy cores (i.e. the addition of excess nucleated emission due to young stars), the potential presence of such features is very unlikely to have biased our other structural parameters.

Overall, the BLRGs as expected display considerably larger nuclear point source contributions than the narrow line radio galaxies, in keeping with the expectations of orientation-based unification schemes where a lesser part of the AGN emission is obscured from view for the former.  The WLRGs have correspondingly weaker AGN emission than BLRGs/NLRGs, and thus the measured nuclear point source contributions are significantly lower than that of either of the other classes on average, and very close to zero. 

Considering purely the NLRGs in our sample, we find a mean unresolved point source contribution of $14.9 \pm 2.9$\%. If we apply only the $n=4$ S\'ersic index results for this type of radio galaxy, which allows for a consistent comparison with other samples, we find a mean nuclear point source contribution of 14.8\% of the total flux (median 10.9\%).

\begin{table*}
\caption{Morphological type and interaction status of the host galaxies broken down as a function of optical class, and interaction status as a function of morphological type. In the first part of this table, we list the distribution of morphological types for the different optical classifications of the host galaxies, both as a percentage of the total number of galaxies of that optical class within our sample, and also as a simple number. In the second part of this table, we list the distribution of the host galaxy interaction status for the different optical classes, again as a percentage of the total number of galaxies of that optical class and as a simple number. In the third part of this table we list the distribution of each interaction status for the different morphological types. Values given are for the percentages and numbers of galaxies of a given morphological type which display each type of interaction.}
\begin{center} 
\begin{tabular}{ccccc}
Optical class vs. Morphological type & NLRGs & BLRGs & WLRGs & All classes\\\hline
Bulges ($n=4,6$)   & 88\% (14)& 50\% (6) & 54\% (7) & 66\% (27)\\
Disks  ($n=1,2$)   & 6\% (1)  & 25\% (3) & 7\% (1) & 12\% (5)\\
Mixed              &  6\% (1) & 35\% (3) & 38\% (5) & 22\% (9)\\
Total              &100\% (16)&100\% (12)&100\% (13)& 100\% (41)\\\\
Optical class vs. Interaction status & NLRGs & BLRGs & WLRGs & All classes\\\hline
Interacting/disturbed& 44\% (7) & 33\% (4) &62\% (8)  &46\% (19)\\
Non-interacting      & 56\% (9) & 67\% (8) &38\% (5)  &54\% (22)\\
Total                &100\% (16)&100\% (12)&100\% (13)&100\% (41)\\\\
Morphological type vs. Interaction status& Bulges ($n=4,6$)&Disks ($n=1,2$)&Mixed & All morphologies\\\hline
Interacting/disturbed & 48\% (13) & 40\% (2) &44\% (4) &46\% (19)\\
Non-interacting       & 52\% (14) & 60\% (3) &56\% (5) &54\% (22)\\
Total                 & 100\% (27)&100\% (5) & 100\% (9) & 100\% (41)\\\\

\end{tabular}                       
\end{center}                        
\end{table*}                       
These results are in very good agreement with others presented in the literature. For example, at higher redshifts ($z \sim 1$) and radio powers, Inskip et al (2005) find typical point source contributions of $\sim 16 \pm 4 \%$ for 6C narrow-line radio galaxies, while Best, Longair and R\"ottgering (1998) find an average value of $\sim 7 \pm 3\%$ for more powerful 3C narrow-line radio galaxies at the same redshift (though it should be noted that the 3C sources are hosted by intrinsically brighter galaxies, and the flux from the unresolved nuclear component is comparable to that observed from the less powerful 6C radio sources).  In the optical, Govoni et al's ground-based study of low redshift radio galaxies also found that the necessary nuclear point source magnitude was not correlated with the host galaxy magnitude, and was about 5-10\% on average.

Higher-resolution infrared studies have been carried out using the Hubble Space Telescope, e.g. the NICMOS observations of low redshift radio galaxies presented by Floyd et al (2008), which include some of the same galaxies.  These observations have the advantage of better separating the unresolved AGN emission from any compact core emission which would not be resolved in ground-based imaging.  As expected, Floyd et al's average detection rate of unresolved cores (in $\sim 40\%$ of their sample objects) is lower than our own (point sources contributing $> 1\%$ of the total flux are observed for 76\% of our sample, while 56\% of our sample objects are best modelled with a point source contribution $>5\%$), illustrating that some of the nuclear point source emission isolated in our structural modelling likely originates from physical process other than solely the AGN, most likely nuclear star formation regions. However, despite this inability to resolve the innermost regions of the host galaxies as accurately as other studies, the close match between the vast majority of effective radii derived from our modelling and other studies in the literature (e.g. Govoni et al 2000; Donzelli et al 2007; Floyd et al 2008 and other references in section 4.2) provides a reassuring confirmation of our modelling results.

\subsection{Galaxy morphologies}
In Table 6, we display information on the distribution of  galaxy morphologies across the different optical classes.  The majority of sources are well described as bulges in all cases, or as a bulge+disk combination. As expected for orientation-based unification scheme, there are no statistically significant differences between the host galaxies of BLRGs and NLRGs, which have an average effective radius of close to 10kpc. While the WLRGs span a similar range of host galaxy sizes, their average effective radius is marginally larger than that of the more powerful BLRG and NLRG sources at $\sim 15$kpc.

Of the forty-one sources in our sample for which morphological modelling could be carried out, twenty-five can be adequately described by either a single $n=4$ S\'ersic profile or the combination of an  $n=4$ S\'ersic profile plus a nuclear point source; i.e.  61\%  of the sample objects are consistent with being de Vaucouleurs elliptical galaxies.  While our best-fit model for the post-merger object PKS2313+03 is also  an $n=4$ S\'ersic profile, it should be noted that there are very large residuals for this fit, and this object would be better described with a more complex morphological description. However, the degeneracies involved prevent us from narrowing the options down to a genuinely best alternative for this source. For the best fit models of the surface profiles of the remaining sixteen sources, two (PKS0349-27 and PKS1559+02) require a steeper $n=6$  S\'ersic index, two (PKS0518-45 and PKS1814-63) a S\'ersic index of $n=2$, and three (PKS0055-01, PKS1547-79 and PKS1549-79) are consistent with exponential disks. The final nine galaxies require a combination of host galaxy components for their best-fit model.  In contrast with the modelling of Donzelli et al (2007), where they find that 45\% of their sample objects require multi-component models, only 22\% of our sample objects fall into this latter category, and of these only half are very inconsistent with a single S\'ersic profile.  Overall, our results are more in keeping with the findings for AGN in the Sloan Digital Sky Survey, where AGN are almost exclusively hosted either by early-type objects or heavily bulge-dominated spirals (e.g. Kauffmann et al 2003).

Powerful radio galaxies hosted by late type/disk galaxies, or galaxies with a {\it substantial} disk component, are clearly a rarity. Of the three sources in our sample best described by pure exponential disk models, one (PKS1547-79) is currently undergoing a major merger, and the other  (PKS1549-79) is a highly-disturbed post-merger ULIRG. With this in mind, an obvious question to ask is whether such sources are more likely to be disturbed/interacting than the more typical elliptical hosts.   We have therefore categorised the galaxies in terms of their interaction status; these data are also included in Table 6. Sources which display multiple nuclei (within 10kpc), clear interactions (not merely spatial proximity) with other galaxies (either bright or faint relative to the radio source host), or noticeably disturbed isophotes are classed as disturbed/interacting. This accounts for 19 of the sample galaxies, while the remaining 22 objects in the sample appear isolated/undisturbed on the basis of these observations.  This is a very preliminary assessment, which provides no handle on interaction timescales and in some cases may miss the lowest surface brightness features of a late-stage merger. A more detailed analysis of interaction signatures is provided by Ramos Almeida et al in their study of the optical images of these sources (RA10, Ramos Almeida et al in prep). However, our basic treatment here does provide a simple, broad overview of the prevalence of clear-cut signs of such activity in radio galaxy hosts.

Considering first the different optical classes of radio galaxies, a higher proportion of WLRG hosts appear to be interacting or disturbed: 62\% compared with  40\% for the NLRG/BLRG. The tendency is contrary to that found by Ramos Almeida et al. (2010) based on optical imaging of the 2Jy sample at $z>0.05$: 55\% for WLRG (though note that the near-IR and optical morphological classifications agree for the objects common to both samples) compared with  94\% for NLRG/BLRG. Given the small number statistics, the rate of morphological disturbance for the WLRG is consistent between the optical and infared studies, and the major difference lies in the much lower rate of interaction detected in the infared observations of the NLRG/BLRG than in optical observations of the same objects. The latter is to be expected, since the optical observations are far more sensitive to subtle, low-surface-brightness signs of galaxy interactions than the infrared observations. Moreover, the NLRG/BLRg are at higher redshifts on average than the WLRG.  Finally, we also look at the breakdown of interaction status across the different morphological types.  Here, we see that roughly 40-50\% of galaxies are either interacting or disturbed in all cases, and that there is no significant variation in the relative numbers of interacting/disturbed galaxies between the different morphological types.

\section{Conclusion}
We have presented near infrared imaging observations for 41 radio galaxies of the 2Jy sample with $0.03 \lta z \lta 0.5$. The sources lie close to the tracks for passively evolving galaxies formed at high redshift on the infrared Hubble diagram.  Through a combination of aperture photometry and 2-d surface profile modelling using \textsc{galfit}, several useful galaxy parameters have been quantified for the sample, including the percentage contamination by a nuclear point source, S\'ersic index and effective radius.  Overall, our structural parameter modelling results are in very good agreement with those in the literature for other radio galaxies at higher and lower redshifts, though a small fraction of the unresolved nuclear emission detected in our objects may be due to unresolved cuspy nuclear starlight profiles rather than AGN emission alone.

Roughly two thirds of the sources in our sample are hosted by massive elliptical galaxies, with a further 20\% being hosted by galaxies best modelled as bulges with an additional disk-component contributing additional flux to the galaxy. Half are clearly either disturbed or interacting with companion objects.   Except at the lowest redshifts (where the minimum intrinsic radio power required for membership in a flux limited sample is lower), disky-type galaxies rarely host the powerful radio sources of samples such as the 2Jy sample, unless other events (such as a major merger) increase the likelihood of a radio source having been triggered.

Although we see no obvious trends between AGN type, recent/ongoing merger activity and a tendency towards galaxies being best modelled by either multi-component or disky models, the physical links between AGN activity and the host galaxy properties may indeed be very indirect, and subject to time delays between mergers/interactions and any subsequently triggered AGN activity. In the second paper of this series, we investigate the case for such links in greater depth.  There, we extend our investigation of the global properties of this sample, tying the newly-measured structural parameters to the other properties of these systems (including the radio source properties, the ongoing visible and obscured star formation, and the interaction status of the host galaxies). The derived galaxy luminosities, morphological parameters and measures of star formation activity are used to fully characterise the scatter on the infrared Hubble diagram. With the aid of literature data, we will also assess the dependence of the host galaxy characteristics on redshift and radio power.  Finally, the combination of optical (from RA10) and near-IR signatures of recent merger activity will be correlated with the overall galaxy morphologies, star formation histories and radio source properties to highlight the links between radio source triggering/AGN activity and the ongoing evolution of the host galaxies, and to develop a more thorough picture of the diversity of radio source triggering mechanisms, and the timelines/chains of events involved.

\section*{Acknowledgments}
KJI is supported through the Emmy Noether programme of the German Science Foundation (DFG). We thank the
ESO technical and support staff for indulging our request for the use of SOFI in the
small field mode, which greatly improved our data quality at the subsequent temporary expense of telescope functionality.  The United Kingdom Infrared Telescope is operated
  by the Joint Astronomy Centre on behalf of the Science and
  Technology Facilities Council of the U.K.  This research has made use of the NASA/IPAC
  Extragalactic Database (NED) which is operated by the Jet Propulsion
  Laboratory, California Institute of Technology, under contract with
  the National Aeronautics and Space Administration.  We would also like to thank the anonymous referee for some swift and very useful suggestions.

\appendix
\section[]{Landscape tables - presented here for astro-ph version...}
\begin{figure*}
\vspace{8.3 in}
\begin{center}
\includegraphics{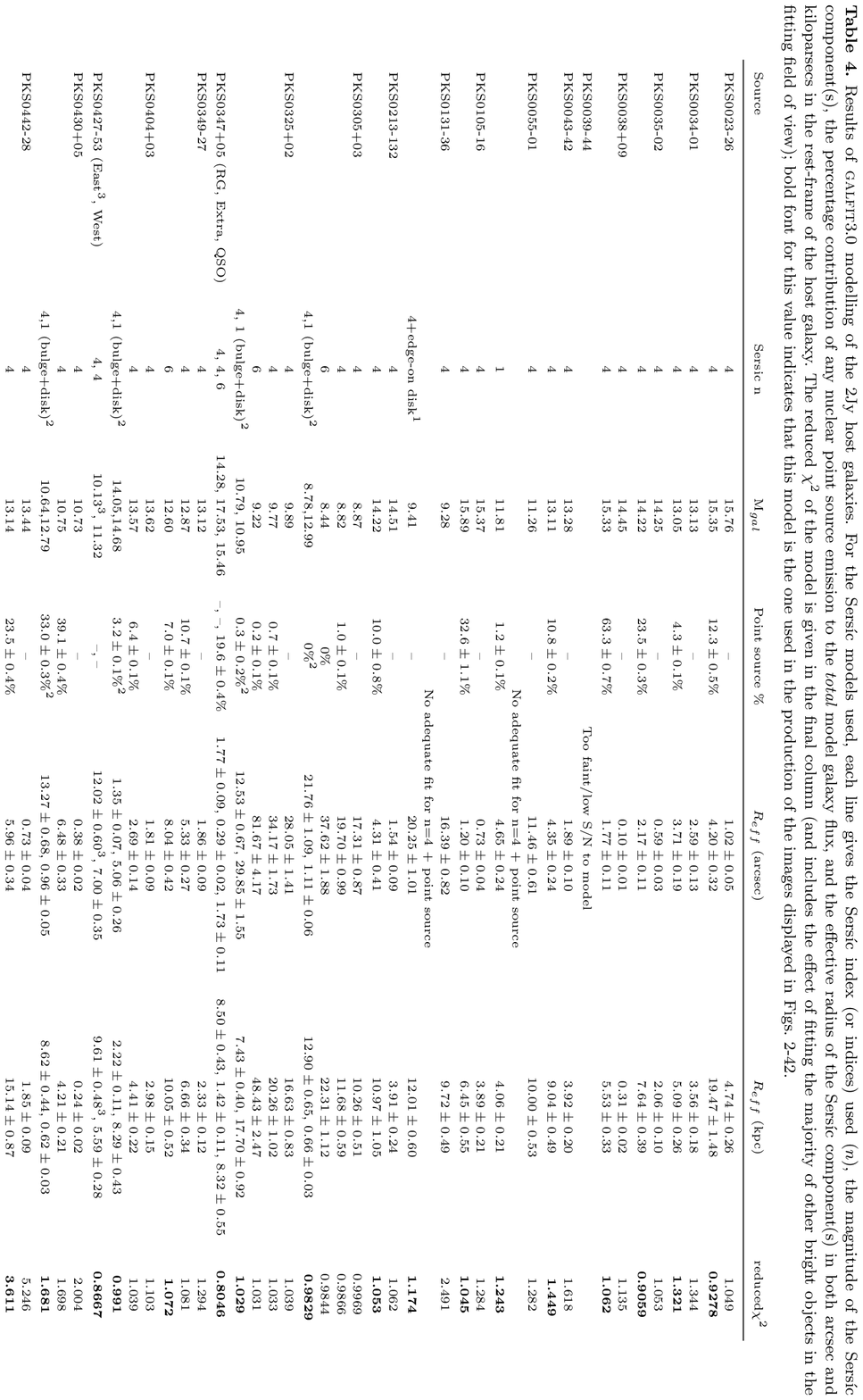}
\includegraphics{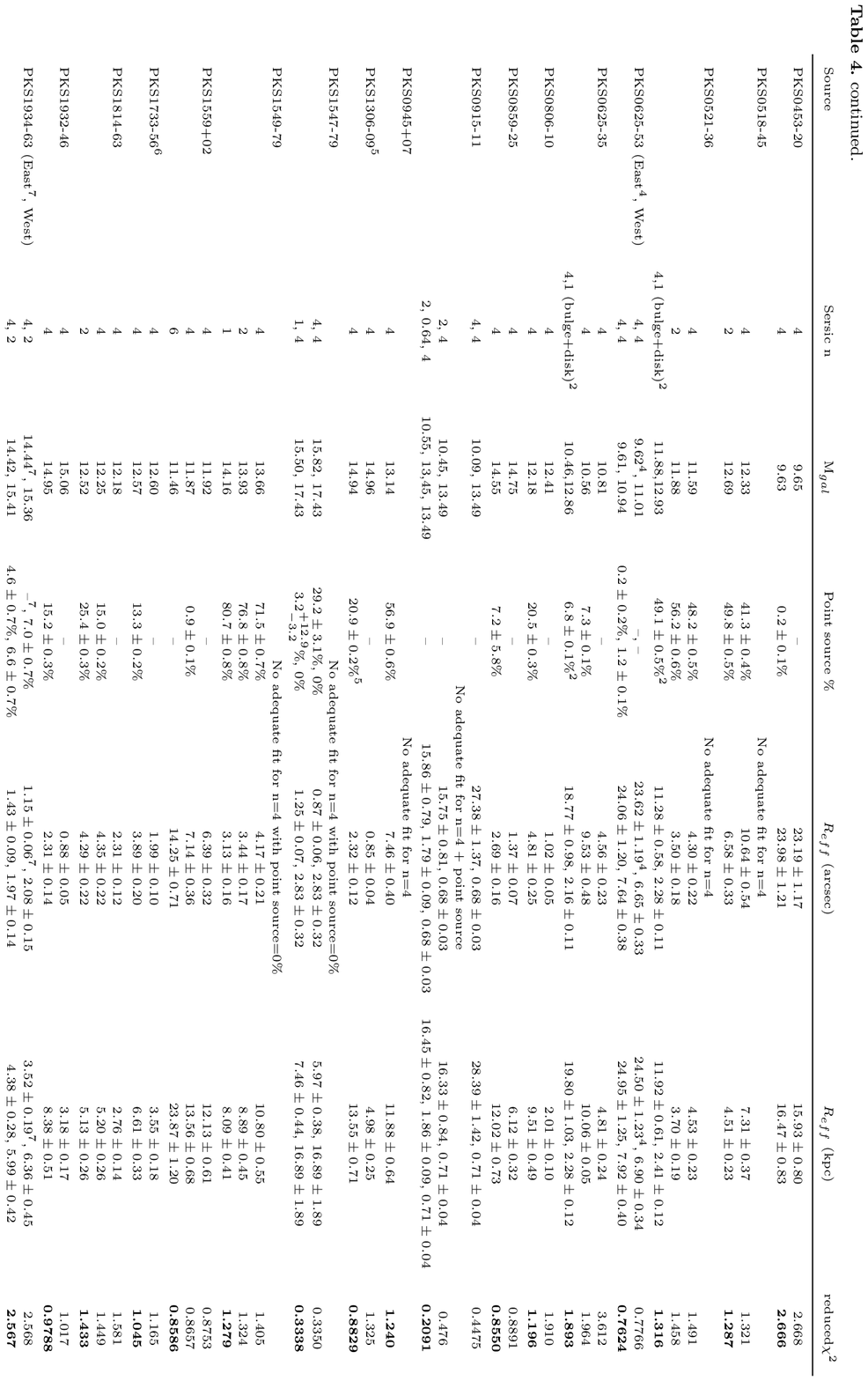}
\end{center}
\caption{
\label{ltabs}}
\end{figure*}

Figure ~\ref{ltabs} displays a shrunken version of the landsape-format table 4, for easy inclusion of this table in the astro-ph version of this paper.

\label{lastpage}

\end{document}